\documentclass{aa}  

\usepackage{graphicx}
\usepackage{txfonts}
\usepackage{comment}
\usepackage[flushleft]{threeparttable} 
\graphicspath{{figures/}}
\usepackage{hyperref}
\usepackage{mathrsfs}
\usepackage{caption}
\begin{document}

   \title{Variability of OB stars from TESS southern Sectors 1-13 and high-resolution IACOB and OWN spectroscopy\thanks{Based on observations made with the Nordic Optical Telescope (FIES), operated by NOTSA, and the Mercator Telescope (HERMES), operated by the Flemish Community, both at the Observatorio del Roque de los Muchachos (La Palma, Spain) of the Instituto de Astrofísica de Canarias. In addition, observations collected with the FEROS spectrograph at the La Silla observatory in the framework of the OWN survey were used. }}

   \author{S. Burssens
          \inst{1}
          \and
          S. Sim\'{o}n-D\'{i}az\inst{2,3}
          \and
          D. M. Bowman\inst{1}
          \and
          G. Holgado\inst{2,3,4}
          \and 
          M. Michielsen\inst{1}
          \and 
          A. de Burgos \inst{2,3,5}
          \and
          N. Castro \inst{6}
          \and 
          R.~H.~Barb\'{a} \inst{7}
          \and
          C. Aerts\inst{1,8,9} \fnmsep
          }

   \institute{Instituut voor Sterrenkunde, KU Leuven, Celestijnenlaan 200D, 3001 Leuven, Belgium \\
              \email{siemen.burssens@kuleuven.be}
         \and
             Instituto de Astrof\'{i}sica de Canarias, 38200 La Laguna, Tenerife, Spain
                \and
            Departamento de Astrof\'{i}sica, Universidad de La Laguna, 38205 La Laguna, Tenerife, Spain
            \and 
            Centro de Astrobiolog\'{i}a, CSIC-INTA, Campus ESAC, Camino bajo del Castillo s/n, E-28692 Madrid, Spain.
            \and
            Nordic Optical Telescope, E-38 711, Bre\~{n}a Baja, La Palma, Spain
            \and
            Leibniz-Institut f\"ur Astrophysik Potsdam (AIP), An der Sternwarte 16, 14482 Potsdam, Germany
            \and 
            Departamento de F\'{i}sica y Astronom\'{i}a, Universidad de La Serena, Avenida Juan Cisternas 1200, La Serena, Chile
            \and 
            Department of Astrophysics, IMAPP, Radboud University Nijmegen, 6500 GL Nijmegen, The Netherlands
            \and 
            Max Planck Institute for Astronomy, K\"{o}nigstuhl 17, 69117 Heidelberg, Germany \\
             }

   \date{Received x xx, xxxx; accepted x x, xxx}

 
  \abstract
   {The lack of high-precision long-term continuous photometric data for large samples of stars has impeded the large-scale exploration of pulsational variability in the OB star regime. As a result, the candidates for in-depth asteroseismic modelling have remained limited to a few dozen dwarfs. The TESS nominal space mission has surveyed the southern sky, including parts of the galactic plane, yielding continuous data across at least 27~d for hundreds of OB stars.}
   {We aim to couple TESS data in the southern sky with ground-based spectroscopy to study the variability in two dimensions, mass and evolution. We focus mainly on the presence of coherent pulsation modes that may or may not be present in the predicted theoretical instability domains and unravel all frequency behaviour in the amplitude spectra of the TESS data.}
   {
   We compose a sample of 98 OB-type stars observed by TESS in Sectors 1-13 and with available multi-epoch, high-resolution spectroscopy gathered by the IACOB and OWN surveys. We present the short-cadence 2-min light curves of dozens of OB-type stars, which have one or more spectra in the IACOB or OWN database. Based on these light curves and their Lomb-Scargle periodograms, we performed variability classification and frequency analysis. We placed the stars in the spectroscopic Hertzsprung-Russell diagram to interpret the variability in an evolutionary context.}
   {We deduce the diverse origins of the mmag-level variability found in all of the 98 OB stars in the TESS data. We find among the sample several new variable stars, including three hybrid pulsators, three eclipsing binaries, high frequency modes in a Be star, and potential heat-driven pulsations in two Oe stars. }
   {We identify stars for which future asteroseismic modelling is possible, provided mode identification is achieved. By comparing the position of the variables to theoretical instability strips, we discuss the current shortcomings in non-adiabatic pulsation theory and the distribution of pulsators in the upper Hertzsprung-Russell diagram.}

    \titlerunning{Variability of OB stars with TESS and IACOB+OWN}
   \maketitle
%

\section{Introduction}\label{sec:introduction}

OB-type stars are key in understanding the chemical, dynamical, and energetic evolution of galaxies (e.g. \citealp{Langer2012}). Despite their importance, large uncertainties regarding their structure and evolution remain and strongly dictate the ultimate fate of these stars \citep{Ekstrom2012, Georgy2012, Chieffi2013, Martins2013, Castro2014}.
Spectroscopic surveys such as the IACOB \citep{Simon-Diaz2011a, Simon-Diaz2015b}, OWN \citep{Barba2010, Barba2014, Barba2017}, GOSSS \citep{Sota2011, Maiz2011}, the two VLT-FLAMES massive star surveys \citep{Evans2005, Evans2011} and MiMeS \citep{Wade2016, Grunhut2017, Petit2019}, have contributed significantly to our understanding of OB-type stars. More specifically, the effects of rotation \citep{Dufton2013, RamirezAgudelo2013, RamirezAgudelo2015, SimonDiaz2014a, Dufton2018, Markova2018}, the high rate of binarity \citep{Sana2012, Sana2014}, magnetic fields \citep{Wade2016, Grunhut2017}, and mass-loss through line-driven winds \citep{Vink2011, Bestenlehner2014, Vink2018} are all important aspects of stellar evolution.

Despite tremendous spectroscopic efforts, direct access to OB-type star interiors is most powerful through asteroseismology. Analysis of stellar pulsations has gained traction across the Hertzsprung-Russell diagram (HRD), allowing for both a direct inference of the internal rotation and chemical stratification from pulsation mode frequency or period spacing patterns in stars born with a convective core \citep{VanReeth2015b, Moravveji2015, Papics2017, Ouazzani2019,  Li2020} and a determination of masses and ages through forward seismic modelling \citep{Handler2006, Briquet2011, Kurtz2014, Moravveji2015, Moravveji2016b, Schmid2016, Buysschaert2018b, Szewczuk2018, Aerts2019b, Mombarg2019}. A limitation is that asteroseismology requires long-term, near-continuous, high-precision data sets. 

The era of studying OB-type star variability has arrived with the advent of high-precision (at $\mu$mag) space missions MOST \citep{Walker2003}, CoRoT \citep{Auvergne2009}, Kepler/K2 \citep{Borucki2010, Howell2014}, BRITE \citep{Weiss2014}, and now the Transiting Exoplanet Survey Satellite (TESS, \citealp{Ricker2014}). The first two TESS sectors have already shown that OB-type stars reveal diverse variability \citep{Pedersen2019a, Handler2019, Bowman2019b}. These studies provide an exploratory evolutionary context and mass range based on distance and brightness measurements assembled by the ESA Gaia mission \citep{Gaia2016, Gaia2018}, which included (large) uncertainties linked to extinction and reddening effects. Coupling the photometric wealth of TESS observations to high-resolution spectroscopy for OB-type stars is the next step in understanding how to improve our knowledge of stellar evolution.

Coherent pulsations in OB-type ($M \gtrsim 3$~M$_{\sun}$) stars\footnote{This includes O- and early B-type main sequence stars and B-type supergiants.} come in two main flavours: gravity (g) modes --- historically associated with the Slowly Pulsating B stars (SPB; \citealp{Waelkens1991, DeCat2002}) --- and pressure (p) modes --- historically associated with $\beta$~Cep type variables \citep{Frost1902, Stankov2005}. These coherent pulsation modes are standing waves for which the dominant restoring forces are buoyancy and the pressure force, respectively \citep{Moskalik1992, Dziembowski1993a, Dziembowski1993b, Gautschy1993}. Using stellar evolution and non-adiabatic pulsational codes, instability regions can be calculated. These regions are sensitive to metallicity \citep{Pamyatnykh1999}, rotation \citep{Szewczuk2017}, opacity, and the chemical mixture \citep{Miglio2007b, Salmon2012, Walczak2015, Moravveji2016a}. Coherent pulsations are also found in Be and Oe stars, which are main sequence stars with spectral types that range between late O and early A and show at least one Balmer line in emission \citep{Porter2003}. The origin of the emission feature is related to episodic mass ejections producing a circumstellar disk, which is induced by a combination of fast rotation and stellar pulsations \citep{Townsend2004, Huat2009, Neiner2013, Kurtz2015, Baade2016, Rivinius2016, Papics2017, Baade2018}.

In addition to coherent pulsations, internal gravity waves (IGWs) are also predicted in massive stars. These are travelling waves  excited at the interface of convective and radiative zones. Numerical simulations of OB-type stars, have shown that the turbulent motions are able to generate IGWs by the convective cores of these stars \citep{Rogers2013, Edelmann2019, Augustson2019, Horst2020}. Every star with a convective core is expected to generate IGWs (see e.g. \citealt{Browning2003, Brun2005} for A-type stars). They are efficient at transporting angular momentum and chemical species within such stars \citep{Rogers2013, Rogers2015, Rogers2017, Aerts2019a} --- see \citet{Aerts2020} for an extensive overview. The recent near-ubiquitous detection of stochastic low frequency variability in space photometry of OB-type stars reveals a frequency spectrum that is consistent with predictions from numerical simulations of IGWs generated by core convection \citep{Aerts2015b, Ramiaramanantsoa2018b, Bowman2019a, Bowman2019b}.  By comparing different samples of Galactic and LMC OB-type stars, \citet{Bowman2019b} highlighted that the morphology of the stochastic low frequency variability in massive stars is connected to the luminosity (i.e.  mass and age) yet insensitive to the metallicity of the star. It is also thought that IGWs are partly responsible, alongside heat-driven modes, for the turbulent motions inducing line-profile variations in OB-type stars \citep{Aerts2009, SimonDiaz2010, Aerts2015b, Tkachenko2014, SimonDiaz2017, Aerts2017b, Godart2017, SimonDiaz2018}. An alternative and non-mutually exclusive explanation for the detected low-frequency power excess could also be sub-surface convection associated with the iron opacity peak \citep{Cantiello2009, Lecoanet2019}, although this is not efficient in stars with masses between 3 and 7 M$_{\sun}$. Sub-surface convection in such mid-B type stars strongly depends on their metallicity \citep{Cantiello2019}, while such stars do reveal low-frequency variability due to a stochastic process \citep{Bowman2019a, Pedersen2020}.

Furthermore, there is the possibility of tidally induced pulsations given that almost all massive stars are born in multiple systems \citep{Sana2012,Sana2014}. These pulsations are driven by the temporary tidal distortion occurring during periastron passage in highly eccentric binary systems \citep{Willems2002, Welsh2011, Thompson2012, Fuller2017}. On the other hand, tides in multiple systems can also potentially affect pulsations intrinsic to the star (e.g. tidally perturbed p modes found by \citealt{Hambleton2013, Bowman2019c}). Tidally induced oscillations have already been discovered in massive OB-type stars (see e.g. \citealt{Pablo2017, Jayasinghe2019}), allowing for new ways to combine the information provided by asteroseismology and binarity.

In this work, we combine high-resolution spectroscopy, from the large-scale spectroscopic surveys IACOB \citep{Simon-Diaz2011a, Simon-Diaz2011b, Simon-Diaz2015b} and OWN \citep{Barba2010, Barba2014, Barba2017}, with high-precision long-term continuous photometry from TESS \citep{Ricker2014} to study variability in OB-type stars. By populating the HRD we constrain the different types of variability seen in main sequence (MS) O and B-stars and B-type supergiants and compare them to theoretically predicted instability regions for coherent pulsation modes.

\section{Methodology}\label{sec:TESSmethod}

TESS is currently observing many stars across the sky in consecutive sectors every month for at least $\sim 27$~d. Data are now available for all sectors (1-13) covered by the nominal TESS mission in the southern hemisphere. Until Sector 6, the number of Galactic OB-type stars observed was limited. In Sector 6 the southern part of the Galactic plane started to be surveyed by TESS. Sectors 6 to 8 thus contain a large variety of OB-type stars ranging from main sequence OB dwarfs to OB supergiants.

\begin{figure}
	\includegraphics[width=1\columnwidth, scale = 1]{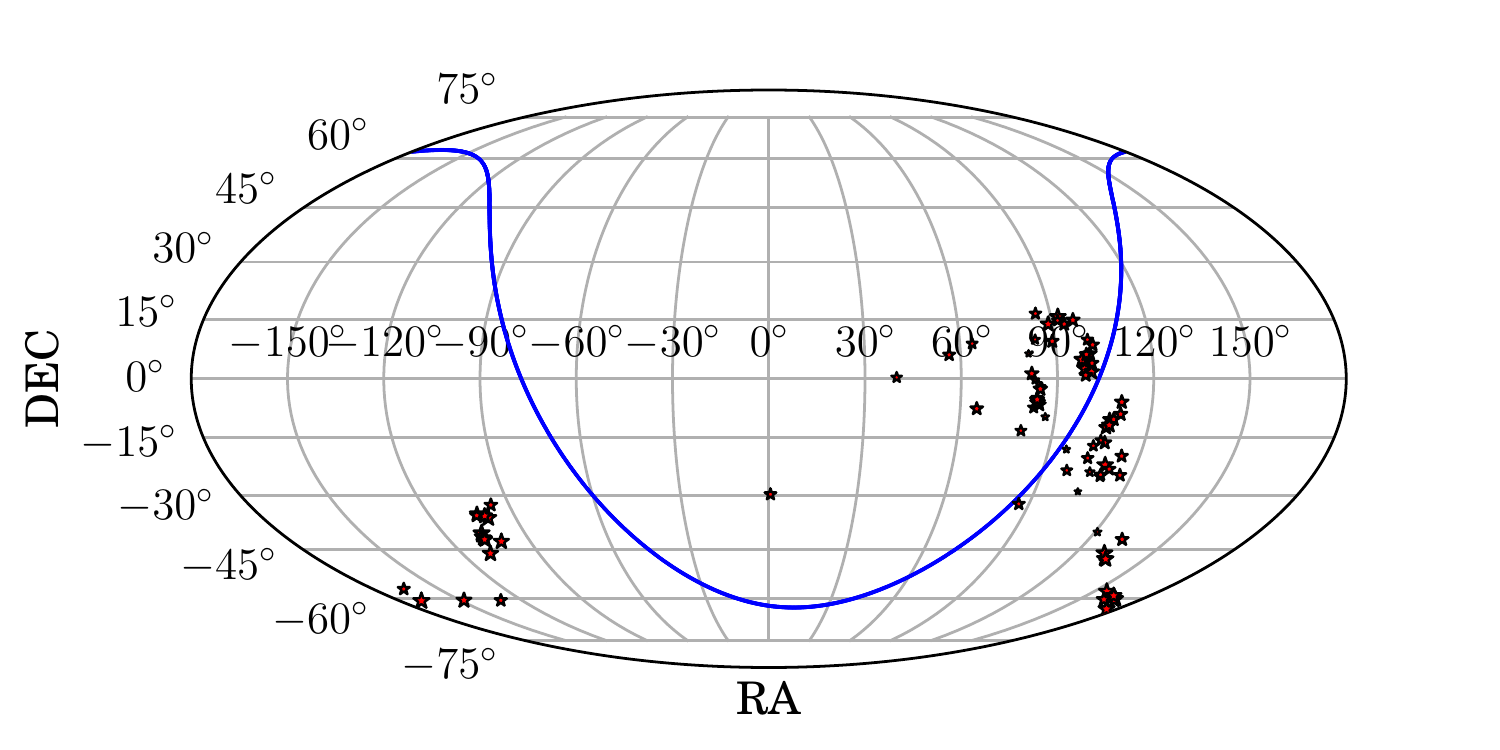}
    \caption{Location of the stars in our sample in galactic coordinates. The galactic plane (in blue) is plotted as reference.}
    \label{fig:moll_overview}
\end{figure}

\begin{figure}
	\includegraphics[width=1.0\columnwidth, scale = 0.5]{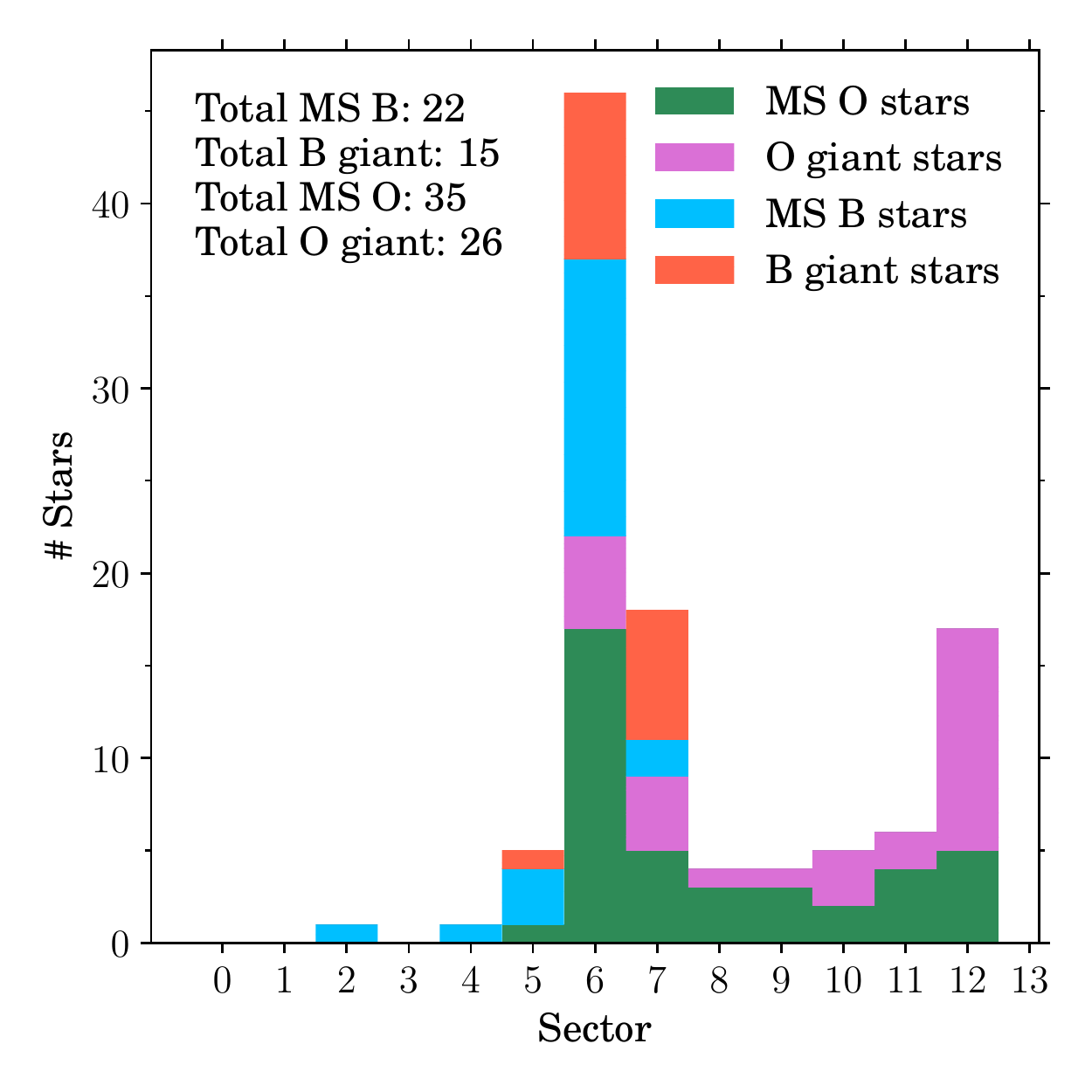}
    \caption{Stacked histogram showing the distribution of the stars observed in each sector, adding up to a total sample size of 98. Different colours indicate different types of stars: MS O stars (V, IV), O giant stars (III, II, I), MS B stars (V, IV) and B giant stars (III, II, I). }
    \label{fig:sector_overview}
\end{figure}

Our sample of stars is based on the list of Galactic OB stars for which the IACOB and OWN projects have, at least, one high resolution spectrum available. Based on the coordinates of each star we scanned the southern hemisphere using the high-precision TESS pointing tool\footnote{\href{https://github.com/christopherburke/tess-point by C. Burke}{https://github.com/christopherburke/tess-point}}. Every star in the IACOB/OWN database with at least one epoch, that falls on one of the TESS cameras was searched for in the TESS Candidate Target List (CTL, \citealp{Stassun2018}), that includes all targets observed at 2-min cadence. Spectral types of B3 and earlier, for the V and IV luminosity classes, and B8 and earlier for luminosity classes I and II, are included here. Dwarf stars earlier than B3 have approximately 5~M$_{\sun}$ or higher, while the supergiant B8 stars and earlier, are expected to be similar mass stars in a more advanced stage of evolution. The galactic distribution of the stars in our sample is shown in Fig.~\ref{fig:moll_overview}. The distribution of our total of 98 stars over the sectors is shown in Fig.~\ref{fig:sector_overview}. 

\subsection{TESS aperture photometry}\label{sec:ap_photometry}

TESS data in short cadence is assembled at the Mikulski Archive for Space Telescopes (MAST\footnote{\href{https://archive.stsci.edu/}{https://archive.stsci.edu/}}) archive as target pixel files (TPF) and light curves (LC). The light curves are available in calibrated (SAP, Simple Aperture Photometry) and pre-conditioned form (PDCSAP, Pre-search Data Conditioning Simple Aperture Photometry). The TESS data reduction pipeline is tailored to stars with potential exoplanets \citep{Jenkins2016}. 

We performed a case-by-case viewing of the aperture mask assigned by the TESS pipeline and the extracted light curve. If the mask selection was poor (i.e. wrong star or large contamination by nearby stars) we extracted an optimised light curve from the TPF following the methodology in \citet{Buysschaert2015, Buysschaert2018a}. Any long-term trends are removed by means of a linear or low-order polynomial fit to the light curve.

\subsection{IACOB Spectroscopy}\label{sec:method_IACOB}

The stars considered in this work cover a wide range in mass ($\approx5-80$~M$_{\sun}$) and evolutionary stage. It is therefore important to locate them as accurately as possible in the HRD for any further interpretation of the detected variability as a function of mass and age.

We use a large database of high resolution spectra of Galactic OB-type stars compiled in the last decade. This mainly refers to the IACOB spectroscopic database, comprising spectra of Northern O and B-type stars obtained with the HERMES \citep{Raskin2011} and FIES \citep{Telting2014} spectrographs attached to the 1.2-m \textit{Mercator} telescope, and the 2.6-m NOT telescope, respectively, but also includes high resolution observations of Southern O-type stars gathered by the OWN survey with the FEROS \citep{Kaufer1997} instrument attached to the ESO/MPG 2.2-m telescope at the ESO La Silla observatory in Chile. For an large number (approximately $85$\%) of the stars under study the multi-epoch character of the observations compiled by these projects has allowed us to probe possible line-profile variability. In addition, we were able to identify stars with strong (variable) stellar winds or other type of circumstellar material which is physically related to the interpretation of the photometric variability. We classify the spectroscopic signatures following \citet{Holgado2018}. The following cases are differentiated:
\begin{itemize}
    \item \textit{C}: constant, two or more spectra show no significant differences in line profile or radial velocity;
    \item \textit{SB1, SB2}: spectroscopic binarity, if one or multiple components are detected respectively, shifted in radial velocity;
    \item \textit{WVa/WVe}: variability seen in the wind lines, that is H$_{\alpha}$ or He~{\sc ii}4686, or both. We differentiate between variability in emission (WVe) and variability in absorption (WVa);
    \item \textit{LPV}: any line profile variations not clearly assignable to any of the above cases;
    \item \textit{S}: only a single spectrum was available.
\end{itemize}

\noindent These signatures are assigned by viewing overplotted spectra. In addition, we calculate the first two lowest-order line-profile moments (see \citealt{Aerts1992}, for a definition) over the multiple epochs, should the visual inspection be unclear. These correspond to equivalent width (EW) and radial velocity (<v>). We note that our variability assignments based on these spectroscopic metrics are mostly qualitative (as a results of the limited number of spectroscopic epochs for some stars) and serve mainly to complement the interpretation of the TESS data. Several peculiar stars are considered separately: the Oe stars or stars showing signatures of a significant magnetic field (see Sect.~\ref{sec:peculiar}). 

Additionally, we make use of parameters from a quantitative spectroscopic analysis to place the stars in the spectroscopic Hertzsprung-Russell diagram\footnote{This diagnostic diagram first proposed by \citet{Langer2014}, and widely used in the literature since then \citep{Castro2014, SimonDiaz2014a, Markova2018, Holgado2018, Castro2018} replaces the stellar luminosity by the parameter $\mathscr{L}:= T_{\rm eff}^{4}/g$, hence removing the uncertainties resulting from the propagation of uncertainties in distance and reddening correction.} (sHRD). The reduction, normalization, analysis and extraction of the spectroscopic parameters are discussed in \citet{Holgado2017, Holgado2018}, and \citet{SimonDiaz2017} for the case of the O-, and B-type stars respectively\footnote{The work by \citet{Holgado2017, Holgado2018} extends and supersedes previous information on the O-type stars included in \citet{SimonDiaz2017}.} (see also \citealt{Castro2012} for the general methodology).  The stars included in our final sample and their spectroscopic properties are given in Table~\ref{table:Star_List}. 
We note that despite using all the spectroscopic variability classifiers indicated above during the analysis process, we do not show in Table~\ref{table:Star_List} the ``C'' and ``S'' labels, and only quote those cases labelled as ``LPV'' in which clear variability is detected.

\setlength{\tabcolsep}{4pt}
\begin{table}
\caption{Overview of the spectroscopic instruments used in this work.}
\begin{center}
\begin{tabular}{ c c c c } 
\hline

Telescope & Instrument & Resolving power &  Range [\AA] \\ \hline 

2.56m~NOT & FIES &46000/25000& 3750-7250 \\ 
1.2m~\textit{Mercator} & HERMES & 85000&3770-9000 \\ 
2.2m~ESO/MPG & FEROS & 48000& 3530-9210 \\ 
 \hline
\end{tabular}
\end{center}
\label{table:telescope}
\end{table}

\setlength{\tabcolsep}{6pt}

\subsection{TESS photometric variability classification}\label{sec:var_class}

We compute Lomb-Scargle (LS) periodograms \citep{Lomb1976, Scargle1982} to obtain an amplitude spectrum for each TESS light curve. Each TESS sector data set contains up to 27~d of near-continuous high-precision photometry and stars in multiple sectors are concatenated. Ten stars in our sample have two available sectors in the 2-min cadence, instead of one. The frequency resolution of the amplitude spectra is defined as the reciprocal of the time-span:  $1/\Delta$T.  The $2$-min cadence of the targets in the TESS CTL yields a Nyquist frequency of $\nu_{\rm Ny}=359.995$~d$^{-1}$. Based on visual inspection of the light curves and their LS-periodograms we provide a variability classification based on the range of frequencies and amplitudes. Additionally all light curves undergo a frequency analysis, following the methodology described in \citet{Burssens2019a}, using the conservative significance criterion that the frequency must have an amplitude with signal-to-noise ratio S/N $\geq 5$. 
We refer to Appendix \ref{sec:appendix_freqanal} for further details.

We differentiate between the following variability types:
\begin{itemize}
    \item \textit{$\beta$~Cep}: coherent p modes with frequencies between approximately $3$ and $20$~d$^{-1}$;
    \item \textit{SPB}: coherent g modes with frequencies below approximately $4$~d$^{-1}$;
    \item \textit{hybrid}: a combination of $\beta$~Cep and SPB; 
    \item  \textit{SLF}: stochastic low-frequency variability with broad amplitude excess, which is different from coherent modes and has periods of several hours to days and amplitudes up to a few mmag;
    \item \textit{EB}: eclipsing binary;
    \item \textit{rot}: rotational variability caused by spots, the stellar wind, or both; 
    \item \textit{cont./poor quality data (PQ)}: Light curve is contaminated by nearby star or instrumental periodicities dominate the LS-periodogram.

\end{itemize}

To assist the classification we estimate the possible rotation frequency range based on the measured v$\,\sin\,i$, an average near-critical rotation velocity v$=450$~km~s$^{-1}$ and an assumed radius. The rotation frequency $\nu_{\rm rot}$ (in d$^{-1}$) is given by:
\begin{equation}
\nu_{\rm rot} \approx 0.02 \rm v_{\rm rot} ({\rm R_{\sun}}/ R),
\end{equation}
where v$_{\rm rot}$ is the equatorial velocity in km~s$^{-1}$ and $ R/ \rm R_{\sun}$ the radius of the star in solar units. The radii are taken from fundamental parameter studies by \citet{Martins2005} for the O-type stars and \citet{Searle2008} for the blue supergiants (who incorporate results from the analysis by \citealt{Crowther2006}). These studies used non-LTE, line-blanketed atmosphere models computed with CMFGEN \citep{Hillier1998}, which were compared to observations of representative stars. The derived fundamental parameters were used to calibrate the fundamental parameters for each spectral type by ways of an interpolation. \citet{Martins2005} report a standard deviation of $10-20$~\%, and \citet{Searle2008} around 10~\% on the calibrated radius.  
For the B dwarfs we use the tabulated values by \citet{Gray2005}.

Disentangling the contributions of rotational modulation and low frequency modes remains difficult as both may be simultaneously present. In this work, we classified as potential rotational variables those stars where the main variability (or a significant contribution) is due to a single frequency and its (sub-)harmonics (following e.g. \citealt{Buysschaert2018a, Sikora2019}). As indicated above rotational variability is expected to be caused by spots, stellar wind, or both.

We provide the results of the variability classification in Table~\ref{table:Star_List}. Uncertain classification is indicated by a question mark. In the case of pulsators this means that the $27$~d TESS data set is insufficient to resolve the complex beating of g modes in the low-frequency regime. Therefore some degeneracy exists between the SLF and the SPB classification. The full results of the frequency analysis are given in Appendix~\ref{sec:appendix_freqanal}. There we provide an overview table (Table~\ref{table:freq_anal}), and full frequency lists of the multi-periodic stars found in the sample (Table~\ref{table:multi_freq_anal}).  

\setlength{\tabcolsep}{6pt}
\begin{table*}
\begin{threeparttable} 
\caption{Overview of OB-type stars with available TESS photometry and IACOB/OWN spectroscopy with spectral type O4 to B3 considered in this work.}

\begin{tabular}{c r l c c c c c c c}
\hline
HD & SpT &  & log\,$T_{\rm eff}$&  log\,$\mathscr{L/L_{\sun}}$&v$\,\sin\,i$ &v$_{\rm mac}$& \# sp. & Sp. Var. Type & TESS Var. Type \\ 
&&&[K] & & [km~s$^{-1}$]&[km~s$^{-1}$]&&\\
\hline
&&&&&&&& \\
 & \multicolumn{9}{c}{O-type dwarfs and subgiants (V and IV)} \\
\cline{2-10}
&&&&&&&& \\
HD\,96715  & O4   & V((f))z        &  4.66   & 4.10  &  59    &  86   & 2   &  ...   &	SLF	        \\
HD\,46223  & O4   & V((f))	   &  4.62   & 4.16  &  60    &  91   & 7   &  ...   &	SLF+SPB?        \\
HD\,155913 & O4.5 & Vn((f))        &  4.63   & 3.88  &  278   &  ...  & 9   &  ...   &	SLF+SPB	        \\
HD\,46150  & O5   & V((f))z  	   &  4.61   & 4.03  &  71    &  94   & 20  &  SB1?  &	SLF	        \\
HD\,90273  & ON7  & V              &  4.59   & 3.95  &  55    &  55   & 1   &  ...   &	SLF+SPB?        \\
HD\,110360 & ON7  & V              &  4.59   & 3.60  &  96    &  86   & 2   &  SB1   &	rot?	        \\
HD\,47839  & O7   & V              &  4.58   & 3.70  &  43    &  65   & 113 &  SB1   &	SLF+$\beta$~Cep \\
HD\,46485  & O7   & V((f))nz~var?  &  4.55   & 3.74  &  322   &  ...  & 3   &  SB1?  &	EB	        \\
HD\,53975  & O7.5 & Vz             &  4.56   & 3.74  &  181   &  ...  & 5   &  SB1   &	SLF+rot?         \\
HD\,41997  & O7.5 & Vn((f))        &  4.55   & 3.85  &  262   &  ...  & 4   &  ...   &	SLF+rot?        \\
HD\,152590 & O7.5 & Vz	           &  4.58   & 3.79  &  48    &  56   & 10  &  SB1   &  EB             \\
HD\,46573  & O7   & V((f))z        &  4.56   & 3.93  &  77    &  81   & 5   &  SB1   &	SLF+rot?        \\
HD\,48279  & O8   & V              &  4.55   & 3.76  &  131   &  74   & 4   &  ...   &	SLF+SPB?        \\
HD\,101191 & O8   & V              &  4.55   & 3.80  &  138   &  ...  & 4   &  ...   &	cont.           \\
HD\,38666  & O9.5 & V              &  4.53   & 3.59  &  111   &  56   & 11  &  ...   &	PQ	        \\
HD\,36512  & O9.7 & V              &  4.51   & 3.44  &  13    &  33   & 23  &  SB1   &	SLF?         \\
&&&&&&&& \\
HD\,123056 & O9.5 & IV(n)          &  4.50   & 3.70  &  193   &  ...  & 9   &  SB1   &	SLF+SPB?      \\
HD\,76556  & O6   & IV(n)((f))p    &  4.58   & 3.80  &  239   &  ...  & 4   &  ...   &	SLF+SPB?      \\
HD\,74920  & O7.5 & IVn((f))       &  4.54   & 3.90  &  291   &  ...  & 1   &  ...   &	SLF+SPB?      \\
HD\,135591 & O8   & IV((f))        &  4.54   & 3.99  &  60    &  60   & 3   &  ...   &	SLF	      \\
HD\,326331 & O8   & IVn((f))       &  4.54   & 3.82  &  332   &  ...  & 6   &  WVa   &	SLF	      \\
HD\,37041  & O9.5 & IVp            &  4.54   & 3.28  &  134   &  ...  & 30  &  SB1   &  SLF?      \\
&&&&&&&& \\
HD\,48099  & \multicolumn{2}{c}{O5\,V((f))z\,+\,O9:\,V}    &  ...    & ...   &  ...   &  ...  & 4  & SB2 & SLF  \\
HD\,159176 & \multicolumn{2}{c}{O7\,V((f))\,+\,O7\,V((f))} &  ...    & ...   &  ...   &  ...  & 4  & SB2 & EB       \\
HD\,54662  & \multicolumn{2}{c}{O7\,Vzvar?}                &  ...    & ...   &  ...   &  ...  & 12 & SB2 & SLF \\
HD\,57236  & \multicolumn{2}{c}{O8.5\,V\,+\,?}             &  ...    & ...   &  ...   &  ...  & 6  & SB2 & SLF+rot  \\
HD\,75759  & \multicolumn{2}{c}{O9\,V\,+\,B0\,V}           &  ...    & ...   &  ...   &  ...  & 6  & SB2 & SLF      \\
HD\,37468  & \multicolumn{2}{c}{O9.5\,V\,+\,B0.5\,V}       &  ...    & ...   &  ...   &  ...  & 80 & SB2 & cont.    \\
&&&&&&&& \\
 & \multicolumn{9}{c}{O-type giants, bright giants and supergiants (III, II and I)} \\
\cline{2-10}
&&&&&&&& \\
HD\,66811  &  O4    & I(n)fp      &  .. .  &  ...  &  198   &  12   & 10  &  ...     &  SLF+rot       \\
HD\,97253    &  O5    & III(f) 	       &  4.59  &   4.16  &  70    &  105   & 2   &   WVa      &  SLF+SPB?     \\
HD\,93843    &  O5    & III(fc)	       &  4.57  &   4.15  &  58    &  120   & 4   &   WVa      &  SLF+rot      \\
HD\,156738   &	O6.5  & III(f)	       &  4.58  &   3.87  &  65    &  103   & 2   &   ...      &  SLF          \\
HD\,36861    &	O8    & III((f))       &  4.55  &   4.06  &  53    &  75    & 881 &   ...      &  SLF          \\
HD\,150574   &	ON9   & III(n)	       &  4.52  &   3.87  &  252   &  ...   & 1   &   ...      &  SLF+rot?     \\
HD\,152247   &	O9.2  & III	       &  4.51  &   3.94  &  82    &  96    & 7   &   SB1      &  SLF+SPB?      \\
HD\,55879    &	O9.7  & III	       &  4.49  &   3.85  &  26    &  60    & 9   &   ...        &  SLF     \\
HD\,154643   &	O9.7  & III	       &  4.49  &   3.85  &  101   &  78    & 5   &   SB1      &  SLF+rot?          \\
&&&&&&&& \\
HD\,152233   &  O6    & II(f)	       &  4.58  &   4.01  &  62    &  105   & 9   &   SB1?     &  cont.        \\
HD\,57061    &	O9    & II	       &  4.51  &   4.01  &  57    &  93    & 9   &   WVa/SB1  &  EB           \\
HD\,36486    &  O9.5  & II~Nwk         &  4.51  &   3.99  &  121   &  96    & 65  &   SB1      &  EB	       \\
&&&&&&&& \\
CPD-47\,2963 &	O5    & Ifc	       &  4.57  &   4.16  &  67    &  110   & 5   &   WVa      &  SLF     \\
HD\,57060    &	O7    & Iafp~var       &  ...   &  ...    &  ...   &  ...   & 12  &   WVe/SB1  &  EB           \\
HD\,156154   &	O7.5  & Ib(f)	       &  4.53  &   4.22  &  62    &  102   & 4   &   WVa      &  SLF+SPB?     \\
HD\,112244   &	O8.5  & Iab(f)p	       &  4.50  &   4.15  &  124   &  80    & 12  &   WVe/SB1? &  SLF          \\
HD\,151804   &	O8    & Iaf	       &  4.45  &   4.33  &  72    &  73    & 2   &   WVe      &  SLF     \\
HD\,47129    &	O8    & fp~var         &  ...   &  ...    &  ...   &  ...   & 5   &   WVe/SB1  &  EB      \\
HD\,303492   &	O8.5  & Iaf	       &  4.45  &   4.29  &  87    &  55    & 4   &   WVe      &  SLF     \\
HD\,152249   &	OC9   & Iab	       &  4.49  &   4.15  &  71    &  70    & 19  &   WVe      &  SLF+SPB?     \\
HD\,152424   &	OC9.2 & Ia	       &  4.48  &   4.14  &  59    &  66    & 5   &   WVe/SB1  &  SLF+SPB?      \\

\hline
\end{tabular}
\begin{tablenotes}
\item \textit{Table continued below.}
\end{tablenotes}

\end{threeparttable}

\label{table:Star_List} 
\end{table*}

\begin{table*}
\centering 
\ContinuedFloat
\caption{Continued.}

\begin{threeparttable}

\begin{tabular}{c r l c c c c c c c} 
\hline

HD & SpT &  & log\,$T_{\rm eff}$&  log\,$\mathscr{L/L_{\sun}}$&v$\,\sin\,i$ &v$_{\rm mac}$& \# sp. & Sp. Var. Type & TESS Var. Type \\
&&&[K] & & [km~s$^{-1}$]&[km~s$^{-1}$]&&\\
\hline
&&&&&&&& \\
 & \multicolumn{9}{c}{O-type giants, bright giants and supergiants (III, II and I) (continued)} \\
\cline{2-10}
HD\,154368   &	O9.5  & Iab	       &  4.48  &   4.28  &  65    &  78    & 5   &   WVe      &  SLF+SPB?     \\
HD\,152003   &	O9.7  & Iab~Nwk        &  4.48  &   4.12  &  65    &  83    & 2   &   WVe      &  SLF+rot?      \\
HD\,152147   &	O9.7  & Ib~Nwk         &  4.48  &   4.04  &  91    &  64    & 4   &   WVa/SB1  &  SLF+SPB?     \\
&&&&&&&& \\
HD\,37043    &	\multicolumn{2}{c}{O9\,III\,+\, B1\,III/IV} &  ...    & ...  &  ...   &  ...   & 99  &   SB2      &  SLF+SPB?     \\
&&&&&&&& \\
 & \multicolumn{9}{c}{Early B-type dwarfs, subgiants and giants (V, IV, and III)} \\
\cline{2-10}
&&&&&&&& \\
HD\,36960  &  B0.5  & V       &  4.46  &   3.31   &   23   &   37  & 4  & ...        &  SLF+SPB?	    \\  
HD\,37042  &  B0.7  & V       &  4.47  &   3.06   &   33   &   13  & 25 & ...        &  SLF?     \\  
HD\,36959  &  B1    & V       &  4.41  &   2.82   &   11   &   14  & 2  & ...        &  cont.     \\  
HD\,43112  &  B1    & V       &  4.41  &   2.95   &    7   &   12  & 1  & ...        &  SLF?     \\  
HD\,37303  &  B1.5  & V       &  4.32  &   2.88	  &  280   &  ...  & 1  & ...        &  hybrid      \\  
HD\,35912  &  B2    & V       &  4.26  &   2.44   &   11   &   21  & 2  & LPV	     &  SPB         \\  
HD\,48977  &  B2.5  & V       &  4.25  &   2.61	  &   26   &	8  & 1 & LPV/SB2?    &  SPB	    \\  
HD\,23466  &  B3    & V       &  4.20  &   2.51	  &   75   &  100  & 1  & ...	     &  EB 	    \\  
&&&&&&&& \\
HD\,34816  &  B0.5  & IV      &  4.46  &   3.22	  &   25   & ...   & 1  & ...        &  rot	    \\  
HD\,46328  &  B0.5  & IV      &  4.40  &   3.28   &    7   &   20  & 2  & LPV/SB1?   &  $\beta$~Cep \\  
HD\,50707  &  B1    & IV      &  4.38  &   3.31   &   29   &   46  & 2  & LPV/SB1?   &  $\beta$~Cep \\  
HD\,37481  &  B1.5  & IV      &  4.34  &   2.78   &   74   &   21  & 2  & ...        &  hybrid       \\  
HD\,16582  &  B2    & IV      &  4.34  &   2.94   &   9    &   19  & 1  & ...	     &  $\beta$~Cep \\  
HD\,37209  &  B2    & IV      &  4.38  &   2.76   &   50   &   15  & 2  & ...	     &  hybrid       \\  
HD\,26912  &  B3    & IV      &  4.20  &   2.61	  &   53   &   30  & 1  & ...        &  SPB	     \\  
HD\,37711  &  B3    & IV      &  4.21  &   2.61   &   68   &   51  & 3  & LPV/SB1?   &  SPB	     \\  
HD\,57539  &  B3    & IV      &  4.13  &   2.54   &   162  &   13  & 1  & ...        &  SPB	     \\     
HD\,41753  &  B3    & IV      &  4.23  &   2.61	  &   25   &   40  & 1  & ...	     &  SPB	     \\  
HD\,224990 &  B5    & IV      &  4.13  &   2.33   &   20   &   40  & 1  & ...	     &  SPB	    \\     
&&&&&&&& \\
HD\,37018  &  \multicolumn{2}{c}{B1\,V\,+\,?}     &  ...    & ...    &  ...   &  ...   & 33 & SB2        &  SLF+SPB?      \\     

&&&&&&&& \\
 & \multicolumn{9}{c}{B-type giants, bright giants and supergiants (III, II, and I)} \\
\cline{2-10}
&&&&&&&& \\
HD\,48434  &  B0   & III     &  4.48  &   3.93   &   48   &   82   & 3   & WVa       &  SLF\\  
HD\,61068  &  B2   & III     &  4.39  &   3.08   &   12   &   23   & 2   & LPV/SB1?  &  $\beta$~Cep \\  
HD\,35468  &  B2   & III     &  4.29  &   2.99   &   53   &   27   & 5   & ...       &  SLF+rot?      \\  
&&&&&&&& \\
HD\,44743  &  B1   & II-III  &  4.37  &   3.20   &   24   &   40   & 6   & LPV/SB1?  &  $\beta$~Cep \\  
HD\,54764  &  B1   & II      &  4.30  &   3.97   &  123   &   87   & 3   & ...       &  SLF+SPB?     \\  
HD\,52089  &  B2   & II      & 4.34 & 3.60 &   26  &  50 & 21  & ...       &  SLF+rot?      \\  
HD\,62747  &  B2   & II      &  4.34  &   3.36   &   98   &   29   & 2   & SB1	     &  EB           \\     
HD\,51309  &  B3   & II      &  4.20  &   3.64   &  27    &   43   & 11  & ...       &  SLF      \\  
HD\,46769  &  B5   & II      & 4.11  &   3.16 &  70    &   23   & 25  & ...       &  SLF+rot?     \\  
HD\,27563  &  B7   & II      &  4.16  &   2.61   &   34   &   26   & 1   & ...       &  SLF+SPB      \\     
HD\,53244  &  B8   & II      &  4.14  &   2.74   &   36   &   21   & 3   & ...       &  SLF+rot      \\  
&&&&&&&& \\
HD\,37128  &  B0   & Ia      &  4.47  &   4.05   &  55    &   85   & 281 & WVe       &  SLF+SPB?      \\   
HD\,38771  &  B0.5 & Ia      &  4.47  &   4.06   &  53    &   83   & 247 & WVe       &  SLF+SPB?      \\   
HD\,53138  &  B3   & Iab     &  4.23  &   4.13   &  37    &   56   & 12  & WVe       &  SLF+SPB?      \\  
HD\,39985  &  B9   & Ib      &  4.11  &   2.34 	 &  26    &  28   & 7   & ...       &  SLF?      \\     

\hline
\end{tabular}
\begin{tablenotes}
\item \textit{Table continued below.}
\end{tablenotes}
\end{threeparttable}

\label{table:Star_List}  
\end{table*}

\begin{table*}
\centering 
\ContinuedFloat
\caption{Continued.}  

\begin{threeparttable}

\begin{tabular}{c r l c c c c c c c} 
\hline

HD & SpT &  & log\,$T_{\rm eff}$&  log\,$\mathscr{L/L_{\sun}}$&v$\,\sin\,i$ &v$_{\rm mac}$& \# sp. & Sp. Var. Type & TESS Var. Type \\
&&&[K] & & [km~s$^{-1}$]&[km~s$^{-1}$]&&\\
\hline
&&&&&&&& \\
 & \multicolumn{9}{c}{Magnetic O- and B-type stars} \\
\cline{2-10}
&&&&&&&& \\
HD\,37022  &  O7    & Vp          &  ...   &  ...  &  23    &  68   & 67  &  ...     &  cont.         \\  
HD\,46056  & O8   & Vn             &  4.55   & 3.58  &  370   &  ...  & 4   &  ...   &	PQ	        \\
HD\,37061  & O9.5 & V            &  4.49   & 3.07  &  210   &  ...  & 4  &  SB1   &  rot      \\
HD\,37742  &  O9.2  & Ib~var~Nwk  &  4.47  & 4.15  &  122   &  97   & 177 &  WVe/SB1  &  SLF          \\
HD\,57682  &  O9.2  & IV          &  4.54  & 3.62  &  12    &  38   & 5   &  ...      &  SLF     \\
HD\,54879  &  O9.7  & V           &  4.52  & 3.16  &  7     &  10   & 8   &  ...     &  SLF?      \\
HD\,37479  &  B2    & Vp          &  ...   & ...   &  155   & 49   & 2   &  ...      &  rot	      \\ 
&&&&&&&& \\
 & \multicolumn{9}{c}{Oe and Be stars} \\
\cline{2-10}
&&&&&&&& \\
HD\,39680  &  O6   & V:[n]pevar  &  ...	  & ...   &  101   &  31   & 4   &  ...	   &  SLF+SPB?	    \\
HD\,45314  &  O9:  & npe         &  ...   &  ...  &  128   &  4	   & 5   &  ...	   &  SLF+SPB?      \\
HD\,58978  &  B0   & IV:e        &  ...   &  ...  &  ...   &  ...  & 2   &  ...     &  $\beta$~Cep   \\  
\hline
\end{tabular}
\begin{tablenotes}

\item \textbf{Header:} \# sp. indicates the number of available spectra, Sp. Var. Type and TESS Var. Type indicate the spectroscopic and photometric variability classifications respectively. We note that the values of log$\mathscr{L/L_{\sun}}$ provided in this table have been computed using gravities already corrected for centrifugal acceleration (following the recipe proposed in \citealt{Herrero1992} and \citealt{Repolust2004}).
\item \textbf{TESS Variability types:} EB~(eclipsing binary), rot~(rotational modulation), SPB~(low frequency pulsation modes), $\beta$~Cep~(high frequency pulsation modes), SLF~(stochastic low frequency signal), PQ~(poor quality data), or cont. (contaminated). 
A question mark indicates that the TESS light curve is insufficient to disentangle the contribution of g modes, rotation effects and SLF.
\item \textbf{Spectroscopic Variability types:} SB1 or SB2~(single or double lined spectroscopic binary), LPV~(line profile variability in photospheric lines), 
WVa or WVe~(Variability of the H$_{\alpha}$ line in absorption or emission).

\end{tablenotes}
\end{threeparttable}
\label{table:Star_List}
\end{table*}

\section{Results}\label{sec:results}

\begin{figure*}
	\includegraphics[width=2\columnwidth, scale = 1]{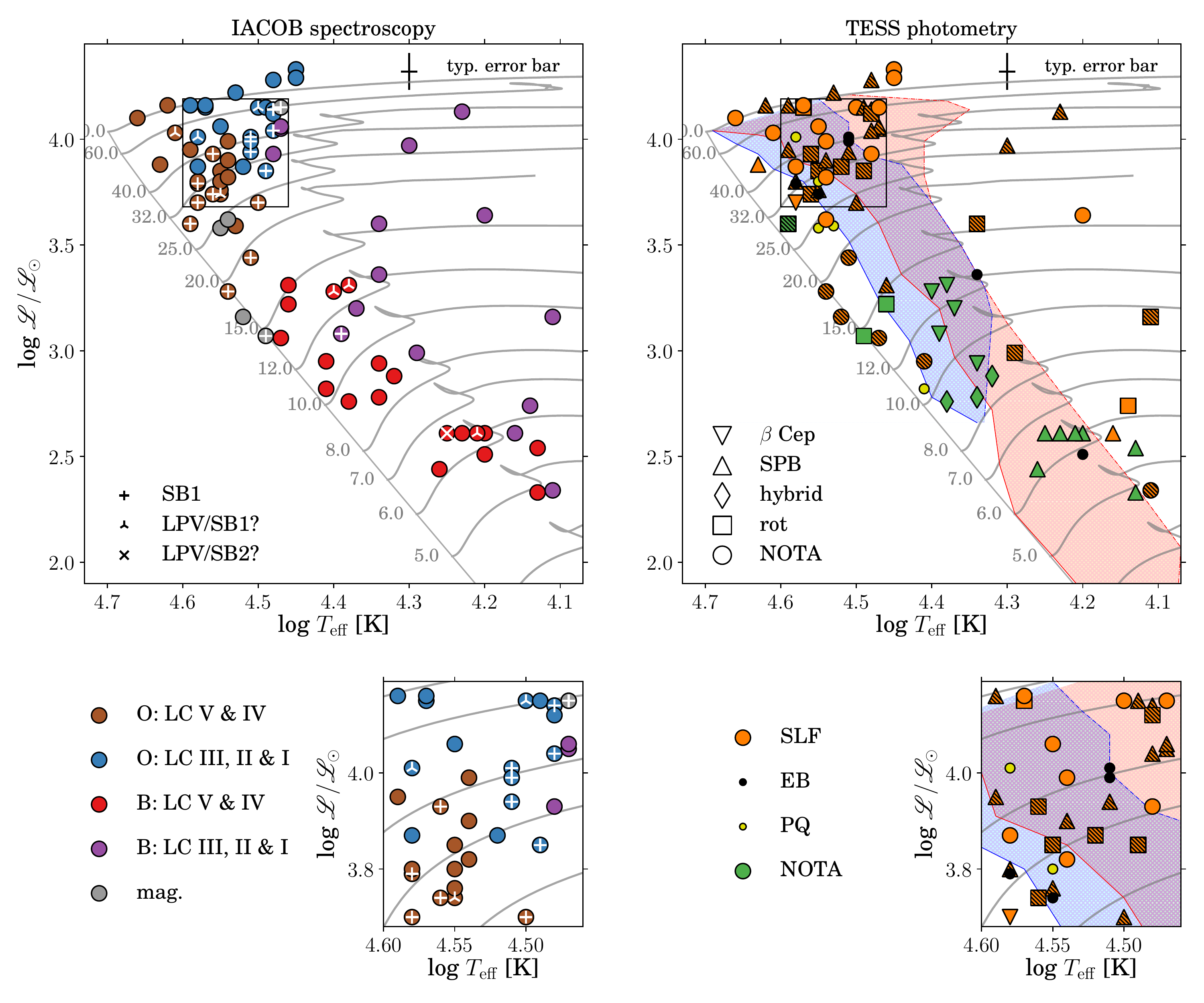}
    \caption{sHRD including stars from Table~\ref{table:Star_List} with determined spectroscopic parameters. On the left panels the stars are sorted following spectral classification criteria (cf. Table~\ref{table:Star_List}). A zoom-in of the dense regions is marked by the black quadrant. Stars identified as single line spectroscopic binaries (SB1) or line profile variables due to the effect of pulsations (LPV/SB1? and LPV/SB2?) are highlighted. 
        On the panels to the right the stars are sorted according to the information provided by the TESS photometric data. The different markers indicate different stable variability seen in the light curves. NOTA indicates "None Of The Above". Hatched symbols mark variable stars where the exact contribution of g modes, wind variability and SLF is difficult to disentangle.
        The colours indicates if the light curve shows stochastic low frequency variability or not (SLF/NOTA), eclipses (EB), or is of poor quality (PQ). Our MESA evolutionary tracks are given in grey on both sides (cf. Sect.~\ref{sec:results}). On the right we also plot our calculated instability strips, blue for the p modes ($\ell=0-2$) and orange for g modes ($\ell=1-2$).}
    \label{fig:overview_sHRD}
\end{figure*}

We populate the sHRD in Fig.~\ref{fig:overview_sHRD} where we combine the TESS photometric variability classification and the parameters from the spectroscopic analysis. On the left panels of Fig.~\ref{fig:overview_sHRD}, we sort the stars according to their spectral type, in the same manner as Table~\ref{table:Star_List}. On the right panels of Fig.~\ref{fig:overview_sHRD}, we sort the stars according to their variability classification. 

To situate the masses and evolutionary stages of the stars we compute evolutionary tracks for M~$\in[3, 4, 5, 6, 7, 8, 10, 12, 15, 20, 25, 32, 40, 50, 60, 80]$~M$_{\sun}$ with the MESA stellar evolution code (\citealp{Paxton2011, Paxton2013, Paxton2015, Paxton2018, Paxton2019}, v.12115), for Z$_{\rm ini}=0.014$, an initial hydrogen abundance X$_{\rm ini}$ = $0.71$, an exponential core overshoot of $f_{\rm ov}$ = $0.02$, envelope mixing coefficient $\log D_{\rm mix}=1.0$, the \citet{Nieva2012} heavy element mixture, and the standard MESA OP opacity tables \citep{Paxton2011}.
Using these tracks and the GYRE stellar oscillation code (\citealp{Townsend2013b, Townsend2018}, v.5.2) we also calculate instability regions along the main sequence for both low-order p modes and high-order g modes for spherical degree $\ell=0-2$ and $\ell=1-2$, respectively. In this we solve for non-adiabatic oscillations (see appendix A in \citealt{Townsend2018}) at spaced intervals of $X_{\rm c} = 0.05$ starting from the Zero-Age Main Sequence (ZAMS) up to core hydrogen depletion, and analyse their stability. We consider azimuthal orders $-\ell \leq m \leq \ell$ and radial orders from $n_{pg}$ from $-50$ to $-1$ for g modes and $n_{pg}$ from $1$ to $5$ for p modes. The results of these computations are presented in the right panel of Fig.~\ref{fig:overview_sHRD}. The blue edges of the instability strips are given by a solid line as they are resolved by our calculations, whereas the red edges are given in dashed lines given that we only compute mode instability up to core hydrogen depletion.
In Sect.~\ref{sec:instability} we examine the effect of metallicity on the instability strips. The detailed MESA and GYRE set-ups are given in Appendix~\ref{sec:appendix_inlists}.

The highest density of stars in the sHRD is found in the O star domain, including MS stars with evolutionary masses in the range $\sim$20\,--\,80~M$_{\sun}$. This is just a consequence of the origin of the compiled spectroscopic data sets being biased towards these stars. Since we are dealing with stars in the TESS Southern sectors, most of our spectra come from the OWN survey, which is mostly concentrated in the monitoring of Southern O-type stars. All the remaining B-type stars are drawn from the Northern IACOB survey, hence resulting in a much smaller overlap with the stars included in TESS Sectors 1\,-\,13. As a result, and despite the large number of B-type supergiants comprising the present version of the IACOB spectroscopic database, the sample of such stars considered in this work is small, with no more than 7 B-type stars apparently located beyond the Terminal-Age Main Sequence (TAMS) for our stellar models with a given metallicity and overshooting. The situation is much better for the case of the B-type stars populating the MS. In this case, we have a relatively good coverage of stars in the mass range 6\,--\,20~M$_{\sun}$.

We highlight in Table~\ref{table:Star_List} (and the left panel of Fig.~\ref{fig:overview_sHRD} for the case of SB1 systems) those stars identified as spectroscopic binaries. In the same table, we identify as separated groups those stars with a confirmed detection of a magnetic field, or which are classified as Oe or Be stars (see Sect.~\ref{sec:peculiar}), and highlight those stars for which we have detected line profile variability in H$_{\alpha}$. All these identifiers add complementary information of interest for the correct interpretation of the TESS light curves.

\setlength{\tabcolsep}{4pt}
\begin{table}
\caption{Global overview of the number of spectroscopic binaries, stars with detected magnetic fields and Oe/Be stars found in our working sample. The numbers in brackets indicates the total number of O- and B-type stars.}\label{table:summary_spvar}
\begin{center}
\begin{tabular}{c c c c c c c} 
\hline
        & SB1 & SB1? & SB2 & SB2? & Mag. & Oe/Be \\
\cline{2-7}
O-type (61)  & 19  &  4       & 7   &   0      &  6   &  2 \\  
B-type (37)  & 2   &  4       & 2   &   1      &  1   &  1 \\  
\hline
\end{tabular}
\end{center}
\end{table}

Table~\ref{table:summary_spvar} provides a global summary of the number of spectroscopic binaries, stars with detected magnetic fields and Oe/Be stars in our working sample. We also note a few cases (identified as LPV/SB1? and LPV/SB2?) in which the detected spectroscopic variability is more likely associated with the effect of $\beta$~Cep- and SPB-type pulsations on the shape of the line profiles\footnote{Those cases labelled as SB1? are most likely binaries; however, our spectroscopic data set does not allow to clearly confirm the binary nature.} than with orbital motion in a single or double line binary system. Also, we note that most of the stars identified as eclipsing binaries (EB) from the inspection of the TESS light curves are labelled as SB1 or SB2 systems. 

In the following sections we provide general remarks on the different classes of variable stars found in the sample. We provide detailed comments on individual objects in Appendix~\ref{sec:appendix_individual}.

\subsection{High frequency pulsators: $\beta$~Cep stars}\label{sec:bcep}

\begin{figure*}
	\includegraphics[width=2\columnwidth, scale = 1]{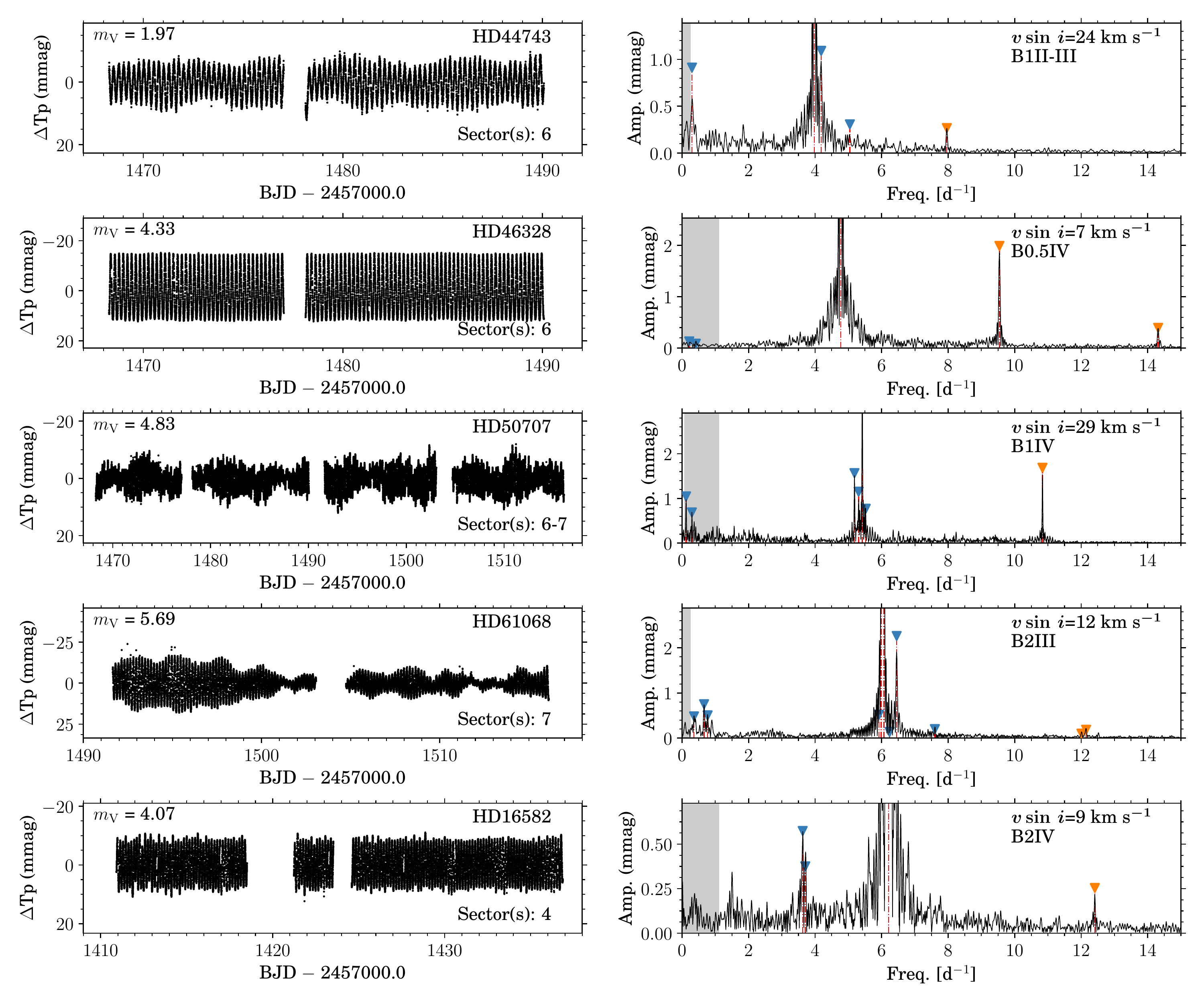}
    \caption{MS B-type stars showing coherent p mode pulsations. Light curves (\textit{left panels}) and their corresponding LS-periodogram (\textit{right panels}) for HD~44743 ($\beta$~CMa), HD~46328 ($\zeta^{1}$~CMa), HD~50707 (15~CMa),  HD~61068 (PT~Pup), and HD~16582 ($\delta$~Cet). No significant variability is seen beyond $15$~d$^{-1}$. The amplitude axis in the LS-periodogram is cut-off at the second most significant peak to highlight the low amplitude frequencies. For HD~61068 we use the third most significant peak.
    We mark significant frequencies with blue markers. Orange markers indicate possible combinations or harmonics of the dominant frequency (in terms of amplitude) and second dominant frequency. The grey band indicates the approximate frequency range where the effects of rotation may be expected if the radius corresponding to the spectral type and v$\,\sin\,i$ are available for that particular star (cf. Sect.~\ref{sec:var_class}).}
    \label{fig:PG_LINE_Bcep}
\end{figure*}

Our sample contains five bright previously studied $\beta$~Cep stars, that is, HD~44743 ($\beta$~CMa), HD~46328 ($\zeta^{1}$~CMa), HD~50707 (15~CMa),  HD~61068 (PT~Pup) and HD~16582 ($\delta$~Cet). We confirm their variable behaviour, typical of the $\beta$~Cep pulsators, see Fig.~\ref{fig:PG_LINE_Bcep}. A more detailed analysis of each TESS data set is given in Appendix~\ref{sec:appendix_Bcep}.

All the stars in this group have a low projected rotational velocity (v$\,\sin\,i$ $\le$ 30~km\,s$^{-1}$). In addition, line profiles of metal lines in these stars are largely affected by pulsational broadening, as indicated by the large v$_{\rm mac}$/v$\,\sin\,i$ ratio (see Table~\ref{table:Star_List}). Interestingly, as previously known, in all those targets from the list above for which we have multi-epoch spectroscopy we detect line-profile variability that could lead to the misinterpretation of the stars as single line spectroscopic binaries, while this variability actually comes from the effect of the pulsation. This highlights the importance of combining TESS photometry and spectroscopy when studying massive stars.

As indicated in the right panel of Fig.~\ref{fig:overview_sHRD}, the five $\beta$\,Cep stars are located well inside the low order p mode instability domain (see also further notes in Sect.~\ref{sec:Discussion}). They are found in the second part of the MS, indicating considerable evolution beyond the ZAMS and their evolutionary masses range between approximately 9 and 14~M$_{\sun}$.

\subsection{Low-frequency pulsators: SPB stars}\label{sec:spb}

\begin{figure*}
	\includegraphics[width=2\columnwidth, scale = 1]{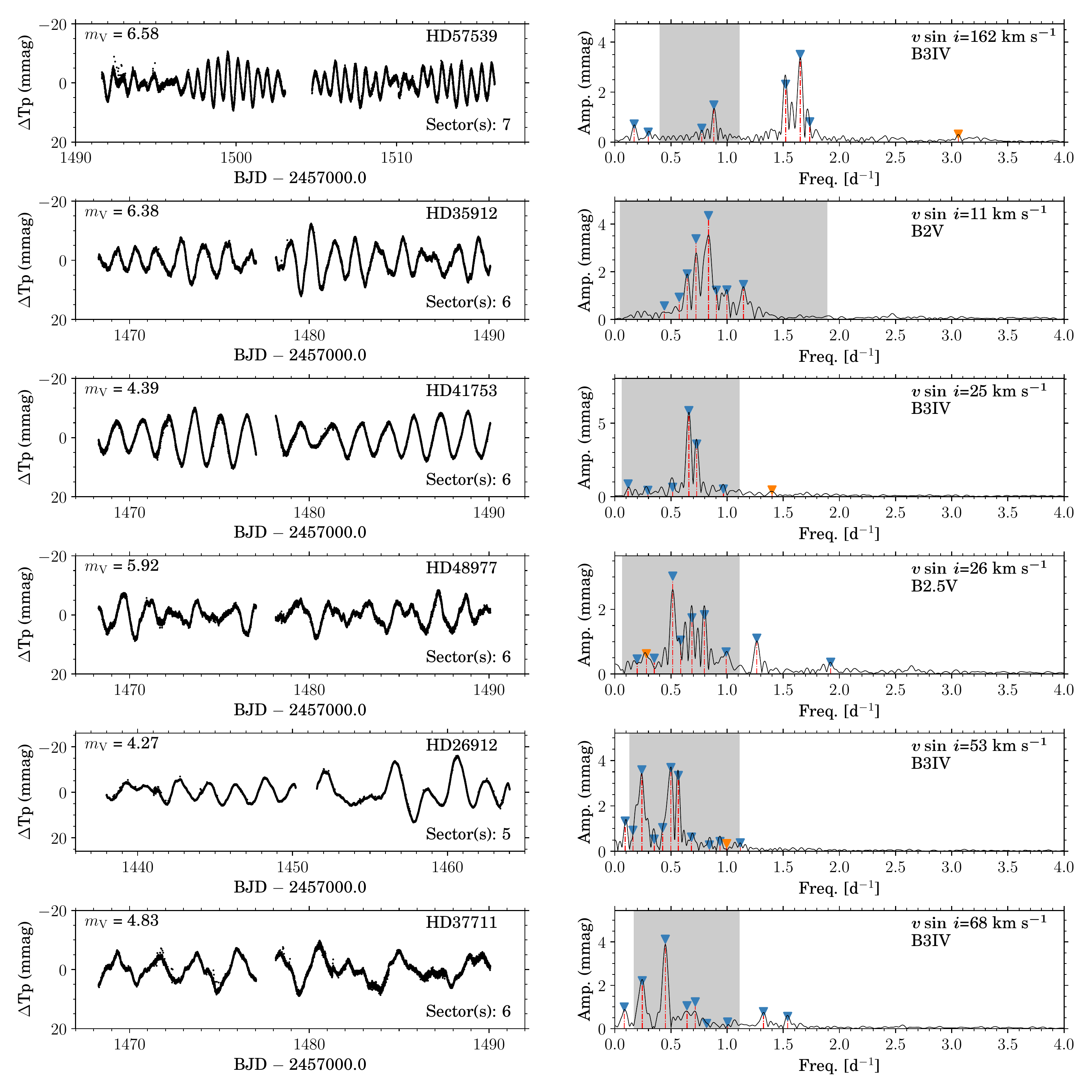}
    \caption{MS B-type stars showing coherent g mode pulsations. Light curves (\textit{left}) and their corresponding LS-periodograms (\textit{right}) for HD~57539, HD~35912, HD~41753, HD~48977, HD~26912 and HD~37711. No significant variability is seen beyond $4$~d$^{-1}$. The markers in the LS-periodograms have the same meaning as in Fig.~\ref{fig:PG_LINE_Bcep}}
    \label{fig:PG_LINE_SPB}
\end{figure*}

In total, there are five previously unknown SPB stars in our sample (HD~57539, HD~35912, HD~41753, HD~26912, and HD~37711), and two known (HD~48977 and HD~224990). The TESS light curves and their periodograms are given in Fig.~\ref{fig:PG_LINE_SPB}. A more detailed analysis of each TESS data set is given in Appendix~\ref{sec:appendix_SPB}.

In comparing the LS-periodograms of the $\beta$~Cep stars (Fig.~\ref{fig:PG_LINE_Bcep}) to the LS-periodograms of the SPB stars (Fig.~\ref{fig:PG_LINE_SPB}), it is clear that single sector TESS data sets are too short to resolve a high number of frequencies in the low frequency regime. For mode identification and asteroseismic modelling stars with a longer time coverage, such as in the TESS continuous viewing zones, are higher priority targets.

All these stars, except HD\,57539, are located inside the instability domain associated with high-order g modes (see right panel in Fig.~\ref{fig:overview_sHRD}, and further notes in Sect.~\ref{sec:Discussion}). As in the case of the five $\beta$\,Cep in our working sample (Sect.~\ref{sec:bcep}) they are also found in the second part of the MS, and have lower masses of approximately \,6\,--\,7~M$_{\sun}$. 

We detected likely signatures of spectroscopic binarity in two of the stars (HD\,37711 and HD\,48977, labelled in Table~\ref{table:Star_List} as LPV/SB1? and LPV/SB2?, respectively) although given the low number of available spectra, we cannot discard that the detected spectroscopic variability is produced by the SPB-type pulsations. We also note that three out of the seven stars have a projected rotational velocity larger than 50~km\,s$^{-1}$. HD~57539 (the star which is located outside of our computed instability strip) has a much larger v$\,\sin\,i$ than the other stars, and the main pulsation frequencies detected in its LS-periodogram are located at higher frequencies. Modelling HD~57539 would be an interesting test for current theoretical understanding of pulsations in fast rotating stars. We provide further notes about this (and the other stars) in Appendix~\ref{sec:appendix_SPB}. Lastly, we note that the macroturbulent broadening contribution (relative to the rotational broadening) seems to be smaller in the SPB stars than in the case of the $\beta$\,Cep stars in our sample, in agreement with \citet{Aerts2014}.

\subsection{Hybrid pulsators}\label{sec:hybrids}

\begin{figure*}
	\includegraphics[width=2\columnwidth, scale = 1]{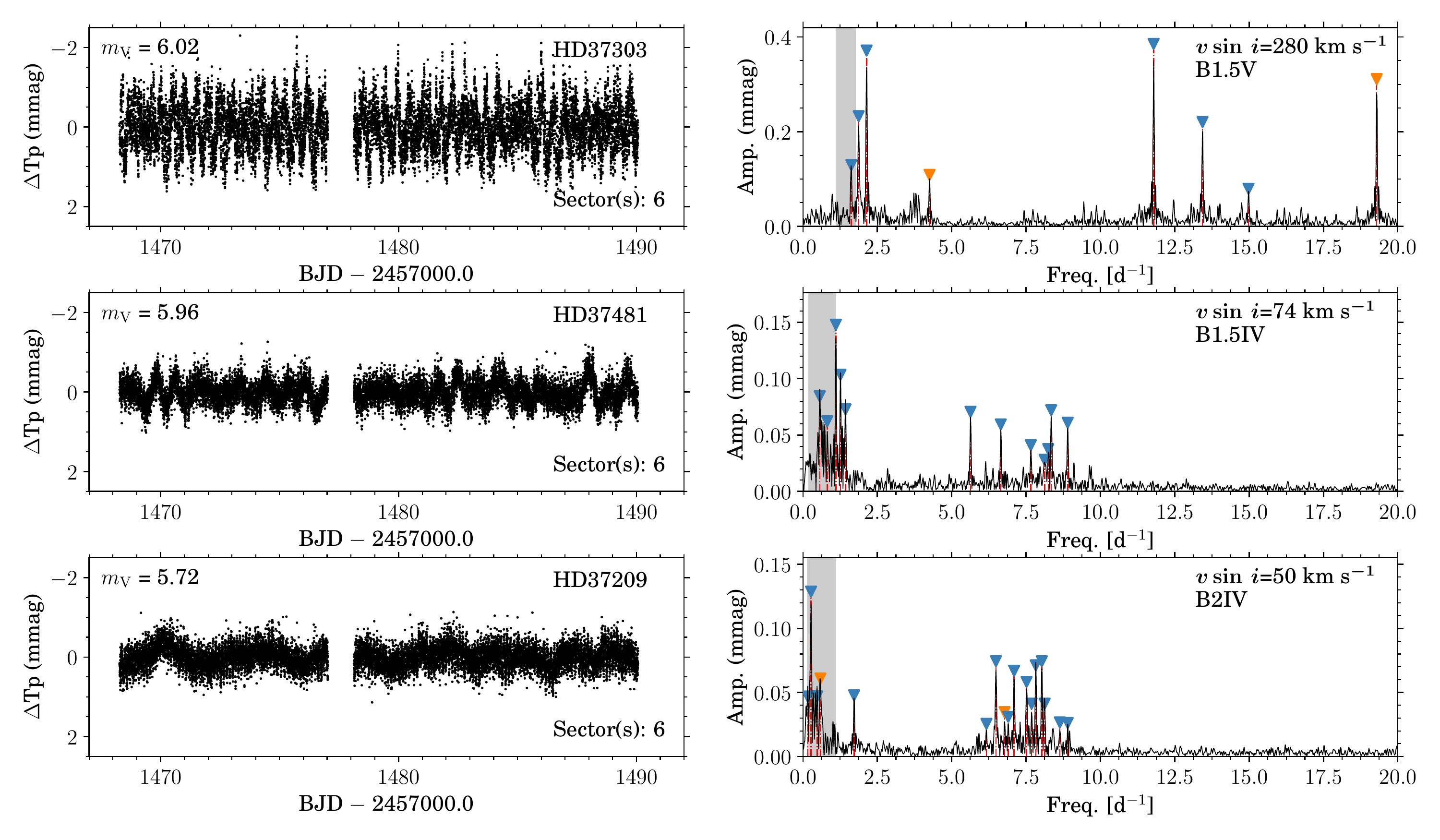}
    \caption{MS B-type stars showing both coherent p and g mode pulsations. Light curves (\textit{left}) and their corresponding LS-periodogram (\textit{right}) for HD~37303, HD~37481, and HD~37209. The markers in the LS-periodograms have the same meaning as in Fig.~\ref{fig:PG_LINE_Bcep}.}
    \label{fig:LC+PG_hybrids}
\end{figure*}

We identify in our sample three previously unknown hybrid pulsators among the MS B-type stars: HD~37303, HD~37481, and HD~37209. We show their light curves and LS-periodograms in Fig.~\ref{fig:LC+PG_hybrids}. A more detailed analysis of each TESS data set is given in Appendix~\ref{sec:appendix_hybrid}. 

As expected, these stars are located in the region between the detected $\beta$\,Cep and SPB stars, with evolutionary masses in the approximate range 8\,--\,10~M$_{\sun}$. HD \,37303 has a relatively large projected rotational velocity (v$\,\sin\,i$\,=\,280~km\,s$^{-1}$), while the other two have more moderate values of v$\,\sin\,i$ ($\approx$\,50\,--75~km\,s$^{-1}$). The macroturbulent broadening contribution (relative to the rotational broadening) is small in these three stars (see Table~\ref{table:Star_List}).

Hybrid pulsators have great asteroseismic potential as they allow us to probe different regions in the star simultaneously. HD~37303 in particular has significant modelling potential given its large number of frequencies and multiplets which provide means of mode identification. However, additional photometry is required to fully resolve the rich spectrum.

\subsection{Peculiar stars}\label{sec:peculiar}
\subsubsection{Be and Oe stars}

\begin{figure*}
	\includegraphics[width=2\columnwidth, scale = 1]{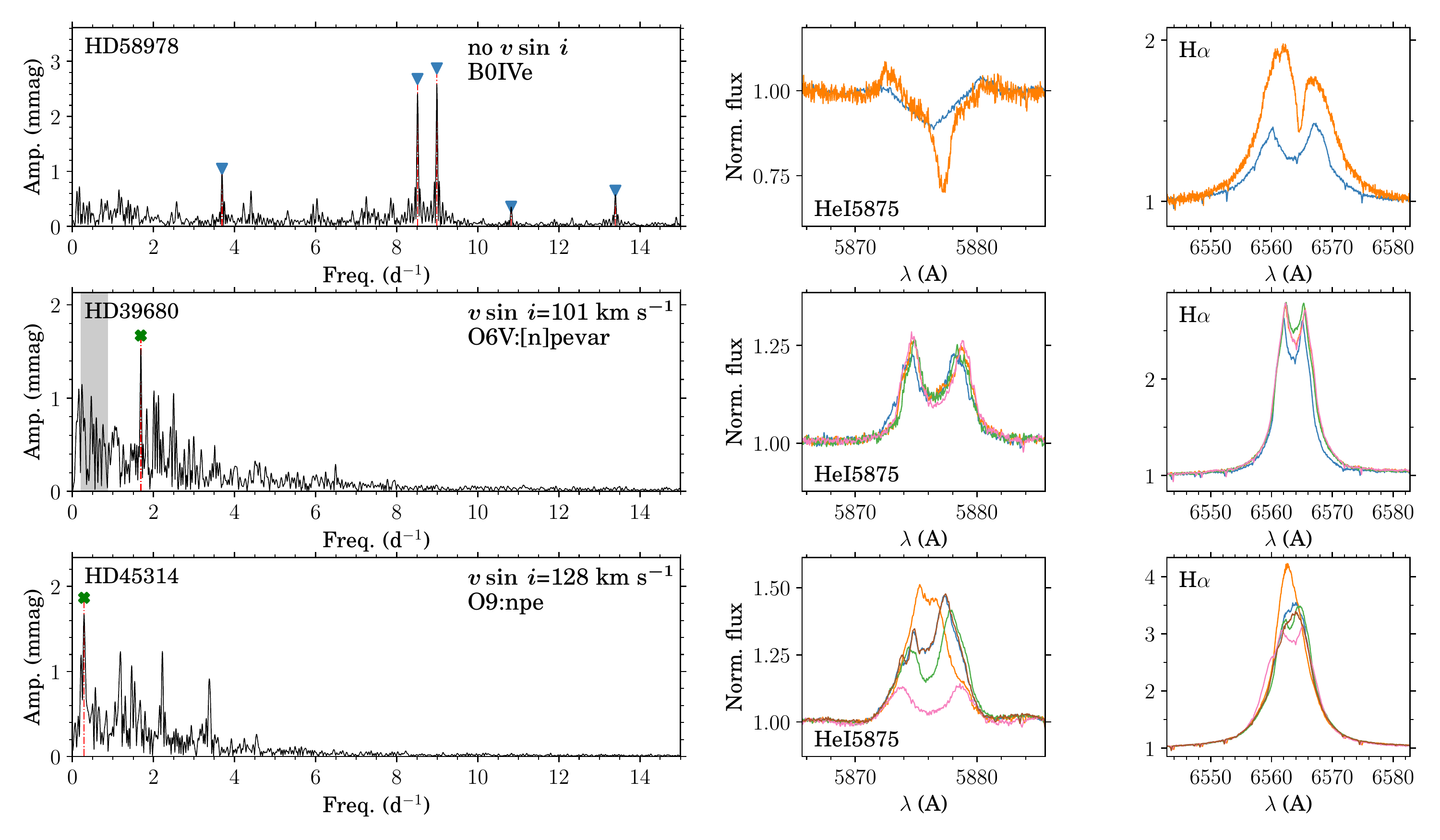}
    \caption{LS-periodogram up to 15~d$^{-1}$ (\textit{left}) and two spectroscopic lines (\textit{right}, He~{\sc i}~$\lambda$5875 and H$_{\alpha}$) for five epochs from days to months to a year, for emission-line stars HD~58978, HD~39680, and HD~45314. The markers in the LS-periodograms have the same meaning as in Fig.~\ref{fig:PG_LINE_Bcep}. If the dominant frequency does not satisfy the significance criterion we mark it with a green cross.}
    \label{fig:PG+lines_Oe}
\end{figure*}

One Be star and two Oe stars are included in our sample: HD~58978, HD~39680, and HD~45314. The TESS LS-periodograms, and the He~{\sc i}~$\lambda$5875 and H$_{\alpha}$ lines are given in Fig.~\ref{fig:PG+lines_Oe}, in the left and right panels respectively. A more detailed analysis of each TESS data set is given in Appendix~\ref{sec:appendix_emission}. 

The newly discovered pulsating Be star HD~58978 has high frequency modes, which would allow for a better understanding of this Be+sdO system. Additionally the two Oe stars (HD~39680 and HD~45314) show variability outside the estimated rotational modulation frequency range which could be due to coherent or non-coherent pulsations in dense frequency groups. These three stars further show variability in the He~{\sc i}~$\lambda$5875 and H$_{\alpha}$ wind lines, as expected from their Oe or Be membership. Frequency groups have also been observed and interpreted as a combination of radial and non-radial pulsations in Oe star $\zeta$~Oph by \citet{Howarth2014a}. These frequency groups are similar to those in Be stars (see i.e. \citealt{Neiner2009, Neiner2012a, Kurtz2015, Semaan2018}, for later B-type stars). The presence of frequency groups in these two Oe stars observed with TESS supports the link between pulsations and the Oe/Be phenomenon.

\subsubsection{Magnetic O- and B-type stars}
We identified stars distributed across the whole sHRD whose TESS light curves show signatures that could be associated with rotational modulation (see square symbols in the right panel of Fig.~\ref{fig:overview_sHRD}). In this section, we briefly discuss those stars known to be magnetic that should contain strong contributions of rotational modulation in their LS-periodogram. 

\begin{figure}
	\includegraphics[width=1\columnwidth, scale = 1]{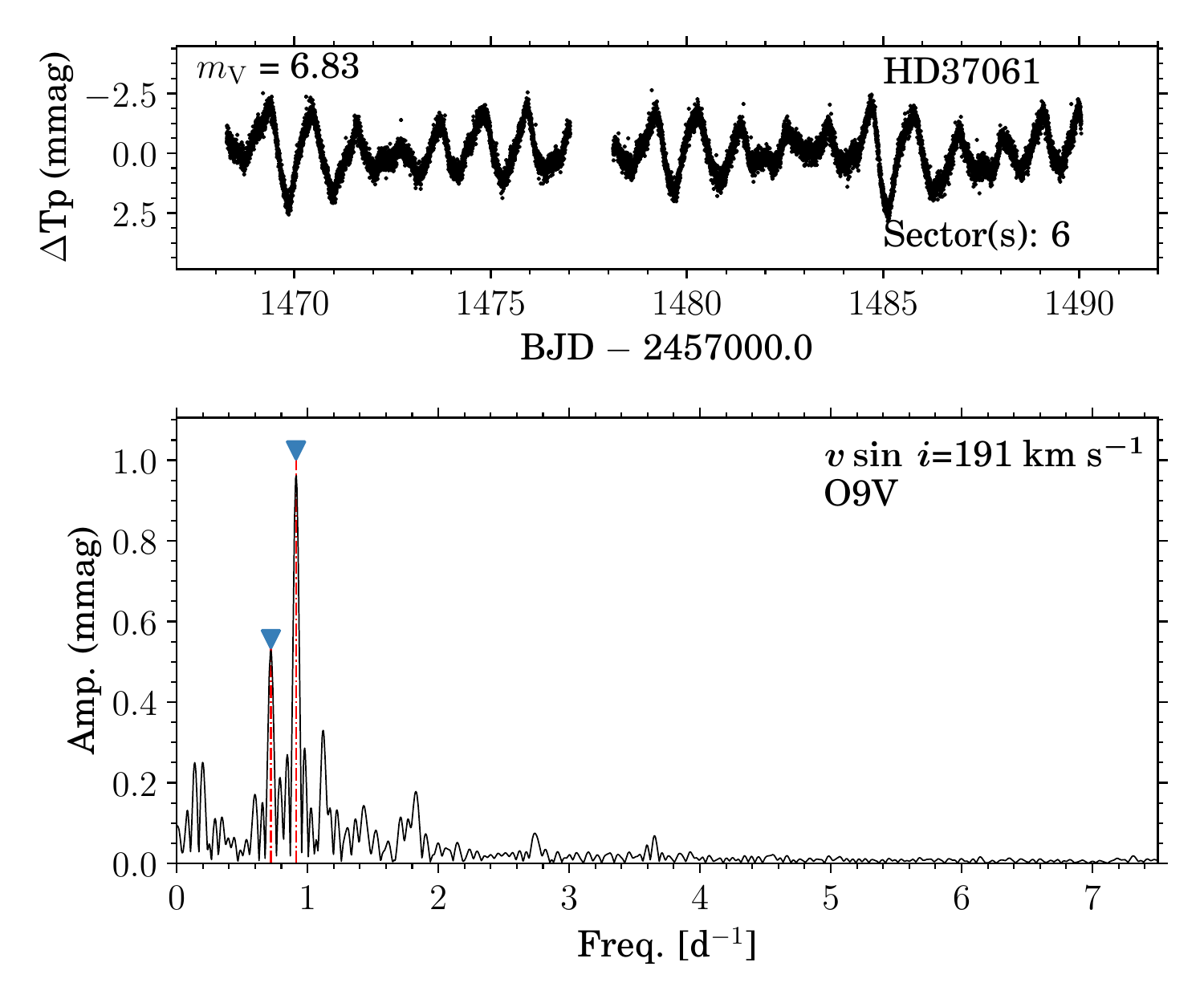}
    \caption{Light curve (\textit{Top}) and LS-periodogram (\textit{Bottom}) of HD~37061, or NU~Ori. The markers in the LS-periodogram have the same meaning as in Fig.~\ref{fig:PG_LINE_Bcep}. The dominant frequency corresponds to the rotational period of NU~Ori~C ($P_{\rm rot}=1.0960(4)$~d).}
    \label{fig:LC+PG_NU_Ori}
\end{figure}

We have seven stars with a clear magnetic field detection in the literature, namely HD~37061 \citep{Shultz2019}, HD~57682 \citep{Grunhut2009, Grunhut2012}, HD~54879 \citep{Castro2015, Shenar2017, Hubrig2019, Wade2020b}, HD~37022 \citep{Donati2002, Wade2006}, HD~37742 \citep{Bouret2008, Blazere2015}, HD~46056 \citep{Grunhut2017, Petit2019}, and HD~37479 \citep{Landstreet1978, Oksala2012, Townsend2013a}. In all LS-periodograms singular frequencies are visible in the low frequency regime (ranging between $0.31$ and $1.67$~d$^{-1}$, with amplitudes between $0.11$ and $6.6$~mmag). The origin of these frequencies is different from star to star. In the case of HD~37061 and HD~37479 they correspond to the rotational period of the star or harmonics hereof (see example in Fig.~\ref{fig:LC+PG_NU_Ori}) and linked to spots induced by magnetic fields on the surface that cause brightness variations due to rotation of the star \citep{DavidUraz2019}. In other cases, for example, HD~54879 and HD~57682, the dominant periodicities do not agree with constraints on the rotational period, hence some other mechanism must be responsible. A more detailed analysis of each TESS data set is given in Appendix~\ref{sec:appendix_magnetic}. We conclude that even a single sector TESS data set can be promising in studying the variability in magnetic OB-type stars in line with recent photometric studies on magnetic A and B-type stars \citep{DavidUraz2019, Sikora2019}, but also magnetic O-type stars \citep{Barron2020}.

\subsection{Binaries}\label{sec:binaries}

\begin{figure*}
	\includegraphics[width=2\columnwidth, scale = 1]{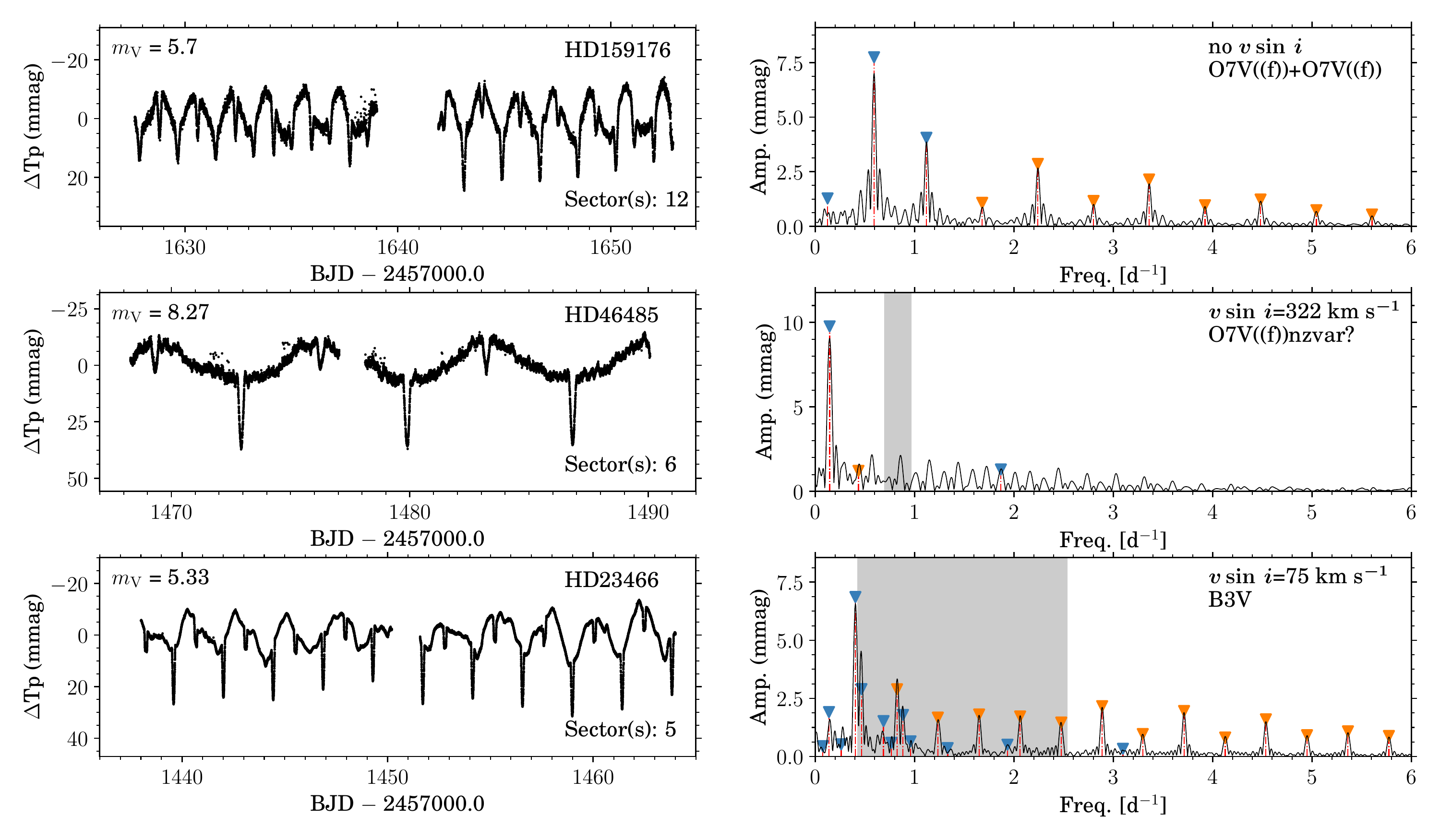}
    \caption{TESS light curve, and LS-periodogram for newly found eclipsing binaries, HD~159176, HD~46485 and HD~23466. The markers in the LS-periodogram have the same meaning as in Fig.~\ref{fig:PG_LINE_Bcep}.}
    \label{fig:LC_LINE_EB}
\end{figure*}

We find nine eclipsing binaries in our sample (see last column in Table~\ref{table:Star_List}), including three new systems. Seven of these include one or two O-type stars and another two include one or two B-type components. All of them have also been identified as SB1 or SB2 systems by means of the available spectroscopy. These EB systems take up about one third of the total sample of spectroscopic binaries comprising our working sample. As the focus of this paper is on the intrinsic variability of OB-type stars we only show the newly eclipsing systems as examples here: HD~159176 (V1036 Sco), HD~46485 and HD~23466. Their TESS light curves, and LS-periodograms are given in Fig.~\ref{fig:LC_LINE_EB}. A detailed analysis of each TESS data set is given in Appendix~\ref{sec:appendix_binaries}.

Only one of the nine stars detected as EB by TESS has been found to be a clear double line spectroscopic binary. The other eight are interesting candidates to be massive star binaries including a OB-type star and compact object companion\footnote{If the presence of a faint lower mass star companion can be discarded.} \citep{Langer2020} and likely progenitors of gravitational wave sources \citep{deMink2015}.

\subsection{Upper-main sequence stars}\label{sec:Ostars}

\begin{figure}
	\includegraphics[width=1\columnwidth, scale = 1]{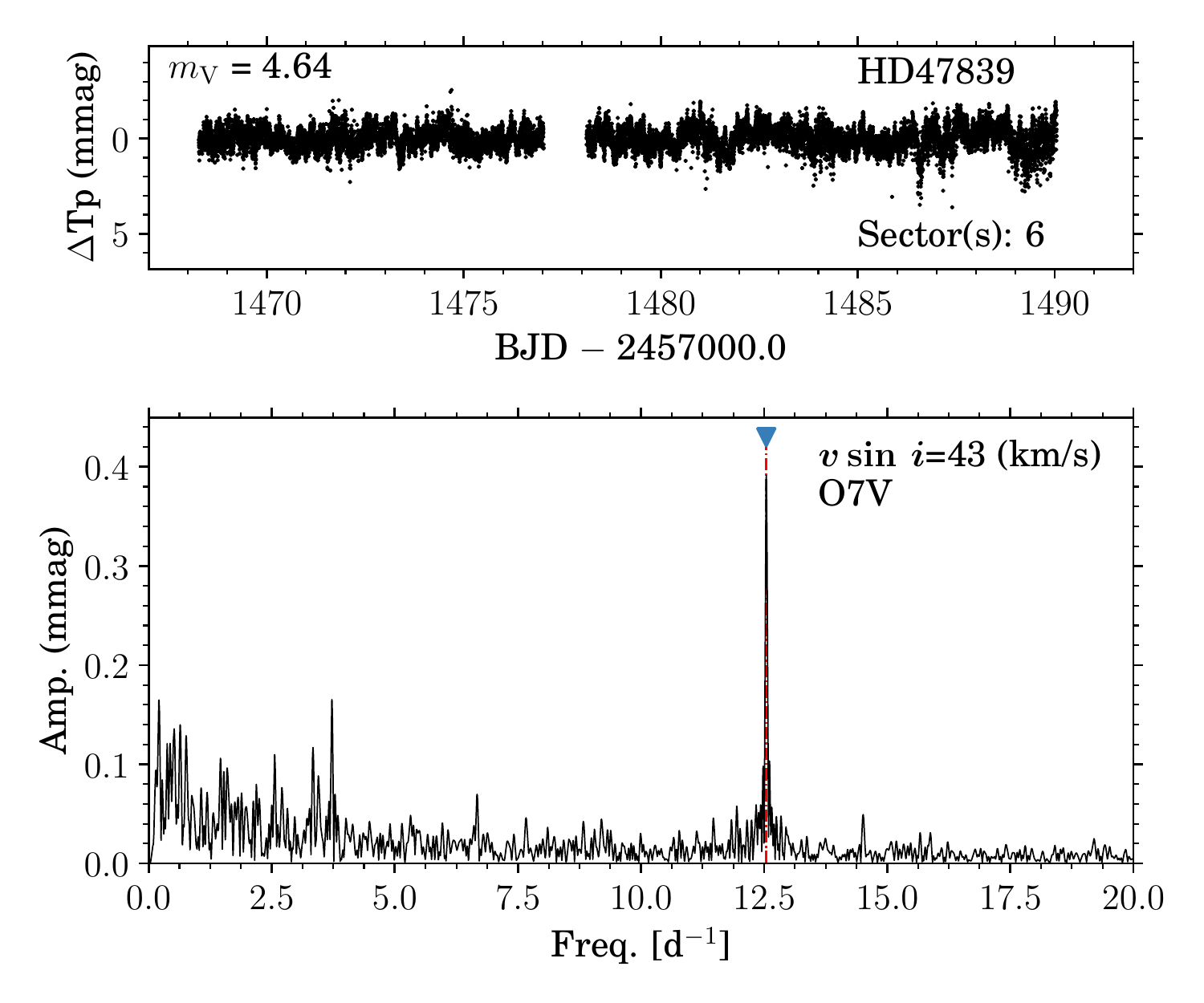}
    \caption{Light curve (\textit{Top}) and LS-periodogram (\textit{Bottom}) of HD~47839, or 15~Mon. The markers in the LS-periodogram have the same meaning as in Fig.~\ref{fig:PG_LINE_Bcep}.}
    \label{fig:LC+PG_15Mon}
\end{figure}

\begin{figure*}
	\includegraphics[width=2\columnwidth, scale = 1]{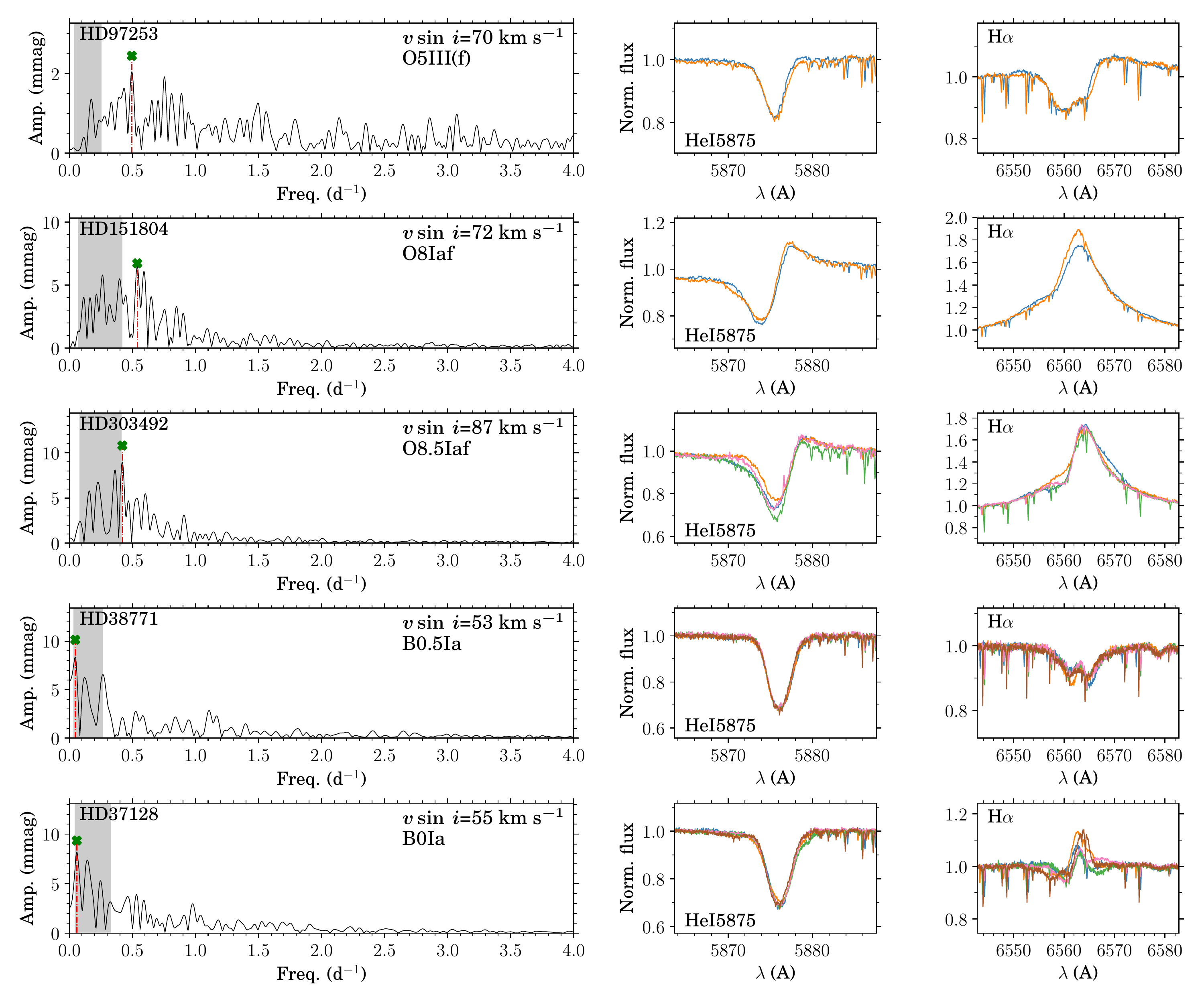}
    \caption{LS-periodogram up to 4~d$^{-1}$ (\textit{left}) and two spectroscopic lines (\textit{right}, He~{\sc i}~$\lambda$5875 and H$_{\alpha}$) for five epochs from days to months to a year, for upper main sequence stars, HD~97253, HD~151804, HD~303492, HD~38771 and HD~37128. The markers in the LS-periodogram have the same meaning as in Fig.~\ref{fig:PG_LINE_Bcep}. If the dominant frequency does not satisfy the significance criterion we mark it with a green cross.}
    \label{fig:PG+lines_OV}
\end{figure*}

As illustrated in Fig.~\ref{fig:overview_sHRD}, most of the stars in our sample identified as {\em classical} $\beta$~Cep, SPB and hybrid pulsators are located on the MS below the 15~M$_{\sun}$ evolutionary tracks, and are mostly classified as B-type dwarfs, subgiants or giants. In this section, we concentrate on the top right part of the sHRD, comprising MS stars with masses above $\gtrapprox$20~M$_{\sun}$ and including all O-type stars of different luminosity classes and two early B-type supergiants (HD\,37128 and HD\,38771).

Out of the 55 O-type stars and early B-type supergiants not labelled as magnetic or Oe stars, seven light curves are either contaminated by nearby stars, saturated or showed signs of instrumental systematics, and seven show clear binary eclipses (see Sect.~\ref{sec:binaries}). In all (but one) of the remaining 41 TESS light curves, we detect signatures of stochastic low-frequency variability (similar to e.g \citealp{Ramiaramanantsoa2018b, Bowman2019a, Bowman2019b, Pedersen2019a}). Similarly to the K2 sample of OB stars studied by \citet{Bowman2019b}, frequencies associated with coherent SPB pulsation modes and rotational modulation are seen in some of them. However, it remains difficult to conclude that these are stable coherent pulsation modes or damped modes (i.e. IGWs) given the limited frequency resolution. In almost none of these stars significant frequencies could be extracted, despite the variability being clearly present. 

The dominant variability in the photometry could also be partly or fully stochastic in nature, that is, a combination of IGWs with a stellar wind, thus highlighting how variability in the photospheres and winds of massive stars are related \citep{Prinja2004, Haucke2018}.
We therefore stress that the interpretation and analysis of the light curves of O-type stars and B supergiants is more complex than in the case of MS B-type stars due to the following reasons.

Firstly, even discarding SB2 and EB systems, a large percentage of the O-type stars in our sample have been identified as SB1. Despite that the companion must be faint, we cannot discard some type of contamination by this lower mass companion in the light curve. This is the case, e.g for HD\,47839 (15\,Mon), one of the O-type stars traditionally used as standard for spectral classification (defining the O7~V spectral class), but which is known to be part of a long period binary system ($\sim$25~years, \citealp{Gies1994}) including a fast rotating B-type star companion. This faint companion is barely detected even in our high-quality spectra, yet the high frequency clearly detected in the LS-periodogram likely corresponds to the B-type companion and not the bright O7~V star (see Fig.~\ref{fig:LC+PG_15Mon} and detailed analysis in Appendix~\ref{sec:appendix_UMS}). We note that up 40\% (11\%) of the O-type stars in our sample have been identified as SB1 (SB2) systems.

Secondly, part of the detected low-frequency variability could be associated with different types of wind variability. O-type stars and luminous B supergiants develop strong winds which leave their imprints in the H$_{\alpha}$ (and some other) lines. As illustrated in Table~\ref{table:Star_List}, many of the O- and early B-type supergiants show a varying H$_{\alpha}$ line in emission, hence implying that the wind is intrinsically varying on time-scales similar to g modes and the existence of large-scale, co-rotating wind structures. These phenomena are expected to leave signatures in the low frequency part of the LS-periodograms. Finally, some of the stars not labelled as spectroscopic binaries may actually be binaries and we did not detect them using our spectroscopic dataset (see e.g. HD~97253 below).

Having this in mind, we show some illustrative examples of the type of variability detected in these type of stars in Fig.~\ref{fig:PG+lines_OV} for stars HD~97253, HD~151804, HD~303492, HD~37128. Stochastic low frequency variability with amplitudes up to 10~mmag is present in all these stars but the morphology differs somewhat from star to star. For example, in the case of O giant HD~97253 and O supergiant HD~151804 the dominant frequency variability is found outside the estimated rotational modulation frequency range, while in the case of B supergiants HD~38771 and HD~37128 the variability occurs on much longer time scales. The spectroscopic lines, in particular H$_{\alpha}$, are also shown to be varying to different degrees for different stars. We come back to this in Sect.~\ref{sec:high_mass_evo}. A detailed analysis of these stars and others in the upper part of the main sequence is presented in Appendix~\ref{sec:appendix_UMS}.

\subsection{Post-main sequence stars}\label{sec:BSG}

\begin{figure*}
	\includegraphics[width=2\columnwidth, scale = 1]{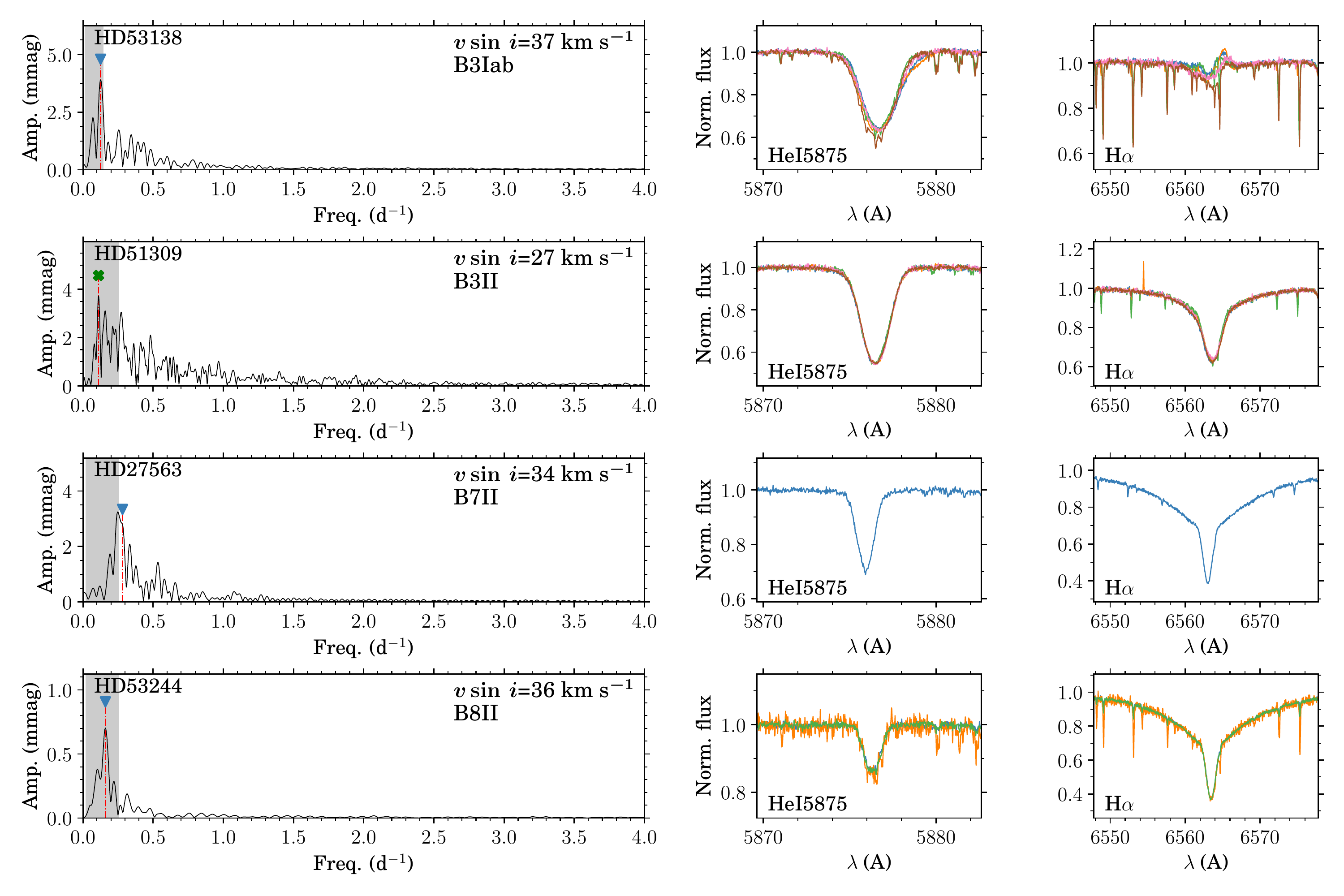}
    \caption{LS-periodogram up to 4~d$^{-1}$ (\textit{left}) and two spectroscopic lines (\textit{right}, He~{\sc i}~$\lambda$5875 and H$_{\alpha}$) for five epochs from days to months to a year, for post-MS stars, HD~53138, HD~51309, HD~27563 and HD~53244. The markers in the LS-periodogram have the same meaning as in Fig.~\ref{fig:PG_LINE_Bcep}. If the dominant frequency does not satisfy the significance criterion we mark it with a green cross.}
    \label{fig:PG_LINE_BSG}
\end{figure*}

There are eight stars (all of them classified as B-type and luminosity class I or II) located beyond the MS\footnote{As resulting from our evolutionary model computations. The location of the TAMS is very sensitive to the input parameters of the stellar model, for example, metallicity, overshooting and rotation.}, see Fig.~\ref{fig:overview_sHRD}: HD~54764, HD~53138, HD~51309, HD~46769, HD~53244, HD~27563, HD~57539, and HD~39985. All of them (except one, HD~57539, the SPB star with a v$\,\sin\,i$\,=\,160~km~s$^{-1}$, see notes about this star in Sect.~\ref{sec:spb}) exhibit stochastic low frequency variability in the LS-periodograms. In addition, we note one or multiple frequencies associated with possible SPB-type coherent modes and rotational modulation. In this aspect, the variability behaviour is very similar to the stars discussed in Sect.~\ref{sec:Ostars}. Likewise for most of these stars no significant frequencies were extracted, complicating the interpretation of the observed variability. We illustrate this by showing several examples in Fig.~\ref{fig:PG_LINE_BSG}. The low-frequency variability in photometry outside of the grey region in Fig.~\ref{fig:PG_LINE_BSG} cannot be caused by rotational modulation, and has been inferred to be caused by IGWs in OB-type stars given its broad frequency range \citep{Blomme2011, Aerts2015b, Bowman2019a, Bowman2019b}. We show one example for comparison, HD~53244, where rotational modulation is the cause of the dominant variability. A detailed analysis of these stars and others is presented in Appendix~\ref{sec:appendix_PMS}.

As a final note in this section, we emphasize that the low number of B-type post-MS stars in our sample is just a consequence of the availability in our spectroscopic database of stars of this type located in the Southern hemisphere. The density of targets in this region is envisaged to considerably increase in further studies of stars in the Northern TESS sectors, cf.\ the number of B-type post-MS stars surveyed by the IACOB project in \citet{SimonDiaz2017}.

\section{Discussion}\label{sec:Discussion}

Despite the relatively short data sets that hinder mode identification, the TESS data are sufficient to characterise different types of pulsational variability. When combined with high-resolution spectroscopy from, for example the IACOB and OWN surveys, there is the  potential to assess stellar variability over a large mass and evolutionary range. In a first instance, we compare the variable stars to theoretical instability strips calculated for early-type stars. In a second instance, we explore how the variability changes from main sequence O-type stars to more evolved B-type supergiants.
 
\subsection{Variables in the instability strip}\label{sec:instability}

\begin{figure*}
	\includegraphics[width=2\columnwidth, scale = 1]{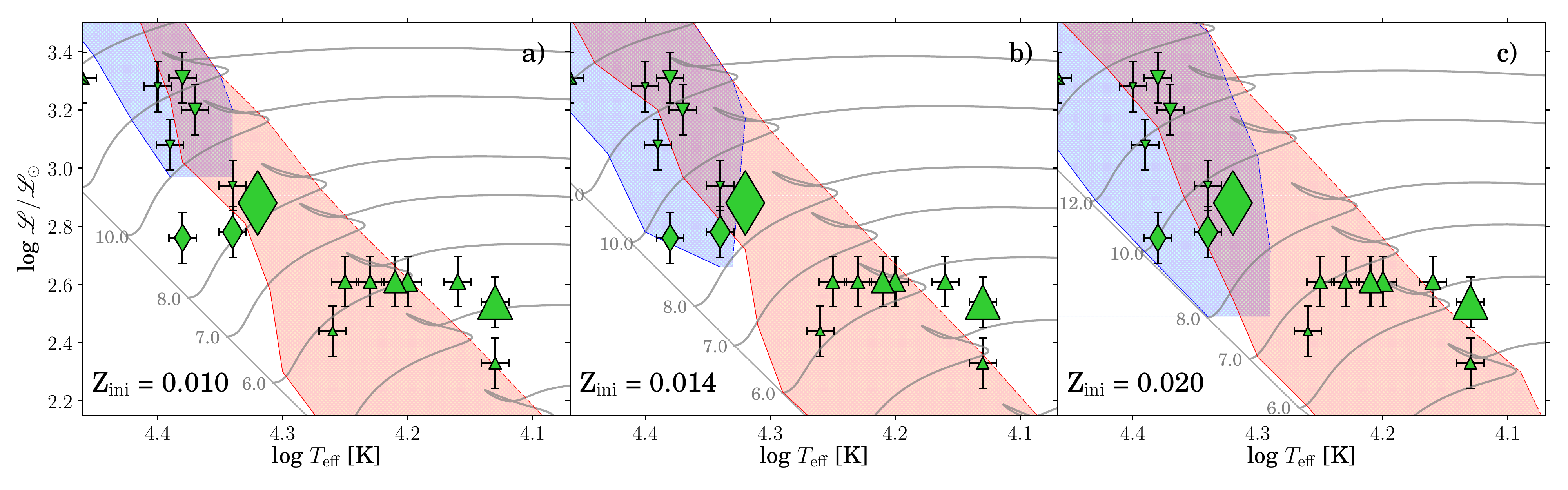}
    \caption{Comparison of the pulsating B-type stars with the theoretical instability strips for $\ell=0-2$ and three different metallicities Z$_{\rm ini}=0.010, 0.014, 0.020$. $\beta$~Cep stars are marked by downward triangles, SPB stars by upward triangles, and hybrids pulsators by diamonds. The size of the symbol is proportional to the projected rotational velocity of the star which ranges between v$\,\sin\,i=7-280$~km~s$^{-1}$. Hatched symbols mark variable stars where the exact contribution of g modes, rotational variability, and SLF is difficult to disentangle.}
    \label{fig:instability_strips_low}
\end{figure*}

We computed non-rotating evolutionary tracks and instability strips for three metallicities:  Z$_{\rm ini}=0.010, 0.014, 0.020$ with stellar evolution code MESA \citep{Paxton2011,Paxton2013,Paxton2015,Paxton2018, Paxton2019} and the non-adiabatic stellar oscillation GYRE code \citep{Townsend2013b, Townsend2018}. In the case of heat-driven pulsators, increasing the iron-group abundances increases the opacity, which also increases the driving of modes \citep{Gautschy1993, Dziembowski1993a, Dziembowski1993b}. As such, the excitation of modes is very sensitive to the heavy element abundance. Overall, our instability strips conform to those in the literature \citep{Pamyatnykh1999, Saio2006, Miglio2007b, Walczak2015, Paxton2015, Godart2017}. In Fig.~\ref{fig:instability_strips_low} we compare the observed positions of the multiperiodic pulsators discussed in Sects.~\ref{sec:bcep},~\ref{sec:spb}, and~\ref{sec:hybrids} to the computed instability strips.

The panels in Fig.~\ref{fig:instability_strips_low} show the p mode instability domain ($n_{pg}>0$) in blue for three different metallicities. The blue edge is confined to later evolutionary stages in the low metallicity regime (a), but extends (b) and finally meets the ZAMS for increasing metallicity (c), as expected \citep{Pamyatnykh1999}. For the stars showing possible p mode pulsations ($\beta$~Cep and hybrids) a metallicity of Z$_{\rm ini}=0.014$ is sufficient to explain the occurrence of high frequency pulsation modes.  The g mode instability domain ($n_{pg}<0$) is shown in orange. The effect of increasing the metallicity is seen as the extension of the instability strip to a wider mass and evolutionary range (e.g. modes in the 7~M$_{\sun}$ ZAMS model are only excited for metallicity Z$_{\rm ini}=0.020$). The g modes in B2\,IV hybrid HD~37209 cannot be explained by any increase in the metallicity. Rotation may be invoked to explain this. A similar argument can be made for SPB star HD~57539 that appears to be post-MS in our evolutionary calculations. In a study by \citet{Szewczuk2017} on the effect of rotation on the g mode instability region, they found that the strip widens and moves to hotter and more massive stars. Moreover, they argue that rotation has a larger effect than the collective influence of initial hydrogen abundance, metallicity, overshooting, and opacity.

\begin{figure*}
	\includegraphics[width=2\columnwidth, scale = 1]{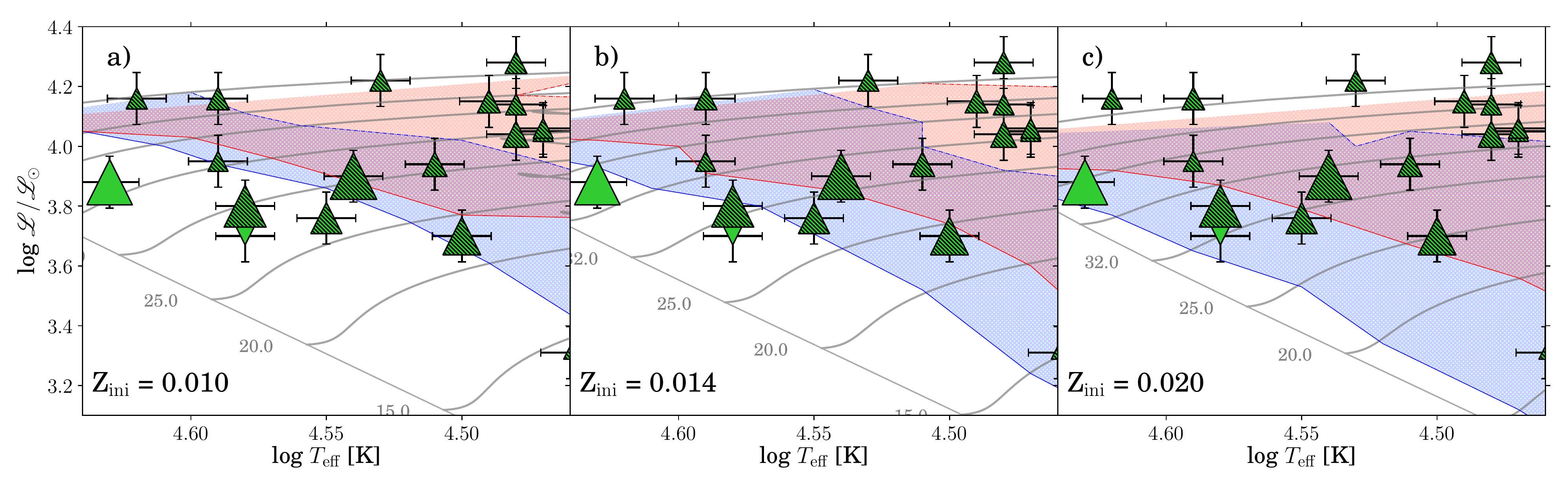}
    \caption{Comparison of the candidate pulsating O-type stars with the theoretical instability strips for $\ell=0-2$ and three different metallicities Z$_{\rm ini}=0.010, 0.014, 0.020$ (same figure style as Fig.~\ref{fig:instability_strips_low}). The size of the symbol is proportional to the projected rotational velocity of the star which ranges between v$\,\sin\,i= 26-291$~km~s$^{-1}$. Hatched symbols mark variable stars where the exact contribution of g modes, rotational variability, and SLF is difficult to disentangle.}
    \label{fig:instability_strips_high}
\end{figure*}

We show the instability strips for different metallicities in the higher mass regime in the upper panels in Fig.~\ref{fig:instability_strips_high}. Again the extent of the strip is similar to what is found in previous work, that is, \citet{Paxton2015} reaching $25$~M$_{\sun}$ or \citet{Godart2017} for Z$_{\rm ini}=0.015$. Compared to the latter, we note that the red edge of our p mode instability regime does not extend all the way to the red edge of the MS, even for high metallicity, because we only consider modes up to $\ell=2$ while these authors computed high-degree modes as well. For visual clarity, we only plot the OB-type stars that are identified as candidate pulsators in addition to stars with stochastic low frequency variability. The effect of increasing the metallicity is similar to what was seen in Fig.~\ref{fig:instability_strips_low}. However, no high frequency modes were detected, except in the case of 15~Mon (HD~47839), where this signal more likely originates from the lower mass B-type companion (see Sect.~\ref{sec:Ostars}).

The blue edge of the g mode instability strip (orange) is less sensitive to the metallicity value (a,b,c) but the blue edge still increases in temperature. Several of the stars that show stochastic low frequency variability, fall out of the instability region for coherent p and g modes in all metallicity regimes. Again rotation may be invoked to explain this but, in addition, we note that hydrodynamical simulations of massive stars suggest that pulsation frequencies of a more coherent nature at predicted eigenfrequencies can also be excited by core convection \citep{Edelmann2019, Horst2020}. In the cases of the more evolved stars, seen in the upper right of each panel in Fig.~\ref{fig:instability_strips_high}, the pulsations are predicted to be g modes, as the red edge of the p mode instability domain does not reach the TAMS. This conforms to the frequency ranges of variability observed in these stars.

For simplicity we only considered three discrete metallicities. This can only give approximate placement of the strips. For a more in-depth analysis a wider variety of input parameters need to be considered, for example, rotation, interior mixing, opacity tables (as was already made clear in the seminal work by \citealt{Pamyatnykh1999} and later by \citealt{Szewczuk2017}). This is beyond the scope of the current paper.

\subsection{High-mass (>25~M$_{\sun}$) evolutionary sequence}\label{sec:high_mass_evo}

\begin{figure*}
	\includegraphics[width=2\columnwidth, scale = 1]{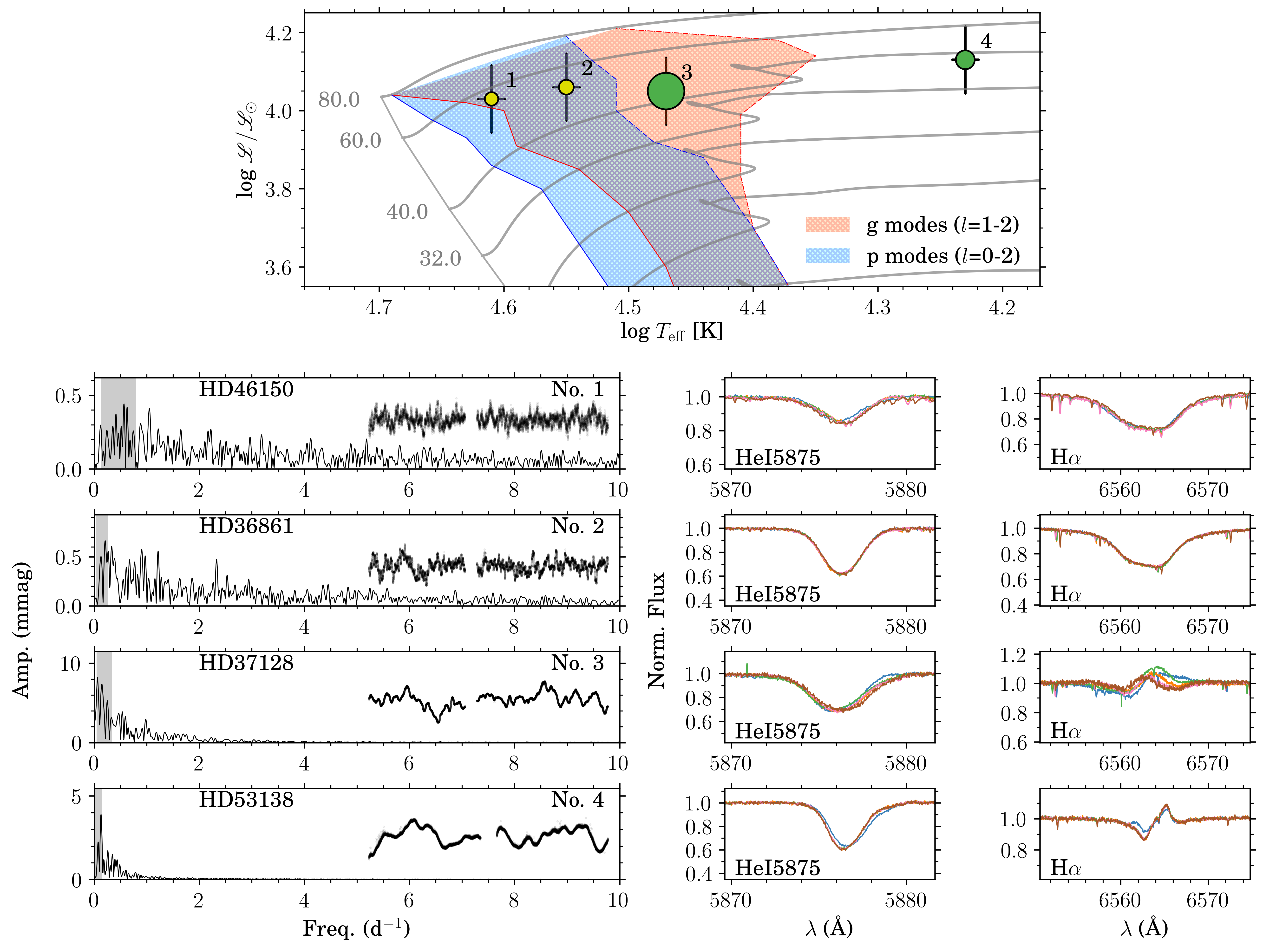}
    \caption{High-mass evolutionary sequence. sHRD (\textit{top panel}) with the size of the symbols scaled to the peak-to-peak value of the light curve (between $7.7$ and $63.8$~mmag). The colour refers to the spectral type, yellow for O and green for B.
    We include the evolutionary tracks in grey and the instability strips for Z$_{\rm in}=0.014$. 
    For each star marked with a number in the top panel, the LS-periodogram (\textit{bottom left panels}), and two spectroscopic lines (He~{\sc i}~$\lambda$5875 and H$_{\alpha}$ over multiple epochs \textit{bottom right panels}) are given. The grey region in the LS-periodogram marks the estimated rotational modulation frequency range (see Sect.~\ref{sec:var_class}). We show the TESS light curve on the inset of each LS-periodogram (not to scale).}
    \label{fig:high_mass_evo}
\end{figure*}

We selected several O and B stars in the high-mass regime to trace the variability for a hypothetical evolving high-mass star. We plot the stars in Fig.~\ref{fig:high_mass_evo}. On the top panel we show the sHRD with four stars following the same approximate evolutionary path between $40-60$~M$_{\sun}$, with different ages. The numbering corresponds to the rows of panels below, where we plot their LS-periodogram and two spectroscopic lines, He~{\sc i}~$\lambda$5875 and H$_{\alpha}$. The light curve of each star is approximately $27$~d. Star~3 and Star~4 show LPV and variable emission in H$_{\alpha}$, indicative of the strong stellar wind. This is interesting with respect to the origin of the variability: clearly various physical mechanisms are responsible, which are linked to mass and age.

An immediate feature of the stars in Fig.~\ref{fig:high_mass_evo} is the decrease in frequency (increase in period) of the dominant variability as one advances on the evolutionary tracks, as seen in the LS-periodogram and the inset light curve (right panels in Fig.~\ref{fig:high_mass_evo}). In Star~1 the stochastic low frequency variability only reaches the instrumental white noise at much higher frequencies ($\gg10$~d$^{-1}$), while in Star~4 the stochastic variability is constrained to lower frequencies. The increasing radius of an evolving star is responsible for the decreasing frequency. Another effect is the increase in the relative amplitude of the variability as the star ages, seen as the markers in the top panel of Fig.~\ref{fig:high_mass_evo} that scale with the peak-to-peak variability in the light curve.

\section{Summary}
By combining TESS data and multi-epoch high-resolution spectroscopy gathered by the IACOB and OWN surveys, we provided a variability study for a sample of OB-type stars in the southern hemisphere based on visual inspections followed by a frequency analysis. We find a large diversity of causes for the observed variability even for stars that reside in an area in the sHRD where this is not expected from non-rotating pulsation models. Each star is interesting in its own right for improving stellar models, and we discussed several examples aided by the available spectroscopic parameters and variability. The low frequency resolution of only one or two TESS Sectors makes the identification of individual pulsation modes, combination frequencies, and rotational multiplets or patterns difficult, especially in the low frequency regime (cf.\ Sects.~\ref{sec:spb} and \ref{sec:hybrids}) and in stars with significant stochastic low frequency variability (cf.\ Sects.~\ref{sec:Ostars} and \ref{sec:BSG}). 

The one-sector TESS light curves are generally sufficient to place constraints on the type of variability. The computed instability strips were not able to fully explain the range of observed variability in OB type stars. The stars falling outside the instability strips are mostly fast-rotating, indicating that instability strips with rotation are needed to make detailed comparisons between such strips and large samples (see also \citealt{Szewczuk2017}). Since the inclusion of rotation in stellar structure models and in pulsation analysis for the modes in the mass regime considered in this study is of major importance (e.g. \citealt{Aerts2019a}) but requires detailed work when modelling stars, we will take up an upgrade of the tracks in Fig.~\ref{fig:overview_sHRD} in future work. Rotation also alters the apparent location of a star in the HRD as the centrifugal force counteracts the gravity, leading to the equator being cooler and having a lower log~$g$. This needs to be accounted for in any spectroscopic analysis\footnote{As is the case in our analysis (see caption of Table~\ref{table:Star_List}).} and illustrates that variability classification of OB-type stars in large numbers is necessary to accurately constrain the observational instability boundaries. The current ongoing TESS observations in the northern sky will surely further aid in this venture. To cover wider ranges in the upper part of the HRD, there is also large potential in extracting OB-type star light curves from the TESS Full Frame Images and combining this with complementary spectroscopy from upcoming all-sky spectroscopic surveys,  such as SDSS-V \citep{Kollmeier2017}.

By tracing the variability for a hypothetical massive star we saw observational confirmation of stellar evolution: the dominant variability decreases in frequency (increases in period), and increases in amplitude, as a star ages. This is a starting point to understand the structure and evolution of massive stars through their photometric variability and pulsations, providing new empirical anchors to the stellar evolutionary theory. In all the newly found potential pulsating stars additional continuous space photometry, multi-colour photometry or high-resolution, multi-epoch spectroscopy is advised to achieve mode identification. To constrain the physics of the interiors of massive stars multidimensional theoretical instability strips (i.e. in rotation, metallicity, overshooting, and envelope mixing) should be considered. However, to achieve this efficiently, the top-down method in this work should be reversed. First, the selection of optimal candidates from long-term continuous photometry (i.e. the TESS Continuous Viewing Zone and Kepler/K2) that allow for mode identification through period or frequency patterns is needed. Only then are these to be followed-up with spectroscopy and asteroseismic modelling. We refer for example, to the ensemble modelling approaches taken for low-mass stars (i.e. \citealt{Chaplin2014, Yu2018}) and intermediate-mass stars (i.e. \citealt{Mombarg2019, Ouazzani2019}), that have allowed for constraints to be placed on interior physics, such as overshooting, angular momentum transport, and chemical mixing. 

\begin{acknowledgements}
SB is thankful to the staff of the IAC for the possibility of and kind hospitality during his research visit there, which made this joint project possible. SB and MM are grateful to Rich Townsend for his inspiring and highly didactical tutorial on mode excitation during the 2019 MESA summer school and for his continuous efforts to upgrade the GYRE pulsation code for the benefit of the asteroseismology community. The MESA developers are also thanked for their efforts. We thank the anonymous referee for helpful comments that have improved the quality of the paper. The research leading to these results has received funding from the European Research Council (ERC) under the European Union's Horizon 2020 research and innovation program (grant agreement No. 670519: MAMSIE). S-SD acknowledges support from the Spanish Government Ministerio de Ciencia, Innovaci\'on y Universidades through grants PGC-2018-091\,3741-B-C22. This research has made use of the \texttt{SIMBAD} database, operated at CDS, Strasbourg, France. Some/all of the data presented in this paper were obtained from the Mikulski Archive for Space Telescopes (MAST). STScI is operated by the Association of Universities for Research in Astronomy, Inc., under NASA contract NAS5-26555. The computational resources and services used in this work were provided by the VSC (Flemish Supercomputer Center), funded by the Research Foundation - Flanders (FWO) and the Flemish Government – department EWI.
\end{acknowledgements}

%
%
\bibliographystyle{aa}
\bibliography{mybib}

\begin{appendix}

\section{Frequency analysis}\label{sec:appendix_freqanal}
For the frequency analysis we perform iterative pre-whitening following the method described in \citet{Degroote2009a} and improved by \citet{Papics2012a}. At each step in this iterative process the LS-periodogram is calculated followed by a search for the frequency with the highest significance. This frequency, together with previous frequencies, is then used to perform a non-linear squares fit on the light curve using a sinusoidal model. Each iteration this model is subtracted from the light curve and the next frequency is searched for. 

\subsection{Significance criterion for TESS single sectors}
A crucial point in any frequency analysis is to consider what determines the signal-to-noise ratio (S/N) of a frequency. A widely used criterion in the field is the S/N $\geq4$ criterion derived empirically by \citet{Breger1993} based on  ground-based observations with a long time base. In this study they concluded that a S/N $\geq4$ corresponds to a 99.9\% confidence that the frequency is real and not due to noise, which was later supported by simulations \citep{Kuschnig1997}. In the context of modern space-based photometric campaigns (specifically Kepler/K2), \citet{Baran2015} revisited this criterion. They concluded that a higher threshold is recommended (S/N $\geq5$) for data sets lasting less than a few months as the typically much larger number of data points compared to ground-based light curves increases the probability of spurious frequency detection. We therefore adopt this more conservative criterion in the frequency analysis of the 1-2 sector TESS data sets.

\begin{figure*}
	\includegraphics[width=2\columnwidth, scale = 1]{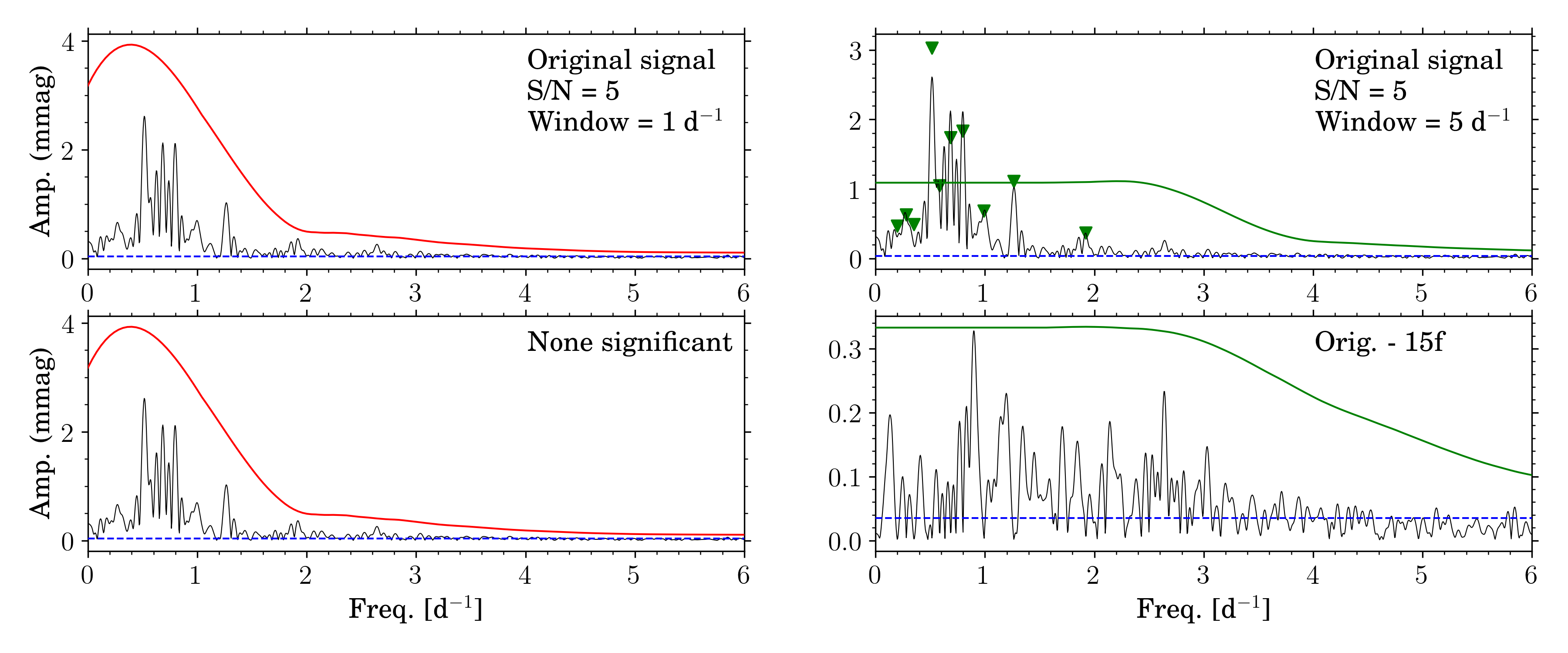}
    \caption{Comparison of the different noise window size calculations for SPB star HD~48977. The left hand side shows the periodogram of the original signal (top) and the residual signal (bottom) after frequency extraction using a significance criterion of S/N $=5$ with the noise calculated in a window of $1$~d$^{-1}$ around the frequency. The right hand side shows the same periodograms but with the noise calculated in a window of $5$~d$^{-1}$ around the frequency. Green markers indicate singificant frequencies. The coloured solid lines represent the significance levels for each periodogram. The blue dashed line indicates the mean noise level over the whole periodogram.}
    \label{fig:Handler_type_fig_HD48977}
\end{figure*}

In photometry, the noise is frequency dependent, consisting of a combination of white noise and instrumental (coloured) noise (see \citealt{Handler2019} for different frequency extraction techniques using TESS data). In Fig.~\ref{fig:Handler_type_fig_HD48977} we compare two different noise calculations for the SPB star HD~48977 \citep{Thoul2013}. In both calculations the noise is calculated in a window around the frequency to be extracted, but this yields different results depending on the size of the window. The high g mode density contributes appreciably to the noise level if a small window size is used and is therefore not appropriate in these cases (left panels of Fig.~\ref{fig:Handler_type_fig_HD48977}). An increase in window size allows for frequency extraction while taking the frequency dependence of the noise into account and thus avoiding over-interpretation of the data (right panels of Fig.~\ref{fig:Handler_type_fig_HD48977}).

\begin{figure*}
	\includegraphics[width=2\columnwidth, scale = 1]{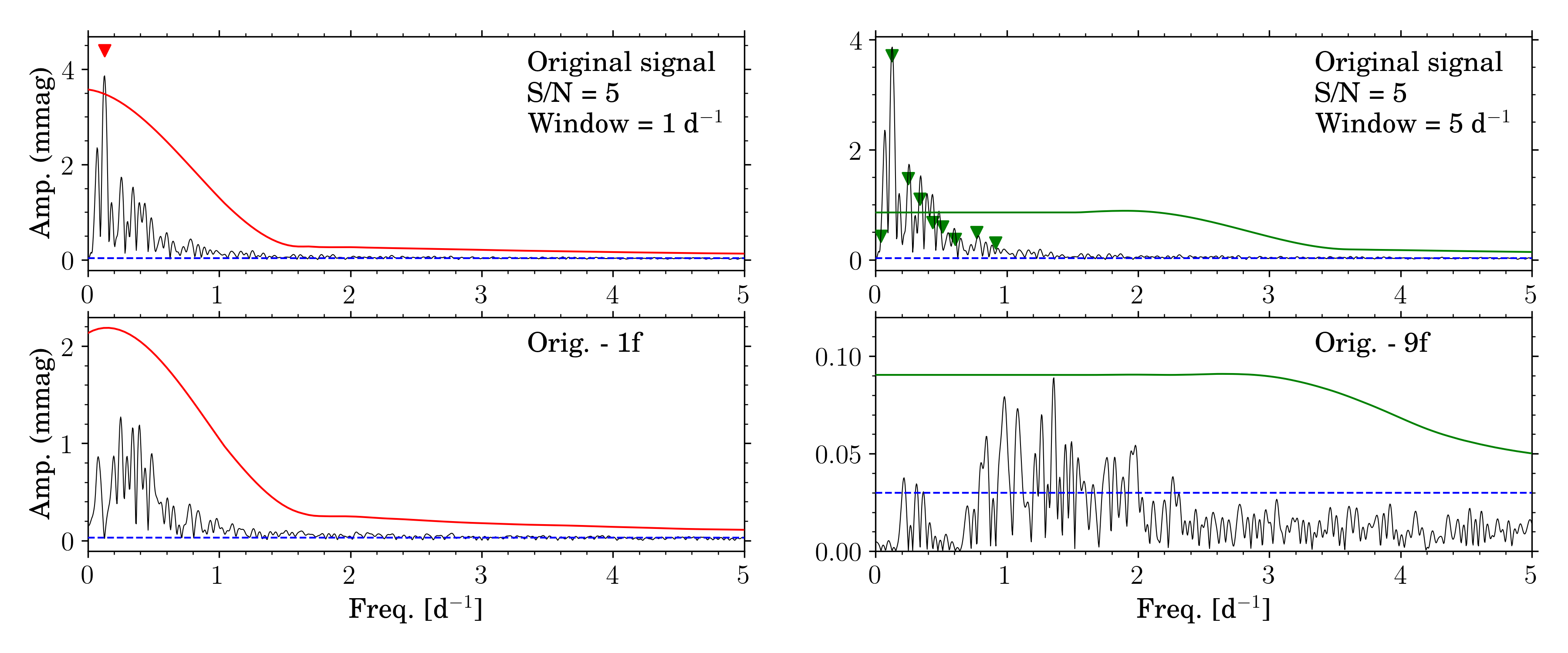}
    \caption{Comparison of the different noise window size calculations for B3\,Iab star HD~53138. Same figure style as Fig.~\ref{fig:Handler_type_fig_HD48977}}
    \label{fig:Handler_type_fig_HD53138}
\end{figure*}

Many of the upper main sequence stars show stochastic low frequency variability, which in the Fourier decomposition of the signal can lead to many closely spaced frequencies. This is shown for B3\,Iab star HD~53138 in Fig.~\ref{fig:Handler_type_fig_HD53138}. In this particular case the noise behaviour is not properly accounted for when choosing a large window size. We therefore employ a window size of $1$~d$^{-1}$ in stars where the stochastic low frequency variability is prominent (generally massive M $>20$ M$_{\sun}$ dwarfs and OB-type giants). In lower mass stars the stochastic low frequency variability is generally of lower amplitude and we employ a window size of $5$~d$^{-1}$.

\subsection{Further processing}

When the significance criterion is reached the frequency list is processed further. This includes removing frequencies that are unresolved to one another, using the Loumos \& Deeming criterion, 1.5/$\Delta$T for a light curve of length $\Delta$~T \citep{Loumos1978}. The time span $\Delta$T~(d) is calculated using the first and last day of observation, not accounting for the intermittent gap that is present in the middle of every light curve. This due to a pause of 1.09~d in data collection during perigree passage while downloading data. We note that Sector 6 provides the shortest light curves in the whole sample, just 21.77~d of data were collected (1/$\Delta$T = 0.046~d$^{-1}$). This is due to the first $\approx3$~d being used to collect calibration data\footnote{see the TESS-DR8 release notes for more information, \href{https://archive.stsci.edu/missions/tess/doc/tess\_drn/
tess\_sector\_06\_drn08\_v02.pdf}{https://archive.stsci.edu/missions/tess/doc/tess\_drn/
tess\_sector\_06\_drn08\_v02.pdf}}.

We further search for potential combination frequencies and harmonics. This involves selecting the two highest amplitude frequencies and looking for other frequencies of the form $n\nu_{i}+m\nu_{j}$ with orders $n,m \in [-3,3]$ in a semi-automatic way. We use the frequency resolution ($\nu_{\rm res}=1/\Delta$T) as the matching criterion, i.e we scan for frequencies $\nu=n\nu_{i}+m\nu_{j} \pm \nu_{\rm res}$. We process these matches further by evaluating the likelihood that higher order combinations are real based on the presence of lower order combinations, given that the chance occurrence of a match increases with the order \citep{Papics2012b}. As mentioned in Sect.~\ref{sec:TESSmethod} any long-term trends were removed by linear or low-order polynomial fits. As such, we caution that frequencies $\nu\lessapprox0.1$~d$^{-1}$ are dubious given that they are sampled only approximately 2 (4) times for a 1 (2) sector TESS data set.

\subsection{Results}
We summarise the global results of the frequency analysis in Table~\ref{table:freq_anal}. In Col.~3 we give the frequency resolution in d$^{-1}$ of the TESS data set. In Col.~4 we indicate the size of the noise window used to calculate the significance levels in the frequency analysis. Cols.~5 and 6 give the dominant frequency and its associated amplitude. The number in brackets in the frequency and amplitude columns gives the error on the last digit. This error is calculated based on the non-linear squares fit, and we further employ a correction factor to compensate for the correlated nature of the data. We use the correction factor by \citet{Schwarzenberg-Czerny2003}, calculated following \citet{Degroote2009a}. The signal-to-noise of that particular frequency is given in Col.~6. In Col.~8 we provide notes for that particular star: \textit{IV} (invalid) if the light curve is of poor quality or contaminated, \textit{HARM} if the periodogram is dominated by harmonics, and an asterisk (*) if more than one frequency was measured. 

The full frequency lists of stars with multiple measured frequencies are given in Table~\ref{table:multi_freq_anal}.  The number in brackets in the frequency, amplitude, and phase columns gives the error on the last digit. In the last column we add notes to highlight potential combinations, harmonics or multiplet memberships. The latter are indicated as follows: tp/tp? (triplet) or mp/mp? (multiplet) followed by a number if more than one was detected.

\begin{table*}
\centering
\caption{Summary of the frequency analysis of the stars considered in this work.}

\begin{threeparttable}\label{table:freq_anal} 

\begin{tabular}{l l c c c c c c c} 
\hline

HD & TIC &  1/$\Delta$T  & Window & $\nu_{\rm dom}$& $A_{\rm dom}$& S/N& Var. Type & Notes\\ 
&&[d$^{-1}$]&[d$^{-1}$]&[d$^{-1}$]&[mmag]&&&\\
\hline
&&&&&&&& \\
\multicolumn{8}{c}{O-type dwarfs and subgiants (V and IV)} \\
HD\,96715	&	306491594	&	0.018	&	1.0	&	0.4780(7)	&	0.33(2)	&	\textbf{3.34}	&	SLF	&		\\
HD\,46223	&	234881667	&	0.046	&	1.0	&	0.181(2)	&	0.80(5)	&	\textbf{3.16}	&	SLF+SPB?	&		\\
HD\,155913	&	216662610	&	0.040	&	1.0	&	1.637(1)	&	1.55(7)	&	\textbf{4.27}	&	SLF+SPB	&		\\
HD\,46150	&	234840662	&	0.046	&	1.0	&	0.563(2)	&	0.44(3)	&	\textbf{2.27}	&	SLF	&		\\
HD\,90273	&	464295672	&	0.041	&	1.0	&	0.581(1)	&	0.67(4)	&	\textbf{2.46}	&	SLF+SPB?	&		\\
HD\,110360	&	433738620	&	0.039	&	1.0	&	1.2372(6)	&	0.77(2)	&	5.59	&	rot?	&		\\
HD\,47839	&	220322383	&	0.046	&	1.0	&	12.5435(7)	&	0.39(1)	&	7.87	&	SLF+$\beta$~Cep	&		\\
HD\,46485	&	234933888	&	0.046	&	5.0	&	0.1453(4)	&	8.9(1)	&	15.43	&	EB	&	HARM, *	\\
HD\,53975	&	148506724	&	0.041	&	1.0	&	0.1700(5)	&	1.19(3)	&	5.13	&	SLF+rot?	&		\\
HD\,41997	&	294114621	&	0.046	&	1.0	&	1.220(2)	&	2.5(2)	&	\textbf{3.19}	&	SLF+rot?	&		\\
HD\,152590	&	341258182	&	0.040	&	5.0	&	0.445(3)	&	25(4)	&	5.95	&	EB	&	HARM, *	\\
HD\,46573	&	234947719	&	0.046	&	1.0	&	0.861(2)	&	0.97(7)	&	\textbf{2.60}	&	SLF+rot?	&		\\
HD\,48279	&	234009943	&	0.046	&	1.0	&	1.213(2)	&	0.68(4)	&	\textbf{3.09}	&	SLF+SPB?	&		\\
HD\,101191	&	319936861	&	0.018	&	1.0	&	0.9553(2)	&	1.98(4)	&	7.28	&	cont.	& IV		\\
HD\,38666	&	100589904	&	0.019	&	1.0	&	0.8349(3)	&	0.162(4)	&	5.62	&	PQ	&IV		\\
HD\,36512	&	34512896	&	0.046	&	1.0	&	0.306(1)	&	0.095(5)	&	\textbf{2.41}	&	SLF?	&		\\
HD\,123056	&	330281456	&	0.039	&	1.0	&	0.4753(7)	&	2.25(7)	&	5.30	&	SLF+SPB?	&		\\
HD\,76556	&	30135331	&	0.041	&	1.0	&	0.235(2)	&	3.0(2)	&	\textbf{3.49}	&	SLF+SPB?	&		\\
HD\,74920	&	430625455	&	0.020	&	1.0	&	1.568(1)	&	1.0(1)	&	\textbf{2.89}	&	SLF+SPB?	&		\\
HD\,135591	&	455675248	&	0.036	&	1.0	&	0.625(2)	&	0.95(8)	&	\textbf{2.88}	&	SLF	&		\\
HD\,326331	&	339568114	&	0.040	&	1.0	&	0.189(2)	&	2.5(3)	&	\textbf{2.33}	&	SLF	&		\\
HD\,37041	&	427395049	&	0.046	&	1.0	&	0.178(1)	&	0.127(5)	&	\textbf{3.05}	&	SLF?	&		\\
HD\,48099	&	231154751	&	0.046	&	1.0	&	0.6559(10)	&	0.98(4)	&	\textbf{3.43}	&	SLF	&		\\
HD\,159176	&	102643159	&	0.040	&	5.0	&	0.5929(5)	&	7.0(2)	&	15.35	&	EB	&	HARM, *	\\
HD\,54662	&	177717406	&	0.041	&	1.0	&	0.710(1)	&	0.19(1)	&	\textbf{2.63}	&	SLF	&		\\
HD\,57236	&	5051420	&	0.041	&	1.0	&	0.7155(6)	&	0.349(9)	&	5.05	&	SLF+rot	&		\\
HD\,75759	&	29207816	&	0.020	&	1.0	&	0.4947(4)	&	0.136(5)	&	\textbf{2.59}	&	SLF	&		\\
HD\,37468	&	11286198$^{\dagger}$	&	0.046	&	1.0	&	1.6788(8)	&	1.48(5)	&	7.85	&	cont.	&		\\

&&&&&&&& \\
\multicolumn{8}{c}{O-type giants, bright giants and supergiants (III, II and I)} \\
HD\,66811	&	133422778	&	0.041	&	1.0	&	0.566(2)	&	3.0(3)	&	\textbf{3.91}	&	SLF+rot	&		\\
HD\,97253	&	467065657	&	0.039	&	1.0	&	0.495(3)	&	2.0(2)	&	\textbf{2.35}	&	SLF+SPB?	&		\\
HD\,93843	&	465012898	&	0.018	&	1.0	&	0.3445(3)	&	4.8(1)	&	7.15	&	SLF+rot	&		\\
HD\,156738	&	195288472	&	0.040	&	1.0	&	0.894(2)	&	1.01(8)	&	\textbf{2.63}	&	SLF	&		\\
HD\,36861	&	436103278	&	0.046	&	1.0	&	0.205(4)	&	0.67(9)	&	\textbf{2.07}	&	SLF	&		\\
HD\,150574	&	234648113	&	0.040	&	1.0	&	0.715(2)	&	2.4(3)	&	\textbf{2.17}	&	SLF+rot?	&		\\
HD\,152247	&	339570292	&	0.040	&	1.0	&	0.1586(9)	&	3.3(1)	&	\textbf{3.79}	&	SLF+SPB?	&		\\
HD\,55879	&	178489528	&	0.041	&	1.0	&	0.222(1)	&	0.39(2)	&	\textbf{2.23}	&	SLF	&		\\
HD\,154643	&	43284243	&	0.040	&	1.0	&	0.109(2)	&	0.98(7)	&	\textbf{2.24}	&	SLF+rot?	&		\\
HD\,152233	&	339565205	&	0.040	&	1.0	&	0.226(1)	&	7.6(4)	&	\textbf{2.45}	&	cont.	&		\\
HD\,57061	&	106347931	&	0.041	&	5.0	&	1.5604(8)	&	19.3(7)	&	20.35	&	EB	&	HARM, *	\\
HD\,36486	&	50743469	&	0.046	&	5.0	&	0.3494(10)	&	8.7(3)	&	14.02	&	EB	&	HARM, *	\\
CPD-47\,2963	&	30653985	&	0.020	&	1.0	&	0.523(1)	&	2.1(2)	&	\textbf{2.87}	&	SLF	&		\\
HD\,57060	&	106349204$^{\dagger}$	&	0.041	&	5.0	&	0.4553(7)	&	186(5)	&	31.61	&	EB	&	HARM, *	\\
HD\,156154	&	152659955	&	0.040	&	1.0	&	0.237(2)	&	5.1(5)	&	\textbf{3.13}	&	SLF+SPB?	&		\\
HD\,112244	&	406050497	&	0.039	&	1.0	&	0.314(4)	&	8(1)	&	\textbf{2.47}	&	SLF	&		\\
HD\,151804	&	337793038	&	0.040	&	1.0	&	0.538(2)	&	6.4(5)	&	\textbf{2.51}	&	SLF	&		\\
HD\,47129	&	220197273	&	0.046	&	5.0	&	1.645(2)	&	18(1)	&	12.35	&	EB	&	HARM, *	\\
HD\,303492	&	459532732	&	0.039	&	1.0	&	0.420(2)	&	9.0(9)	&	\textbf{3.15}	&	SLF	&		\\
HD\,152249	&	339567904	&	0.040	&	1.0	&	0.455(2)	&	5.1(5)	&	\textbf{2.59}	&	SLF+SPB?	&		\\
HD\,152424	&	247267245	&	0.040	&	1.0	&	0.423(4)	&	8(2)	&	\textbf{2.25}	&	SLF+SPB?	&		\\
HD\,154368	&	41792209	&	0.040	&	1.0	&	0.788(2)	&	4.4(4)	&	\textbf{2.53}	&	SLF+SPB?	&		\\
HD\,152003	&	338640317	&	0.040	&	1.0	&	0.112(2)	&	9.4(9)	&	\textbf{2.86}	&	SLF+rot?	&		\\
HD\,152147	&	246953610	&	0.040	&	1.0	&	0.398(2)	&	8.4(9)	&	\textbf{2.69}	&	SLF+SPB?	&		\\
HD\,37043	&	427395774	&	0.046	&	1.0	&	0.505(4)	&	1.6(3)	&	\textbf{2.06}	&	SLF+SPB?	&		\\

\hline
\end{tabular}
\begin{tablenotes}
\item \textit{Table continued below.}
\end{tablenotes}

\end{threeparttable}

\end{table*}

\begin{table*}
\centering 
\ContinuedFloat
\caption{Continued.}\label{table:freq_anal}  

\begin{threeparttable}

\begin{tabular}{l l c c c c c c c} 
\hline
HD & TIC & 1/$\Delta$T  & Window & $\nu_{\rm dom}$  & $A_{\rm dom}$ & S/N& Var. Type & Notes\\ 
&&[d$^{-1}$]&[d$^{-1}$]&[d$^{-1}$]&[mmag]&&&\\
\hline
&&&&&&&& \\
\multicolumn{8}{c}{Early B-type dwarfs, subgiants and giants (V, IV)} \\
HD\,36960	&	427373484	&	0.046	&	1.0	&	1.0024(9)	&	0.111(4)	&	\textbf{3.46}	&	SLF+SPB?	&		\\
HD\,37042	&	427395058	&	0.046	&	1.0	&	0.201(1)	&	0.167(9)	&	\textbf{2.78}	&	SLF?	&		\\
HD\,36959	&	427373476	&	0.046	&	1.0	&	0.299(1)	&	0.53(3)	&	\textbf{3.17}	&	cont.	&IV		\\
HD\,43112	&	434384707	&	0.046	&	1.0	&	0.311(2)	&	0.036(3)	&	\textbf{3.28}	&	SLF?	&		\\
HD\,37303	&	332856560	&	0.046	&	5.0	&	11.7889(8)	&	0.35(1)	&	15.29	&	hybrid	&	*	\\
HD\,35912	&	464839773	&	0.046	&	5.0	&	0.835(1)	&	4.0(2)	&	14.08	&	SPB	&	*	\\
HD\,48977	&	202148345$^{\dagger}$	&	0.046	&	5.0	&	0.516(2)	&	2.8(2)	&	12.0	&	SPB	&	*	\\
HD\,23466	&	426588729	&	0.038	&	5.0	&	0.405(1)	&	6.2(4)	&	13.69	&	EB	&	HARM, *	\\
HD\,34816	&	442871031	&	0.038	&	1.0	&	0.7923(2)	&	1.56(1)	&	9.49	&	rot	&	*	\\
HD\,46328	&	47763235$^{\dagger}$	&	0.046	&	5.0	&	4.7715(1)	&	12.91(8)	&	30.65	&	$\beta$~Cep	&	*	\\
HD\,50707	&	78897024	&	0.021	&	5.0	&	5.4188(2)	&	3.87(8)	&	30.64	&	$\beta$~Cep	&	*	\\
HD\,37481	&	332913301	&	0.046	&	5.0	&	1.1052(8)	&	0.134(4)	&	8.74	&	hybrid	&	*	\\
HD\,16582	&	328161938	&	0.039	&	5.0	&	6.20598(9)	&	8.20(3)	&	27.88	&	$\beta$~Cep	&	*	\\
HD\,37209	&	388935529	&	0.046	&	5.0	&	0.2709(9)	&	0.117(4)	&	11.02	&	hybrid	&	*	\\
HD\,26912	&	283793973	&	0.038	&	5.0	&	0.500(3)	&	3.4(4)	&	15.43	&	SPB	&	*	\\
HD\,37711	&	59215060	&	0.046	&	5.0	&	0.4503(7)	&	3.8(1)	&	18.57	&	SPB	&	*	\\
HD\,57539	&	10176636	&	0.041	&	5.0	&	1.6518(7)	&	3.2(1)	&	16.16	&	SPB	&	*	\\
HD\,41753	&	151464886$^{\dagger}$	&	0.046	&	5.0	&	0.6600(8)	&	5.3(2)	&	23.80	&	SPB	&	*	\\
HD\,224990	&	313934087	&	0.036	&	5.0	&	0.5733(3)	&	0.504(7)	&	19.72	&	SPB	&	*	\\
HD\,37018	&	427393354	&	0.046	&	1.0	&	0.8320(8)	&	0.148(4)	&	\textbf{4.35}	&	SLF+SPB?	&		\\

&&&&&&&& \\
\multicolumn{8}{c}{B-type giants, bright giants and supergiants (III, II, and I)} \\
HD\,48434	&	234052684$^{\dagger}$	&	0.046	&	1.0	&	0.034(4)	&	4.7(7)	&	\textbf{2.63}	&	SLF	&		\\
HD\,61068	&	349043273	&	0.041	&	5.0	&	6.072(1)	&	5.5(3)	&	18.98	&	$\beta$~Cep	&	*	\\
HD\,35468	&	365572007	&	0.046	&	1.0	&	0.2761(9)	&	0.195(7)	&	\textbf{2.94}	&	SLF+rot?	&		\\
HD\,44743	&	34590771$^{\dagger}$	&	0.046	&	5.0	&	3.9784(1)	&	5.86(3)	&	25.78	&	$\beta$~Cep	&	*	\\
HD\,54764	&	95513457	&	0.041	&	1.0	&	0.490(1)	&	4.9(3)	&	\textbf{3.10}	&	SLF+SPB?	&		\\
HD\,52089	&	63198307	&	0.021	&	1.0	&	0.2340(5)	&	0.71(3)	&	\textbf{3.71}	&	SLF+rot?	&		\\
HD\,62747	&	126586580	&	0.041	&	5.0	&	0.510(2)	&	12(1)	&	10.75	&	EB	&	HARM, *	\\
HD\,51309	&	146908355	&	0.021	&	1.0	&	0.1112(7)	&	3.8(2)	&	\textbf{3.51}	&	SLF	&		\\
HD\,46769	&	281148636	&	0.046	&	1.0	&	0.2249(9)	&	0.082(3)	&	\textbf{4.05}	&	SLF+rot?	&		\\
HD\,27563	&	37777866	&	0.038	&	1.0	&	0.2820(6)	&	2.78(8)	&	5.2	&	SLF+SPB	&		\\
HD\,53244	&	148109427	&	0.041	&	1.0	&	0.1600(2)	&	0.755(8)	&	7.12	&	SLF+rot	&		\\
HD\,37128	&	66651575$^{\dagger}$	&	0.046	&	1.0	&	0.059(3)	&	9(1)	&	\textbf{3.21}	&	SLF+SPB?	&		\\
HD\,38771	&	427451176$^{\dagger}$	&	0.046	&	1.0	&	0.046(2)	&	8.5(7)	&	\textbf{4.04}	&	SLF+SPB?	&		\\
HD\,53138	&	80466973	&	0.041	&	1.0	&	0.1258(4)	&	3.99(7)	&	5.62	&	SLF+SPB?	&		\\
HD\,39985	&	102281507	&	0.046	&	1.0	&	0.2864(8)	&	0.102(3)	&	\textbf{4.20}	&	SLF?	&		\\

&&&&&&&& \\
\multicolumn{8}{c}{Magnetic O- and B-type stars} \\
HD\,37022	&	427394772	&	0.046	&	1.0	&	0.307(2)	&	0.97(8)	&	\textbf{2.41}	&	cont.	&IV		\\
HD\,46056	&	234834992	&	0.046	&	1.0	&	2.063(1)	&	0.30(2)	&	\textbf{3.69}	&	PQ	&IV		\\
HD\,37061	&	427393920	&	0.046	&	1.0	&	0.9124(4)	&	0.93(1)	&	5.88	&	rot	& *		\\
HD\,37742	&	11360636$^{\dagger}$	&	0.046	&	1.0	&	0.280(2)	&	4.4(4)	&	\textbf{2.69}	&	SLF	&		\\
HD\,57682	&	187458882	&	0.041	&	1.0	&	0.2870(8)	&	0.40(1)	&	\textbf{3.13}	&	SLF	&		\\
HD\,54879	&	177860391	&	0.041	&	1.0	&	1.002(1)	&	0.112(5)	&	\textbf{4.18}	&	SLF?	&		\\
HD\,37479	&	11286209	&	0.046	&	1.0	&	1.6789(9)	&	6.5(2)	&	8.11	&	rot	&	*	\\

&&&&&&&& \\
\multicolumn{8}{c}{Oe and Be stars} \\
HD\,39680	&	91791971	&	0.046	&	1.0	&	1.687(2)	&	1.5(1)	&	\textbf{3.41}	&	SLF+SPB?	&		\\
HD\,45314	&	438306275	&	0.046	&	1.0	&	0.287(2)	&	1.7(1)	&	\textbf{3.34}	&	SLF+SPB?	&		\\
HD\,58978	&	139385056	&	0.041	&	1.0	&	8.991(1)	&	2.6(2)	&	5.92	&	$\beta$~Cep	&	*	\\

\hline
\end{tabular}
\begin{tablenotes}
\item \textbf{Col.~2:} $\dagger$ indicates that the light curve was extracted using our own method.
\item \textbf{Col.~7:} numbers marked in bold indicate stars for which the dominant frequency is below the significance level.
\item \textbf{Col.~8:} Var. Type has same meaning as in Table~\ref{table:Star_List}.

\end{tablenotes}

\end{threeparttable}

\end{table*}

\begin{table*}
\centering 

\caption{Detailed frequency analysis of multiperiodic pulsators. Errors on the last value are given in the brackets.}\label{table:multi_freq_anal}  

\begin{threeparttable}

\begin{tabular}{c c c c c c c} 
\hline
HD & Freq.& $\nu$  & $A$  & $\phi$  &  S/N&  Notes\\
&		 \# 	&[d$^{-1}$]		&[mmag]		&	[-0.5, 0.5]	& &		\\
\hline
&		&		&		&		&		&		\\
\multicolumn{7}{c}{$\beta$~Cep stars} \\
HD\,44743	&	$\nu_{1}$	&	3.9784(1)	&	5.86(3)	&	-0.147(5)	&	25.8	&		\\
	&	$\nu_{2}$	&	4.1832(5)	&	0.99(2)	&	0.26(2)	&	12.1	&		\\
	&	$\nu_{3}$	&	0.2986(6)	&	0.83(2)	&	0.39(2)	&	6.3	&		\\
	&	$\nu_{4}$	&	5.046(2)	&	0.28(2)	&	0.32(7)	&	6.3	&		\\
	&	$\nu_{5}$	&	7.962(2)	&	0.24(2)	&	-0.43(8)	&	6.3	&	$2\nu_{1}$	\\
	&		&		&		&		&		&		\\
HD\,46328	&	$\nu_{1}$	&	4.7715(1)	&	12.91(8)	&	0.373(6)	&	30.6	&		\\
	&	$\nu_{2}$	&	9.5430(1)	&	1.808(9)	&	0.023(5)	&	30.4	&	$2\nu_{1}$	\\
	&	$\nu_{3}$	&	14.3144(4)	&	0.358(5)	&	-0.14(1)	&	27.4	&	$3\nu_{1}$	\\
	&	$\nu_{4}$	&	0.2166(9)	&	0.118(4)	&	0.10(4)	&	6.1	&		\\
	&	$\nu_{5}$	&	19.085(1)	&	0.080(5)	&	-0.31(6)	&	6.1	&	$4\nu_{1}$	\\
	&	$\nu_{6}$	&	0.414(1)	&	0.076(4)	&	-0.19(5)	&	5.1	&		\\
	&	$\nu_{7}$	&	23.860(7)	&	0.017(5)	&	-0.5(3)	&	6.1	&	$5\nu_{1}$	\\
	&		&		&		&		&		&		\\
HD\,50707	&	$\nu_{1}$	&	5.4188(2)	&	3.87(8)	&	0.24(2)	&	30.6	&	tp	\\
	&	$\nu_{2}$	&	10.8370(5)	&	1.53(6)	&	0.47(4)	&	27.8	&	$2\nu_{1}$	\\
	&	$\nu_{3}$	&	5.1842(4)	&	1.42(5)	&	0.29(4)	&	16.9	&	tp	\\
	&	$\nu_{4}$	&	5.3084(5)	&	1.05(5)	&	-0.19(4)	&	14.9	&	tp	\\
	&	$\nu_{5}$	&	0.1208(5)	&	0.95(4)	&	0.30(4)	&	9.7	&		\\
	&	$\nu_{6}$	&	5.5229(7)	&	0.70(4)	&	-0.26(6)	&	9.6	&		\\
	&	$\nu_{7}$	&	0.2950(6)	&	0.62(3)	&	0.27(6)	&	7.2	&		\\
	&	$\nu_{8}$	&	21.672(9)	&	0.05(4)	&	-0.2(8)	&	9.7	&	$4\nu_{1}$	\\
	&	$\nu_{9}$	&	16.25(1)	&	0.04(4)	&	-0.2(9)	&	7.2	&	$3\nu_{1}$	\\
	&		&		&		&		&		&		\\
HD\,61068	&	$\nu_{1}$	&	6.072(1)	&	5.5(3)	&	-0.34(5)	&	19.0	&		\\
	&	$\nu_{2}$	&	5.9875(6)	&	5.4(1)	&	0.02(3)	&	21.4	&		\\
	&	$\nu_{3}$	&	6.4530(5)	&	2.06(5)	&	0.29(2)	&	18.9	&		\\
	&	$\nu_{4}$	&	0.6615(8)	&	0.68(3)	&	-0.45(4)	&	8.6	&		\\
	&	$\nu_{5}$	&	5.922(1)	&	0.47(3)	&	-0.08(6)	&	8.6	&		\\
	&	$\nu_{6}$	&	0.768(1)	&	0.46(2)	&	0.37(5)	&	5.8	&		\\
	&	$\nu_{7}$	&	0.359(1)	&	0.44(2)	&	-0.35(5)	&	5.6	&		\\
	&	$\nu_{8}$	&	7.602(3)	&	0.18(3)	&	-0.3(1)	&	5.8	&		\\
	&	$\nu_{9}$	&	12.149(4)	&	0.17(3)	&	-0.4(2)	&	8.6	&	$2\nu_{1}$	\\
	&	$\nu_{10}$	&	6.240(4)	&	0.13(2)	&	-0.3(2)	&	5.8	&		\\
	&	$\nu_{11}$	&	12.010(6)	&	0.09(3)	&	-0.2(3)	&	5.8	&	$2\nu_{2}$	\\
	&		&		&		&		&		&		\\
\multicolumn{7}{c}{SPB stars} \\
HD\,35912	&	$\nu_{1}$	&	0.835(1)	&	4.0(2)	&	-0.15(5)	&	14.1	&	mp?	\\
	&	$\nu_{2}$	&	0.723(1)	&	3.1(2)	&	0.39(5)	&	12.7	&	mp?	\\
	&	$\nu_{3}$	&	0.645(1)	&	1.73(9)	&	0.03(5)	&	8.6	&	mp?	\\
	&	$\nu_{4}$	&	1.146(1)	&	1.33(7)	&	0.36(5)	&	7.1	&		\\
	&	$\nu_{5}$	&	0.998(2)	&	1.12(7)	&	-0.16(6)	&	7.7	&		\\
	&	$\nu_{6}$	&	0.904(1)	&	1.11(5)	&	-0.06(4)	&	7.9	&	mp?	\\
	&	$\nu_{7}$	&	0.574(1)	&	0.84(4)	&	-0.04(5)	&	6.4	&	mp?	\\
	&	$\nu_{8}$	&	0.442(2)	&	0.52(4)	&	-0.20(7)	&	5.7	&		\\
	&	$\nu_{9}$	&	4.345(6)	&	0.20(5)	&	-0.1(2)	&	5.8	&		\\
	&		&		&		&		&		&		\\
HD\,26912	&	$\nu_{1}$	&	0.500(3)	&	3.4(4)	&	-0.5(1)	&	15.4	&		\\
	&	$\nu_{2}$	&	0.241(2)	&	3.3(3)	&	0.31(8)	&	15.2	&		\\
	&	$\nu_{3}$	&	0.566(1)	&	3.0(2)	&	-0.02(7)	&	15.5	&		\\
	&	$\nu_{4}$	&	0.093(1)	&	1.21(7)	&	-0.46(6)	&	10.2	&		\\
	&	$\nu_{5}$	&	0.427(1)	&	0.95(6)	&	-0.14(6)	&	10.1	&		\\
	&	$\nu_{6}$	&	0.162(1)	&	0.84(4)	&	-0.19(5)	&	9.4	&		\\
	&	$\nu_{7}$	&	0.683(1)	&	0.57(3)	&	0.15(6)	&	7.3	&		\\
	&	$\nu_{8}$	&	0.351(1)	&	0.49(3)	&	0.06(5)	&	7.1	&		\\

	&		&		&		&		&		&		\\
\hline
\end{tabular}
\begin{tablenotes}
\item \textit{Table continued below.}
\end{tablenotes}

\end{threeparttable}
\end{table*}

\begin{table*}
\centering 
\ContinuedFloat
\caption{Continued.}\label{table:multi_freq_anal}  

\begin{threeparttable}

\begin{tabular}{c c c c c c c} 
\hline
HD & Freq.  & $\nu$  & $A$ & $\phi$ &  S/N&  Notes\\
&		 \# 	&[d$^{-1}$]		&[mmag]		&	[-0.5, 0.5]	& &		\\
\hline
&		&		&		&		&		&		\\
\multicolumn{7}{c}{SPB stars (continued)} \\
HD\,26912	&	$\nu_{9}$	&	0.937(2)	&	0.39(3)	&	0.49(8)	&	7.1	&		\\
	&	$\nu_{10}$	&	1.118(1)	&	0.34(2)	&	-0.11(7)	&	6.4	&		\\
	&	$\nu_{11}$	&	0.9970(10)	&	0.31(1)	&	0.46(5)	&	6.0	&	$2\nu_{1}$	\\
	&	$\nu_{12}$	&	0.842(2)	&	0.26(2)	&	-0.12(8)	&	6.7	&		\\
	&		&		&		&		&		&		\\
HD\,57539	&	$\nu_{1}$	&	1.6518(7)	&	3.2(1)	&	-0.27(3)	&	16.2	&		\\
	&	$\nu_{2}$	&	1.5213(7)	&	2.10(7)	&	-0.24(3)	&	13.5	&		\\
	&	$\nu_{3}$	&	0.8823(7)	&	1.35(4)	&	-0.04(3)	&	9.6	&		\\
	&	$\nu_{4}$	&	1.7363(8)	&	0.73(3)	&	0.10(4)	&	7.1	&		\\
	&	$\nu_{5}$	&	0.1729(8)	&	0.65(2)	&	0.37(3)	&	6.8	&		\\
	&	$\nu_{6}$	&	0.7750(9)	&	0.50(2)	&	0.09(4)	&	5.7	&		\\
	&	$\nu_{7}$	&	0.298(1)	&	0.36(2)	&	0.46(5)	&	5.0	&		\\
	&	$\nu_{8}$	&	3.059(2)	&	0.28(2)	&	-0.48(8)	&	5.5	&	$2\nu_{2}$	\\
	&		&		&		&		&		&		\\
HD\,48977	&	$\nu_{1}$	&	0.516(2)	&	2.8(2)	&	0.01(6)	&	12.0	&		\\
	&	$\nu_{2}$	&	0.798(1)	&	1.67(10)	&	-0.39(6)	&	12.0	&		\\
	&	$\nu_{3}$	&	0.687(1)	&	1.58(7)	&	-0.18(4)	&	10.5	&		\\
	&	$\nu_{4}$	&	1.265(1)	&	1.01(5)	&	-0.13(5)	&	7.5	&		\\
	&	$\nu_{5}$	&	0.586(1)	&	0.95(4)	&	0.17(5)	&	6.7	&		\\
	&	$\nu_{6}$	&	0.991(1)	&	0.62(3)	&	0.23(5)	&	5.9	&		\\
	&	$\nu_{7}$	&	0.282(1)	&	0.57(3)	&	-0.17(5)	&	6.2	&	$\nu_{2}-\nu_{1}$	\\
	&	$\nu_{8}$	&	0.3516(10)	&	0.44(2)	&	-0.23(4)	&	5.3	&		\\
	&	$\nu_{9}$	&	0.200(1)	&	0.42(2)	&	-0.28(5)	&	5.7	&		\\
	&	$\nu_{10}$	&	1.923(2)	&	0.33(2)	&	0.35(6)	&	5.0	&		\\
	&		&		&		&		&		&		\\
HD\,37711	&	$\nu_{1}$	&	0.4503(7)	&	3.8(1)	&	0.27(3)	&	18.6	&		\\
	&	$\nu_{2}$	&	0.245(1)	&	2.01(9)	&	-0.39(4)	&	12.2	&		\\
	&	$\nu_{3}$	&	0.717(1)	&	1.12(5)	&	0.10(4)	&	6.9	&		\\
	&	$\nu_{4}$	&	0.643(1)	&	0.96(4)	&	-0.09(4)	&	7.5	&		\\
	&	$\nu_{5}$	&	0.0846(8)	&	0.91(3)	&	-0.15(3)	&	7.1	&		\\
	&	$\nu_{6}$	&	1.324(2)	&	0.71(5)	&	0.45(8)	&	6.2	&	$3\nu_{1}$	\\
	&	$\nu_{7}$	&	1.539(1)	&	0.51(3)	&	0.14(5)	&	7.1	&		\\
	&	$\nu_{8}$	&	1.005(1)	&	0.28(1)	&	0.21(5)	&	5.4	&		\\
	&	$\nu_{9}$	&	0.8178(10)	&	0.218(8)	&	0.22(4)	&	5.1	&		\\
	&		&		&		&		&		&		\\
HD\,41753	&	$\nu_{1}$	&	0.6600(8)	&	5.3(2)	&	0.25(3)	&	23.8	&		\\
	&	$\nu_{2}$	&	0.7297(3)	&	3.25(4)	&	-0.16(1)	&	17.8	&		\\
	&	$\nu_{3}$	&	0.1189(9)	&	0.79(3)	&	0.47(4)	&	7.9	&		\\
	&	$\nu_{4}$	&	0.515(1)	&	0.58(3)	&	0.04(4)	&	7.4	&		\\
	&	$\nu_{5}$	&	0.967(1)	&	0.47(2)	&	-0.21(4)	&	6.3	&		\\
	&	$\nu_{6}$	&	1.401(1)	&	0.42(2)	&	-0.09(4)	&	6.2	&	$\nu_{1}+\nu_{2}$	\\
	&	$\nu_{7}$	&	0.2946(10)	&	0.39(1)	&	0.01(4)	&	6.4	&		\\
	&		&		&		&		&		&		\\
HD\,224990	&	$\nu_{1}$	&	0.5733(3)	&	0.504(7)	&	-0.15(1)	&	19.7	&		\\
	&	$\nu_{2}$	&	1.1712(5)	&	0.174(5)	&	-0.33(3)	&	9.9	&		\\
	&	$\nu_{3}$	&	0.9027(5)	&	0.163(4)	&	-0.41(2)	&	9.8	&		\\
	&	$\nu_{4}$	&	0.754(1)	&	0.053(3)	&	-0.14(6)	&	5.1	&		\\
	&		&		&		&		&		&		\\
\multicolumn{7}{c}{Hybrid stars} \\
HD\,37303	&	$\nu_{1}$	&	11.7889(8)	&	0.35(1)	&	0.01(3)	&	15.3	&		\\
	&	$\nu_{2}$	&	2.1388(6)	&	0.338(8)	&	-0.15(2)	&	11.5	&		\\
	&	$\nu_{3}$	&	19.292(1)	&	0.28(1)	&	0.18(5)	&	15.3	&	$2\nu_{1}-2\nu_{2}$	\\
	&	$\nu_{4}$	&	1.8722(8)	&	0.211(7)	&	-0.27(3)	&	8.6	&		\\
	&	$\nu_{5}$	&	13.433(1)	&	0.20(1)	&	-0.15(5)	&	11.5	&		\\
	&	$\nu_{6}$	&	1.625(1)	&	0.117(6)	&	-0.02(5)	&	5.7	&		\\
	&	$\nu_{7}$	&	4.256(2)	&	0.099(6)	&	0.25(6)	&	5.3	&	$2\nu_{2}$	\\

	&		&		&		&		&		&		\\
\hline
\end{tabular}
\begin{tablenotes}
\item \textit{Table continued below.}
\end{tablenotes}

\end{threeparttable}
\end{table*}

\begin{table*}
\centering 
\ContinuedFloat
\caption{Continued.}\label{table:multi_freq_anal}  

\begin{threeparttable}

\begin{tabular}{c c c c c c c} 
\hline
HD & Freq.& $\nu$  & $A$  & $\phi$ &  S/N&  Notes\\ 
&		 \# 	&[d$^{-1}$]		&[mmag]		&	[-0.5, 0.5]	& &		\\
\hline
&		&		&		&		&		&		\\
\multicolumn{7}{c}{Hybrid stars (continued)} \\
HD\,37303	&	$\nu_{8}$	&	14.980(2)	&	0.072(6)	&	-0.46(9)	&	5.7	&		\\
	&	$\nu_{9}$	&	34.271(3)	&	0.070(8)	&	0.1(1)	&	8.6	&		\\
	&	$\nu_{10}$	&	23.631(3)	&	0.049(6)	&	-0.3(1)	&	5.7	&		\\
	&	$\nu_{11}$	&	26.309(3)	&	0.047(6)	&	-0.1(1)	&	5.7	&		\\
	&		&		&		&		&		&		\\
HD\,37481	&	$\nu_{1}$	&	1.1052(8)	&	0.134(4)	&	0.07(3)	&	8.7	&	mp1?	\\
	&	$\nu_{2}$	&	1.260(1)	&	0.094(4)	&	-0.19(4)	&	7.0	&	mp1?	\\
	&	$\nu_{3}$	&	0.563(1)	&	0.077(3)	&	-0.30(5)	&	7.0	&	$0.5\nu_{1}$	\\
	&	$\nu_{4}$	&	1.432(1)	&	0.066(3)	&	0.48(5)	&	5.9	&	mp1?	\\
	&	$\nu_{5}$	&	8.346(2)	&	0.065(5)	&	0.17(7)	&	8.7	&	mp2?	\\
	&	$\nu_{6}$	&	5.632(2)	&	0.064(5)	&	-0.12(7)	&	8.7	&		\\
	&	$\nu_{7}$	&	0.815(1)	&	0.056(3)	&	-0.29(5)	&	5.4	&		\\
	&	$\nu_{8}$	&	8.901(2)	&	0.055(4)	&	-0.25(8)	&	8.8	&		\\
	&	$\nu_{9}$	&	6.651(2)	&	0.054(4)	&	-0.16(8)	&	8.8	&		\\
	&	$\nu_{10}$	&	7.661(3)	&	0.037(4)	&	0.47(10)	&	7.0	&		\\
	&	$\nu_{11}$	&	8.246(2)	&	0.034(3)	&	-0.36(10)	&	5.8	&	mp2?	\\
	&	$\nu_{12}$	&	8.125(3)	&	0.025(3)	&	0.3(1)	&	5.4	&	mp2?	\\
	&		&		&		&		&		&		\\
HD\,37209	&	$\nu_{1}$	&	0.2709(9)	&	0.117(4)	&	0.23(3)	&	11.0	&		\\
	&	$\nu_{2}$	&	6.488(1)	&	0.068(4)	&	0.17(5)	&	6.6	&		\\
	&	$\nu_{3}$	&	8.031(1)	&	0.068(4)	&	0.06(5)	&	6.3	&		\\
	&	$\nu_{4}$	&	7.827(1)	&	0.065(4)	&	0.08(6)	&	6.2	&		\\
	&	$\nu_{5}$	&	7.098(1)	&	0.061(3)	&	0.16(6)	&	6.9	&		\\
	&	$\nu_{6}$	&	0.583(1)	&	0.055(3)	&	-0.36(5)	&	6.2	&	$2\nu_{1}$	\\
	&	$\nu_{7}$	&	7.517(2)	&	0.053(3)	&	0.48(6)	&	6.9	&		\\
	&	$\nu_{8}$	&	1.717(2)	&	0.043(3)	&	-0.10(7)	&	6.1	&		\\
	&	$\nu_{9}$	&	0.478(2)	&	0.043(3)	&	-0.35(6)	&	5.4	&		\\
	&	$\nu_{10}$	&	0.159(2)	&	0.043(3)	&	0.47(7)	&	5.7	&		\\
	&	$\nu_{11}$	&	8.126(2)	&	0.037(3)	&	-0.30(9)	&	6.9	&		\\
	&	$\nu_{12}$	&	7.686(2)	&	0.037(3)	&	-0.19(9)	&	6.9	&		\\
	&	$\nu_{13}$	&	6.779(3)	&	0.031(3)	&	0.3(1)	&	6.9	&	$\nu_{1}+\nu_{2}$	\\
	&	$\nu_{14}$	&	6.901(3)	&	0.028(3)	&	0.06(10)	&	5.3	&		\\
	&	$\nu_{15}$	&	8.637(3)	&	0.024(3)	&	0.4(1)	&	5.3	&		\\
	&	$\nu_{16}$	&	8.900(3)	&	0.024(3)	&	0.0(1)	&	5.3	&		\\
	&	$\nu_{17}$	&	6.168(3)	&	0.023(3)	&	0.1(1)	&	5.3	&		\\
	&		&		&		&		&		&		\\
\multicolumn{7}{c}{Be/Oe stars} \\
HD\,58978	&	$\nu_{1}$	&	8.991(1)	&	2.6(2)	&	-0.30(6)	&	5.9	&		\\
	&	$\nu_{2}$	&	8.5139(9)	&	2.4(1)	&	-0.26(4)	&	8.6	&		\\
	&	$\nu_{3}$	&	3.691(5)	&	1.0(2)	&	0.4(2)	&	5.9	&		\\
	&	$\nu_{4}$	&	13.395(7)	&	0.6(2)	&	0.0(3)	&	5.9	&		\\
	&	$\nu_{5}$	&	10.82(1)	&	0.3(2)	&	0.5(6)	&	5.9	&		\\
	&		&		&		&		&		&		\\
\multicolumn{7}{c}{Magnetic stars} \\
HD\,37061	&	$\nu_{1}$	&	0.9124(4)	&	0.93(1)	&	-0.19(2)	&	5.9	&		\\
	&	$\nu_{2}$	&	0.7203(5)	&	0.51(1)	&	0.15(2)	&	5.0	&		\\
	&		&		&		&		&		&		\\
HD\,37479	&	$\nu_{1}$	&	1.6789(9)	&	6.5(2)	&	-0.49(4)	&	8.1	&	$2\nu_{2}$	\\
	&	$\nu_{2}$	&	0.8399(7)	&	5.2(1)	&	0.21(3)	&	8.1	&		\\
	&	$\nu_{3}$	&	2.5195(4)	&	4.32(7)	&	0.39(2)	&	8.0	&	$3\nu_{2}$	\\
	&	$\nu_{4}$	&	4.199(3)	&	3.0(3)	&	-0.4(1)	&	8.1	&	$5\nu_{2}$	\\
	&	$\nu_{5}$	&	3.3585(1)	&	2.89(2)	&	-0.307(6)	&	8.5	&	$4\nu_{2}$	\\
	&	$\nu_{6}$	&	5.878(3)	&	1.2(1)	&	-0.2(1)	&	8.0	&	$7\nu_{2}$	\\
	&	$\nu_{7}$	&	5.038(6)	&	1.2(3)	&	-0.2(2)	&	8.1	&	$6\nu_{2}$	\\
	&	$\nu_{8}$	&	7.557(8)	&	0.5(2)	&	-0.1(3)	&	8.0	&	$9\nu_{2}$	\\
	&		&		&		&		&		&		\\

\hline
\end{tabular}
\begin{tablenotes}
\item \textit{Table continued below.}
\end{tablenotes}

\end{threeparttable}
\end{table*}

\begin{table*}
\centering 
\ContinuedFloat
\caption{Continued.}\label{table:multi_freq_anal}  

\begin{threeparttable}

\begin{tabular}{c c c c c c c} 
\hline
HD & Freq.& $\nu$ & $A$ & $\phi$ &  S/N&  Notes\\
&		 \# 	&[d$^{-1}$]		&[mmag]		&	[-0.5, 0.5]	& &		\\
\hline
&		&		&		&		&		&		\\
\multicolumn{7}{c}{Eclipsing binaries} \\
HD\,159176	&	$\nu_{1}$	&	0.5929(5)	&	7.0(2)	&	0.06(2)	&	15.4	&		\\
	&	$\nu_{2}$	&	1.1200(6)	&	3.7(1)	&	-0.02(3)	&	11.4	&	$2\nu_{\rm orb}$	\\
	&	$\nu_{3}$	&	2.2413(7)	&	2.60(8)	&	-0.31(3)	&	9.4	&	$4\nu_{\rm orb}$	\\
	&	$\nu_{4}$	&	3.3612(7)	&	1.95(6)	&	0.41(3)	&	9.8	&	$6\nu_{\rm orb}$	\\
	&	$\nu_{5}$	&	0.1251(6)	&	1.16(3)	&	-0.36(3)	&	5.0	&		\\
	&	$\nu_{6}$	&	4.483(1)	&	1.11(6)	&	0.11(5)	&	8.2	&	$8\nu_{\rm orb}$	\\
	&	$\nu_{7}$	&	2.8019(10)	&	1.05(5)	&	-0.11(4)	&	7.3	&	$5\nu_{\rm orb}$	\\
	&	$\nu_{8}$	&	1.6810(9)	&	0.98(4)	&	0.19(4)	&	6.8	&	$3\nu_{\rm orb}$	\\
	&	$\nu_{9}$	&	3.923(1)	&	0.88(5)	&	-0.42(6)	&	6.3	&	$7\nu_{\rm orb}$	\\
	&	$\nu_{10}$	&	5.045(2)	&	0.67(5)	&	0.29(8)	&	6.9	&	$9\nu_{\rm orb}$	\\
	&	$\nu_{11}$	&	5.604(2)	&	0.49(5)	&	-0.2(1)	&	7.3	&	$10\nu_{\rm orb}$	\\
	&	$\nu_{12}$	&	6.165(3)	&	0.31(5)	&	0.0(1)	&	6.8	&	$11\nu_{\rm orb}$	\\
	&	$\nu_{13}$	&	6.724(4)	&	0.21(4)	&	0.4(2)	&	6.0	&	$12\nu_{\rm orb}$	\\
	&		&		&		&		&		&		\\
HD\,46485	&	$\nu_{1}$	&	0.1453(4)	&	8.9(1)	&	-0.43(2)	&	15.4	&	$\nu_{\rm orb}$	\\
	&	$\nu_{2}$	&	1.869(1)	&	1.18(6)	&	-0.45(5)	&	5.0	&		\\
	&	$\nu_{3}$	&	0.436(1)	&	1.11(5)	&	0.22(5)	&	5.3	&	$3\nu_{\rm orb}$	\\
	&		&		&		&		&		&		\\
HD\,23466	&	$\nu_{1}$	&	0.405(1)	&	6.2(4)	&	-0.22(6)	&	13.7	&		\\
	&	$\nu_{2}$	&	0.466(2)	&	2.6(2)	&	0.46(9)	&	9.0	&		\\
	&	$\nu_{3}$	&	0.825(3)	&	2.6(3)	&	0.0(1)	&	9.4	&	$2\nu_{\rm orb}$	\\
	&	$\nu_{4}$	&	2.889(4)	&	2.0(4)	&	-0.4(2)	&	8.8	&	$7\nu_{\rm orb}$	\\
	&	$\nu_{5}$	&	3.713(5)	&	1.8(4)	&	0.4(2)	&	7.6	&	$9\nu_{\rm orb}$	\\
	&	$\nu_{6}$	&	0.139(2)	&	1.7(1)	&	0.30(8)	&	9.4	&		\\
	&	$\nu_{7}$	&	1.650(2)	&	1.6(2)	&	-0.28(10)	&	8.3	&	$4\nu_{\rm orb}$	\\
	&	$\nu_{8}$	&	0.884(1)	&	1.62(8)	&	-0.21(5)	&	10.4	&		\\
	&	$\nu_{9}$	&	2.063(2)	&	1.6(2)	&	-0.1(1)	&	6.7	&	$5\nu_{\rm orb}$	\\
	&	$\nu_{10}$	&	1.235(2)	&	1.5(1)	&	0.19(9)	&	8.9	&	$3\nu_{\rm orb}$	\\
	&	$\nu_{11}$	&	4.536(5)	&	1.4(4)	&	0.1(3)	&	8.0	&	$11\nu_{\rm orb}$	\\
	&	$\nu_{12}$	&	0.688(1)	&	1.38(7)	&	-0.03(5)	&	8.0	&		\\
	&	$\nu_{13}$	&	2.477(3)	&	1.3(2)	&	0.4(2)	&	7.2	&	$6\nu_{\rm orb}$	\\
	&	$\nu_{14}$	&	5.362(8)	&	1.0(4)	&	-0.2(4)	&	8.4	&	$13\nu_{\rm orb}$	\\
	&	$\nu_{15}$	&	3.297(5)	&	0.9(2)	&	0.1(2)	&	7.9	&	$8\nu_{\rm orb}$	\\
	&	$\nu_{16}$	&	4.950(8)	&	0.8(3)	&	0.4(4)	&	8.5	&	$12\nu_{\rm orb}$	\\
	&	$\nu_{17}$	&	5.77(1)	&	0.8(4)	&	0.1(5)	&	8.4	&	$14\nu_{\rm orb}$	\\
	&	$\nu_{18}$	&	4.127(7)	&	0.7(2)	&	-0.3(3)	&	6.4	&	$10\nu_{\rm orb}$	\\
	&	$\nu_{19}$	&	6.60(1)	&	0.6(4)	&	-0.2(6)	&	8.5	&	$16\nu_{\rm orb}$	\\
	&	$\nu_{20}$	&	6.19(1)	&	0.6(4)	&	-0.5(7)	&	8.5	&	$15\nu_{\rm orb}$	\\
	&	$\nu_{18}$	&	0.960(1)	&	0.59(4)	&	-0.37(6)	&	5.6	&		\\
	&	$\nu_{19}$	&	0.766(1)	&	0.53(3)	&	0.15(7)	&	5.2	&		\\
	&	$\nu_{20}$	&	0.2618(10)	&	0.47(2)	&	0.40(5)	&	5.2	&		\\
	&	$\nu_{21}$	&	1.935(2)	&	0.46(4)	&	-0.04(9)	&	5.3	&		\\
	&	$\nu_{22}$	&	0.077(1)	&	0.40(3)	&	-0.06(6)	&	5.1	&		\\
	&	$\nu_{23}$	&	7.43(2)	&	0.4(4)	&	-0.5(10)	&	8.5	&	$18\nu_{\rm orb}$	\\
	&	$\nu_{24}$	&	7.01(2)	&	0.4(4)	&	0(1)	&	8.5	&	$17\nu_{\rm orb}$	\\
	&	$\nu_{25}$	&	1.334(2)	&	0.33(2)	&	0.24(7)	&	5.0	&		\\
	&	$\nu_{26}$	&	3.097(3)	&	0.30(4)	&	-0.3(1)	&	5.5	&		\\
	&	$\nu_{27}$	&	7.84(3)	&	0.3(4)	&	0(1)	&	8.5	&	$19\nu_{\rm orb}$	\\
	&	$\nu_{28}$	&	11.96(4)	&	0.2(4)	&	0(2)	&	7.6	&	$29\nu_{\rm orb}$	\\
	&	$\nu_{29}$	&	11.14(4)	&	0.2(4)	&	0(2)	&	7.6	&	$27\nu_{\rm orb}$	\\
	&	$\nu_{30}$	&	8.66(4)	&	0.2(3)	&	0(2)	&	9.4	&	$21\nu_{\rm orb}$	\\
	&	$\nu_{31}$	&	12.79(5)	&	0.2(4)	&	0(2)	&	7.6	&	$31\nu_{\rm orb}$	\\
	&	$\nu_{32}$	&	10.31(5)	&	0.2(4)	&	0(2)	&	9.4	&	$25\nu_{\rm orb}$	\\
	&	$\nu_{33}$	&	8.25(5)	&	0.2(3)	&	0(2)	&	9.4	&	$20\nu_{\rm orb}$	\\
	&	$\nu_{34}$	&	13.20(7)	&	0.1(4)	&	0(3)	&	7.6	&	$32\nu_{\rm orb}$	\\
	&	$\nu_{35}$	&	14.02(7)	&	0.1(4)	&	0(3)	&	7.6	&	$34\nu_{\rm orb}$	\\

	&		&		&		&		&		&		\\

\hline
\end{tabular}
\begin{tablenotes}
\item \textit{Table continued below.}
\end{tablenotes}

\end{threeparttable}
\end{table*}

\begin{table*}
\centering 
\ContinuedFloat
\caption{Continued.}\label{table:multi_freq_anal}  

\begin{threeparttable}

\begin{tabular}{c c c c c c c} 
\hline
HD & Freq.& $\nu$ & $A$  & $\phi$ &  S/N&  Notes\\ 
&		 \# 	&[d$^{-1}$]		&[mmag]		&	[-0.5, 0.5]	& &		\\
\hline
&		&		&		&		&		&		\\
\multicolumn{7}{c}{Eclipsing binaries (continued)} \\
HD\,23466	&	$\nu_{36}$	&	9.90(6)	&	0.1(3)	&	0(3)	&	9.4	&	$24\nu_{\rm orb}$	\\
	&	$\nu_{37}$	&	9.49(6)	&	0.1(3)	&	0(3)	&	9.4	&	$23\nu_{\rm orb}$	\\
	&	$\nu_{38}$	&	9.08(5)	&	0.1(2)	&	0(3)	&	6.4	&	$22\nu_{\rm orb}$	\\
	&	$\nu_{39}$	&	13.61(8)	&	0.1(4)	&	0(4)	&	8.4	&	$33\nu_{\rm orb}$	\\
	&	$\nu_{40}$	&	10.72(6)	&	0.1(2)	&	0(3)	&	6.4	&	$26\nu_{\rm orb}$	\\
	&	$\nu_{41}$	&	18.56(7)	&	0.1(3)	&	0(3)	&	6.4	&	$45\nu_{\rm orb}$	\\
	&	$\nu_{42}$	&	19.39(7)	&	0.1(2)	&	0(3)	&	6.4	&	$47\nu_{\rm orb}$	\\
	&	$\nu_{43}$	&	12.38(7)	&	0.1(2)	&	0(3)	&	6.4	&	$30\nu_{\rm orb}$	\\
	&	$\nu_{44}$	&	14.85(7)	&	0.1(3)	&	0(4)	&	6.4	&	$36\nu_{\rm orb}$	\\
	&	$\nu_{45}$	&	17.74(8)	&	0.1(2)	&	0(4)	&	6.4	&	$43\nu_{\rm orb}$	\\
	&	$\nu_{46}$	&	20.21(9)	&	0.1(2)	&	0(4)	&	6.4	&	$49\nu_{\rm orb}$	\\
	&	$\nu_{47}$	&	17.3(1)	&	0.0(2)	&	0(5)	&	8.3	&	$42\nu_{\rm orb}$	\\
	&	$\nu_{48}$	&	14.43(10)	&	0.0(2)	&	0(5)	&	7.2	&	$35\nu_{\rm orb}$	\\
	&	$\nu_{49}$	&	16.9(1)	&	0.0(2)	&	0(5)	&	8.3	&	$41\nu_{\rm orb}$	\\
	&	$\nu_{50}$	&	15.26(5)	&	0.03(8)	&	0(2)	&	8.0	&	$37\nu_{\rm orb}$	\\
	&	$\nu_{51}$	&	20.6(1)	&	0.0(2)	&	0(6)	&	8.3	&	$50\nu_{\rm orb}$	\\
	&	$\nu_{52}$	&	16.09(4)	&	0.03(6)	&	0(2)	&	7.3	&	$39\nu_{\rm orb}$	\\
	&	$\nu_{53}$	&	16.50(6)	&	0.03(8)	&	0(3)	&	8.0	&	$40\nu_{\rm orb}$	\\
	&	$\nu_{54}$	&	25.98(5)	&	0.03(6)	&	0(2)	&	7.3	&	$63\nu_{\rm orb}$	\\
	&	$\nu_{55}$	&	18.15(6)	&	0.03(7)	&	0(3)	&	8.3	&	$44\nu_{\rm orb}$	\\
	&	$\nu_{56}$	&	21.03(5)	&	0.03(6)	&	0(2)	&	7.3	&	$51\nu_{\rm orb}$	\\
	&	$\nu_{57}$	&	25.16(5)	&	0.02(6)	&	0(3)	&	7.3	&	$61\nu_{\rm orb}$	\\
	&	$\nu_{58}$	&	23.93(6)	&	0.02(6)	&	0(3)	&	7.3	&	$58\nu_{\rm orb}$	\\
	&	$\nu_{59}$	&	24.75(6)	&	0.02(6)	&	0(3)	&	7.3	&	$60\nu_{\rm orb}$	\\
	&	$\nu_{60}$	&	21.45(6)	&	0.02(6)	&	0(3)	&	7.3	&	$52\nu_{\rm orb}$	\\
	&	$\nu_{61}$	&	19.80(7)	&	0.02(6)	&	0(3)	&	7.3	&	$48\nu_{\rm orb}$	\\
	&		&		&		&		&		&		\\
HD\,152590	&	$\nu_{1}$	&	0.445(3)	&	25(4)	&	0.0(1)	&	6.0	&	$2\nu_{\rm orb}$	\\
	&	$\nu_{2}$	&	0.889(3)	&	20(2)	&	-0.3(1)	&	5.7	&	$4\nu_{\rm orb}$	\\
	&	$\nu_{3}$	&	1.560(3)	&	17(2)	&	-0.3(1)	&	5.6	&	$7\nu_{\rm orb}$	\\
	&	$\nu_{4}$	&	1.113(3)	&	17(2)	&	0.0(1)	&	6.8	&	$5\nu_{\rm orb}$	\\
	&	$\nu_{5}$	&	0.218(2)	&	16(2)	&	-0.3(1)	&	6.5	&		\\
	&	$\nu_{6}$	&	2.005(4)	&	16(3)	&	0.4(2)	&	5.3	&	$9\nu_{\rm orb}$	\\
	&	$\nu_{7}$	&	0.668(2)	&	16(1)	&	0.27(8)	&	7.1	&	$3\nu_{\rm orb}$	\\
	&	$\nu_{8}$	&	1.336(2)	&	14(1)	&	0.44(9)	&	7.4	&	$6\nu_{\rm orb}$	\\
	&	$\nu_{9}$	&	2.451(3)	&	14(2)	&	0.1(2)	&	6.2	&	$11\nu_{\rm orb}$	\\
	&	$\nu_{10}$	&	1.781(2)	&	10.7(9)	&	0.14(8)	&	7.1	&	$8\nu_{\rm orb}$	\\
	&	$\nu_{11}$	&	2.896(2)	&	10.4(9)	&	-0.14(9)	&	6.5	&	$13\nu_{\rm orb}$	\\
	&	$\nu_{12}$	&	2.227(1)	&	8.6(5)	&	-0.19(6)	&	8.8	&	$10\nu_{\rm orb}$	\\
	&	$\nu_{13}$	&	2.673(2)	&	7.8(8)	&	0.5(1)	&	6.9	&	$12\nu_{\rm orb}$	\\
	&	$\nu_{14}$	&	3.341(3)	&	7.1(8)	&	-0.4(1)	&	6.4	&	$15\nu_{\rm orb}$	\\
	&	$\nu_{15}$	&	3.118(2)	&	6.8(7)	&	0.20(10)	&	7.7	&	$14\nu_{\rm orb}$	\\
	&	$\nu_{16}$	&	3.564(2)	&	5.5(5)	&	-0.09(9)	&	9.0	&	$16\nu_{\rm orb}$	\\
	&	$\nu_{17}$	&	3.787(2)	&	5.0(5)	&	0.26(9)	&	9.6	&	$17\nu_{\rm orb}$	\\
	&	$\nu_{18}$	&	4.011(3)	&	3.6(4)	&	-0.4(1)	&	7.7	&	$18\nu_{\rm orb}$	\\
	&	$\nu_{19}$	&	4.232(3)	&	3.2(4)	&	0.0(1)	&	8.4	&	$19\nu_{\rm orb}$	\\
	&	$\nu_{20}$	&	0.347(2)	&	2.3(2)	&	0.44(9)	&	7.1	&		\\
	&	$\nu_{18}$	&	4.680(3)	&	1.8(2)	&	-0.4(1)	&	6.1	&	$21\nu_{\rm orb}$	\\
	&	$\nu_{19}$	&	7.127(3)	&	1.5(2)	&	0.1(1)	&	5.2	&	$32\nu_{\rm orb}$	\\
	&	$\nu_{20}$	&	6.682(3)	&	1.5(2)	&	0.4(1)	&	5.2	&	$30\nu_{\rm orb}$	\\
	&	$\nu_{21}$	&	4.457(3)	&	1.5(2)	&	0.3(2)	&	6.4	&	$20\nu_{\rm orb}$	\\
	&	$\nu_{22}$	&	6.458(3)	&	1.4(2)	&	0.0(1)	&	5.2	&	$29\nu_{\rm orb}$	\\
	&	$\nu_{23}$	&	6.904(3)	&	1.4(2)	&	-0.3(1)	&	5.2	&	$31\nu_{\rm orb}$	\\
	&	$\nu_{24}$	&	7.573(3)	&	1.4(2)	&	-0.2(1)	&	5.2	&	$34\nu_{\rm orb}$	\\
	&	$\nu_{25}$	&	7.350(3)	&	1.4(2)	&	0.4(1)	&	5.2	&	$33\nu_{\rm orb}$	\\
	
	&		&		&		&		&		&		\\

\hline
\end{tabular}
\begin{tablenotes}
\item \textit{Table continued below.}
\end{tablenotes}

\end{threeparttable}
\end{table*}
\begin{table*}
\centering 
\ContinuedFloat
\caption{Continued.}\label{table:multi_freq_anal}  

\begin{threeparttable}

\begin{tabular}{c c c c c c c} 
\hline

HD & Freq. & $\nu$  & $A$ & $\phi$ &  S/N&  Notes\\ 
&		 \# 	&[d$^{-1}$]		&[mmag]		&	[-0.5, 0.5]	& &		\\
\hline
&		&		&		&		&		&		\\
\multicolumn{7}{c}{Eclipsing binaries (continued)} \\
HD\,152590	&	$\nu_{26}$	&	6.238(3)	&	1.3(2)	&	-0.3(1)	&	5.3	&	$28\nu_{\rm orb}$	\\
	&	$\nu_{27}$	&	7.795(3)	&	1.2(2)	&	0.1(1)	&	5.2	&	$35\nu_{\rm orb}$	\\
	&	$\nu_{28}$	&	8.019(3)	&	1.2(2)	&	0.5(1)	&	5.2	&	$36\nu_{\rm orb}$	\\
	&	$\nu_{29}$	&	6.014(3)	&	1.1(2)	&	0.3(1)	&	5.2	&	$27\nu_{\rm orb}$	\\
	&	$\nu_{30}$	&	5.789(3)	&	1.1(2)	&	0.0(1)	&	5.3	&	$26\nu_{\rm orb}$	\\
	&	$\nu_{31}$	&	8.241(4)	&	1.0(2)	&	-0.2(2)	&	5.2	&	$37\nu_{\rm orb}$	\\
	&	$\nu_{32}$	&	5.123(4)	&	0.9(2)	&	0.4(2)	&	5.3	&	$23\nu_{\rm orb}$	\\
	&	$\nu_{33}$	&	8.464(4)	&	0.8(2)	&	0.2(2)	&	5.2	&	$38\nu_{\rm orb}$	\\
	&	$\nu_{34}$	&	8.688(5)	&	0.7(2)	&	-0.5(2)	&	5.2	&	$39\nu_{\rm orb}$	\\
	&	$\nu_{35}$	&	8.910(6)	&	0.6(2)	&	-0.1(3)	&	5.2	&	$40\nu_{\rm orb}$	\\
	&	$\nu_{36}$	&	5.344(6)	&	0.6(2)	&	0.3(3)	&	5.3	&	$24\nu_{\rm orb}$	\\
	&	$\nu_{37}$	&	9.133(7)	&	0.5(1)	&	0.3(3)	&	5.3	&	$42\nu_{\rm orb}$	\\
	&	$\nu_{38}$	&	9.359(8)	&	0.4(2)	&	-0.5(3)	&	5.3	&	$43\nu_{\rm orb}$	\\
	&	$\nu_{39}$	&	9.80(1)	&	0.3(2)	&	0.3(5)	&	5.3	&	$45\nu_{\rm orb}$	\\
	&	$\nu_{40}$	&	9.58(1)	&	0.3(2)	&	0.0(6)	&	5.3	&	$44\nu_{\rm orb}$	\\
	&	$\nu_{41}$	&	10.25(2)	&	0.2(2)	&	0.1(8)	&	5.3	&	$46\nu_{\rm orb}$	\\
	&		&		&		&		&		&		\\
HD\,57061	&	$\nu_{1}$	&	1.5604(8)	&	19.3(7)	&	-0.28(4)	&	20.4	&	$2\nu_{\rm orb}$	\\
	&	$\nu_{2}$	&	3.121(3)	&	5.0(6)	&	0.2(1)	&	10.8	&	$4\nu_{\rm orb}$	\\
	&	$\nu_{3}$	&	0.278(3)	&	3.6(4)	&	-0.3(1)	&	6.9	&		\\
	&	$\nu_{4}$	&	0.771(3)	&	3.0(4)	&	-0.4(1)	&	6.2	&		\\
	&	$\nu_{5}$	&	0.175(3)	&	2.2(3)	&	0.2(2)	&	5.1	&		\\
	&	$\nu_{6}$	&	4.680(7)	&	2.0(6)	&	-0.3(3)	&	6.5	&	$6\nu_{\rm orb}$	\\
	&		&		&		&		&		&		\\
HD\,36486	&	$\nu_{1}$	&	0.3494(10)	&	8.7(3)	&	-0.48(4)	&	14.0	&	$2\nu_{\rm orb}$	\\
	&	$\nu_{2}$	&	0.523(2)	&	4.6(4)	&	-0.49(9)	&	9.5	&	$3\nu_{\rm orb}$	\\
	&	$\nu_{3}$	&	0.691(1)	&	4.0(2)	&	-0.17(6)	&	10.4	&	$4\nu_{\rm orb}$	\\
	&	$\nu_{4}$	&	0.184(1)	&	3.5(2)	&	0.18(5)	&	10.0	&		\\
	&	$\nu_{5}$	&	1.220(1)	&	2.8(1)	&	0.05(5)	&	9.2	&	$7\nu_{\rm orb}$	\\
	&	$\nu_{6}$	&	0.874(1)	&	2.1(1)	&	-0.27(5)	&	7.2	&	$5\nu_{\rm orb}$	\\
	&	$\nu_{7}$	&	1.038(2)	&	1.6(1)	&	0.13(6)	&	6.1	&	$6\nu_{\rm orb}$	\\
	&	$\nu_{8}$	&	1.571(2)	&	1.10(9)	&	0.27(8)	&	5.1	&	$9\nu_{\rm orb}$	\\
	&		&		&		&		&		&		\\
HD\,57060	&	$\nu_{1}$	&	0.4553(7)	&	186(5)	&	-0.31(3)	&	31.6	&	$2\nu_{\rm orb}$	\\
	&	$\nu_{2}$	&	0.913(2)	&	35(3)	&	0.16(8)	&	16.1	&	$4\nu_{\rm orb}$	\\
	&	$\nu_{3}$	&	1.367(3)	&	16(2)	&	-0.4(2)	&	9.1	&	$6\nu_{\rm orb}$	\\
	&	$\nu_{4}$	&	0.687(4)	&	13(2)	&	-0.1(2)	&	8.7	&		\\
	&	$\nu_{5}$	&	0.106(4)	&	10(2)	&	0.1(2)	&	6.1	&		\\
	&	$\nu_{6}$	&	0.851(5)	&	8(2)	&	-0.5(2)	&	5.3	&		\\
	&	$\nu_{7}$	&	0.241(7)	&	7(2)	&	0.3(3)	&	5.7	&		\\
	&	$\nu_{8}$	&	0.310(4)	&	7(1)	&	-0.3(2)	&	5.0	&		\\
	&	$\nu_{9}$	&	1.819(5)	&	5(1)	&	0.0(2)	&	5.0	&	$8\nu_{\rm orb}$	\\
	&		&		&		&		&		&		\\
	HD\,47129	&	$\nu_{1}$	&	1.645(2)	&	18(1)	&	-0.37(8)	&	12.3	&		\\
	&	$\nu_{2}$	&	0.811(3)	&	9(1)	&	-0.2(1)	&	9.5	&		\\
	&	$\nu_{3}$	&	0.626(3)	&	7.5(10)	&	0.1(1)	&	7.3	&	9$\nu_{\rm orb}$	\\
	&	$\nu_{4}$	&	0.696(3)	&	6.1(7)	&	0.2(1)	&	5.6	&	10$\nu_{\rm orb}$ 	\\
	&	$\nu_{5}$	&	0.289(4)	&	4.9(8)	&	-0.2(2)	&	6.4	&		\\
	&	$\nu_{6}$	&	0.130(4)	&	4.3(6)	&	-0.2(1)	&	5.7	&		\\
	&	$\nu_{7}$	&	1.105(5)	&	3.1(6)	&	0.2(2)	&	5.4	&	16$\nu_{\rm orb}$	\\
	&	$\nu_{8}$	&	0.520(3)	&	2.9(3)	&	0.1(1)	&	5.1	&		\\
	&	$\nu_{9}$	&	3.293(9)	&	2.3(9)	&	-0.1(4)	&	5.5	&	$2\nu_{1}$	\\
	&		&		&		&		&		&		\\
HD\,62747	&	$\nu_{1}$	&	0.510(2)	&	12(1)	&	-0.49(9)	&	10.7	&	$2\nu_{\rm orb}$	\\
	&	$\nu_{2}$	&	0.257(2)	&	11.0(9)	&	-0.15(8)	&	11.8	&	$\nu_{\rm orb}$	\\

	&		&		&		&		&		&		\\

\hline
\end{tabular}
\begin{tablenotes}
\item \textit{Table continued below.}
\end{tablenotes}

\end{threeparttable}
\end{table*}
\begin{table*}
\centering 
\ContinuedFloat
\caption{Continued.}\label{table:multi_freq_anal}  

\begin{threeparttable}

\begin{tabular}{c c c c c c c} 
\hline
HD & Freq. & $\nu$ & $A$  & $\phi$ &  S/N&  Notes\\ 
&		 \# 	&[d$^{-1}$]		&[mmag]		&	[-0.5, 0.5]	& &		\\
\hline
&		&		&		&		&		&		\\
\multicolumn{7}{c}{Eclipsing binaries (continued)} \\
HD\,62747	&	$\nu_{3}$	&	1.018(2)	&	7.7(6)	&	-0.21(8)	&	9.4	&	$4\nu_{\rm orb}$	\\
	&	$\nu_{4}$	&	1.527(2)	&	6.0(5)	&	0.05(9)	&	8.5	&	$6\nu_{\rm orb}$	\\
	&	$\nu_{5}$	&	0.764(2)	&	5.1(4)	&	0.18(8)	&	8.8	&	$3\nu_{\rm orb}$	\\
	&	$\nu_{6}$	&	2.036(2)	&	4.7(5)	&	0.33(10)	&	8.0	&	$8\nu_{\rm orb}$	\\
	&	$\nu_{7}$	&	1.786(2)	&	3.7(3)	&	-0.24(9)	&	7.5	&	$7\nu_{\rm orb}$	\\
	&	$\nu_{8}$	&	1.273(1)	&	3.6(2)	&	0.47(7)	&	8.9	&	$5\nu_{\rm orb}$	\\
	&	$\nu_{9}$	&	2.545(2)	&	3.1(3)	&	-0.4(1)	&	8.0	&	$10\nu_{\rm orb}$	\\
	&	$\nu_{10}$	&	2.294(2)	&	2.2(2)	&	0.00(8)	&	6.4	&	$9\nu_{\rm orb}$	\\
	&	$\nu_{11}$	&	3.055(3)	&	1.8(2)	&	-0.1(1)	&	6.1	&	$12\nu_{\rm orb}$	\\
	&	$\nu_{12}$	&	2.799(2)	&	1.7(2)	&	0.33(9)	&	5.4	&	$11\nu_{\rm orb}$	\\
	&		&		&		&		&		&		\\

\hline
\end{tabular}
\begin{tablenotes}
\item \textbf{Col.~7:} $\nu_{\rm orb}$ indicates the orbital frequency, which is not always directly measured but derived from harmonics. In the case of HD\,47129, $\nu_{\rm orb}$ refers to the orbital frequency measured by \citet{Mahy2011}. See Appendix~\ref{sec:appendix_individual} for more details.
\end{tablenotes}

\end{threeparttable}
\end{table*}

\section{Comments on individual objects}\label{sec:appendix_individual}
In this section, we present the results of the frequency analysis described in Appendix~\ref{sec:appendix_freqanal} and discuss each object individually. Multi-epoch spectroscopy (if available) is consulted when relevant. References are made to spectral type designation and previous work that may be of interest to the variability classification\footnote{We also refer the reader to the various papers of the GOSSS and MONOS project \citep{Sota2011, Sota2014, Maiz2016, Maiz2019} for additional information about the O-type stars}. In referring to frequencies found in previous work we use $f$, and for frequencies derived here we use $\nu$. For each star a figure showing the TESS light curve and its LS-periodogram is provided. The figure style is similar to Figs.~\ref{fig:PG_LINE_Bcep}, \ref{fig:PG_LINE_SPB}, \ref{fig:LC+PG_hybrids}, \ref{fig:PG+lines_Oe}, \ref{fig:LC_LINE_EB}, \ref{fig:PG+lines_OV}, and \ref{fig:PG_LINE_BSG} in the main text.

\subsection{High frequency pulsators}\label{sec:appendix_Bcep}

\begin{figure*}
	\includegraphics[width=2\columnwidth, scale = 1]{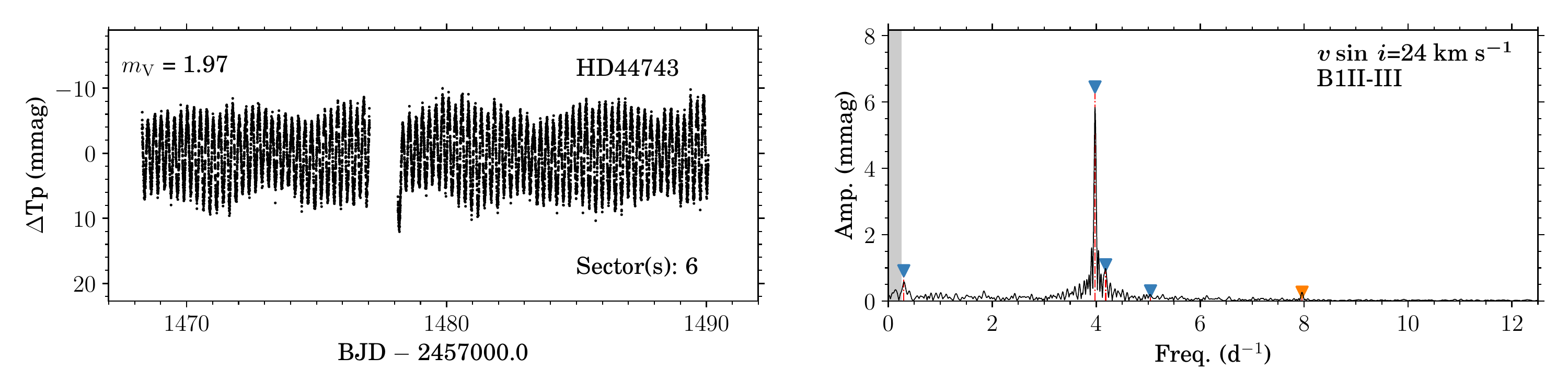}
    \caption{TESS light curve and LS-periodogram of HD\,44743. Same figure style as Fig~\ref{fig:PG_LINE_Bcep}.}
    \label{fig:LC+PG_HD44743}
\end{figure*}

HD~44743 ($\beta$~CMa, B1\,II-III, \citealp{Lesh1968}) has a weak magnetic field \citep{Hubrig2009} and \citet{Fossati2015} used this property to derive a rotation period of $P_{\rm rot}=13.6\pm1.2$~d. $\beta$~CMa has three identified pulsation modes which allowed asteroseismic modelling ($f_{1}=3.979$, $f_{2}=3.995$, and $f_{3}=4.184$~d$^{-1}$) \citep{Shobbrook1973a, Aerts1994, Shobbrook2006, Mazumdar2006}. From the one-sector TESS data set we recover two of these known pulsation modes, $\nu_{1}=3.9784(1)$ and $\nu_{2}=4.1832(5)$~d$^{-1}$. In addition, we detect two new frequencies ($\nu_{3}=0.2986(6)$ and $\nu_{4}=5.046(2)$~d$^{-1}$) and one harmonic of $\nu_{1}$.

\begin{figure*}
	\includegraphics[width=2\columnwidth, scale = 1]{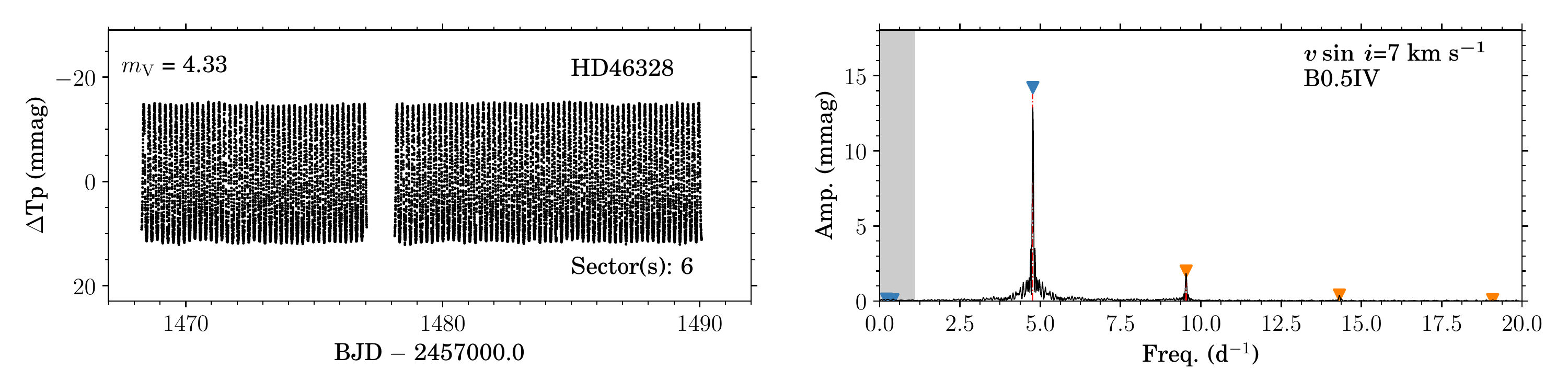}
    \caption{TESS light curve and LS-periodogram of HD\,46328. Same figure style as Fig~\ref{fig:PG_LINE_Bcep}.}
    \label{fig:LC+PG_HD46328}
\end{figure*}

HD~46328 ($\xi^{1}$~CMa, B0.5\,IV, \citealp{Lesh1968}) is a well-studied high-amplitude $\beta$~Cep star, with a single pulsation period of $P=0.209$~d \citep{McNamara1955, Shobbrook1973b, Heynderickx1992, Heynderickx1994, Saesen2006, Shultz2017, Shultz2018b, Wade2020a}. Its one-sector TESS data set was most recently studied by \citet{Wade2020a}, who recover this period again and use it to constrain the period change over the last 100~yr. They also note the possible presence of g modes. Our analysis conforms to these results ($\nu_{1}=4.7715(1)$~d$^{-1}$ or $P=0.20958(4)$~d). We also find two low amplitude frequencies ($\nu_{4}=0.2166(9)$ and $\nu_{6}=0.414(1)$~d$^{-1}$) that could be g modes.

\begin{figure*}
	\includegraphics[width=2\columnwidth, scale = 1]{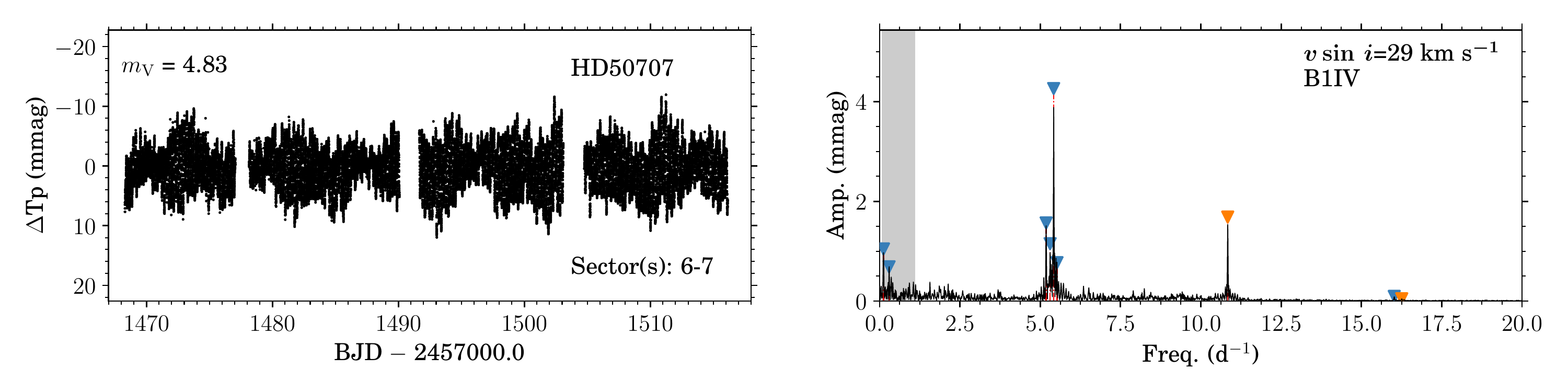}
    \caption{TESS light curve and LS-periodogram of HD\,50707. Same figure style as Fig~\ref{fig:PG_LINE_Bcep}.}
    \label{fig:LC+PG_HD50707}
\end{figure*}

HD~50707 (15~CMa, B1\,IV,  \citealp{Lesh1968}) has four identified pulsation modes ($f_{1}=5.4187(2)$, $f_{2}=5.1831(3)$, $f_{3}=5.5212(8)$ and $f_{4}=5.3085(12)$~d$^{-1}$) \citep{Shobbrook1973b, Heynderickx1992, Shobbrook2006}. It was shown by \citet{Shobbrook2006} that $f_{2}$ is a radial mode and that $f_{1}$, $f_{3}$, $f_{4}$ are part of the same non-radial multiplet. We recover these frequencies in the two-sector TESS data set. In addition, we find two low frequencies ($\nu_{5}=0.1208(5)$ and $\nu_{7}=0.2950(6)$~d$^{-1}$) which could be either g modes, or due to rotational modulation.

\begin{figure*}
	\includegraphics[width=2\columnwidth, scale = 1]{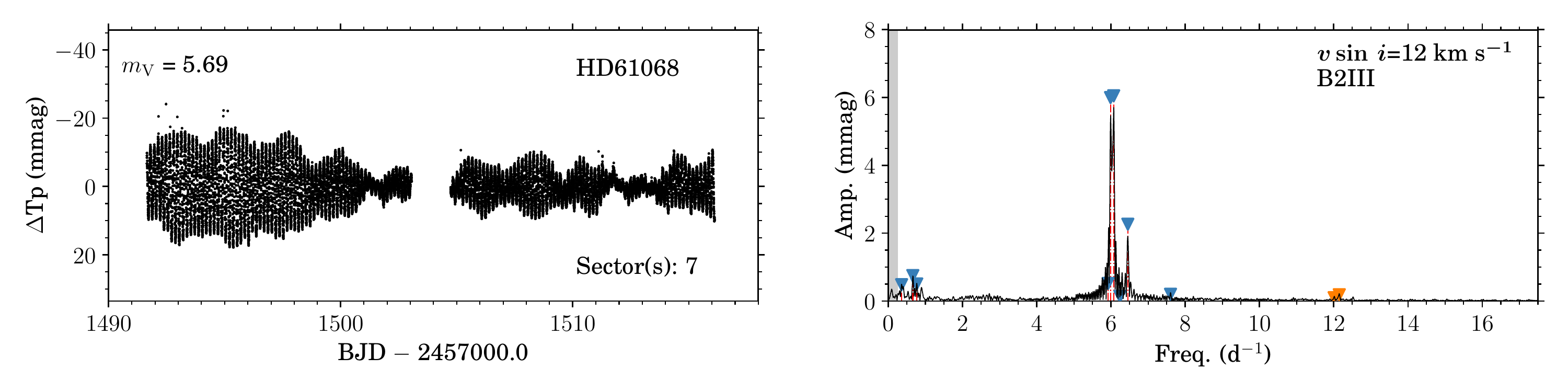}
    \caption{TESS light curve and LS-periodogram of HD\,61068. Same figure style as Fig~\ref{fig:PG_LINE_Bcep}.}
    \label{fig:LC+PG_HD61068}
\end{figure*}

HD~61068 (PT~Pup, B2\,III, \citealp{Lesh1968}) has two reported frequencies that vary in relative amplitude on long time scales \citep{Lesh1979, Shobbrook1981, Heynderickx1992, Maisonneuve2011}. We recover these two frequencies (within 1$\sigma$) in the one-sector TESS data set ($\nu_{1}=6.072(1)$ and $\nu_{2}=5.9875(6)$~d$^{-1}$). In addition, we find seven new frequencies, and harmonics of the previously reported frequencies. Three low frequencies are noted that could indicate the presence of g modes as they fall outside of the estimated rotational modulation frequency range. The dominant frequencies $\nu_{1}$ and $\nu_{2}$ are almost equal in amplitude, in contrast to previous findings. This amplitude variation may be related to beating effects given their close proximity. \citet{Heynderickx1992} and \citet{Maisonneuve2011} identified $\nu_{2}$ as a non-radial mode, either $\ell=1$ or $2$.

\begin{figure*}
	\includegraphics[width=2\columnwidth, scale = 1]{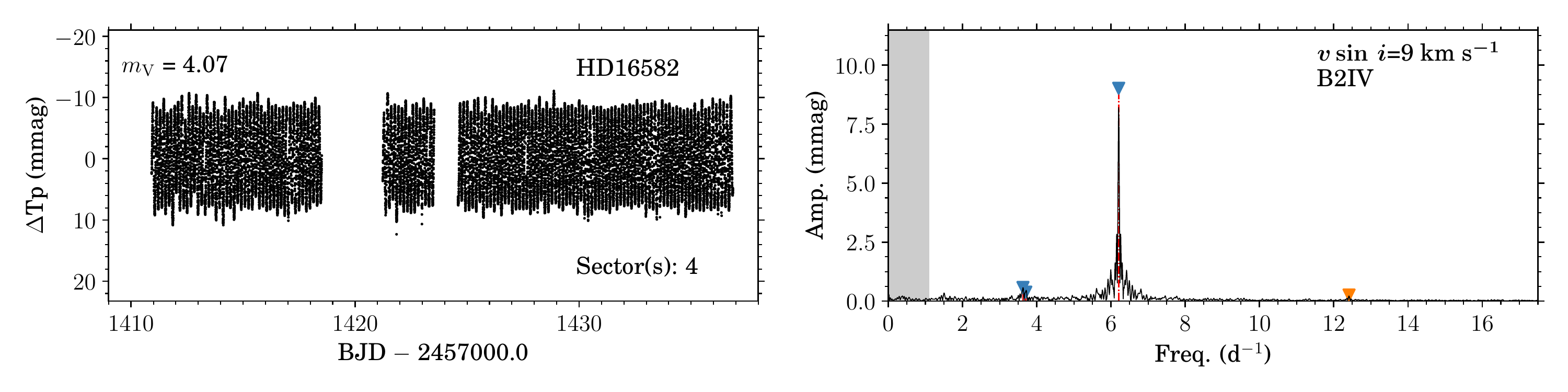}
    \caption{TESS light curve and LS-periodogram of HD\,16582. Same figure style as Fig~\ref{fig:PG_LINE_Bcep}.}
    \label{fig:LC+PG_HD16582}
\end{figure*}

HD~16582 ($\delta$~Cet, B2\,IV, \citealp{Lesh1968}) has long been considered a mono-periodic $\beta$~Cep star with $f=6.205886$~d$^{-1}$ \citep{McNamara1953, McNamara1955, Mohan1981, Peters1987, Jerzykiewicz1988, Aerts2006}. \citet{Aerts2006} detected an additional three frequencies ($f_{2}=3.737(2)$, $f_{3}=3.673(2)$ and $f_{4}=0.318(2)$~d$^{-1}$) and the harmonic of the dominant frequency, using $18.7$~d of MOST data. We extract four frequencies from the one-sector TESS data set including the large amplitude dominant frequency ($\nu_{1}=6.20598(9)$~d$^{-1}$) and one of its harmonics. The lower frequency modes are measured as $\nu_{2}=3.628(1)$~d$^{-1}$ and $\nu_{3}=3.705(2)$~d$^{-1}$. If these are the same modes as measured by \citet{Aerts2006} this indicates a significant decrease in frequency with time. The lower frequency g mode ($f_{4}$) is not significant in the TESS data. Pulsation periods of $\beta$~Cep stars are known to decrease in frequency with time due to stellar evolution (see e.g \citealt{Eggleton1973, Jerzykiewicz1999, Neilson2015}). However, the frequency changes in $\nu_{2}$ and $\nu_{3}$ (approximately $0.045$ and $0.032$~d$^{-1}$ respectively) are much larger than those predicted by stellar evolution models on the short timescale of approximately 13~yr (e.g \citealt{Eggleton1973, Neilson2015}). Conversely, we measure no difference within 3$\sigma$ between our measurement of the dominant frequency and the last measurement by \citet{Aerts2006}, which is in line with period change estimates for $\delta$~Cet \citep{Jerzykiewicz1999}. Follow-up measurements of $\nu_{2}$ and $\nu_{3}$ with high-precision photometry are necessary to verify the stability of these frequencies. 

\subsection{Low-frequency pulsators: SPB stars}\label{sec:appendix_SPB}

\begin{figure*}
	\includegraphics[width=2\columnwidth, scale = 1]{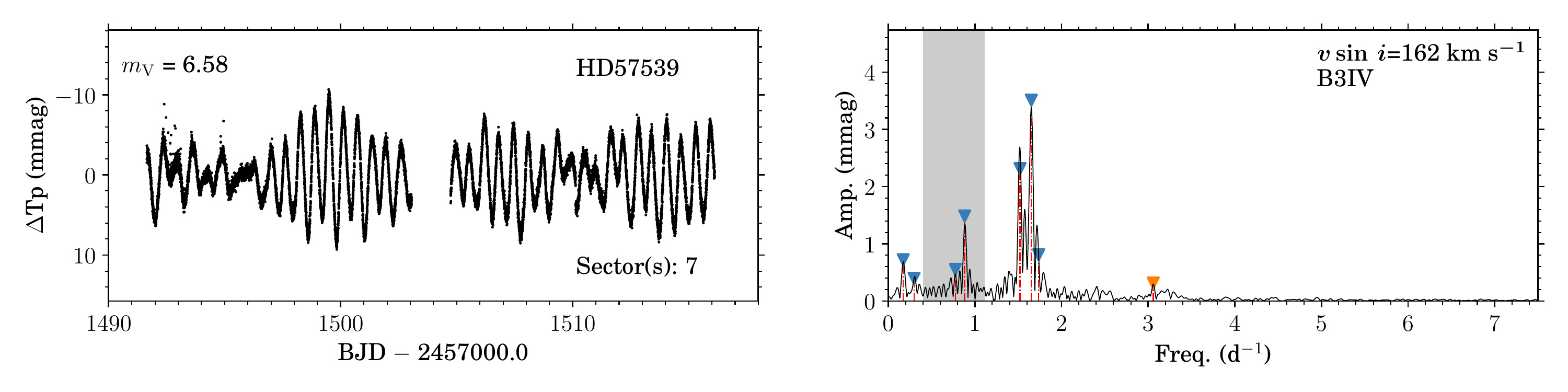}
    \caption{TESS light curve and LS-periodogram of HD\,57539. Same figure style as Fig~\ref{fig:PG_LINE_Bcep}.}
    \label{fig:LC+PG_HD57539}
\end{figure*}

HD~57539 (B3\,IV, \citealp{Houk1999}) is not known to be variable in photometry (i.e. ASAS-3, \citealp{Fremat2006, Gutierrez2007}). We extract seven frequencies from the one-sector TESS data set, including a harmonic of the second dominant frequency. The dominant frequency is surrounded by many low-S/N frequencies which cannot be resolved in a data set of this length. Two frequencies ($\nu_{3}=0.8823(7)$ and $\nu_{6}=0.7750(9)$~d$^{-1}$) are found in the estimated rotational modulation frequency range (v$\,\sin\,i$=280~km~s$^{-1}$) and could be used to constrain the envelope rotation of HD~57539. However, this complicates the explanation of the lowest frequencies. These could be instrumental artefacts, or if they are due to rotation indicate an overestimation of the projected rotational velocity. The spectroscopic parameters of HD~57539 indicate it is evolved and thus falls outside our calculated instability strips for the MS (see also Sect.~\ref{sec:instability}). This is likely because of an extended MS due to fast rotation (see e.g. \citealt{Brott2011}), and hence an extended instability domain. All the above makes HD~57539 an interesting candidate for follow-up, and modelling of fast-rotating pulsating stars in general.

\begin{figure*}
	\includegraphics[width=2\columnwidth, scale = 1]{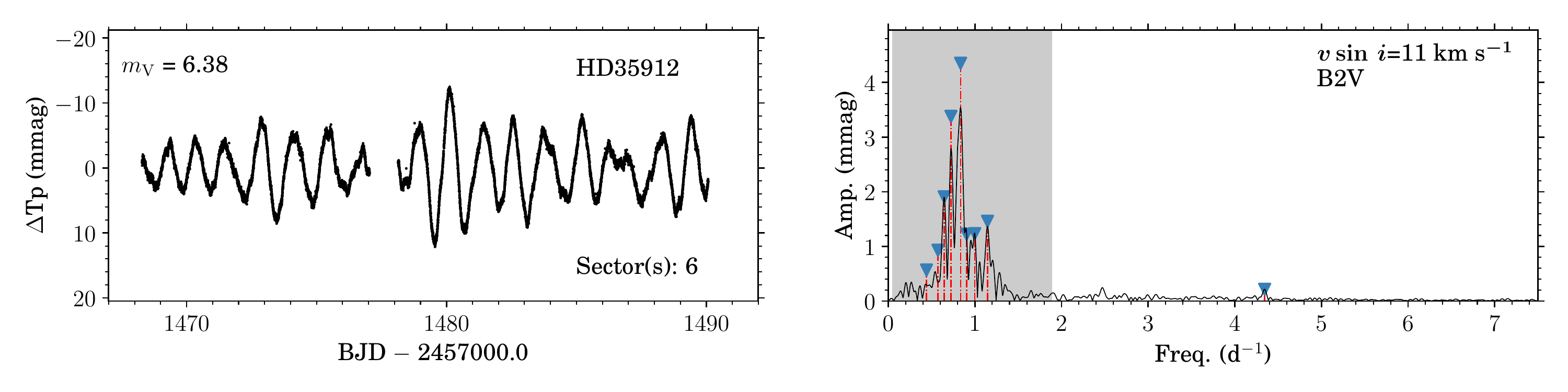}
    \caption{TESS light curve and LS-periodogram of HD\,35912. Same figure style as Fig~\ref{fig:PG_LINE_Bcep}.}
    \label{fig:LC+PG_HD35912}
\end{figure*}

HD~35912 (B2\,V, \citealp{Lesh1968}) has a potential weak magnetic field \citep{Bychkov2003}. Using this property \citet{Bychkov2005} derived a rotation period of $P_{\rm rot} = 0.89786$~d. We measure nine frequencies in the one-sector TESS data set. We explain the frequencies as heat-driven g modes given the large amplitudes. The fourth dominant frequency could correspond to the rotation period, $\nu_{4}=1.146(1)$~d$^{-1}$ ($P=0.8726(8)$~d).  We note equal spacing between five frequencies ($\nu_{7},\nu_{3},\nu_{2}$, and $\nu_{1},\nu_{5}$) but they are unlikely to correspond to rotational splitting, given the rotation period derived by \citet{Bychkov2005} unless the star exhibits strong differential radial rotation \citep{VanReeth2018}. Given the combination of a dense spectrum and low frequency resolution they may also be chance occurrences. Only additional long-term photometric data will be able to resolve this.

\begin{figure*}
	\includegraphics[width=2\columnwidth, scale = 1]{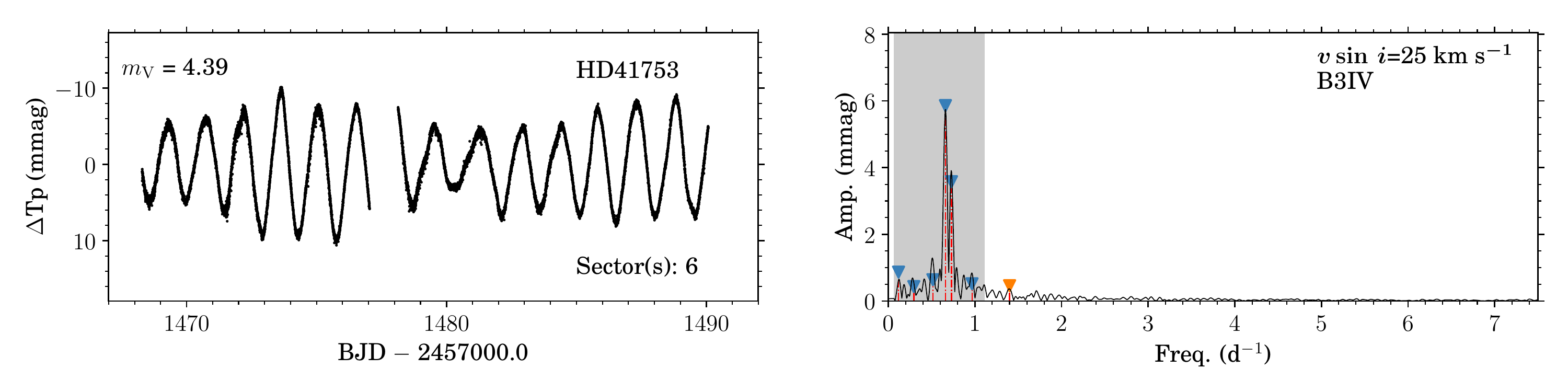}
    \caption{TESS light curve and LS-periodogram of HD\,41753. Same figure style as Fig~\ref{fig:PG_LINE_Bcep}.}
    \label{fig:LC+PG_HD41753}
\end{figure*}

HD~41753 ($\mu$~Ori, B3\,IV, \citealp{Lesh1968}) is part of a binary system with a period of 131.2~d \citep{Eggleton2008}. We extract seven frequencies from the one-sector TESS data set, including a potential combination of the dominant frequencies $\nu_{6}=\nu_{1}+\nu_{2}$. We do not detect any patterns or splittings that allow for mode identification, which is expected given the frequency resolution of this short data set. We attribute the variability to g modes ($0.5<\nu<2.5$~d$^{-1}$) and rotational modulation ($<0.5$~d$^{-1}$) given the frequency range.

\begin{figure*}
	\includegraphics[width=2\columnwidth, scale = 1]{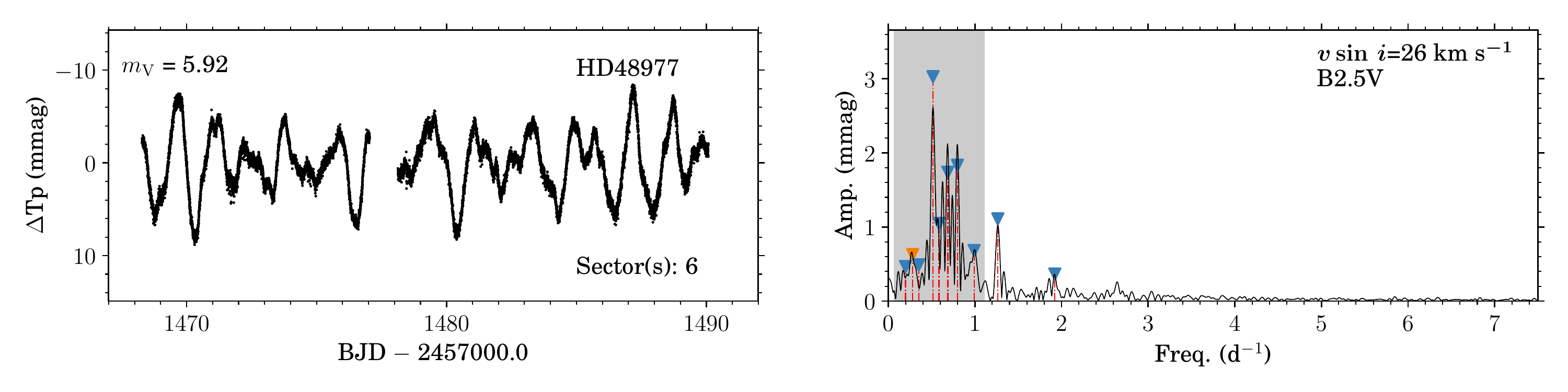}
    \caption{TESS light curve and LS-periodogram of HD\,48977. Same figure style as Fig~\ref{fig:PG_LINE_Bcep}.}
    \label{fig:LC+PG_HD48977}
\end{figure*}

HD~48977 (B2.5\,V, \citealp{Lesh1968}) is a known spectroscopic binary \citep{Karlsson1969} and SPB star \citep{Koen2002, Stankov2005, Telting2006, Thoul2013, Bowman2019a}.  We extract ten frequencies from the TESS data set (fourth row in Fig.~\ref{fig:PG_LINE_SPB}), which is significantly less than in the 25~d CoRoT light curve ($55$ by \citealp{Thoul2013}). The amplitude of the dominant frequency in the CoRoT light curve is $A=2025\pm125$~ppm \citep{Thoul2013}. We find an amplitude of $A=2582\pm185$~ppm (or $2.8(2)$~mmag), which means that this mode has a larger amplitude in the redder passband compared to the blue passband of CoRoT. The difference in CoRoT and TESS pulsation amplitudes could be caused by amplitude modulation or beating among unresolved frequencies. \citet{Thoul2013} noted that the frequency precision of the CoRoT light curve (25~d) was insufficient for asteroseismic modelling, and this proves to be true for the TESS light curve as well. This is the general case for all SPB stars discussed in this section.

\begin{figure*}
	\includegraphics[width=2\columnwidth, scale = 1]{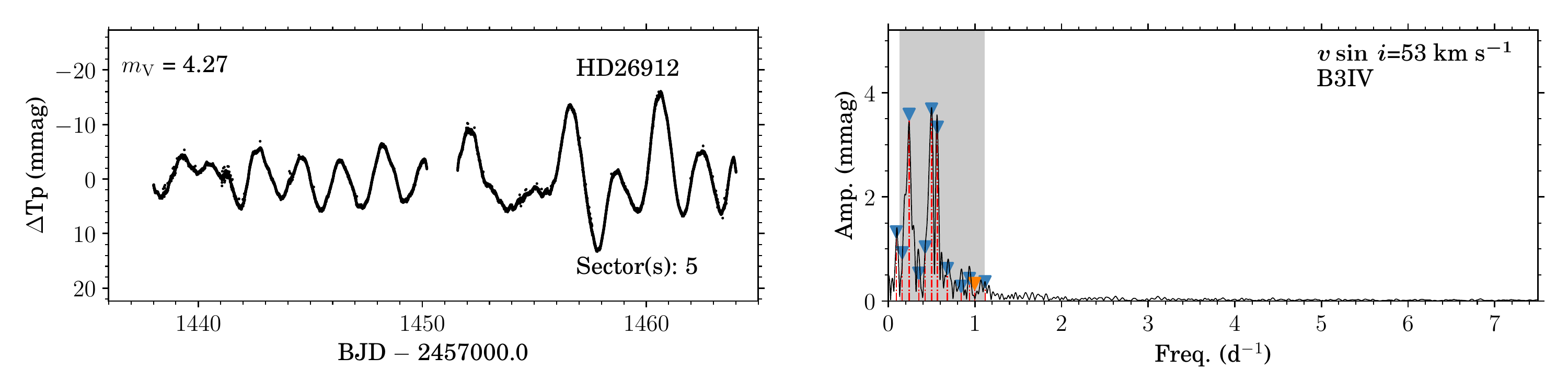}
    \caption{TESS light curve and LS-periodogram of HD\,26912. Same figure style as Fig~\ref{fig:PG_LINE_Bcep}.}
    \label{fig:LC+PG_HD26912}
\end{figure*}

HD~26912 (B3\,IV, \citealp{Lesh1968}) was classified as a constant star by \citet{Percy1977}, however they only investigated short-term variability. We extract twelve frequencies from the TESS data set in the frequency range typical for SPB stars. No patterns or multiplets are immediately apparent inhibiting mode identification.

\begin{figure*}
	\includegraphics[width=2\columnwidth, scale = 1]{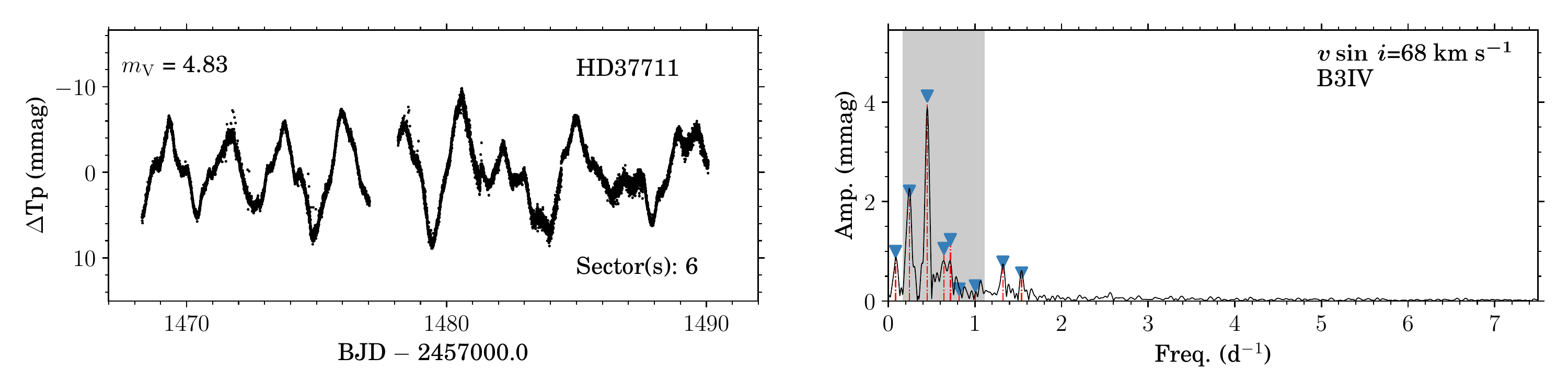}
    \caption{TESS light curve and LS-periodogram of HD\,37711. Same figure style as Fig~\ref{fig:PG_LINE_Bcep}.}
    \label{fig:LC+PG_HD37711}
\end{figure*}

HD~37711 (B3\,IV, \citealp{Lesh1968}) is a known visual binary \citep{Eggleton2008}. We extract nine frequencies from the one-sector TESS data set. Given the estimated rotational modulation frequency range the observed variability is likely a combination of rotation and heat-driven g modes.

\begin{figure*}
	\includegraphics[width=2\columnwidth, scale = 1]{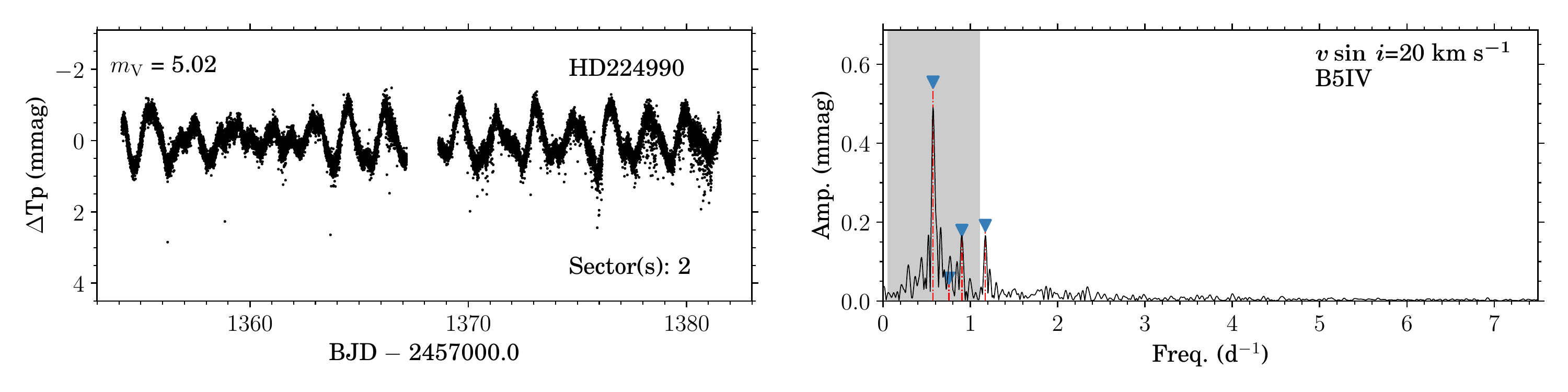}
    \caption{TESS light curve and LS-periodogram of HD\,224990. Same figure style as Fig~\ref{fig:PG_LINE_Bcep}.}
    \label{fig:LC+PG_HD224990}
\end{figure*}

HD~224990 (also known as $\zeta$~Scl) is a spectroscopic binary \citep{Eggleton2008}. Based on its one-sector TESS data set it was classified as SPB by \citet{Pedersen2019a} and \citet{Balona2019b}. We confirm this classification although we remark that its peak-to-peak amplitude is noticeably smaller than SPB stars of similar magnitude considered in this work. We extract four frequencies. 

\subsection{Hybrid pulsators}\label{sec:appendix_hybrid}

\begin{figure*}
	\includegraphics[width=2\columnwidth, scale = 1]{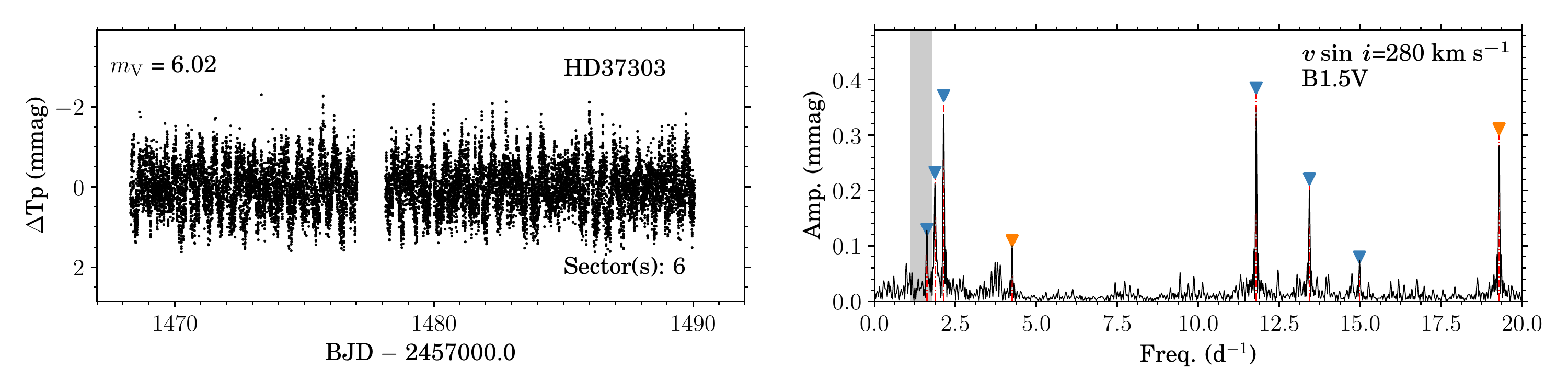}
    \caption{TESS light curve and LS-periodogram of HD\,37303. Same figure style as Fig~\ref{fig:PG_LINE_Bcep}.}
    \label{fig:LC+PG_HD37303}
\end{figure*}

HD~37303 (B1.5\,V, \citealp{Lesh1968}) is known to be variable on a short time-scale \citep{Percy1977}. It was considered SB2 by \citet{Morell1991} but this could not be confirmed by \citet{Eggleton2008}. We extract nine frequencies, both in the low and high frequency regime up to $35$~d$^{-1}$. The dominant frequency is indicative of a young $\beta$~Cep star. 
Many other lower-amplitude frequencies ($4<$ S/N $<5$) are part of multiplets, that is, quadruplets at approximately 8 and 10~d$^{-1}$ and the sextuplet at 16~d$^{-1}$. They are not all equally spaced within each multiplet but two  spacings repeat: a large one $0.24-0.30$~d$^{-1}$ and a small one $0.16-0.17$~d$^{-1}$. Based on our estimated rotational modulation frequency range (v$\,\sin\,i=280$~km~s$^{-1}$), the rotational frequency is between $1.11<\nu_{\rm rot}<1.78$~d$^{-1}$. This excludes rotational splitting as the underlying mechanism. Moreover, uniform rotational multiplets are expected to break down at this rotational velocity \citep{Reese2006}. Nonetheless, the rich frequency spectrum and frequency spacings make HD~37303 a most interesting hybrid candidate for future asteroseismic modelling, provided that the modes can be identified.

\begin{figure*}
	\includegraphics[width=2\columnwidth, scale = 1]{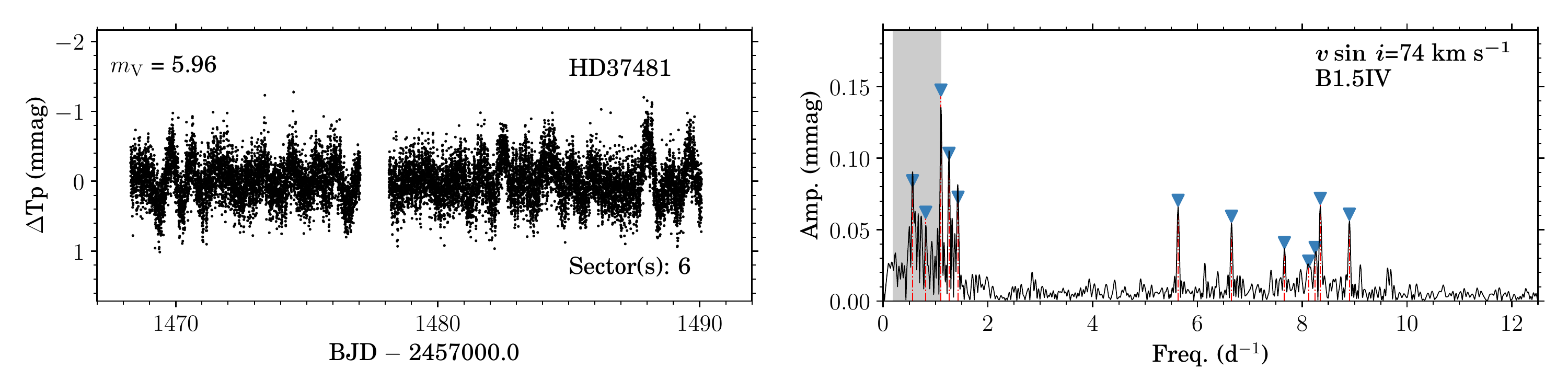}
    \caption{TESS light curve and LS-periodogram of HD\,37481. Same figure style as Fig~\ref{fig:PG_LINE_Bcep}.}
    \label{fig:LC+PG_HD37481}
\end{figure*}

HD~37481 is a B1.5\,IV star \citep{Lesh1968}. We extract twelve frequencies, four of which in the low frequency regime ($<2$~d$^{-1}$), making HD~37481 a hybrid pulsator. The low frequencies form a triplet: $\nu_{1}=1.1052(8), \nu_{2}=1.260(1)$, and $\nu_{4}=1.432(1)$~d$^{-1}$, with a spacing of approximately $0.16$~d$^{-1}$.  A possible triplet is also seen in the high frequency regime: $\nu_{12}=8.125(3)$, $\nu_{11}=8.246(2)$, and $\nu_{5}=8.346(2)$~d$^{-1}$, with a smaller average spacing of approximately $0.11$~d$^{-1}$. However, the presence of (unresolved) residual variability does not allow us to conclude unambiguously that these are in fact rotational triplets. Before mode identification is possible, more photometric data are required to deduce additional significant frequencies.

\begin{figure*}
	\includegraphics[width=2\columnwidth, scale = 1]{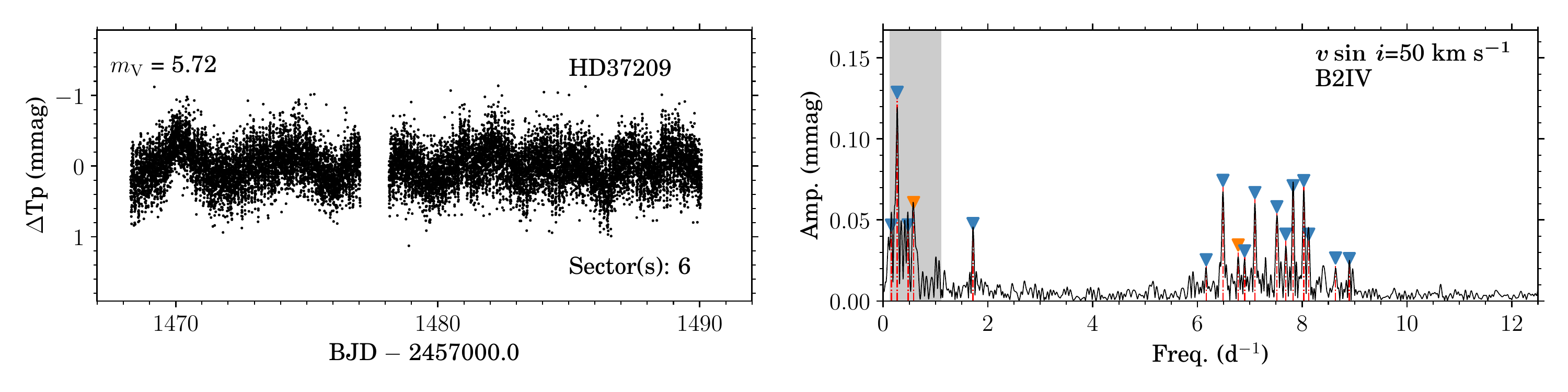}
    \caption{TESS light curve and LS-periodogram of HD\,37209. Same figure style as Fig~\ref{fig:PG_LINE_Bcep}.}
    \label{fig:LC+PG_HD37209}
\end{figure*}

HD~37209 (B2\,IV, \citealp{Lesh1968}) was suspected to be variable on a short time scale by \citet{Percy1977}, but this could not be confirmed by \citet{Telting2006}. It is a confirmed spectroscopic binary with a companion star at a distance of 5.380$\arcsec$ \citep{Eggleton2008}. 
We extract seventeen frequencies from the TESS data set, both in the low frequency ($<2$~d$^{-1}$) and high frequency regime ($6<\nu<9$~d$^{-1}$). Based on our estimated rotational modulation frequency range (the grey area in the bottom right panel Fig.~\ref{fig:LC+PG_hybrids}), it is possible that the dominant low frequency variability is due to rotational modulation. However, the presence of an additional low frequency outside this region is indicative of heat-driven g modes. The high frequencies form a dense frequency group, and the residuals indicate that more variability is present. HD~37209 is clearly a hybrid pulsator but despite the high number of significant frequencies we cannot identify patterns or splittings for mode identification.

\subsection{Peculiar stars}\label{sec:appendix_peculiar}
\subsubsection{Be and Oe stars}\label{sec:appendix_emission}

\begin{figure*}
	\includegraphics[width=2\columnwidth, scale = 1]{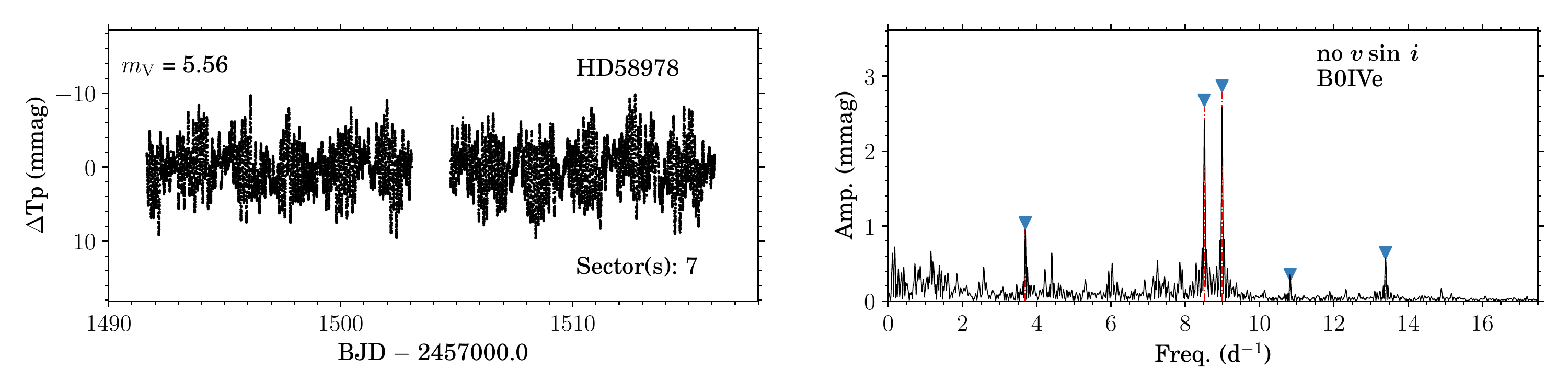}
    \caption{TESS light curve and LS-periodogram of HD\,58978. Same figure style as Fig~\ref{fig:PG_LINE_Bcep}.}
    \label{fig:LC+PG_HD58978}
\end{figure*}

HD~58978 (FY CMa, B0\,IV:e, \citealp{Morgan1955}) is a Be star with a hot sdO sub-dwarf companion \citep{Rivinius2004, Peters2008} with low frequency ($<0.38$~d$^{-1}$) periodicities in HIPPARCOS photometry \citep{Lefevre2009}. We extract five frequencies from the one-sector TESS data set, none of which have been previously reported. FY CMa is a fast rotator as a result of the past mass transfer spin-up \citep{Peters2008}. It remains to be seen what the variability contribution of the subdwarf O-type star is, given that hot subdwarfs are known to exhibit heat-driven modes \citep{Charpinet1997, Green2003, Kilkenny2010, Bloemen2014, Randall2016}. The origin of the dominant variability is still likely from the Be star given that the light contribution of the subdwarf O star is approximately two orders of magnitude lower \citep{Peters2008}. We do not detect any patterns or spacings but the discovery of high frequency modes makes HD~58978 an interesting candidate for asteroseismic modelling. That is, if the modes can be identified which in the case of this emission-line star is likely only possible with additional continuous photometry.

\begin{figure*}
	\includegraphics[width=2\columnwidth, scale = 1]{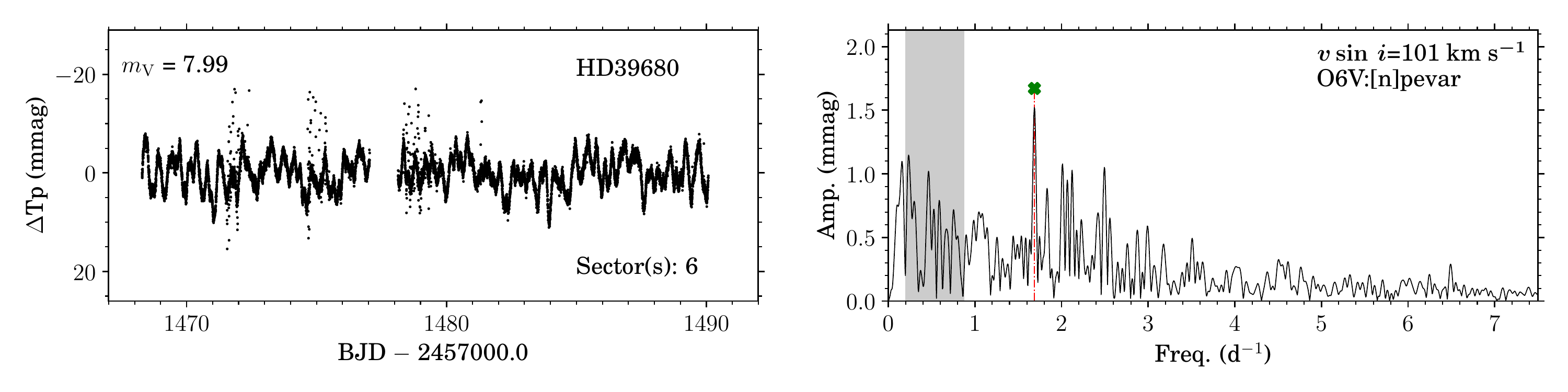}
    \caption{TESS light curve and LS-periodogram of HD\,39680. Same figure style as Fig~\ref{fig:PG_LINE_Bcep}.}
    \label{fig:LC+PG_HD39680}
\end{figure*}

HD~39680 (O6\,V:[n]pevar, \citealp{Sota2011}) is a known emission-line star \citep{Conti1974, Frost1976, Negueruela2004}. This emission is also seen in the multi-epoch IACOB/OWN spectroscopy. Both He~{\sc i}~$\lambda$5875 and H$_{\alpha}$ reveal double peak emission, which is variable over the five randomly selected epochs. No frequencies were extracted from the one-sector TESS data set, but a dense frequency grouping $1.5<\nu<2.7$~d$^{-1}$ is noticed. This is reminiscent of those seen in pulsating Be stars (i.e. \citealt{Huat2009}).  The lowest frequencies could be due to rotational modulation but the large presence of variability outside the estimated rotational modulation frequency range indicates the presence of an additional mechanism, which is likely connected with pulsations or waves. Longer photometric time series data are required to verify the stability of these modes. Nonetheless, studying pulsations in these stars is crucial to understanding the possible connection between pulsations and outflow events.

\begin{figure*}
	\includegraphics[width=2\columnwidth, scale = 1]{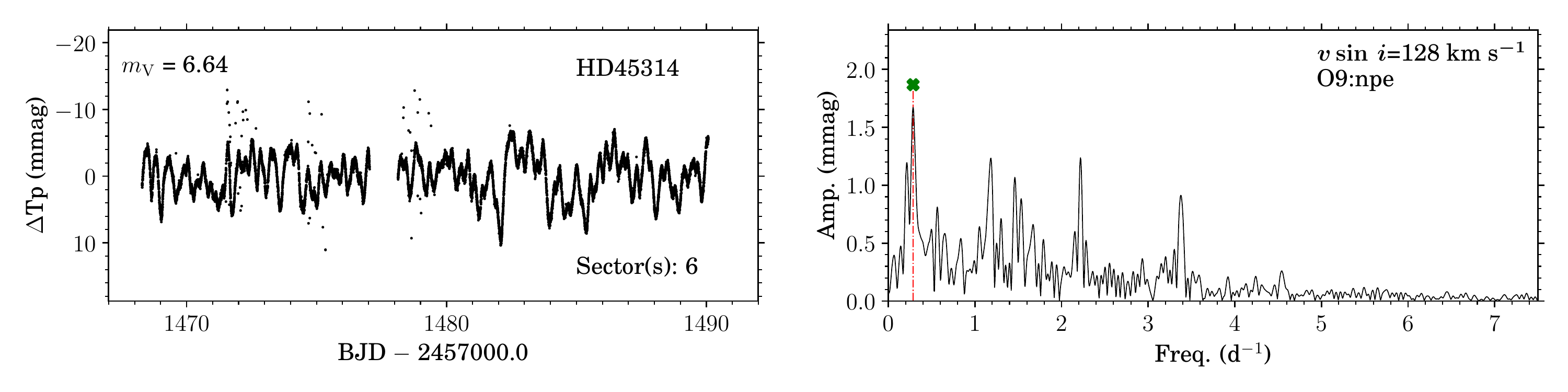}
    \caption{TESS light curve and LS-periodogram of HD\,45314. Same figure style as Fig~\ref{fig:PG_LINE_Bcep}.}
    \label{fig:LC+PG_HD45314}
\end{figure*}

HD~45314 (O9\,:npe, \citealp{Sota2011}) is classified as a $\gamma$~Cas star in the literature  because of its unusually hard and bright X-ray emission \citep{Rauw2013, Rauw2018}. The one-sector TESS data set shows predominantly stochastic low frequency variability, but coherent frequencies are visible in the range $1.5<\nu<4.5$~d$^{-1}$. Given the available data set it remains unclear whether these are stable or stochastic in nature. 
Both He~{\sc i}~$\lambda$5875 and H$_{\alpha}$ are seen in emission, transitioning between single and multi-peaked which was also seen by \citet{Rauw2018}.

\subsubsection{Magnetic O- and B-type stars}\label{sec:appendix_magnetic}
\begin{figure*}
	\includegraphics[width=2\columnwidth, scale = 1]{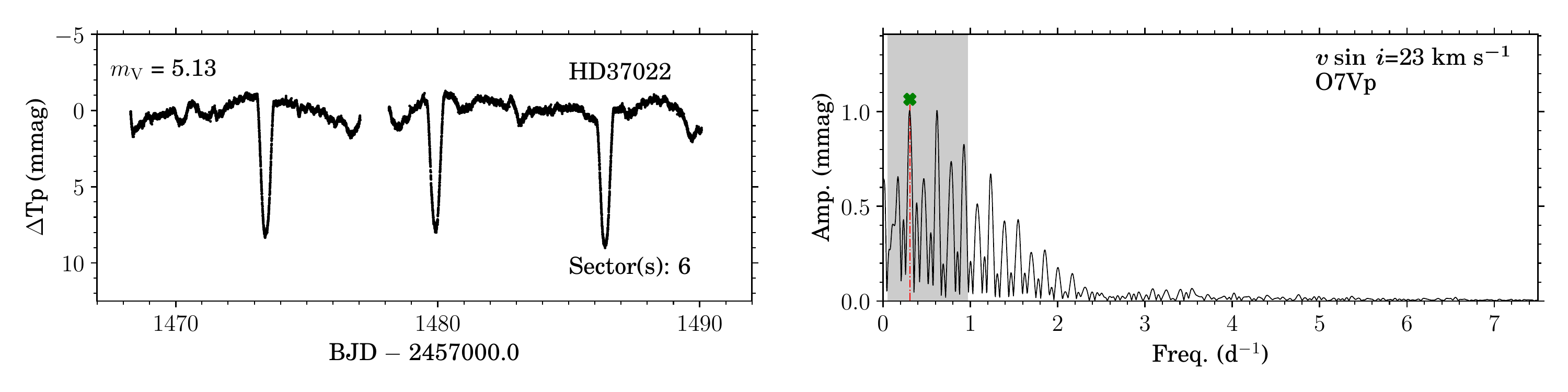}
    \caption{TESS light curve and LS-periodogram of HD\,37022. Same figure style as Fig~\ref{fig:PG_LINE_Bcep}.}
    \label{fig:LC+PG_HD37022}
\end{figure*}

HD~37022 ($\theta^{1}$~Ori~C, O7\,Vp, \citealp{Sota2011}) is a  magnetic star  \citep{Donati2002, Wade2006}. The TESS pixels are contaminated by the nearby eclipsing binary $\theta^{1}$~Ori~B \citep{Windemuth2013}, which dominates the light contribution. We do note variability outside the eclipses which might come from any of the components of $\theta^{1}$~Ori.

\begin{figure*}
	\includegraphics[width=2\columnwidth, scale = 1]{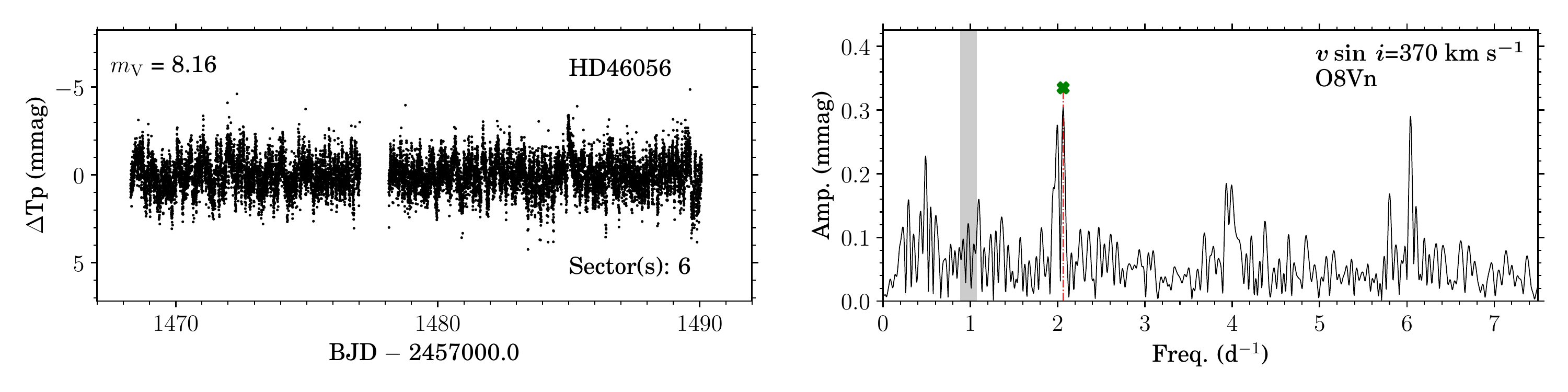}
    \caption{TESS light curve and LS-periodogram of HD\,46056. Same figure style as Fig~\ref{fig:PG_LINE_Bcep}.}
    \label{fig:LC+PG_HD46056}
\end{figure*}

HD~46056 (O8\,Vn, \citealp{Sota2011}) has a detectable magnetic field \citep{Grunhut2017, Petit2019}. The one-sector TESS data set is nearly constant and dominated by low amplitude periodicities. The origin is unclear and could be instrumental.

\begin{figure*}
	\includegraphics[width=2\columnwidth, scale = 1]{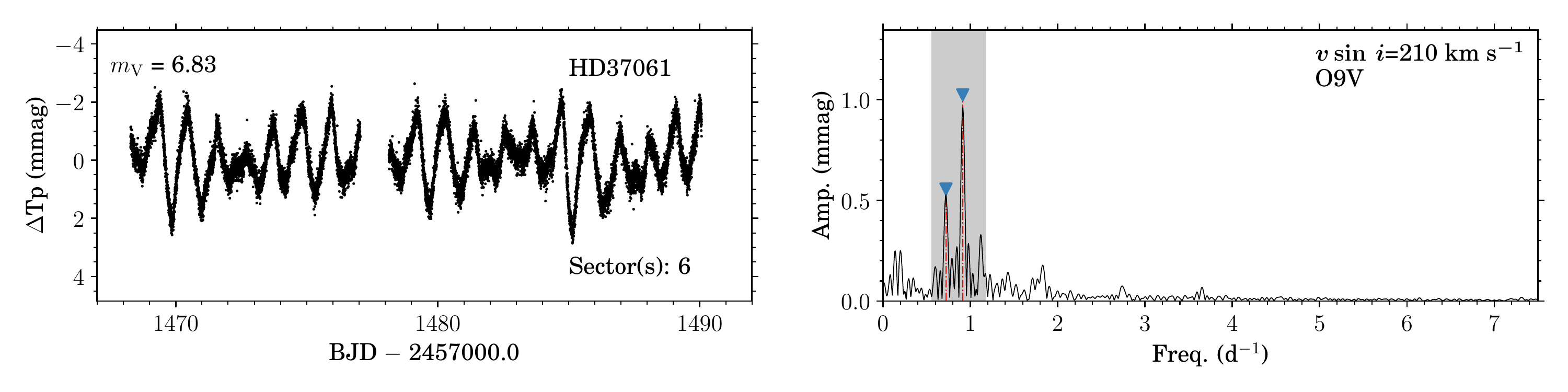}
    \caption{TESS light curve and LS-periodogram of HD\,37061. Same figure style as Fig~\ref{fig:PG_LINE_Bcep}.}
    \label{fig:LC+PG_HD37061}
\end{figure*}

HD~37061 (NU~Ori$_{1}$, O9.5\,V) is a well-studied triple system \citep{Kohler2006, Grellmann2013} with a strongly magnetic B star \citep{Petit2008, Shultz2017, Shultz2019}. We find it as SB1 in our data set. Given the large TESS pixel size all components are present in the mask (separation $<1\arcsec$) and hence in the one-sector TESS data set. The measured large-scale magnetic field of approximately $8$~kG was thought to stem from the primary Aa by \citet{Petit2008} but this was contested by \citet{Shultz2019}, who show that it is the C component instead. These authors also derive a rotational period $P_{\rm rot}=1.09468(7)$~d.
We extract two frequencies, $\nu_{1}=0.9124(4)$ and $\nu_{2}=0.7203(5)$~d$^{-1}$. The dominant frequency (or $P_{1}=1.0960(4)$~d) is consistent with the rotational period measured by \citet{Shultz2019} for NU~Ori C. Interestingly, $\nu_{1}$ is the central frequency of a triplet: $\nu_{1},\nu_{2}$ and $\nu\approx1.118$~d$^{-1}$, with an average splitting of $\Delta \nu \approx 0.199$~d$^{-1}$. This could be the result of an inclination effect ($i_{\rm rot}= 38\pm5^{\circ}$, \citealp{Shultz2019}) but this cannot be confirmed given the available data.

\begin{figure*}
	\includegraphics[width=2\columnwidth, scale = 1]{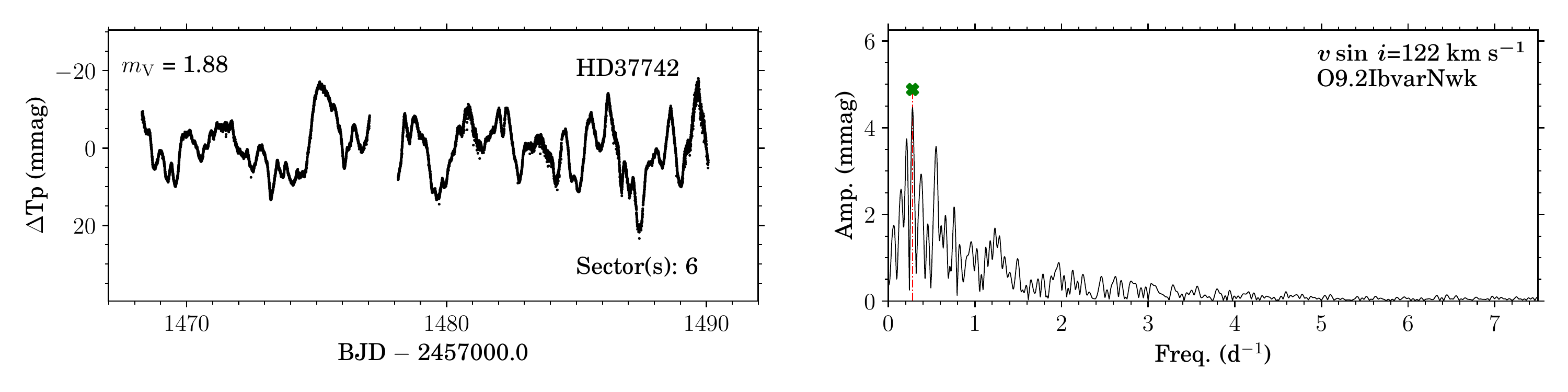}
    \caption{TESS light curve and LS-periodogram of HD\,37742. Same figure style as Fig~\ref{fig:PG_LINE_Bcep}.}
    \label{fig:LC+PG_HD37742}
\end{figure*}

HD~37742 ($\zeta$~Ori~A, O9.2\,Ib~var~Nwk, \citealp{Sota2014}) is a wide-period binary star \citep{Mason2001, Turner2008}. A magnetic field was detected by \citet{Bouret2008}, which was used by \citet{Blazere2015} to derive the rotation period ($P_{\rm rot} = 6.829$~d) and constrain the magnetic configuration. An analysis of the BRITE photometry by \citet{Buysschaert2017} led to the recovery of the rotation period ($P_{\rm rot} = 6.82 \pm 0.18$~d). In addition, a variation with period $P=10.0 \pm 0.3$~d was detected and attributed to the circumstellar environment given its clear presence in the wind lines but absence in the photospheric lines. The dominant type of variability in the one-sector TESS light curve is stochastic low frequency variability.

\begin{figure*}
	\includegraphics[width=2\columnwidth, scale = 1]{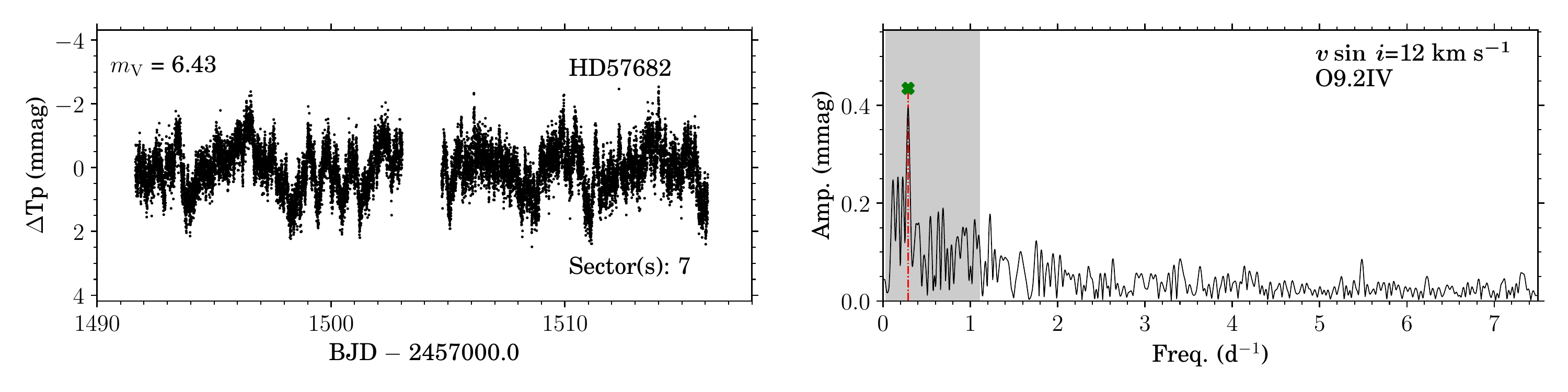}
    \caption{TESS light curve and LS-periodogram of HD\,57682. Same figure style as Fig~\ref{fig:PG_LINE_Bcep}.}
    \label{fig:LC+PG_HD57682}
\end{figure*}

HD~57682 (O9.2\,IV, \citealp{Sota2014}) has a detectable magnetic field \citep{Grunhut2009, Grunhut2012}. Using this property \citet{Grunhut2012} derived a rotation period of $63.5708 \pm 0.0057$~d and were able to constrain the magnetic field. The one-sector TESS data set shows low amplitude stochastic low frequency variability. The dominant frequency ($\nu\approx0.287$~d$^{-1}$ or $P\approx3.48$~d, at S/N $=3.1$) does not agree with the period found by \citet{Grunhut2012} emphasising the stochastic nature of the variability in this star.

\begin{figure*}
	\includegraphics[width=2\columnwidth, scale = 1]{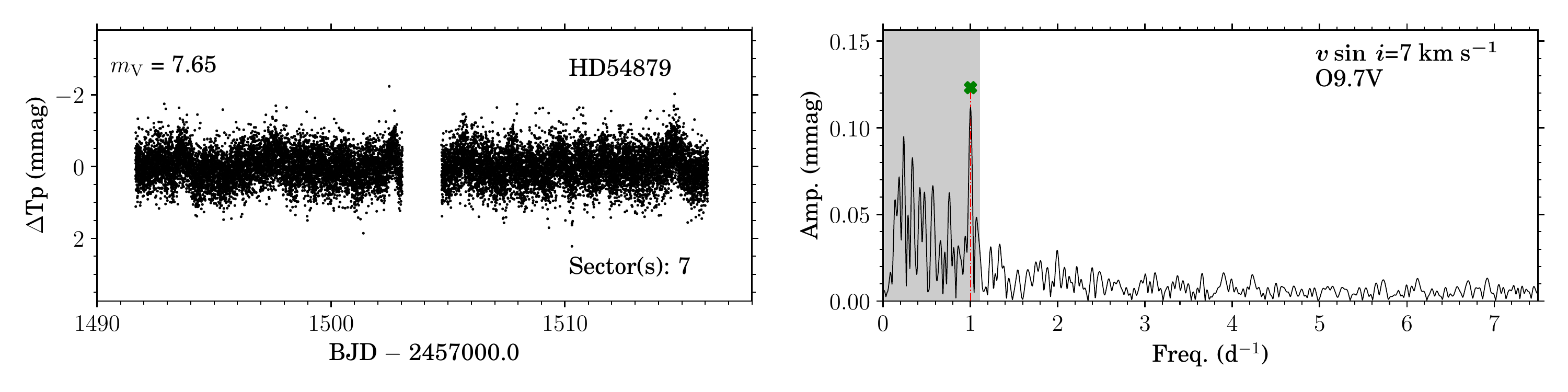}
    \caption{TESS light curve and LS-periodogram of HD\,54879. Same figure style as Fig~\ref{fig:PG_LINE_Bcep}.}
    \label{fig:LC+PG_HD54879}
\end{figure*}

HD~54879 (O9.7\,V, \citealp{Sota2011}) is a magnetic star \citep{Castro2015}. Subsequent studies have shown that the magnetic and spectroscopic variability implies a long rotation period \citep{Shenar2017}. A recent sudden change in magnetic field, accompanied by radial velocity variations on the time-scale of days was claimed by \citet{Hubrig2019} but this has been contested by \citet{Wade2020b}. The one-sector TESS data set shows low amplitude stochastic variability.

\begin{figure*}
	\includegraphics[width=2\columnwidth, scale = 1]{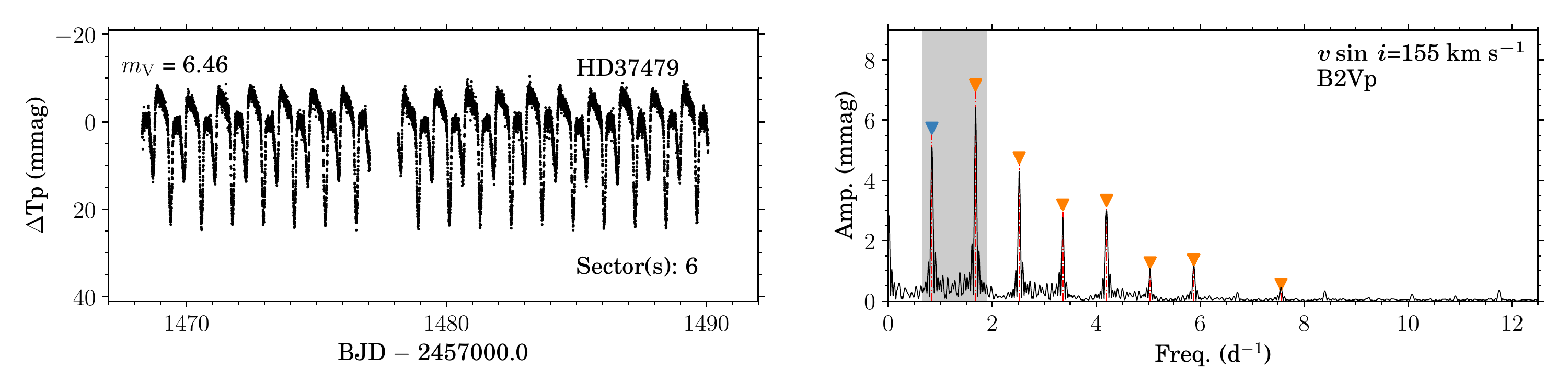}
    \caption{TESS light curve and LS-periodogram of HD\,37479. Same figure style as Fig~\ref{fig:PG_LINE_Bcep}.}
    \label{fig:LC+PG_HD37479}
\end{figure*}

HD~37479 ($\sigma$~Ori~E, B2\,Vp, \citealp{Lesh1968}) has a magnetic field that was first detected by \citet{Landstreet1978}. Many attempts have been made at modelling its complex magnetic configuration, with a rigidly rotating magnetosphere matching the observations best \citep{Townsend2005, Oksala2012, Townsend2013a, Oksala2015}. The morphology of the one-sector TESS data set is similar in all aspects to the MOST data set of similar length studied by \citet{Townsend2013a}. We extract eight frequencies, all harmonics of the rotational frequency ($\nu_{2}=0.8399(7)$~d$^{-1}$ or $P_{\rm rot}=1.19(1)$~d).

\subsection{Binaries}\label{sec:appendix_binaries}

\begin{figure*}
	\includegraphics[width=2\columnwidth, scale = 1]{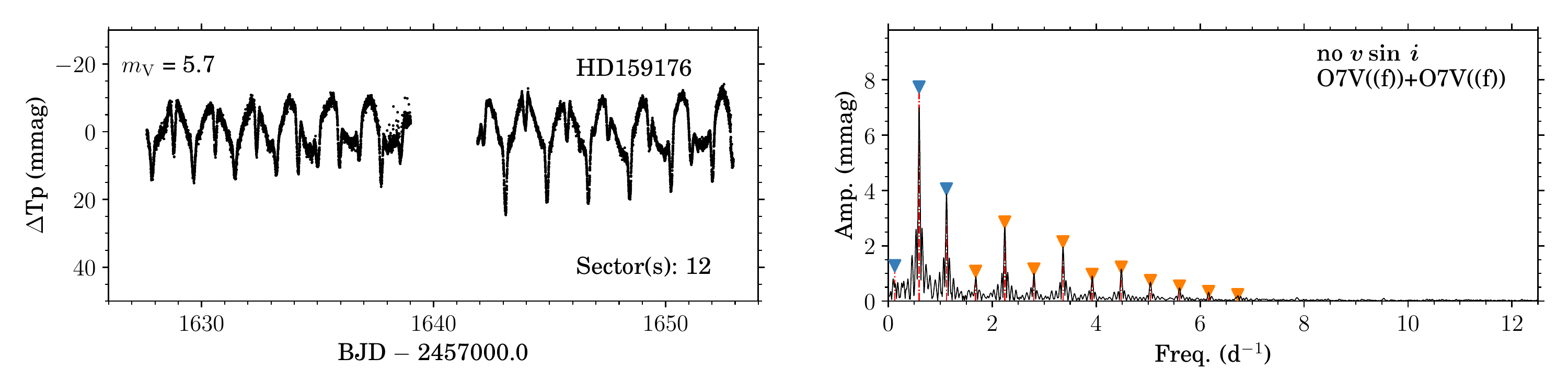}
    \caption{TESS light curve and LS-periodogram of HD\,159176. Same figure style as Fig~\ref{fig:PG_LINE_Bcep}.}
    \label{fig:LC+PG_HD159176}
\end{figure*}

HD~159176 (O7\,V((f)) + O7\,V((f)) is a known massive SB2 system with seven more fainter lower mass components detected \citep{Mason1998, Linder2007, Sana2014}. Its light curve shows complex behaviour with ellipsoidal variations which have been reported in the past \citep{Thomas1975, Linder2007}. The second dominant frequency, $\nu_{2}=1.1200(6)$~d$^{-1}$ is twice the orbital frequency and has a series of harmonics. This yields a period of $P_{\rm orb}=1.786(1)$~d. The dominant frequency $\nu_{1}=0.5929(5)$~d$^{-1}$ corresponds to the out of eclipse periodic signal and could be due to asynchronous rotation of one of the stars.
 
 \begin{figure*}
	\includegraphics[width=2\columnwidth, scale = 1]{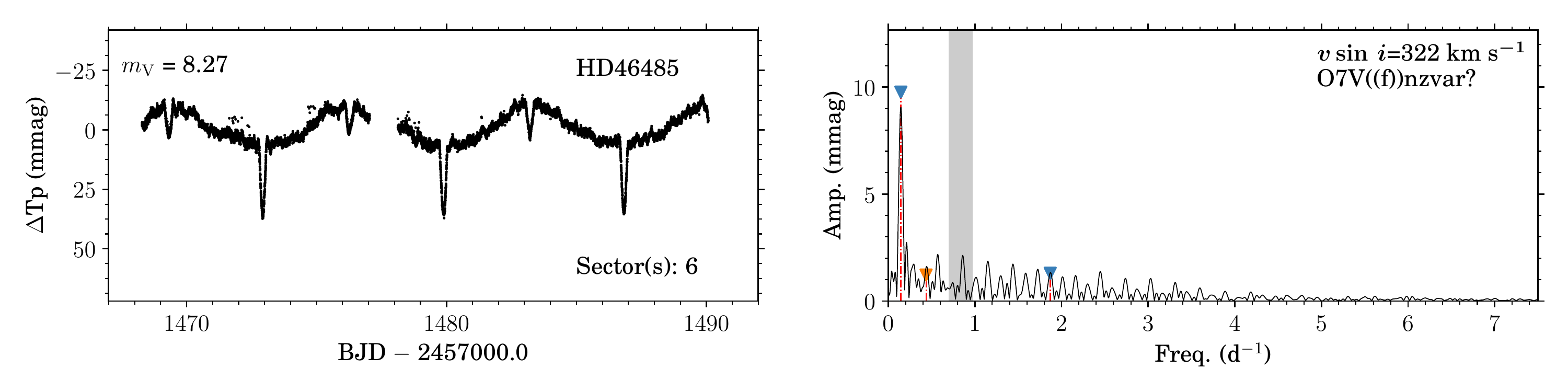}
    \caption{TESS light curve and LS-periodogram of HD\,46485. Same figure style as Fig~\ref{fig:PG_LINE_Bcep}.}
    \label{fig:LC+PG_H46485}
\end{figure*}

HD46485 (O7 V((f))nz var?, \citealp{Sota2014}) is an eclipsing binary system whose spectroscopic binary nature still remains unconfirmed due to the fast rotating nature of its main component (v$\,\sin\,i$\,=\,322 km~s$^{-1}$) \citep{Sana2014, Cazorla2017a}. We find it as a single-lined spectroscopic binary in the IACOB/OWN spectra, these lines being very broad. The TESS data set shows periodic light variations in sync with the eclipses. These are caused by a reflection effect due to a possible large difference of temperature between components. We extract the orbital frequency, $\nu=0.1453(4)$~d$^{-1}$, and one harmonic although many more are visible. After removing numerous multiples of the orbital frequency some variability remains in the periodogram, which could be related to rotational modulation of the main, fast rotating component given the estimated rotational modulation frequency range.

 \begin{figure*}
	\includegraphics[width=2\columnwidth, scale = 1]{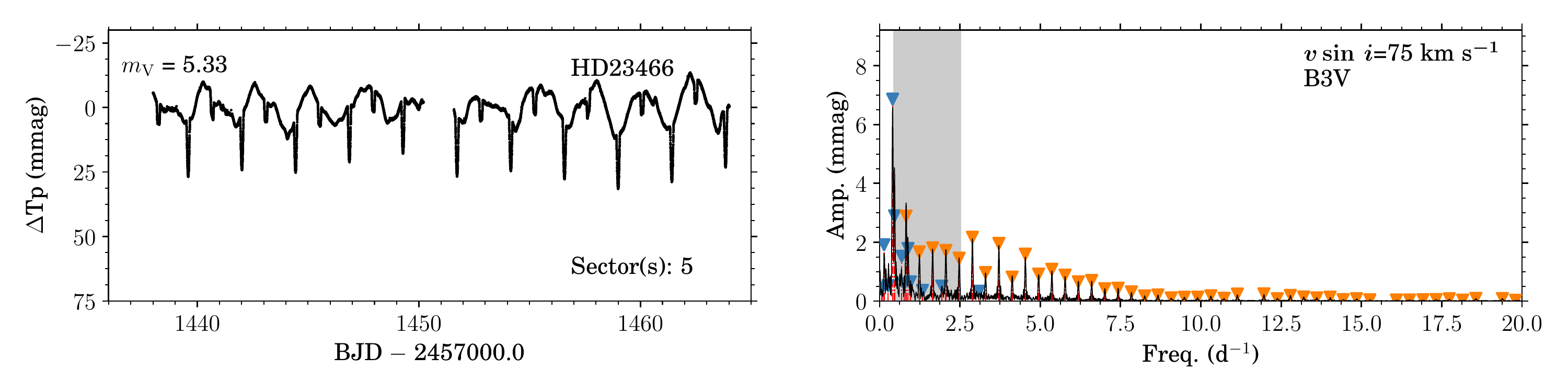}
    \caption{TESS light curve and LS-periodogram of HD\,23466. Same figure style as Fig~\ref{fig:PG_LINE_Bcep}.}
    \label{fig:LC+PG_H23466}
\end{figure*}

HD~23466 (B3\,V) is a known spectroscopic binary \citep{Pourbaix2004, Eggleton2008}. We only see lines from one of the components in our spectra but we cannot discard the star to be a SB2 since we have only epoch available. The one-sector TESS data set shows clear eclipses and periodic variability out of eclipse. The complex variability is difficult to interpret in the LS-periodogram given that the two dominant frequencies ($\nu_{1}= 0.405(1)$ and $\nu_{2}=0.466(2)$~d$^{-1}$) do not represent the orbital frequency. Rather, it seems that $\nu_{3}=0.825(3)$~d$^{-1}$ is twice the orbital frequency yielding a period of $P_{\rm orb}=2.442(8)$~d. Long time-scale variability occurs in between these eclipses that could be due to coherent/non-coherent pulsations or wind variability. However, after pre-whitening all harmonics of the orbital frequency the residual frequencies remain insignificant.

 \begin{figure*}
	\includegraphics[width=2\columnwidth, scale = 1]{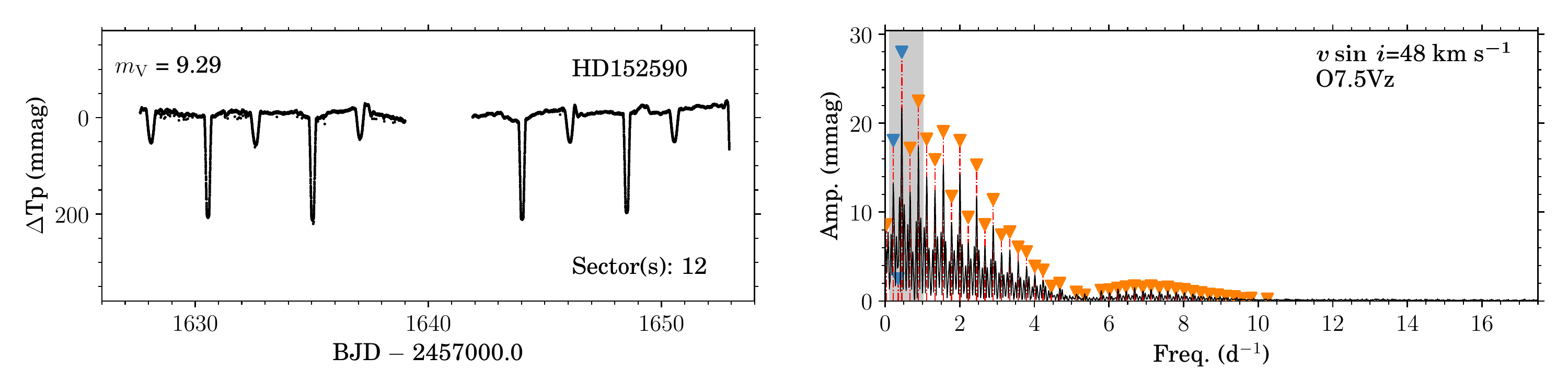}
    \caption{TESS light curve and LS-periodogram of HD\,152590. Same figure style as Fig~\ref{fig:PG_LINE_Bcep}.}
    \label{fig:LC+PG_HD152590}
\end{figure*}

HD~152590 (V1297~Sco, O7.5\,Vz, \citealp{Sota2014}) is a known eclipsing binary with a period of 4.487~d \citep{Otero2004}. We find it as SB1 in the IACOB/OWN spectroscopy. The one-sector TESS data set shows eclipses with a period of $P_{\rm orb}=4.59(4)$~d. The LS-periodogram is dominated by a long series of harmonics of $\nu_{1}$ (twice the orbital frequency) that fades away around $5$~d$^{-1}$ but resurfaces between $6<\nu<10$~d$^{-1}$.

 \begin{figure*}
	\includegraphics[width=2\columnwidth, scale = 1]{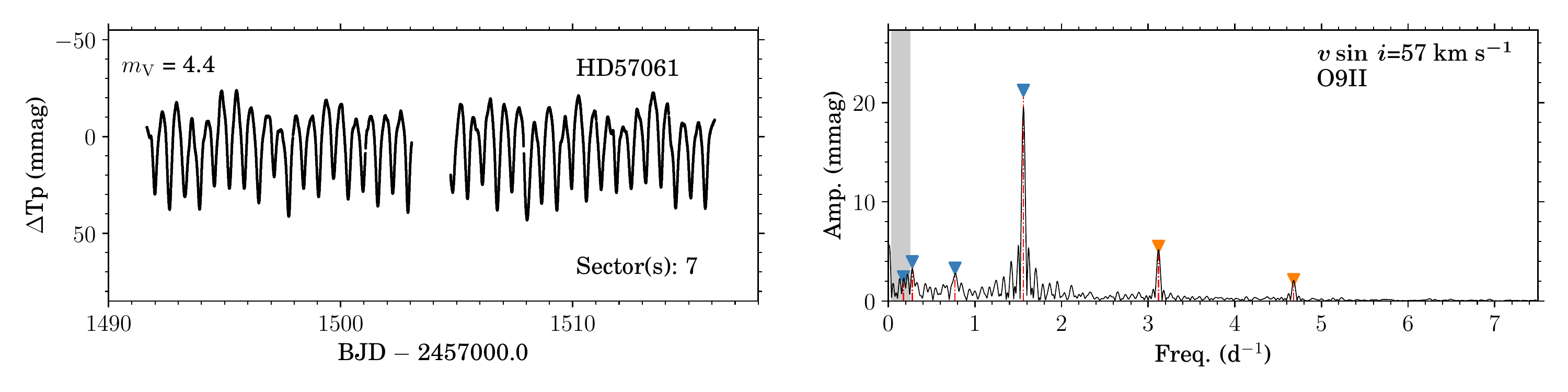}
    \caption{TESS light curve and LS-periodogram of HD\,57061. Same figure style as Fig~\ref{fig:PG_LINE_Bcep}.}
    \label{fig:LC+PG_HD57061}
\end{figure*}

HD~57061 ($\tau$~CMa) is a complex multiple system with at least three close visual components \citep{Sota2014}. It was found as a single-lined binary with a period of $154.9$~d by \citet{Stickland1998} and as an eclipsing binary with a period of $1.282$~d by \citet{vanLeeuwen1997}. We only detect one component in the IACOB/OWN spectroscopy, in addition to wind variability. We find it as an eclipsing binary in the one-sector TESS data set with a period $P_{\rm orb}=1.2818(6)$~d, which agrees with the period found by \citet{vanLeeuwen1997}. Long-term trends are seen, but it is unclear whether these are physical or instrumental in origin.

 \begin{figure*}
	\includegraphics[width=2\columnwidth, scale = 1]{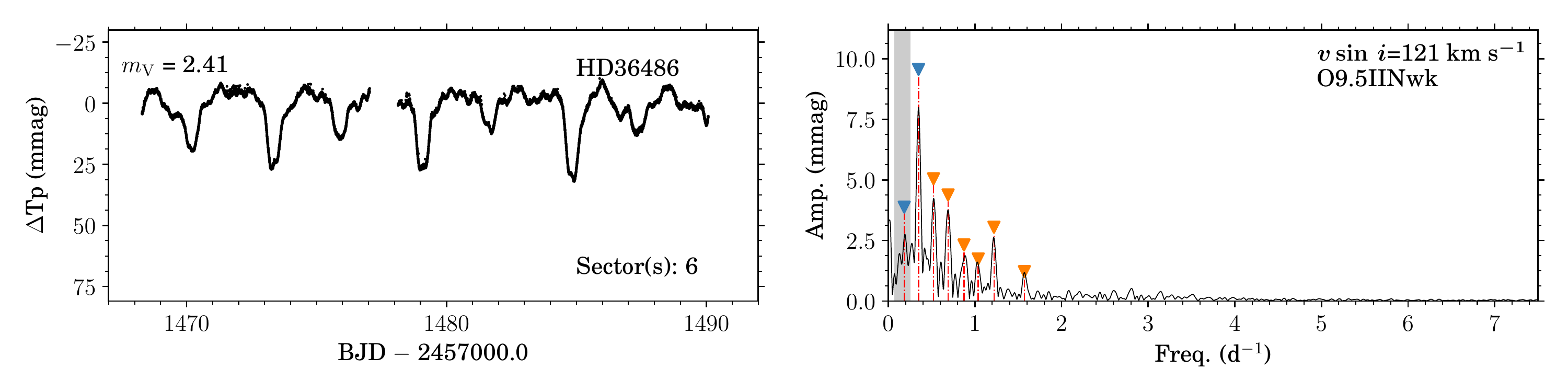}
    \caption{TESS light curve and LS-periodogram of HD\,36486. Same figure style as Fig~\ref{fig:PG_LINE_Bcep}.}
    \label{fig:LC+PG_HD36486}
\end{figure*}

HD~36486 ($\delta$~Ori, O9.5\,II~Nwk, \citealp{Sota2011}) is a multiple system composed of a close inner binary and two more distant companions \citep{Harvin2002, Tokovinin2014}. The system has been studied extensively over multiple wavelengths \citep{Corcoran2015, Pablo2015, Shenar2015, Nichols2015, Richardson2015}. All components are close enough to contribute light to the TESS pixels. We only detect one component in the IACOB/OWN spectroscopy. The one-sector TESS light curve shows eclipses with a period $P=5.72(2)$~d, which agrees with previous findings by \citet{Harvin2002} and \citet{Mayer2010} for the inner eclipsing binary $\delta$~Ori Aa1 and $\delta$~Ori Aa2. Out of eclipse stochastic low frequency variability seems to be present. 

 \begin{figure*}
	\includegraphics[width=2\columnwidth, scale = 1]{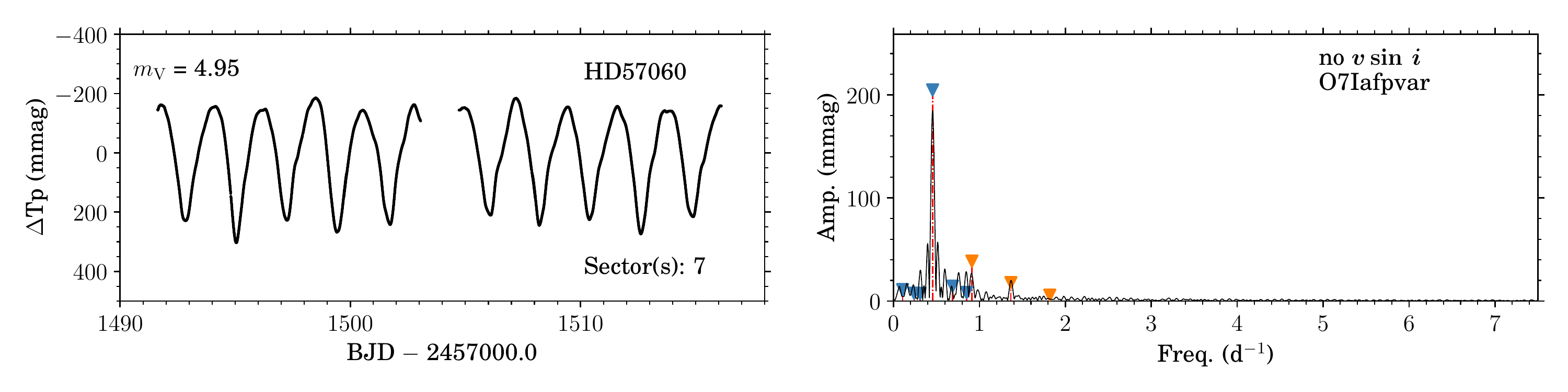}
    \caption{TESS light curve and LS-periodogram of HD\,57060. Same figure style as Fig~\ref{fig:PG_LINE_Bcep}.}
    \label{fig:LC+PG_HD57060}
\end{figure*}

HD~57060 (29~CMa or UW CMa, O7\,Iafpvar, \citealp{Sota2014}) is a known spectroscopic and eclipsing contact binary \citep{Leung1978, Bagnuolo1994, Antokhina2011, Pablo2018}. We find it as SB1 in the IACOB/OWN spectroscopy, in addition to wind variability in emission. The one-sector TESS data set shows large amplitude eclipses with a period of $P=4.40(6)$~d, which agrees with previous findings in HIPPARCOS \citep{Antokhina2011} and BRITE photometry \citep{Pablo2018}. Some low frequencies are detected which are likely related to the asymmetry in the eclipse branches \citep{Antokhina2011, Pablo2018}.

 \begin{figure*}
	\includegraphics[width=2\columnwidth, scale = 1]{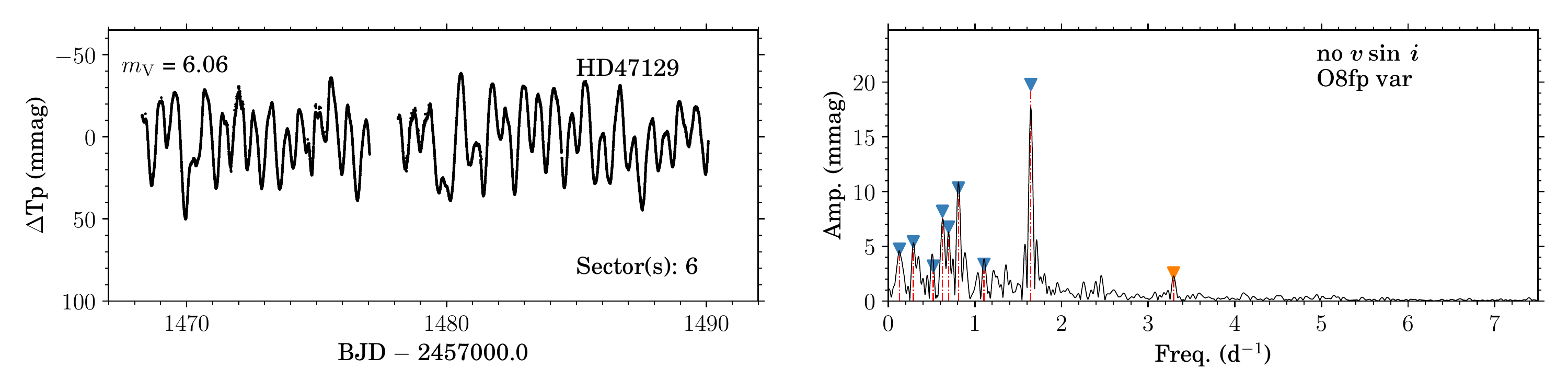}
    \caption{TESS light curve and LS-periodogram of HD\,47129. Same figure style as Fig~\ref{fig:PG_LINE_Bcep}.}
    \label{fig:LC+PG_HD47129}
\end{figure*}

HD~47129 (Plaskett's star, O8\,fp~var, \citealp{Sota2011}), is a well-known SB2 binary system. It was suggested by \citet{Linder2008} to be in a post-Roche lobe overflow phase, with a period $P_{\rm orb}=14.39625$~d. We find it as SB1 in the IACOB/OWN data set. Wind variability in emission is also noted. The CoRoT data was studied in-depth by \citet{Mahy2011}, who find two dominant frequencies ($f_{1}=0.823$ and $f_{2}=0.069$~d$^{-1}$) each with a series of harmonics. The lower frequency $f_{2}$ is attributed to the orbital frequency of the system but the high frequency $f_{1}$ could not be explained. \citet{Grunhut2013} measured a strong magnetic field in the rapidly rotating O companion. We find a high frequency in the one-sector TESS data set ($\nu_{2}=0.811(3)$~d$^{-1}$), which is close to the value found by \citet{Mahy2011} but not within $3\sigma$. We do however detect two harmonics of the frequency found by \citet{Mahy2011} $\nu_{1}=2f_{1}$ and $\nu_{10}=4f_{1}$. Several other frequencies are extracted, some of which are harmonics of the orbital frequency measured by \citet{Mahy2011}. In the last column of Table~\ref{table:multi_freq_anal} these are indicated, with $\nu_{\rm orb}$ being $f_{2}=0.069$~d$^{-1}$ measured by \citet{Mahy2011}.

 \begin{figure*}
	\includegraphics[width=2\columnwidth, scale = 1]{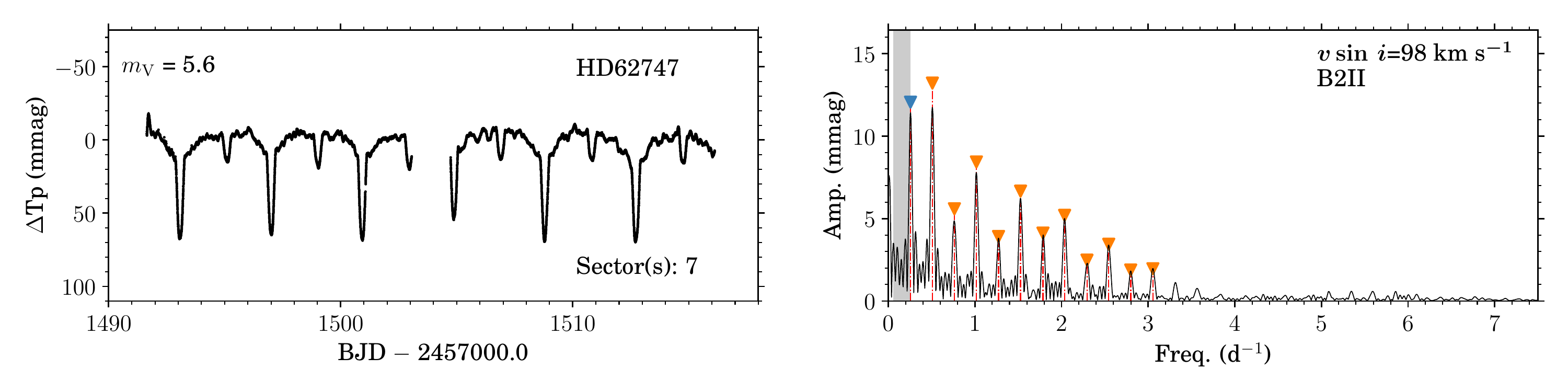}
    \caption{TESS light curve and LS-periodogram of HD\,62747. Same figure style as Fig~\ref{fig:PG_LINE_Bcep}.}
    \label{fig:LC+PG_HD62747}
\end{figure*}

HD~62747 (V390~Pup, B2\,II, \citealp{Houk1988}), is an eclipsing binary \citep{Malkov2006, Eggleton2008, Lefevre2009}. A period of $P=3.928$~d was found in HIPPARCOS data \citep{Lefevre2009}. \citet{Telting2006} detected line-profile variation in several photospheric lines sensitive to pulsations. We find it as SB1 in the IACOB/OWN data. The one-sector TESS data set reveals an orbital frequency $\nu_{2}=0.257(2)$~d$^{-1}$ (or $P_{\rm orb}=3.89(3)$~d) with a series of harmonics up to approximately $3$~d$^{-1}$. Low frequency variability seems to be present out of eclipse but no frequencies are found to be significant.

\subsection{Upper-main sequence stars}\label{sec:appendix_UMS}

 \begin{figure*}
	\includegraphics[width=2\columnwidth, scale = 1]{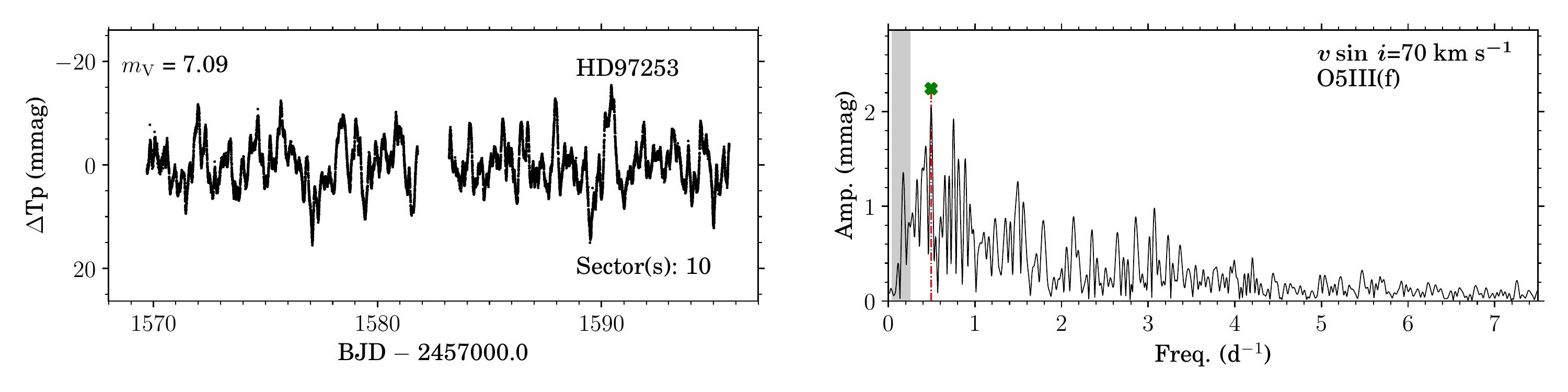}
    \caption{TESS light curve and LS-periodogram of HD\,97253. Same figure style as Fig~\ref{fig:PG_LINE_Bcep}.}
    \label{fig:LC+PG_HD97253}
\end{figure*}

HD~97253 (O5\,III(f), \citealp{Sota2014}) is a known spectroscopic binary \citep{Gies1987, Mason2009, Chini2012, Sana2014}. We find wind variability in absorption in the IACOB/OWN spectroscopy. The TESS light curve shows clear stochastic low frequency variability. 

 \begin{figure*}
	\includegraphics[width=2\columnwidth, scale = 1]{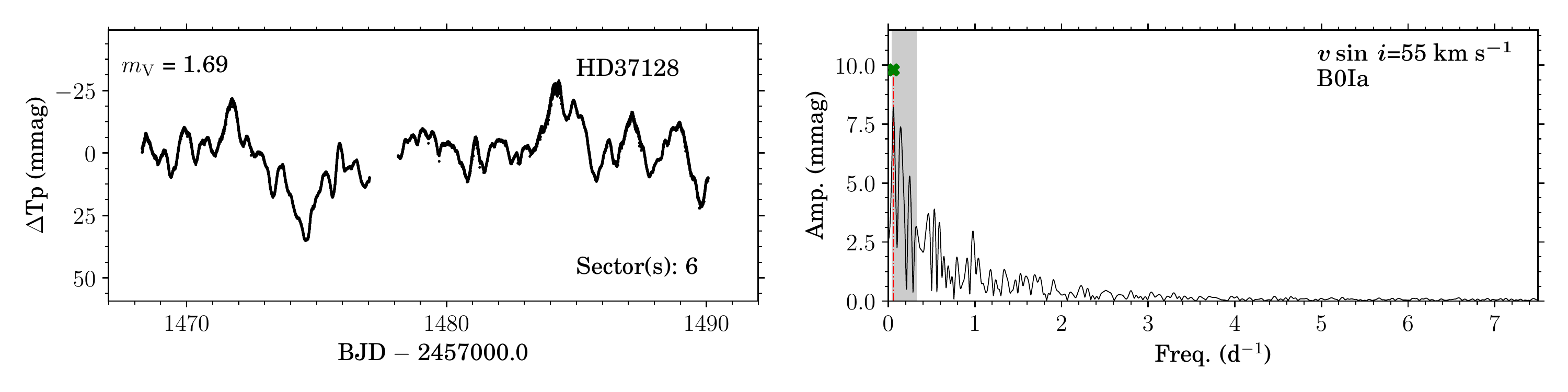}
    \caption{TESS light curve and LS-periodogram of HD\,37128. Same figure style as Fig~\ref{fig:PG_LINE_Bcep}.}
    \label{fig:LC+PG_HD37128}
\end{figure*}

HD~37128 ($\epsilon$~Ori, B0\,Ia \citealp{Lesh1968}) is a blue supergiant with a well-studied variable wind \citep{Blomme2002, Prinja2004, Crowther2006, David-Uraz2014, Puebla2016, Krticka2018}. \citet{Prinja2004} find three spectroscopic periodicities (1.9, 6.5, and 9.5~d) but could not fully constrain their origin. The one-sector TESS data set shows stochastic low frequency variability. Given the estimated rotational modulation frequency range, the dominant frequencies could be related to rotational effects. H$_{\alpha}$ shows variable emission in all (randomly selected) spectroscopic epochs, showing either a variable wind or a static wind combined with rotation. The He~{\sc i}~$\lambda$5875 line shows line profile variability, that is likely to have a similar origin as the variability in the photometry. This line is formed partially in the wind such that we can expect the stellar wind of $\epsilon$~Ori to play a significant role in its variability, much like in B1Ia star HD~2905 \citep{SimonDiaz2018}, and in O9.5Iab star HD~188209 \citep{Aerts2017a}.

 \begin{figure*}
	\includegraphics[width=2\columnwidth, scale = 1]{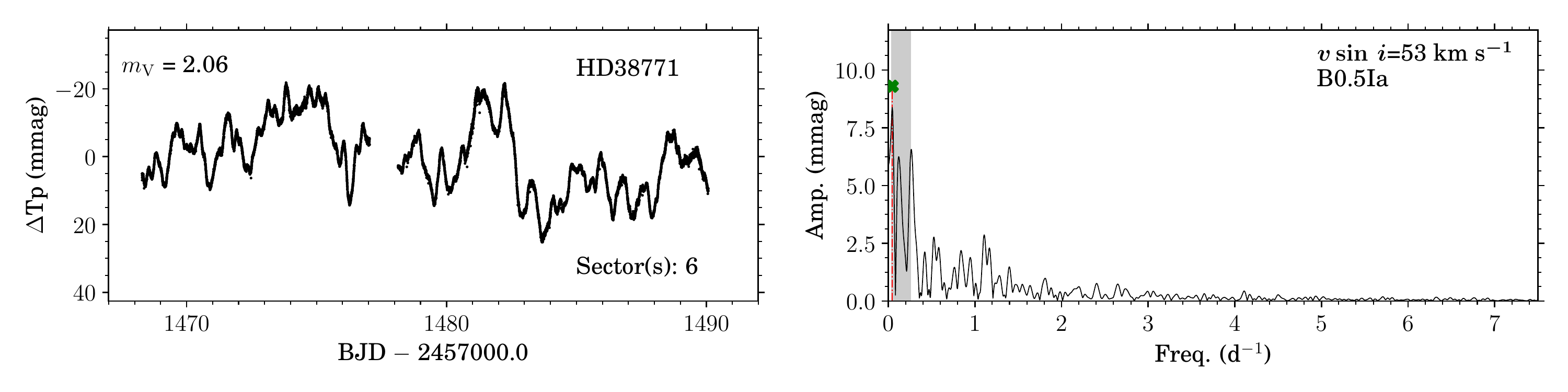}
    \caption{TESS light curve and LS-periodogram of HD\,38771. Same figure style as Fig~\ref{fig:PG_LINE_Bcep}.}
    \label{fig:LC+PG_HD38771}
\end{figure*}

HD~38771, ($\kappa$~Ori, B0.5\,Ia, \citealp{Lesh1968}) is a bright blue supergiant with a strongly variable wind (\citealt{Haucke2018}, and references therein) and a variable magnetic field \citep{Nerney1980}. \citet{Morel2004} finds two photometric periodicities ($f\approx0.21$ and $f\approx0.96$~d$^{-1}$). We note variability in the wind lines in our spectroscopic data set. The dominant variability type in the one-sector TESS data set is stochastic low frequency variability. The dominant frequencies fall inside the estimated rotational modulation frequency range and could therefore be related to (quasi-)cyclical wind variations.

 \begin{figure*}
	\includegraphics[width=2\columnwidth, scale = 1]{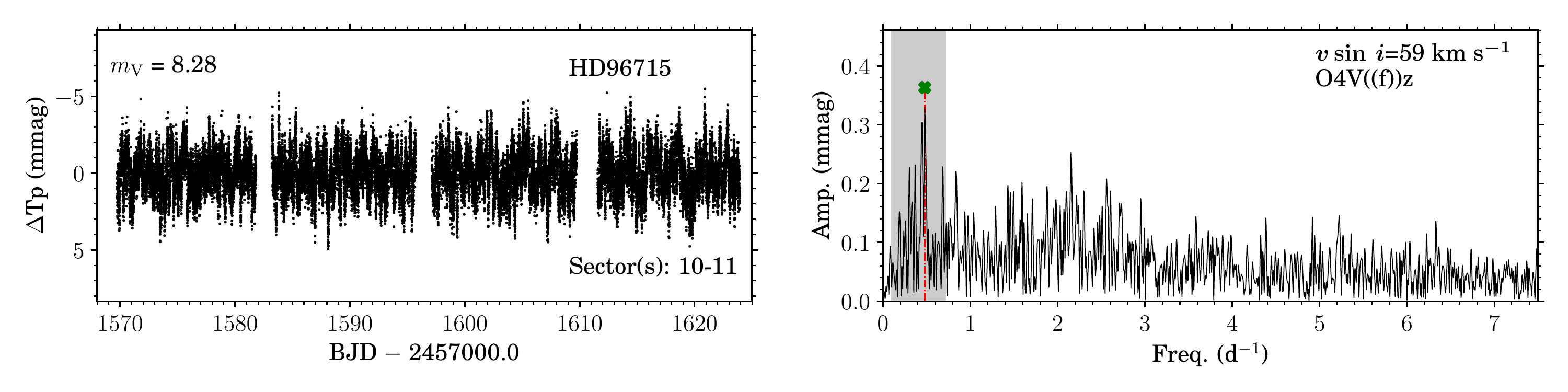}
    \caption{TESS light curve and LS-periodogram of HD\,96715. Same figure style as Fig~\ref{fig:PG_LINE_Bcep}.}
    \label{fig:LC+PG_HD96715}
\end{figure*}

HD~96715 is an O4\,V((f))z star \citep{Sota2014}. The two-sector TESS data set shows stochastic low frequency variability.

 \begin{figure*}
	\includegraphics[width=2\columnwidth, scale = 1]{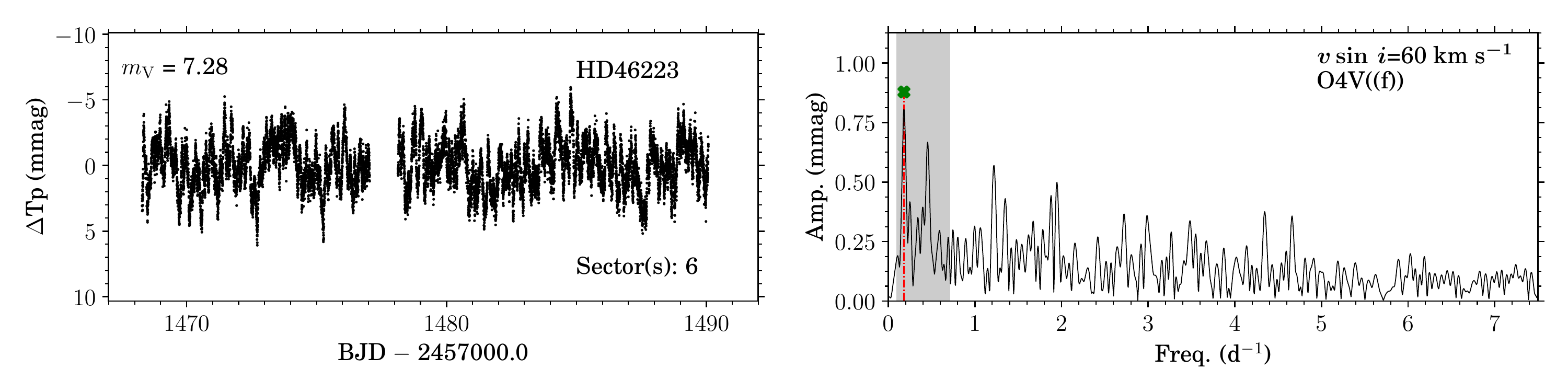}
    \caption{TESS light curve and LS-periodogram of HD\,46223. Same figure style as Fig~\ref{fig:PG_LINE_Bcep}.}
    \label{fig:LC+PG_HD46223}
\end{figure*}

HD~46223 is an O4\,V((f)) star \citep{Sota2011}. The star is characterised by stochastic low frequency variability in its 34~d CoRoT light curve, but no coherent pulsation frequencies were detected \citep{Blomme2011, Bowman2019a}. The variability is again dominantly stochastic in the one-sector TESS data set. This variability has been interpreted as a sign of IGWs triggered by core convection \citep{Aerts2015b, Bowman2019a}.

 \begin{figure*}
	\includegraphics[width=2\columnwidth, scale = 1]{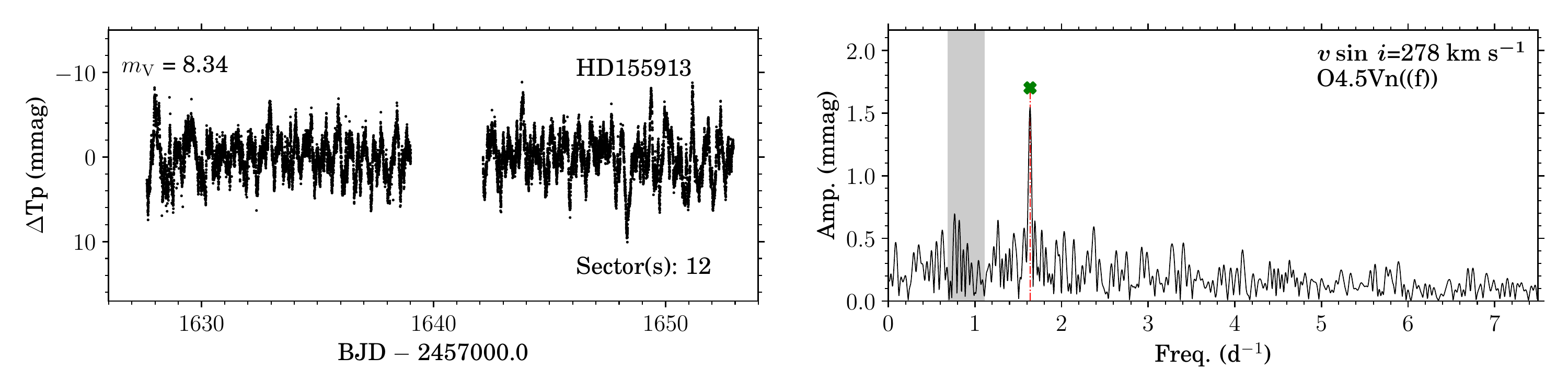}
    \caption{TESS light curve and LS-periodogram of HD\,155913. Same figure style as Fig~\ref{fig:PG_LINE_Bcep}.}
    \label{fig:LC+PG_HD155913}
\end{figure*}

HD~155913 is an O4.5\,Vn((f)) star \citep{Sota2014}. The periodogram of the one-sector TESS data shows stochastic low frequency variability. The dominant variability is too high to be related to rotational period (even with a projected rotational velocity of v$\,\sin\,i=278$~km~s$^{-1}$), and could be pulsational in origin. 

 \begin{figure*}
	\includegraphics[width=2\columnwidth, scale = 1]{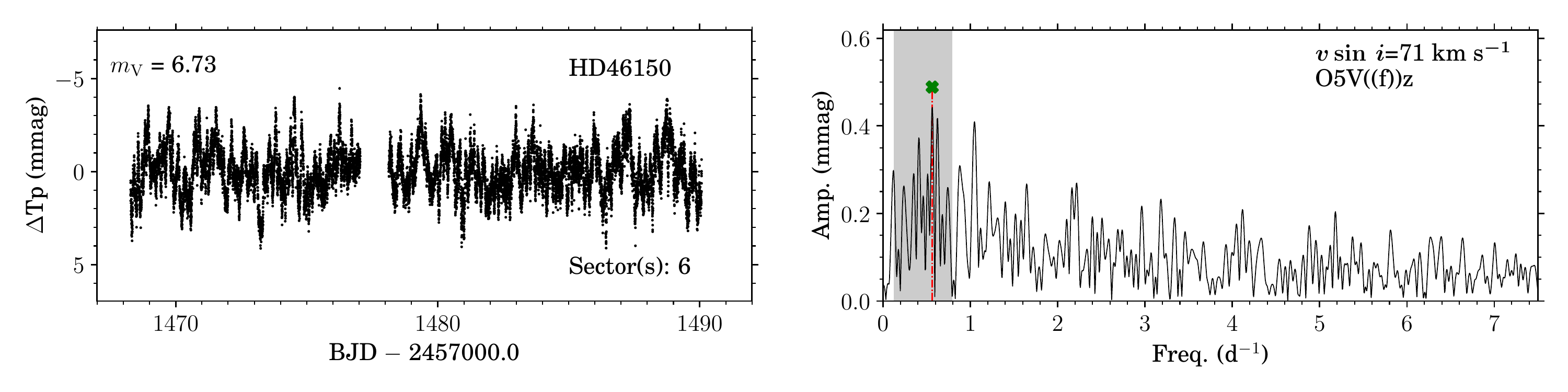}
    \caption{TESS light curve and LS-periodogram of HD\,46150. Same figure style as Fig~\ref{fig:PG_LINE_Bcep}.}
    \label{fig:LC+PG_HD46150}
\end{figure*}

HD~46150 is O5\,V((f))z star \citep{Sota2011}. Its variability was first analysed by \citet{Blomme2011} who noted considerable stochastic low frequency variability in the 34~d CoRoT data set. The variability has been interpreted as a sign of IGWs excited by core convection \citep{Aerts2015b, Bowman2019a}. The one-sector TESS data set shows stochastic low frequency variability over a broad range which agrees with previous findings.

 \begin{figure*}
	\includegraphics[width=2\columnwidth, scale = 1]{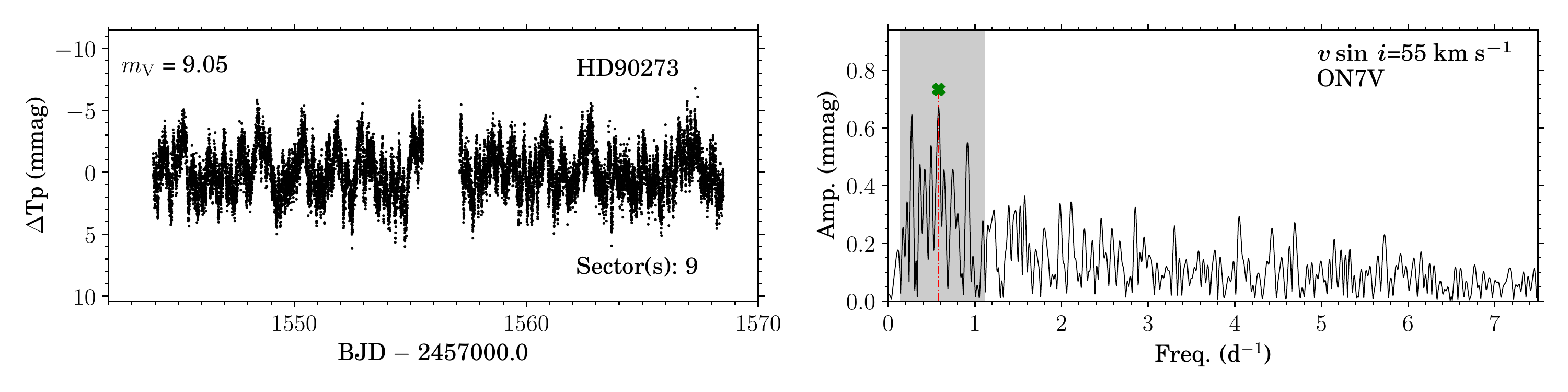}
    \caption{TESS light curve and LS-periodogram of HD\,90273. Same figure style as Fig~\ref{fig:PG_LINE_Bcep}.}
    \label{fig:LC+PG_HD90273}
\end{figure*}

HD~90273 is an ON7\,V((f)) star \citep{Maiz2016}. The one-sector TESS data set shows a broad range of stochastic low frequency variability.

 \begin{figure*}
	\includegraphics[width=2\columnwidth, scale = 1]{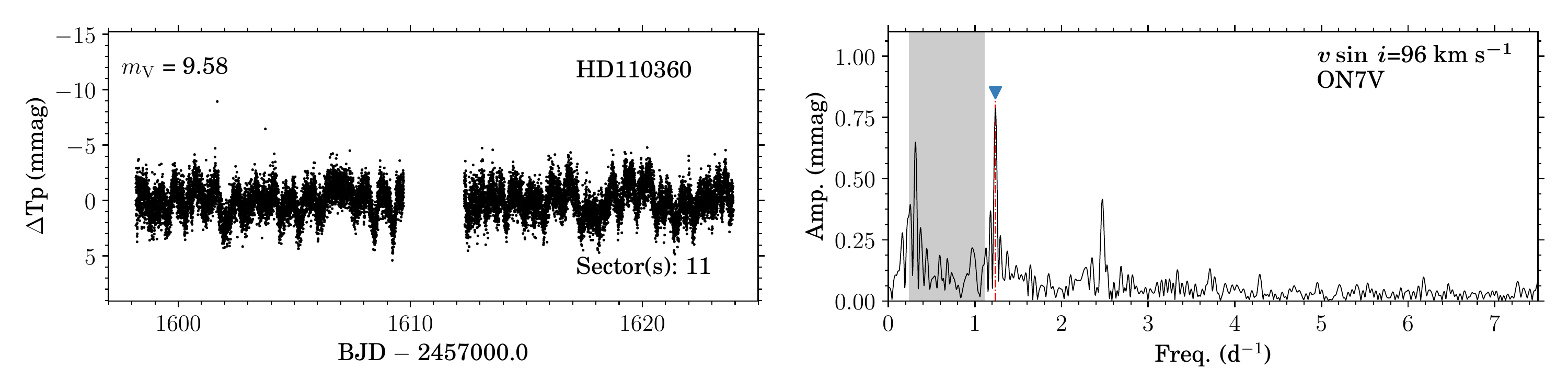}
    \caption{TESS light curve and LS-periodogram of HD\,110360. Same figure style as Fig~\ref{fig:PG_LINE_Bcep}.}
    \label{fig:LC+PG_HD110360}
\end{figure*}

HD~110360 is an ON7\,V star \citep{Maiz2016}. One significant frequency is detected in the one-sector TESS data set, $\nu=1.2372(6)$~d$^{-1}$.  The detected frequency is much higher than the upper bound of the predicted rotational frequency range (v$\,\sin\,i=96$~km~s$^{-1}$).

 \begin{figure*}
	\includegraphics[width=2\columnwidth, scale = 1]{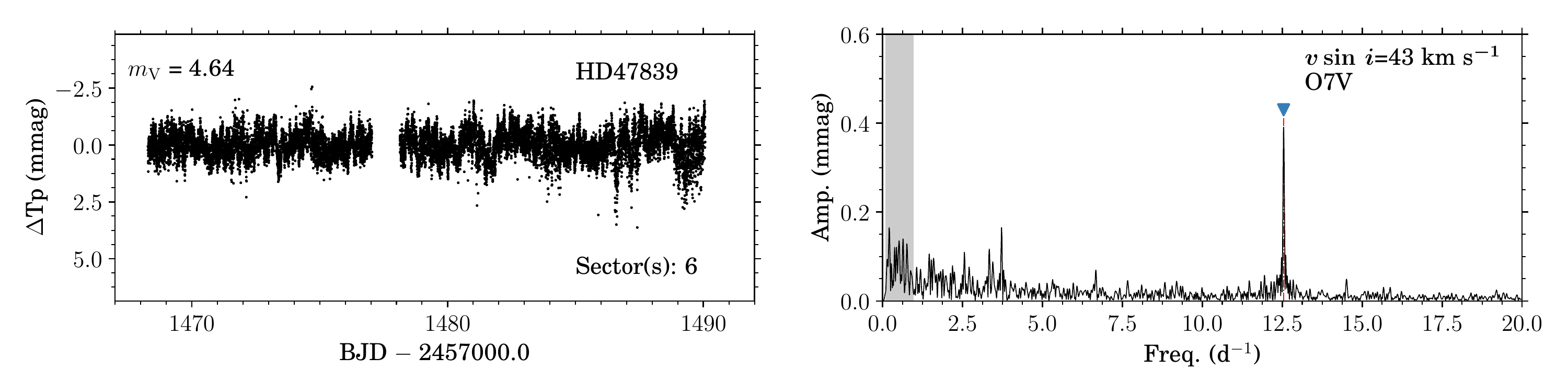}
    \caption{TESS light curve and LS-periodogram of HD\,47839. Same figure style as Fig~\ref{fig:PG_LINE_Bcep}.}
    \label{fig:LC+PG_HD47839}
\end{figure*}

HD~47839 (15~Mon) is a well-known long-period binary system composed of a O7\,V star and a fast-rotating B1.5\,V star \citep{Gies1987, Maiz2018a}. A third wider component is also confirmed \citep{Maiz2010, Maiz2018a}. All components of the system contribute light in the TESS pixels. As alluded to in Sect.~\ref{sec:BSG}, the measured high frequency mode ($\nu=12.5435(7)$~d$^{-1}$) in the one-sector TESS data set likely belongs to the fast-rotating B\,1.5/2V star in the system. Several other nearby stars are blended into the TESS pixels, including potential short-period pulsators, for example, HD~261937, HD~261938, and BD+10~1222, but given their increased magnitudes they are less likely to be the source of the measured frequency.

\begin{figure*}
	\includegraphics[width=2\columnwidth, scale = 1]{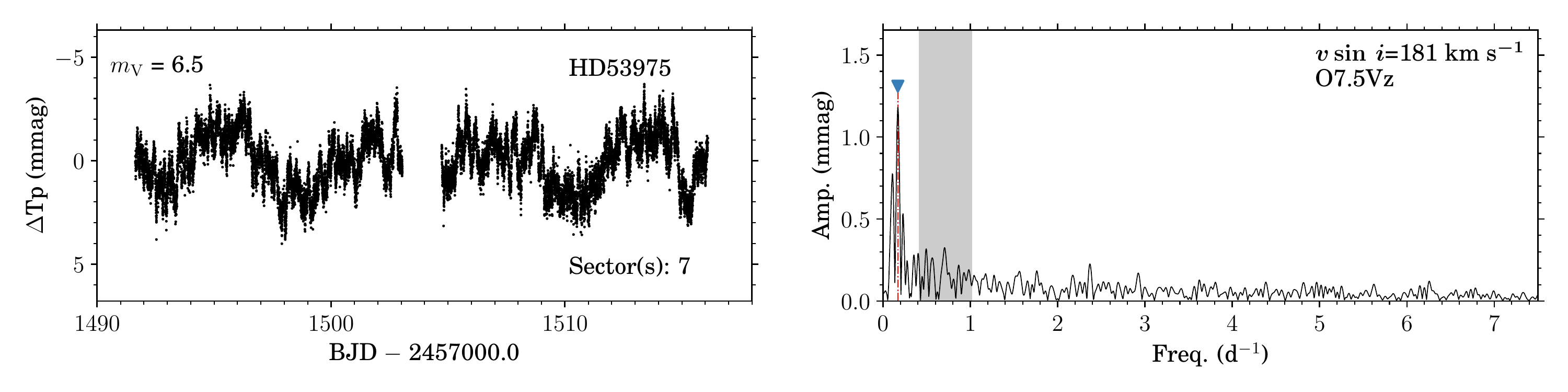}
    \caption{TESS light curve and LS-periodogram of HD\,53975. Same figure style as Fig~\ref{fig:PG_LINE_Bcep}.}
    \label{fig:LC+PG_HD53975}
\end{figure*}

HD~53975 (O7.5\,Vz, \citealp{Sota2011}) is a spectroscopic binary but only a single component has been detected \citep{Gies1994, Maiz2018a}. A low frequency ($\nu=0.1700(5)$~d$^{-1}$) is measured in the one-sector TESS data set, causing a large cyclical light variation. It is below the low boundary of the estimated rotational modulation frequency range (v$\,\sin\,i=181$~km~s$^{-1}$). The frequency is similar to the binary period measured by \citet{Gies1994}, $P=6.0173$~d, and the dominant variability is likely due to a reflection effect. Stochastic low frequency variability is also present.

\begin{figure*}
	\includegraphics[width=2\columnwidth, scale = 1]{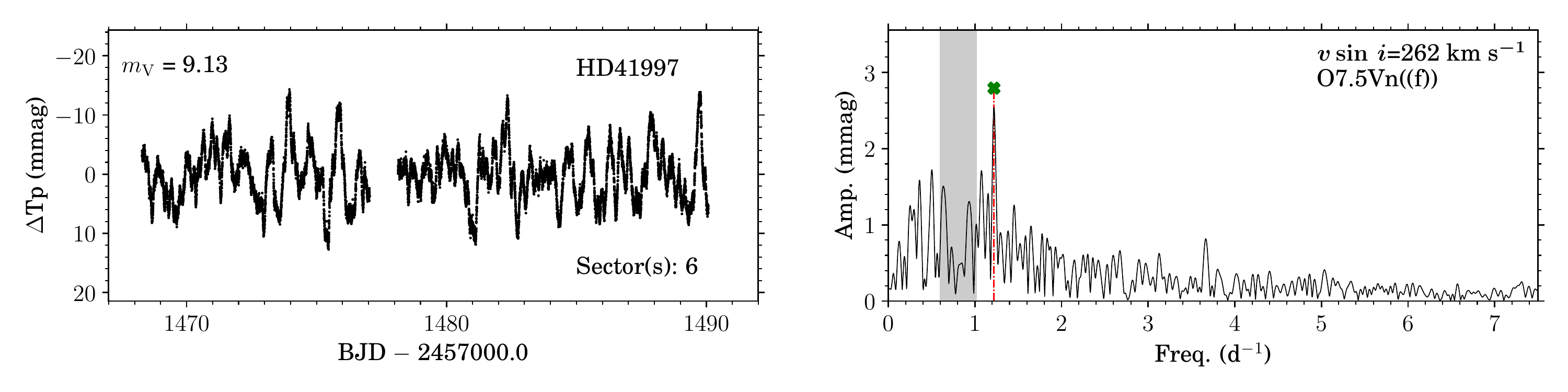}
    \caption{TESS light curve and LS-periodogram of HD\,41997. Same figure style as Fig~\ref{fig:PG_LINE_Bcep}.}
    \label{fig:LC+PG_HD41997}
\end{figure*}

HD~41997 (O7.5\,Vn((f)), \citealp{Sota2011}) is a runaway star \citep{Maiz2018b}. Stochastic low frequency variability is the dominant source of variability in the light curve.

\begin{figure*}
	\includegraphics[width=2\columnwidth, scale = 1]{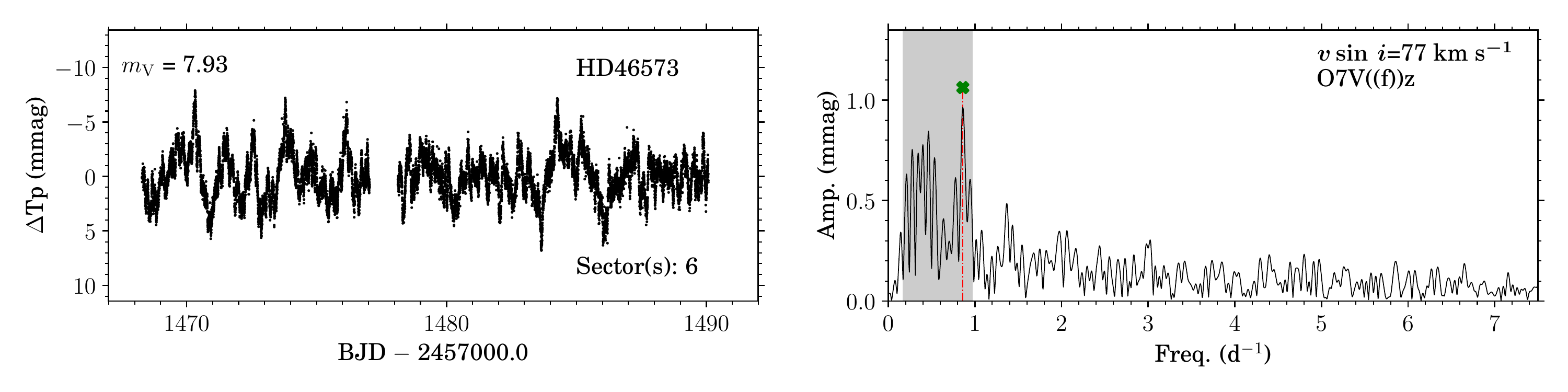}
    \caption{TESS light curve and LS-periodogram of HD\,46573. Same figure style as Fig~\ref{fig:PG_LINE_Bcep}.}
    \label{fig:LC+PG_HD46573}
\end{figure*}

HD~46573 (O7\,V((f))z, \citealp{Sota2011}) is a young star in the Monoceros OB2 association \citep{Mahy2012, Martins2012}. It is a confirmed spectroscopic binary but no companion has been detected \citep{Mason2009, Martins2012, Sana2012}. The one-sector TESS data set shows stochastic low frequency variability.

\begin{figure*}
	\includegraphics[width=2\columnwidth, scale = 1]{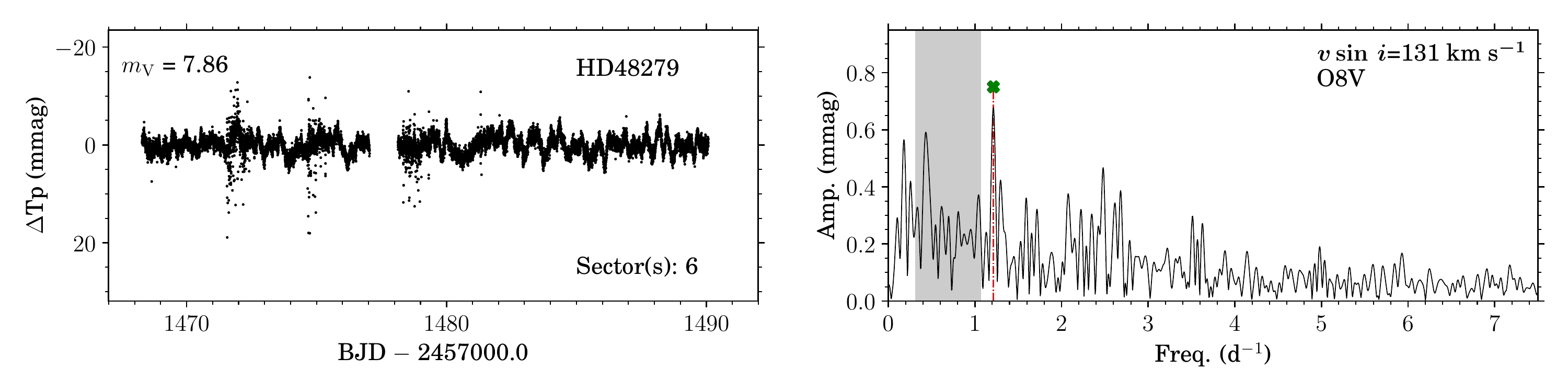}
    \caption{TESS light curve and LS-periodogram of HD\,48279. Same figure style as Fig~\ref{fig:PG_LINE_Bcep}.}
    \label{fig:LC+PG_HD48279}
\end{figure*}

HD~48279 (O8\,V, \citealp{Lesh1968}) is a runaway star \citep{Gvaramadze2012}. It was classified as a binary system by \citet{Meisel1968}, but this remains unconfirmed. \citet{Mahy2009} argued that the likelihood was low based on the measured constant radial velocity, which was later supported by \citet{Martins2015b}. The dominant type of variability in the one-sector TESS data set is stochastic low frequency variability.

\begin{figure*}
	\includegraphics[width=2\columnwidth, scale = 1]{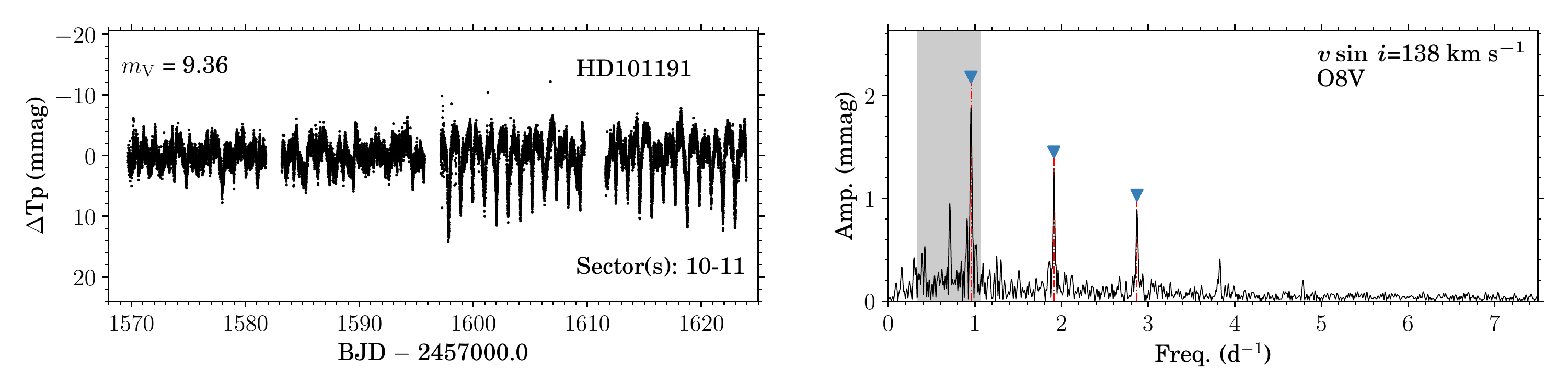}
    \caption{TESS light curve and LS-periodogram of HD\,101191. Same figure style as Fig~\ref{fig:PG_LINE_Bcep}.}
    \label{fig:LC+PG_HD101191}
\end{figure*}

HD~101191 (O8\,V, \citealp{Sota2014}) is a spectroscopic binary \citep{Mason2009, Chini2012}. The TESS pixels are contaminated by nearby eclipsing binary HD~101205 \citep{Balona1992, Mayer1992} which dominates the signal.

\begin{figure*}
	\includegraphics[width=2\columnwidth, scale = 1]{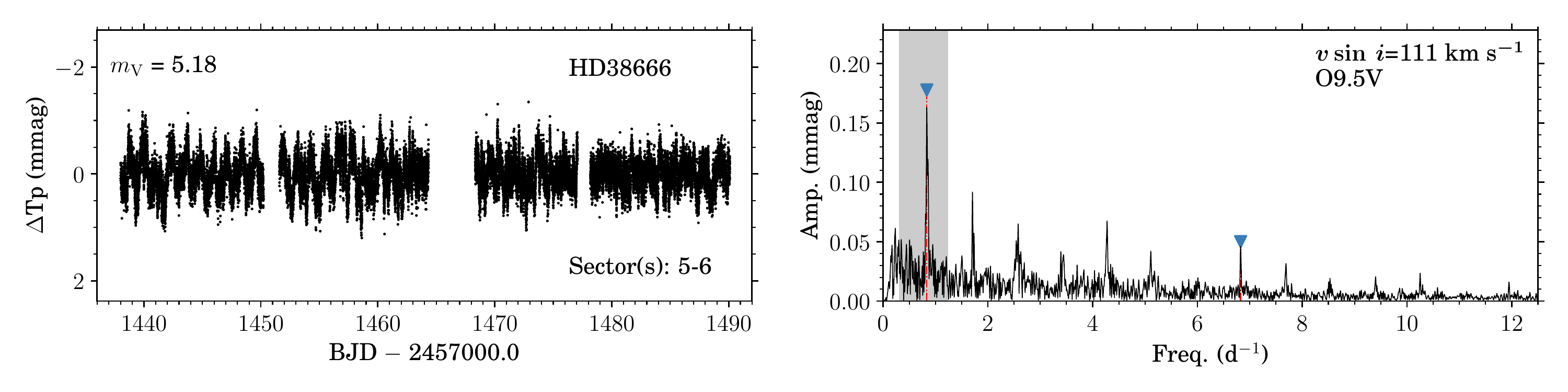}
    \caption{TESS light curve and LS-periodogram of HD\,38666. Same figure style as Fig~\ref{fig:PG_LINE_Bcep}.}
    \label{fig:LC+PG_HD38666}
\end{figure*}

HD~38666 ($\mu$~Col, O9.5\,V, \citealp{Sota2014}) is a well-known run-away star, thought to be the product of a binary supernova ejection scenario in the Orion Nebula \citep{Blaauw1954, Hoogerwerf2000, Hoogerwerf2001}. The periodogram of the TESS light curve shows a low amplitude dominant frequency ($\nu=0.8349(3)$~d$^{-1}$) with many harmonics up to $12$~d$^{-1}$. The origin of this harmonic comb is likely related to instrumental effects.

\begin{figure*}
	\includegraphics[width=2\columnwidth, scale = 1]{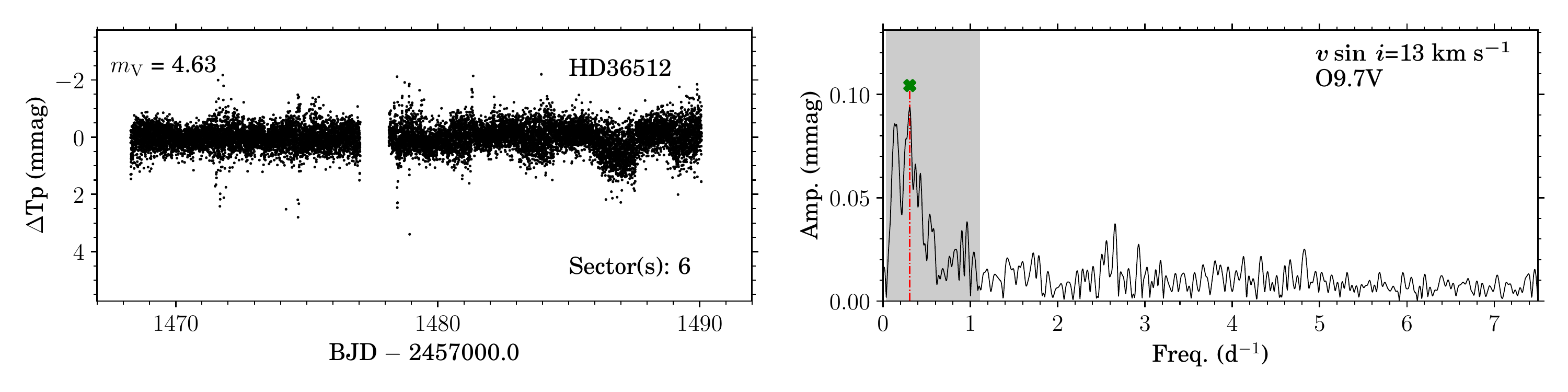}
    \caption{TESS light curve and LS-periodogram of HD\,36512. Same figure style as Fig~\ref{fig:PG_LINE_Bcep}.}
    \label{fig:LC+PG_HD36512}
\end{figure*}

HD~36512 ($\upsilon$~Ori) is an O9.7\,V star \citep{Sota2011}. \citet{Smith1981} detected  of non-radial pulsations based on line-profile variations over a period of 3~yr of time series spectroscopy. They report a period of $\sim0.5$~d, and during one instance a period of approximately $1$~d. \citet{Balona1985} detected two frequencies, 3.762~d$^{-1}$ and 2.995~d$^{-1}$, and therefore classified it as a $\beta$~Cep variable. The one-sector TESS data set shows stochastic low frequency variability but of much lower amplitude than similar stars of this spectral type. We note that the TESS light curve of HD~36512 shows instrumental effects related to the 2.5~d momentum wheel dumps (Fig~\ref{fig:LC+PG_HD36512}).

\begin{figure*}
	\includegraphics[width=2\columnwidth, scale = 1]{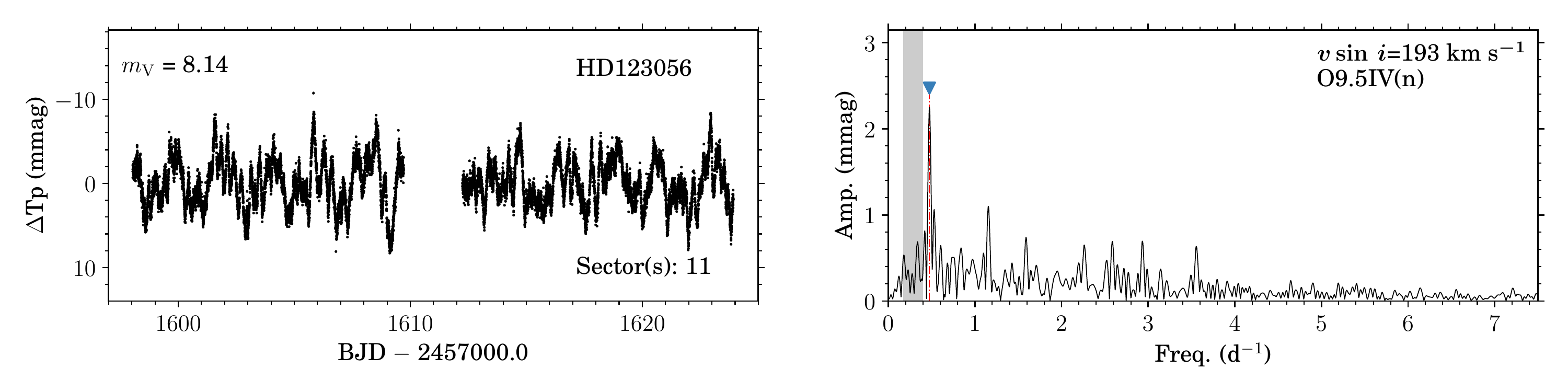}
    \caption{TESS light curve and LS-periodogram of HD\,123056. Same figure style as Fig~\ref{fig:PG_LINE_Bcep}.}
    \label{fig:LC+PG_HD123056}
\end{figure*}

HD~123056 (O9.5\,IV(n), \citealp{Sota2014}) is a multiple system with three confirmed components, a high-mass O star orbiting in long period ($\sim1314$~d) around a short-period inner pair ($<2$~d) of early B stars \citep{Mayer2017}. It is found to be SB1 in the IACOB/OWN spectroscopy. The one-sector TESS data set shows stochastic low frequency variability, and the dominant frequency satisfies our significance criterion, $\nu=0.4753(7)$~d$^{-1}$. This approximately corresponds to the short-period found by \citet{Mayer2017} but from the light curve it is unclear whether this dominant periodicity is due to binarity.

\begin{figure*}
	\includegraphics[width=2\columnwidth, scale = 1]{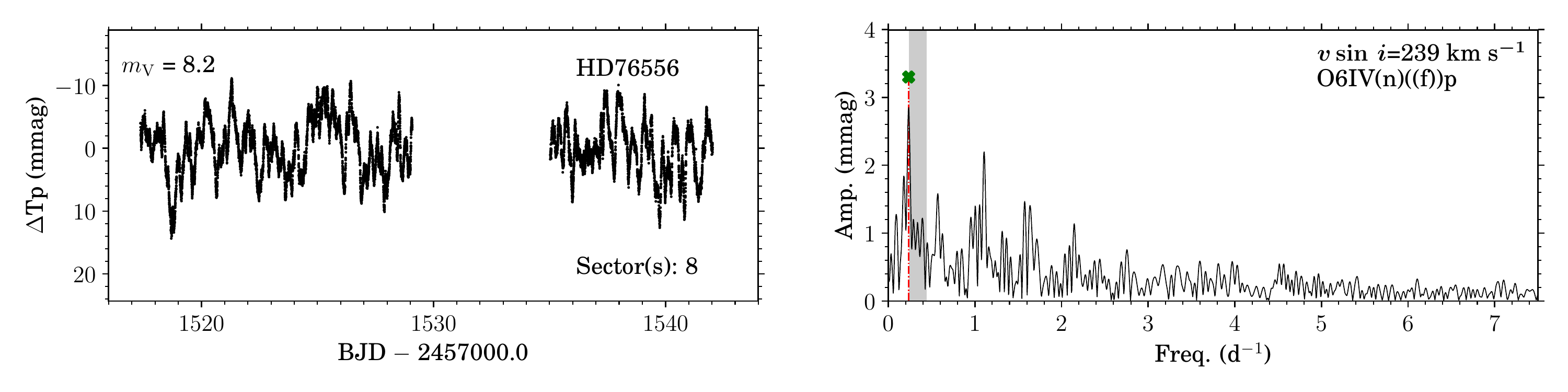}
    \caption{TESS light curve and LS-periodogram of HD\,76556. Same figure style as Fig~\ref{fig:PG_LINE_Bcep}.}
    \label{fig:LC+PG_HD76556}
\end{figure*}

HD~76556 (O6\,IV(n)((f))p, \citealp{Sota2014}) is a long-period double-lined spectroscopic binary \citep{Chini2012}. We attribute the dominant variability type in the one-sector TESS data set to stochastic low frequency variability.

\begin{figure*}
	\includegraphics[width=2\columnwidth, scale = 1]{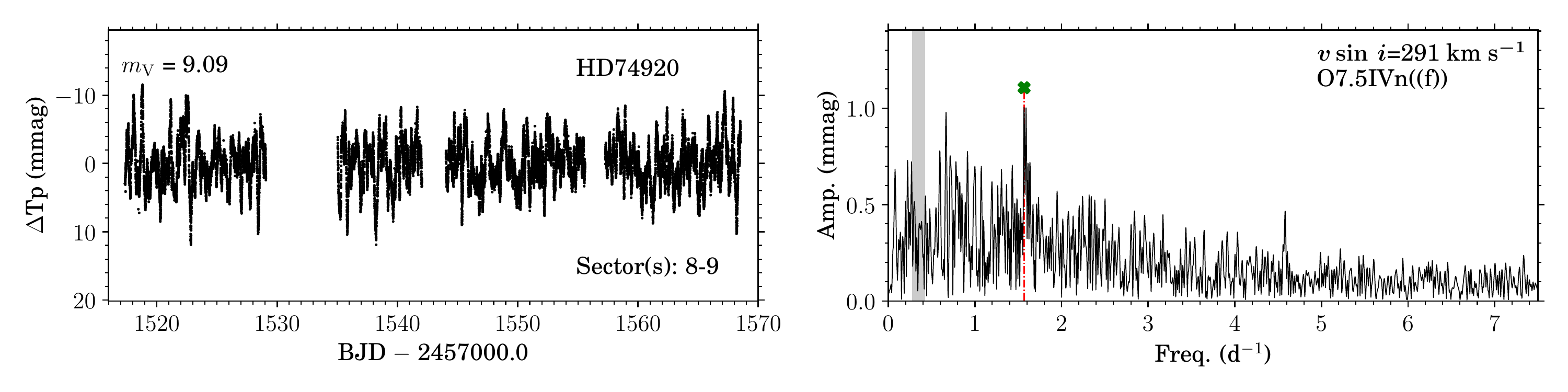}
    \caption{TESS light curve and LS-periodogram of HD\,74920. Same figure style as Fig~\ref{fig:PG_LINE_Bcep}.}
    \label{fig:LC+PG_HD74920}
\end{figure*}

HD~74920 (O7.5\,IVn((f)), \citealp{Sota2014}) is a potential runaway candidate \citep{Maiz2018a}. Stochastic low frequency variability is the dominant type of variability in the two-sector TESS data set.

\begin{figure*}
	\includegraphics[width=2\columnwidth, scale = 1]{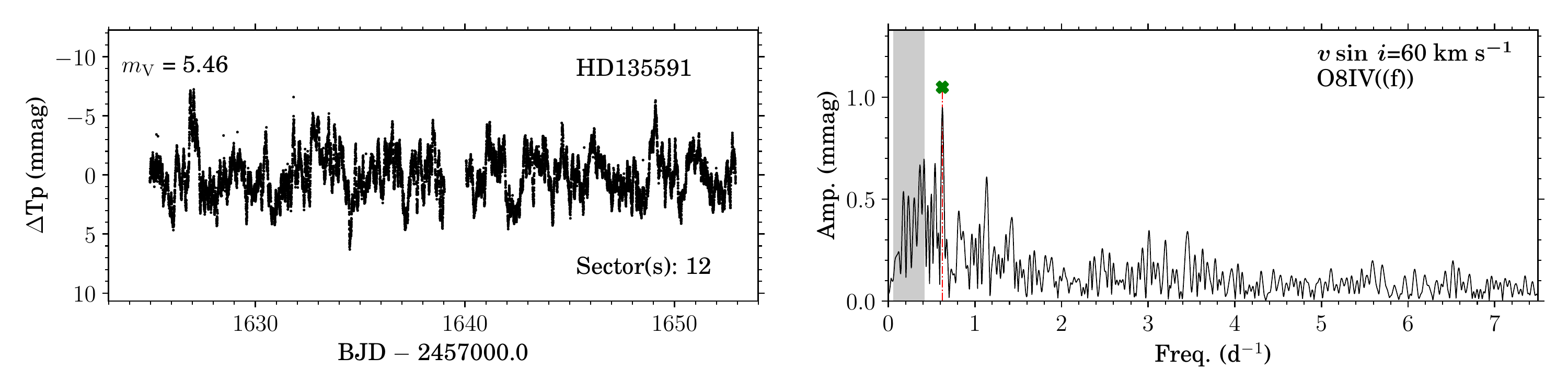}
    \caption{TESS light curve and LS-periodogram of HD\,135591. Same figure style as Fig~\ref{fig:PG_LINE_Bcep}.}
    \label{fig:LC+PG_HD135591}
\end{figure*}

HD~135591 (O8\,IV((f)), \citealp{Sota2014}) is a visual binary system \citep{Eggleton2008, Sana2014}. The dominant type of variability in the one-sector TESS data set is stochastic low frequency variability.

\begin{figure*}
	\includegraphics[width=2\columnwidth, scale = 1]{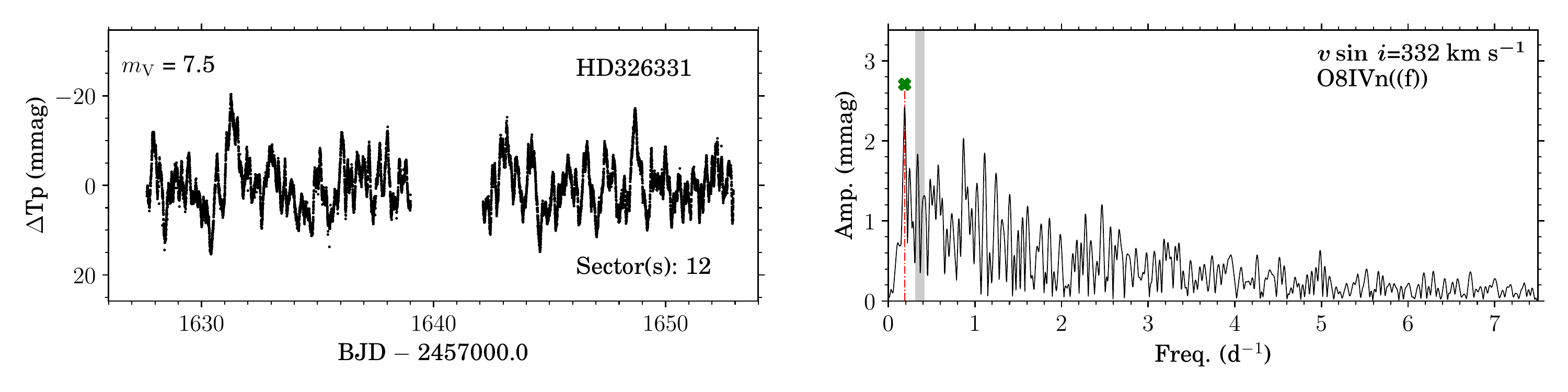}
    \caption{TESS light curve and LS-periodogram of HD\,326331. Same figure style as Fig~\ref{fig:PG_LINE_Bcep}.}
    \label{fig:LC+PG_HD326331}
\end{figure*}

HD~326331 is an O8\,IVn((f)) star \citep{Sota2014}. It was found to be a broad-line fast rotator with LPV by \citet{Sana2008}. A companion was found by \citet{Mason1998}. The one-sector TESS data set shows stochastic low frequency. We find wind variability in the IACOB/OWN spectroscopy, which could be partly responsible for (or caused by) the observed photometric variability.

\begin{figure*}
	\includegraphics[width=2\columnwidth, scale = 1]{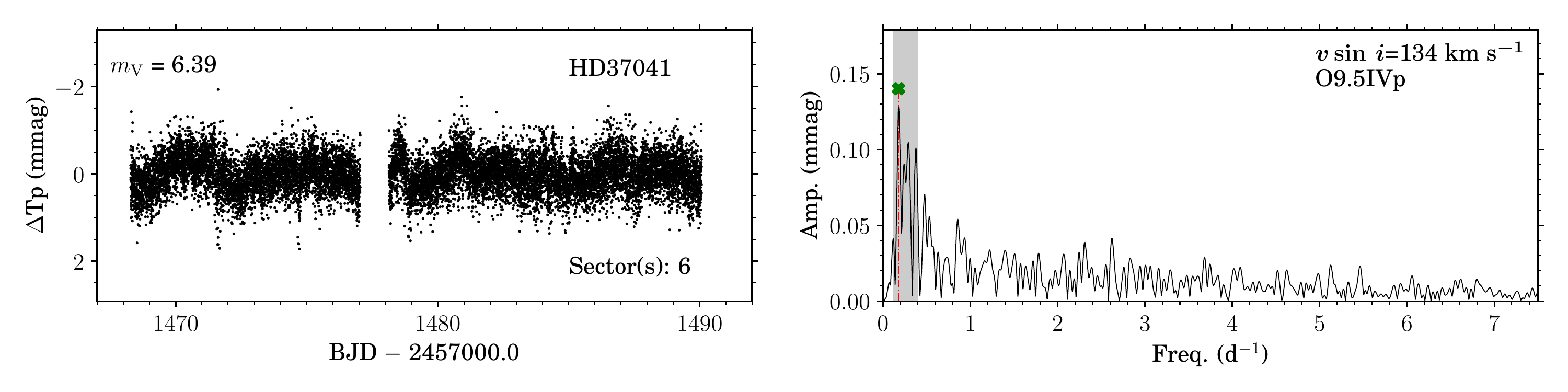}
    \caption{TESS light curve and LS-periodogram of HD\,37041. Same figure style as Fig~\ref{fig:PG_LINE_Bcep}.}
    \label{fig:LC+PG_HD37041}
\end{figure*}

HD~37041 (O9.5\,IVp,  \citealp{Sota2011}) was found to be a SB2 with a secondary B0.5\,V star by \citet{Maiz2019}. We find it as SB1 in the IACOB/OWN spectroscopy. The one-sector TESS data set shows stochastic low frequency variability.

\begin{figure*}
	\includegraphics[width=2\columnwidth, scale = 1]{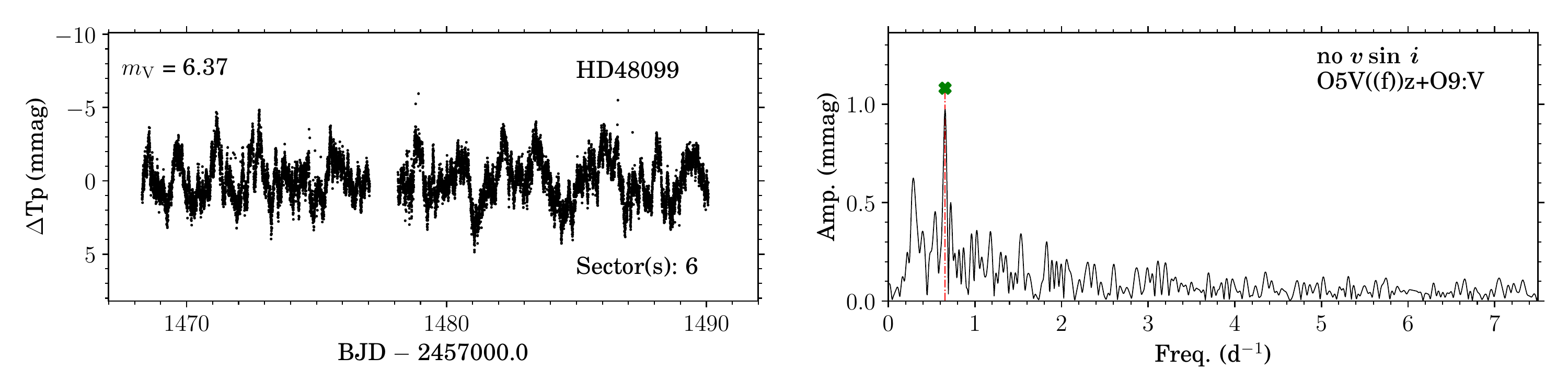}
    \caption{TESS light curve and LS-periodogram of HD\,48099. Same figure style as Fig~\ref{fig:PG_LINE_Bcep}.}
    \label{fig:LC+PG_HD48099}
\end{figure*}

HD~48099 is a double O-star system (O5V((f))z+O9:V, \citealp{Sota2014}) with a 3.078098~d period first measured by \citet{Stickland1998}. The one-sector TESS data set is characterised by stochastic low frequency variability.

\begin{figure*}
	\includegraphics[width=2\columnwidth, scale = 1]{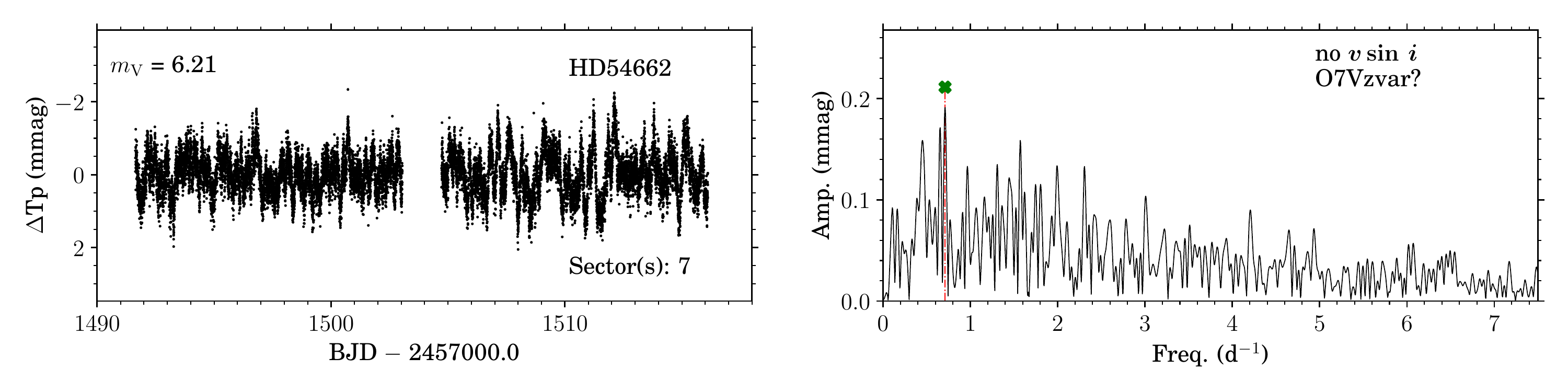}
    \caption{TESS light curve and LS-periodogram of HD\,54662. Same figure style as Fig~\ref{fig:PG_LINE_Bcep}.}
    \label{fig:LC+PG_HD54662}
\end{figure*}

HD~54662 (O7\,Vzvar?, \citealp{Sota2014}) is a binary system with a full interferometric orbital solution \citep{Sana2014, LeBouquin2017}. It is found as SB2 in the IACOB/OWN spectroscopy. A recent investigation by \citet{Mossoux2018} found an orbital period of 2103.4~d, an eccentricity of 0.11, and a mass ratio of 0.84. The one-sector TESS data set is dominated by stochastic low frequency variability with likely contributions from both components in this system.

\begin{figure*}
	\includegraphics[width=2\columnwidth, scale = 1]{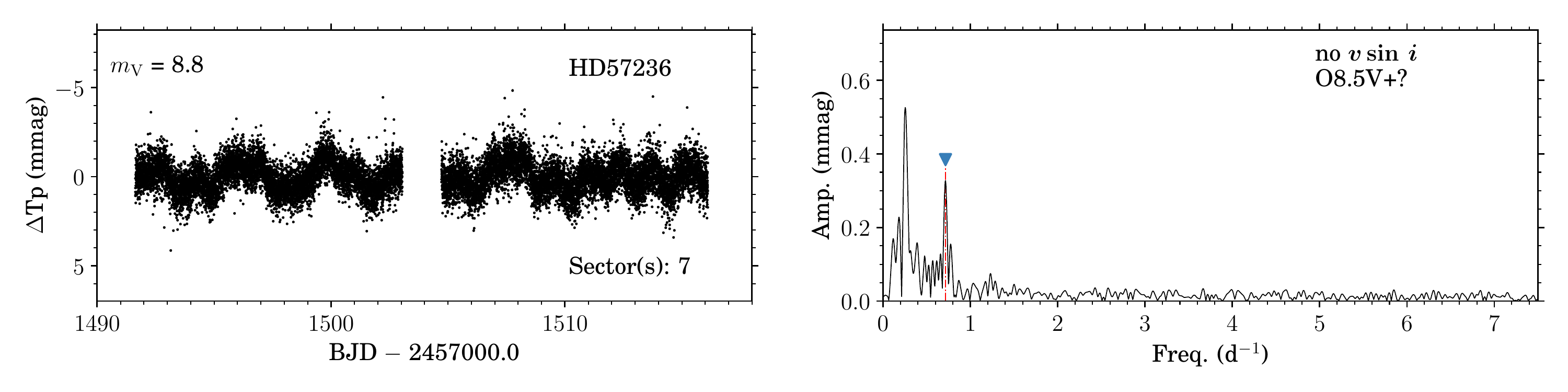}
    \caption{TESS light curve and LS-periodogram of HD\,57236. Same figure style as Fig~\ref{fig:PG_LINE_Bcep}.}
    \label{fig:LC+PG_HD57236}
\end{figure*}

HD~57236 is an O8\,V star \citep{Sota2014}, which we find as SB2 in the IACOB/OWN spectroscopy. We find stochastic low frequency in the TESS data set and only one frequency that satisfies our significance criterion, $\nu=0.7155(6)$~d$^{-1}$.

\begin{figure*}
	\includegraphics[width=2\columnwidth, scale = 1]{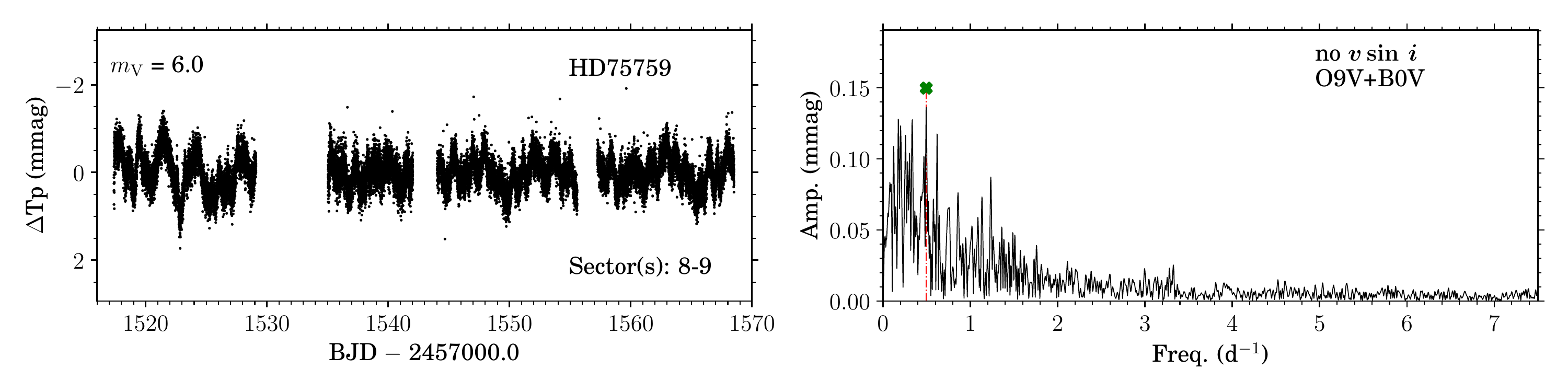}
    \caption{TESS light curve and LS-periodogram of HD\,75759. Same figure style as Fig~\ref{fig:PG_LINE_Bcep}.}
    \label{fig:LC+PG_HD75759}
\end{figure*}

HD~75759 (O9\,V$+$B0\,V, \citealp{Sota2014}) is a binary system with a period of about $P\approx30$~d \citep{Sana2014}. We detect stochastic low frequency variability in the two-sector TESS data set.

\begin{figure*}
	\includegraphics[width=2\columnwidth, scale = 1]{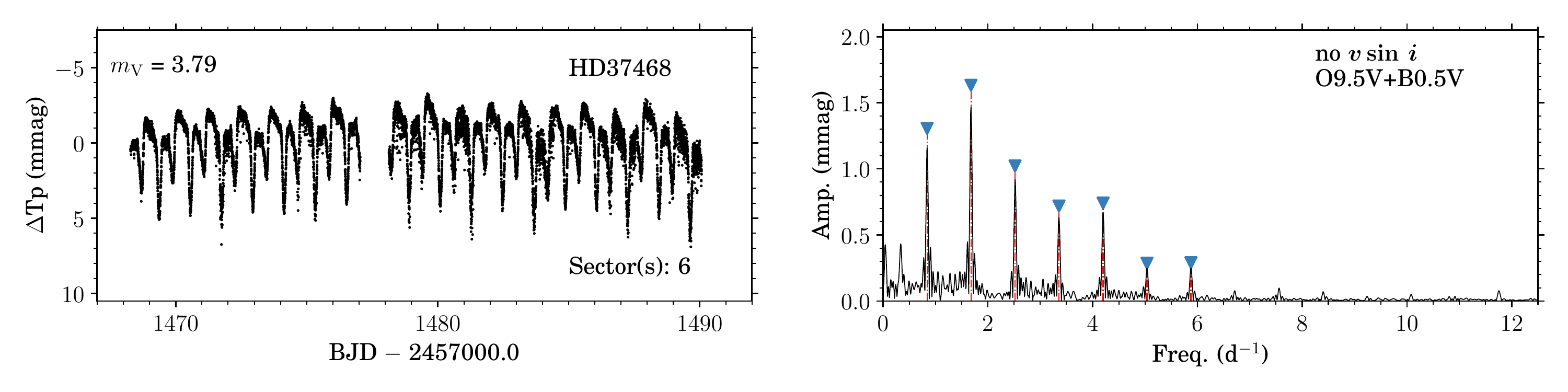}
    \caption{TESS light curve and LS-periodogram of HD\,37468. Same figure style as Fig~\ref{fig:PG_LINE_Bcep}.}
    \label{fig:LC+PG_HD37468}
\end{figure*}

HD 37468 refers to $\sigma$\,Ori\,AaAbB hierarchical massive triple system, including one late-O dwarf and two early-B dwarfs, as described by \citet{Simon-Diaz2011c, SimonDiaz2015a, Maiz2018a}. Furthermore, this system is part of the so-called $\sigma$\,Orionis cluster which also includes another 2 early B-type stars within 1$\arcsec$. Hence, the TESS pixels do not only include multiple stars, but are also heavily contaminated by the close-by magnetic B2\,Vp star HD~37479 ($\sigma$~Ori E), discussed in Appendix~\ref{sec:appendix_magnetic}.

\begin{figure*}
	\includegraphics[width=2\columnwidth, scale = 1]{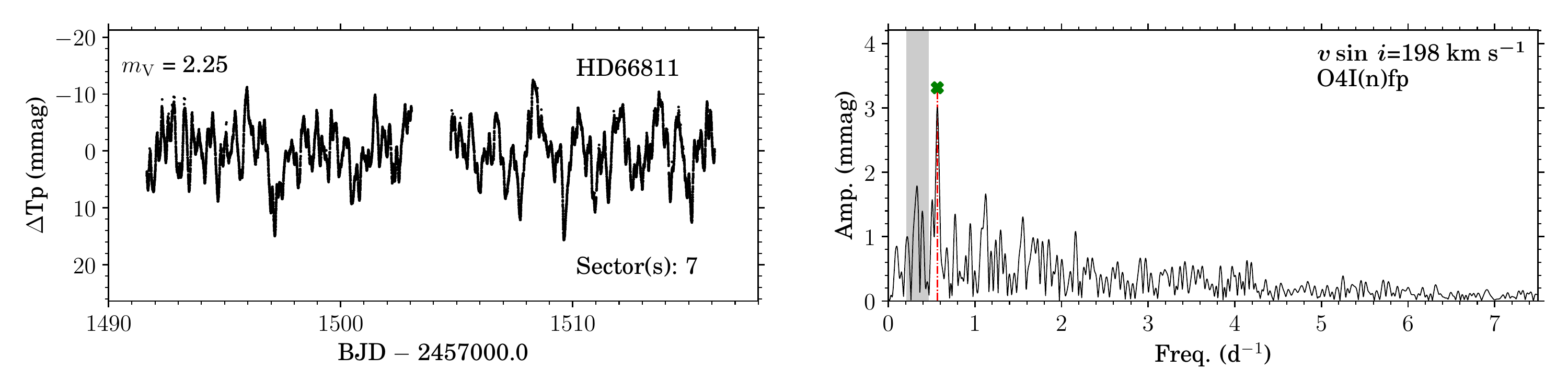}
    \caption{TESS light curve and LS-periodogram of HD\,66811. Same figure style as Fig~\ref{fig:PG_LINE_Bcep}.}
    \label{fig:LC+PG_HD66811}
\end{figure*}

HD~66811 ($\zeta$~Pup, O4\,I(n)fp, \citealp{Sota2014} is known to be variable. Using 5.5~months of BRITE photometry and multi-epoch spectroscopy \citet{Ramiaramanantsoa2018a} unveil two dominant types of variability: a 1.78~d period that is attributed to rotational modulation caused by spots on the surface, and a stochastic component that is attributed to wind clumping which is already occurring at the base of the wind.  The 1.78~d period is recovered in the one-sector TESS data set ($\nu\approx0.566$~d$^{-1}$, or $P\approx1.77$~d), albeit with a S/N $=3.9$. We also note the presence of stochastic low frequency variability.

\begin{figure*}
	\includegraphics[width=2\columnwidth, scale = 1]{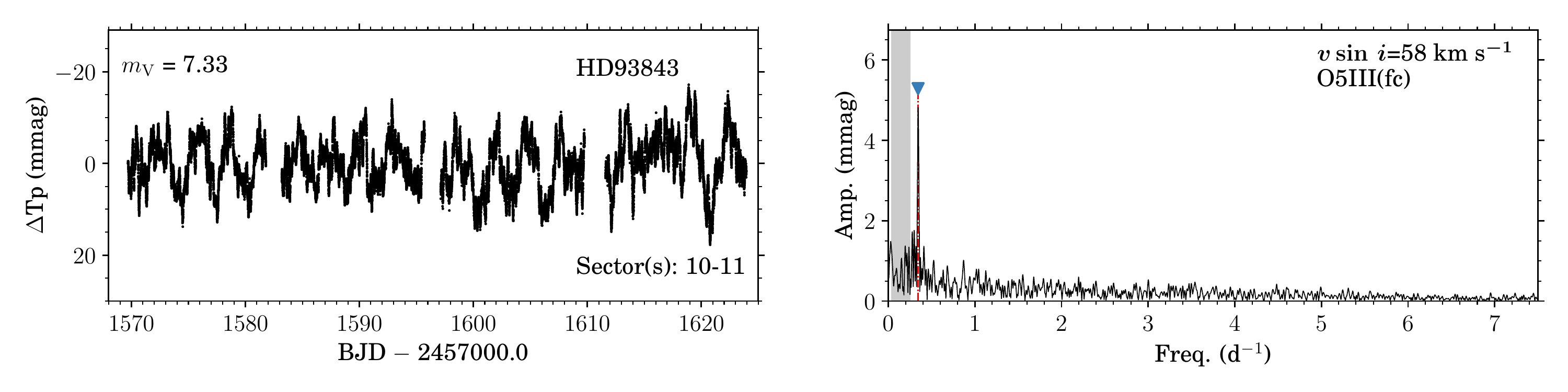}
    \caption{TESS light curve and LS-periodogram of HD\,93843. Same figure style as Fig~\ref{fig:PG_LINE_Bcep}.}
    \label{fig:LC+PG_HD93843}
\end{figure*}

HD~93843 is an O5\,III(fc) star \citep{Sota2014}. \citet{Prinja1998} find repeating structure in the wind, with a period of $\sim7.1$~d at intermediate and high velocities in the wind. We similarly find a variable wind in the IACOB/OWN spectroscopy. The two-sector TESS data set reveals stochastic low frequency variability and a high S/N singular frequency ($\nu=0.3445(3)$~d$^{-1}$). Hence we do not recover the periodicity found by \citet{Prinja1998}. This indicates either a structural change in the wind over the last twenty years or some other cause, for example, co-rotating magnetic spots on the surface given that HD~93843 is a magnetic star \citep{Hubrig2013}. The latter would also agree with the estimated rotational modulation frequency range (v$\,\sin\,i=58$~km~s$^{-1}$).

\begin{figure*}
	\includegraphics[width=2\columnwidth, scale = 1]{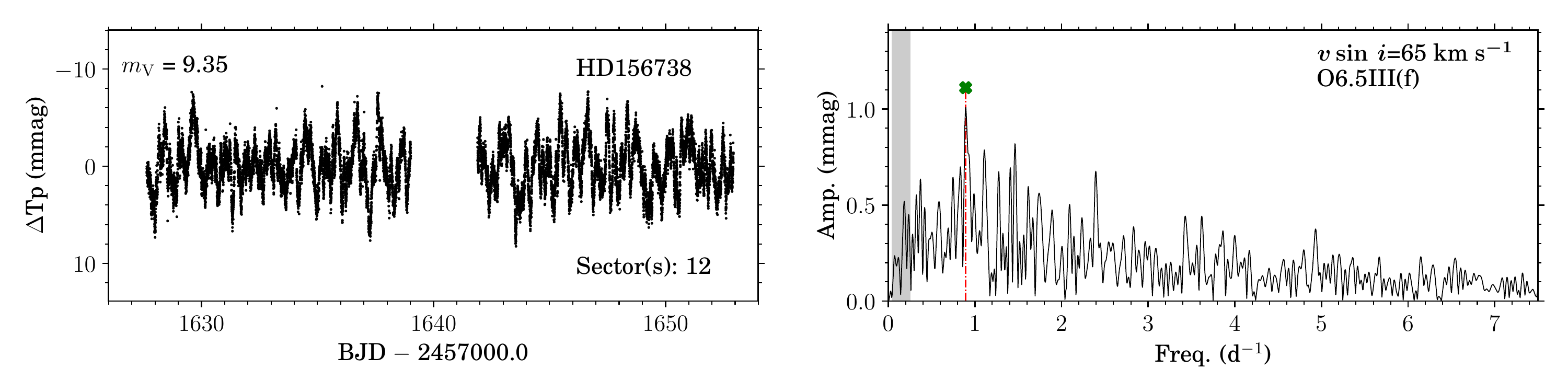}
    \caption{TESS light curve and LS-periodogram of HD\,156738. Same figure style as Fig~\ref{fig:PG_LINE_Bcep}.}
    \label{fig:LC+PG_HD156738}
\end{figure*}

HD~156738 (O6.5\,III(f), \citealp{Sota2014}) is a visual binary consisting of a close pair \citep{Sana2014}. The one-sector TESS data set is characterised by stochastic low frequency variability, with contributions from both stars.

\begin{figure*}
	\includegraphics[width=2\columnwidth, scale = 1]{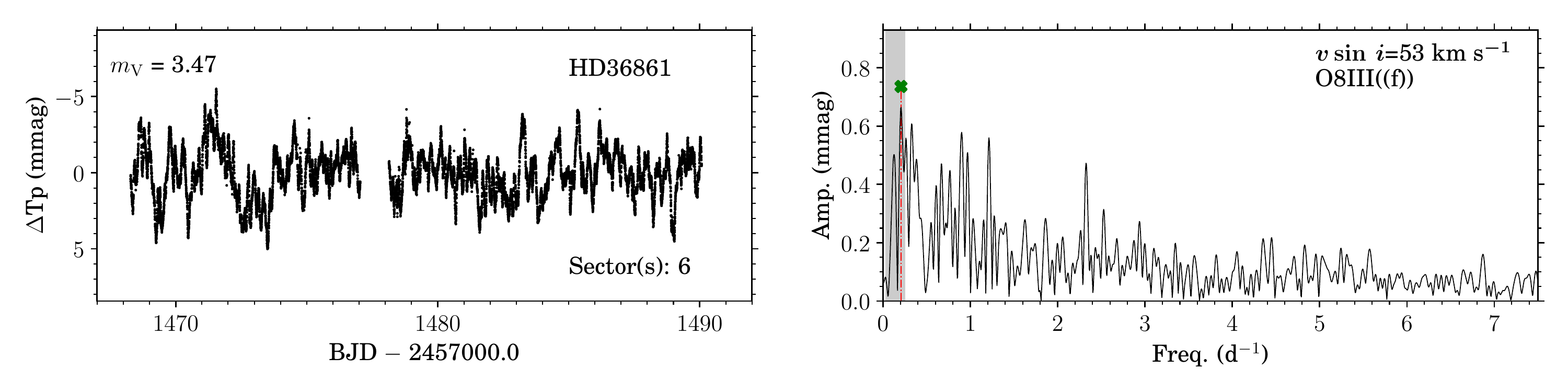}
    \caption{TESS light curve and LS-periodogram of HD\,36861. Same figure style as Fig~\ref{fig:PG_LINE_Bcep}.}
    \label{fig:LC+PG_HD36861}
\end{figure*}

HD~36861 ($\lambda$~Ori A, O8\,III((f)), \citealp{Sota2011}) is a binary system with the secondary early-B star separated by 4.342$\arcsec$ and has a magnitude difference $\Delta m = 1.91$ in the $z$ band \citep{Maiz2010, Sota2011}. The dominant variability type in the one-sector TESS data set is stochastic low frequency variability.

\begin{figure*}
	\includegraphics[width=2\columnwidth, scale = 1]{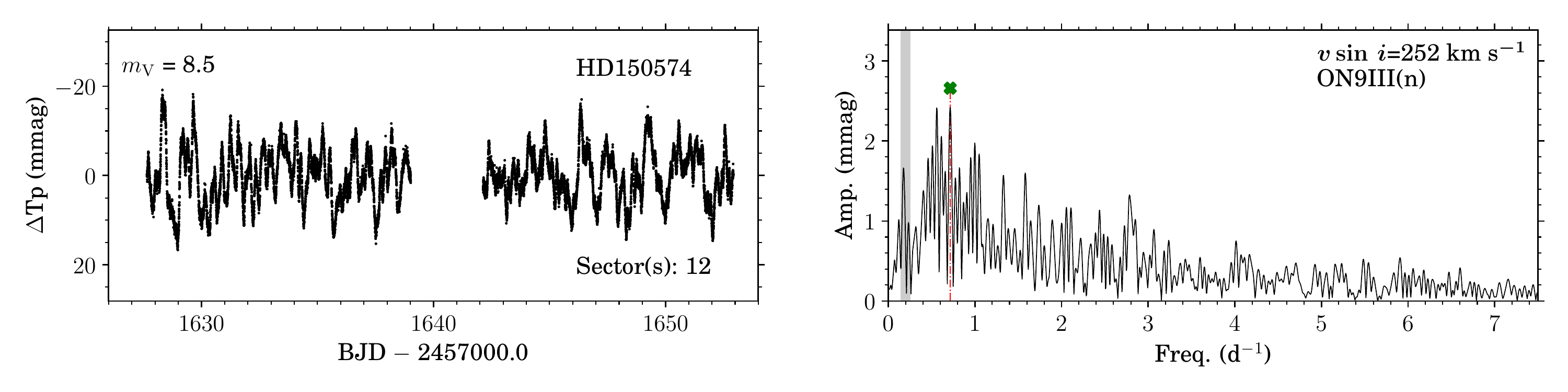}
    \caption{TESS light curve and LS-periodogram of HD\,150574. Same figure style as Fig~\ref{fig:PG_LINE_Bcep}.}
    \label{fig:LC+PG_HD150574}
\end{figure*}

HD~150574 (ON9\,III(n), \citealp{Sota2014}) is a high proper-motion star \citep{Gaia2016}. The one-sector TESS data set reveals stochastic low frequency variability.

\begin{figure*}
	\includegraphics[width=2\columnwidth, scale = 1]{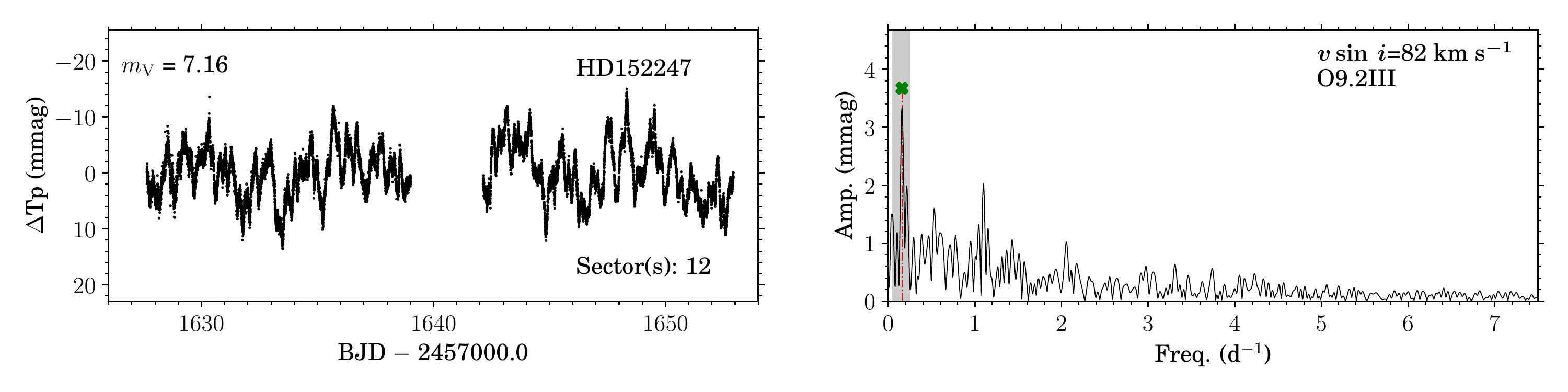}
    \caption{TESS light curve and LS-periodogram of HD\,152247. Same figure style as Fig~\ref{fig:PG_LINE_Bcep}.}
    \label{fig:LC+PG_HD152247}
\end{figure*}

HD~152247 (O9.2\,III, \citealp{Sota2014}) is long-period SB2 system \citep{Chini2012, Sana2012}. The orbit was constrained by \citet{LeBouquin2017}. We find it as SB1 in the IACOB/OWN spectroscopy. The dominant type of variability in the one-sector TESS data set is found to be stochastic low frequency variability but potentially coherent frequencies are noted, both in and out of the estimated rotational modulation frequency range (v$\,\sin\,i=82$~km~s$^{-1}$). 

\begin{figure*}
	\includegraphics[width=2\columnwidth, scale = 1]{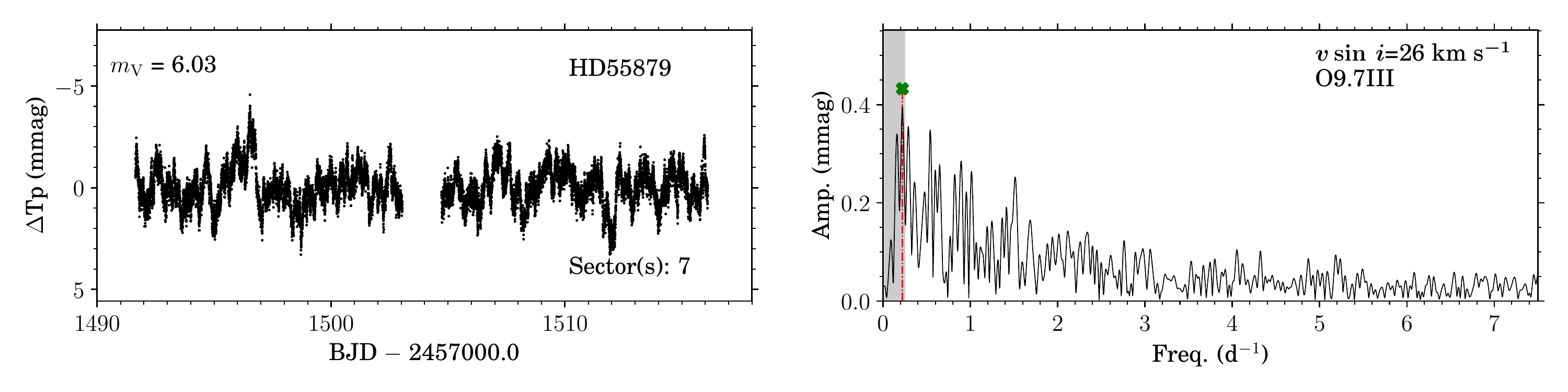}
    \caption{TESS light curve and LS-periodogram of HD\,55879. Same figure style as Fig~\ref{fig:PG_LINE_Bcep}.}
    \label{fig:LC+PG_HD55879}
\end{figure*}

HD~55879 is an O9.7\,III giant star \citep{Sota2011}. The one-sector TESS data set shows stochastic low frequency variability.

\begin{figure*}
	\includegraphics[width=2\columnwidth, scale = 1]{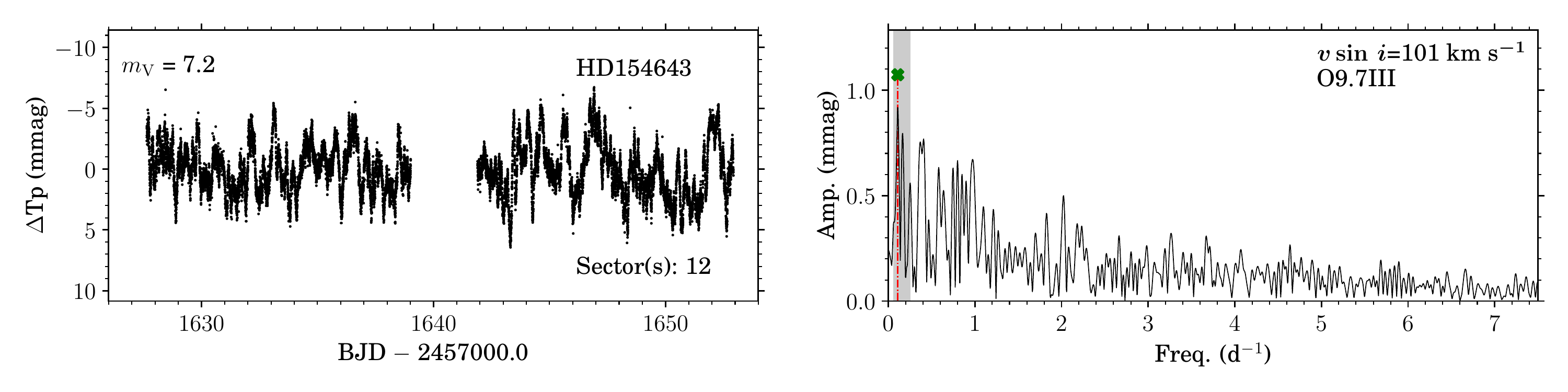}
    \caption{TESS light curve and LS-periodogram of HD\,154643. Same figure style as Fig~\ref{fig:PG_LINE_Bcep}.}
    \label{fig:LC+PG_HD154643}
\end{figure*}

HD~154643 is an O9.7\,III giant star \citep{Sota2014}. We find it as a SB1 in the IACOB/OWN spectroscopy. The one-sector TESS data set is dominated by stochastic low frequency variability.

\begin{figure*}
	\includegraphics[width=2\columnwidth, scale = 1]{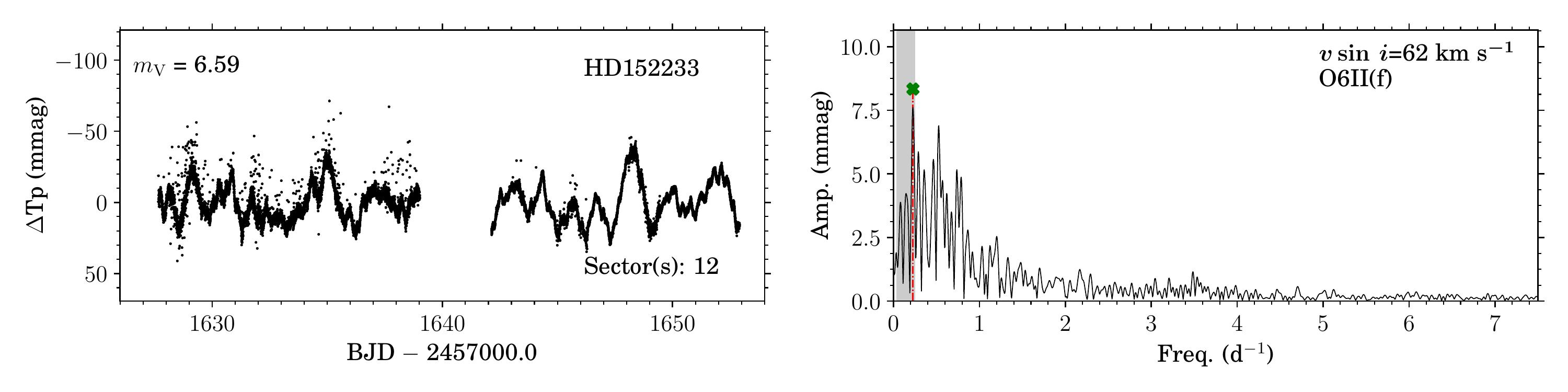}
    \caption{TESS light curve and LS-periodogram of HD\,152233. Same figure style as Fig~\ref{fig:PG_LINE_Bcep}.}
    \label{fig:LC+PG_HD152233}
\end{figure*}

HD~152233 is an O6\,II(f) giant star \citep{Sota2014}. It is found in a crowded region and the TESS pixels are contaminated by nearby stars, i.e bright B0.5\,Ia star HD~152234 (at a distance of $\sim80\arcsec$ vs. the TESS pixel size of $\sim20\arcsec$). This star is brighter than HD~152233 (5.45 vs. 6.59~mag in Johnson~V) and is thus the dominant source of the stochastic low frequency variability seen in the one-sector TESS data set.

\begin{figure*}
	\includegraphics[width=2\columnwidth, scale = 1]{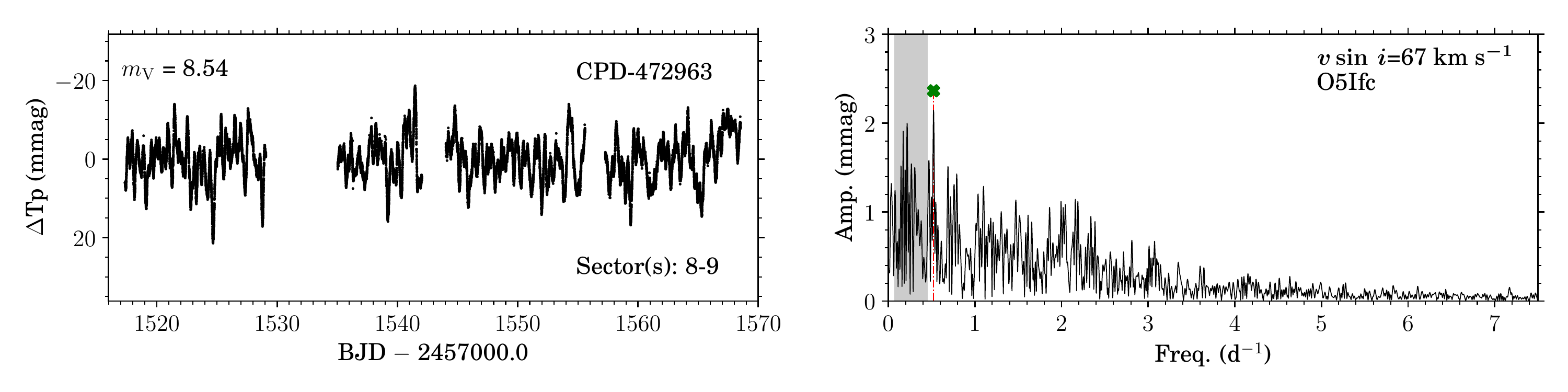}
    \caption{TESS light curve and LS-periodogram of CPD-47~2963. Same figure style as Fig~\ref{fig:PG_LINE_Bcep}.}
    \label{fig:LC+PG_CPD-472963}
\end{figure*}

CPD-47~2963 (O5\,Ifc supergiant, \citealp{Sota2014}) is a spectroscopic binary \citep{Sota2014, Sana2014, LeBouquin2017}. We find wind variability in absorption in the IACOB/OWN spectroscopy. The dominant type of variability in the two-sector TESS data set is stochastic low frequency variability. 

\begin{figure*}
	\includegraphics[width=2\columnwidth, scale = 1]{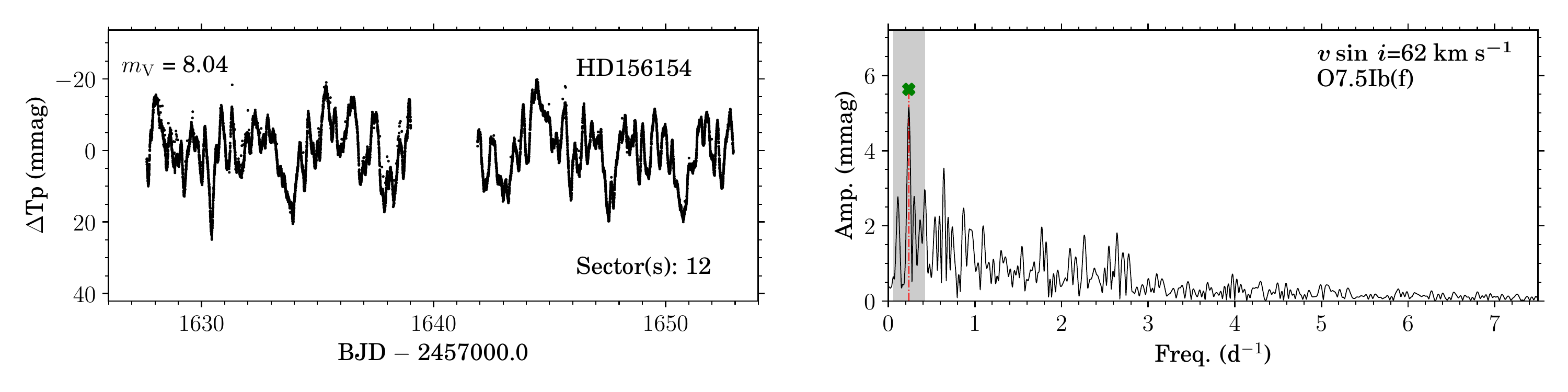}
    \caption{TESS light curve and LS-periodogram of HD\,156154. Same figure style as Fig~\ref{fig:PG_LINE_Bcep}.}
    \label{fig:LC+PG_HD156154}
\end{figure*}

HD~156154 is an O7.5\,Ib(f) supergiant star \citep{Sota2014}. We find signatures of a variable wind in certain absorption lines in the IACOB/OWN spectroscopy. The one-sector TESS light curve is characterised by stochastic low frequency variability.

\begin{figure*}
	\includegraphics[width=2\columnwidth, scale = 1]{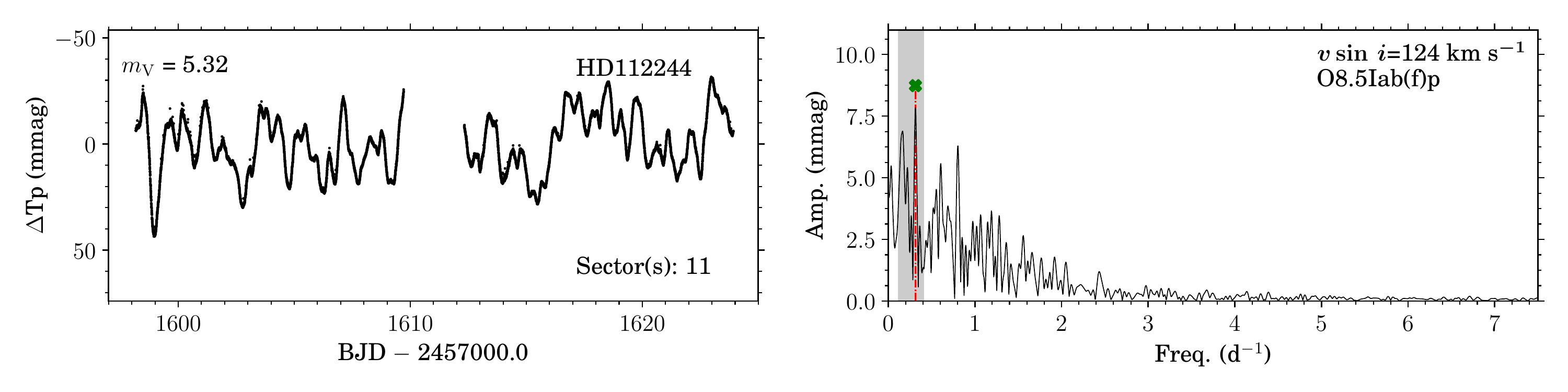}
    \caption{TESS light curve and LS-periodogram of HD\,112244. Same figure style as Fig~\ref{fig:PG_LINE_Bcep}.}
    \label{fig:LC+PG_HD112244}
\end{figure*}

HD~112244 (O8.5\,Iab(f)p, \citealp{Sota2014}) is a spectroscopic binary with a period of $\sim27.665$~d \citep{Chini2012, Mayer2017}. The IACOB/OWN spectra also show RV shifts that could indicate SB1, and we note signatures of a structured wind. The one-sector TESS data show predominantly stochastic low frequency variability.

\begin{figure*}
	\includegraphics[width=2\columnwidth, scale = 1]{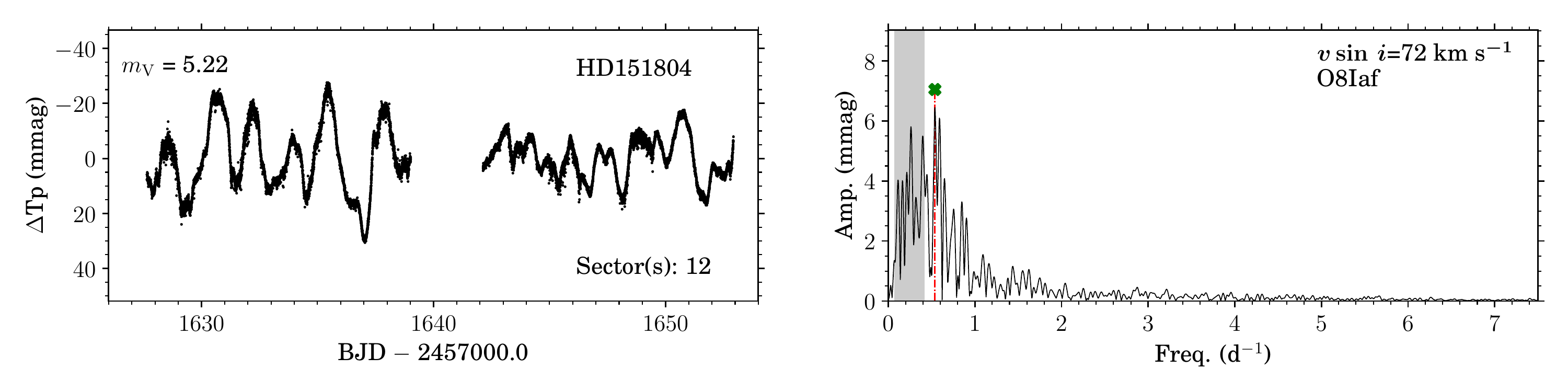}
    \caption{TESS light curve and LS-periodogram of HD\,151804. Same figure style as Fig~\ref{fig:PG_LINE_Bcep}.}
    \label{fig:LC+PG_HD151804}
\end{figure*}

HD~151804 (O8\,Iaf, \citealp{Sota2014}) has a variable wind \citep{Fullerton1992, prinja1996, Crowther1997} and is known to show photometric variations \citep{Bok1966, Vangenderen1989}. We also find signatures of a variable wind in the IACOB/OWN spectroscopy. The one-sector data set shows stochastic low frequency variability.

\begin{figure*}
	\includegraphics[width=2\columnwidth, scale = 1]{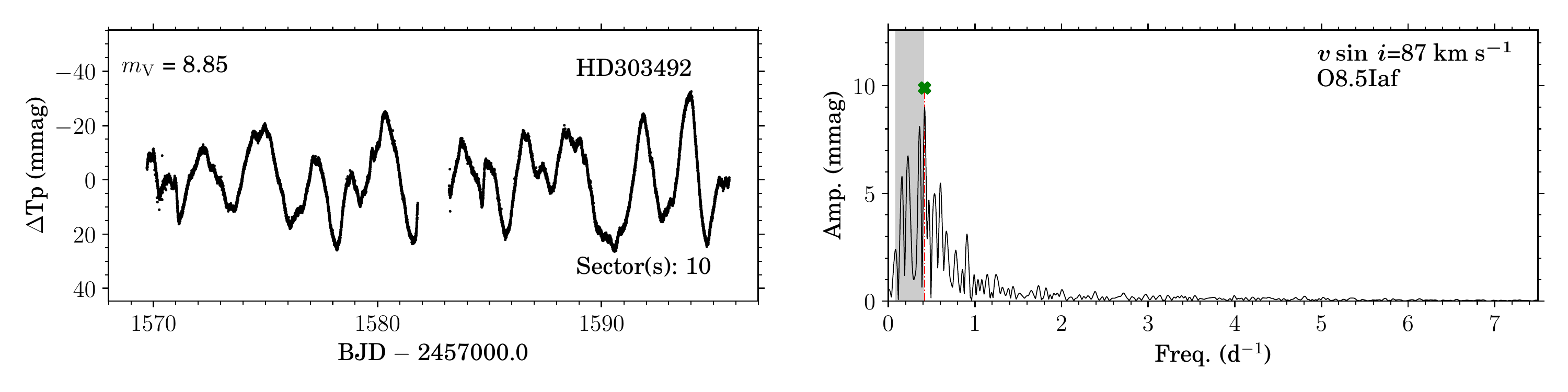}
    \caption{TESS light curve and LS-periodogram of HD\,303492. Same figure style as Fig~\ref{fig:PG_LINE_Bcep}.}
    \label{fig:LC+PG_HD303492}
\end{figure*}

HD~303492 (O8.5Iaf, \citealp{Sota2014}) is a known visual binary \citep{Sana2014}. The one-sector TESS data set is dominated by stochastic low frequency variability. We also find signatures of wind variability in the IACOB/OWN spectroscopy.

\begin{figure*}
	\includegraphics[width=2\columnwidth, scale = 1]{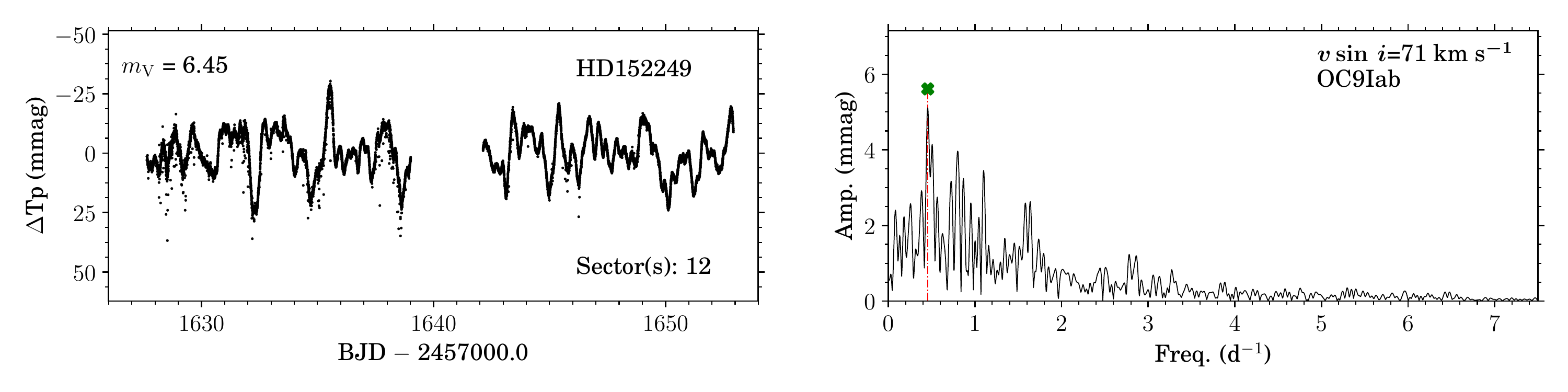}
    \caption{TESS light curve and LS-periodogram of HD\,152249. Same figure style as Fig~\ref{fig:PG_LINE_Bcep}.}
    \label{fig:LC+PG_HD152249}
\end{figure*}

HD~152249 is an OC9\,Iab supergiant star \citep{Sota2014}. It is found in a crowded region and several nearby stars likely bleed into the TESS pixels, that is, CD-41~11037, CD-41~11027A, and CD-41~11027B. Nonetheless, HD~152249 is brighter by at least $1.5$~mag in Johnson~V, indicating that it is the dominant contributor to the light curve variability. We find stochastic low frequency as the dominant type of variability, and we note wind variability in the multi-epoch IACOB/OWN spectroscopy, which may be related to the former.

\begin{figure*}
	\includegraphics[width=2\columnwidth, scale = 1]{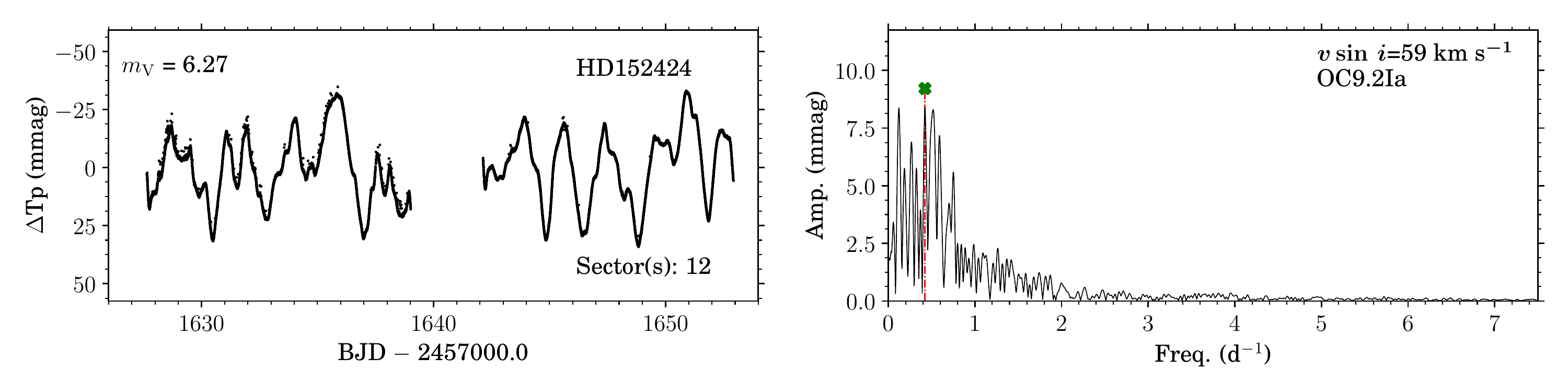}
    \caption{TESS light curve and LS-periodogram of HD\,152424. Same figure style as Fig~\ref{fig:PG_LINE_Bcep}.}
    \label{fig:LC+PG_HD152424}
\end{figure*}

HD~152424 (OC9.2\,Ia, \citealp{Sota2014}) is a spectroscopic binary \citep{Chini2012}. We find it as SB1 in the IACOB/OWN spectroscopy. We further note the presence of a variable wind. We find stochastic low frequency variability as the dominant variability type in the one-sector TESS photometry.

\begin{figure*}
	\includegraphics[width=2\columnwidth, scale = 1]{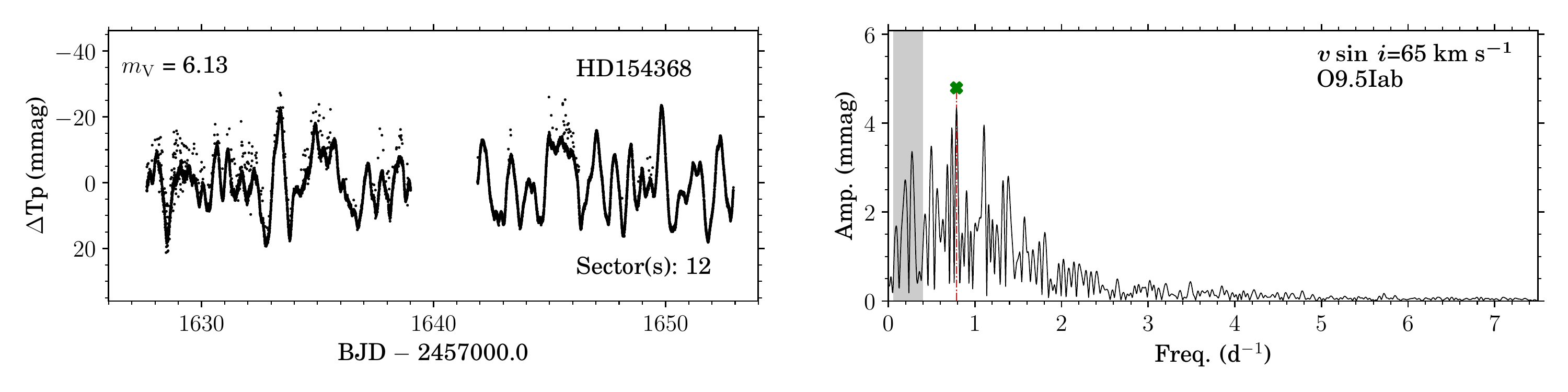}
    \caption{TESS light curve and LS-periodogram of HD\,154368. Same figure style as Fig~\ref{fig:PG_LINE_Bcep}.}
    \label{fig:LC+PG_HD154368}
\end{figure*}

HD~154368 (V1074~Sco, O9.5\,Iab \citealp{Sota2014}) was found to be an eclipsing binary with a period of 16.1~d by \citet{Mason1998} but this could not be confirmed by \citet{Sana2014}. We find wind variability in emission in the IACOB/OWN spectroscopy.  
The dominant type of variability found in the one-sector TESS data set is stochastic low frequency variability.

\begin{figure*}
	\includegraphics[width=2\columnwidth, scale = 1]{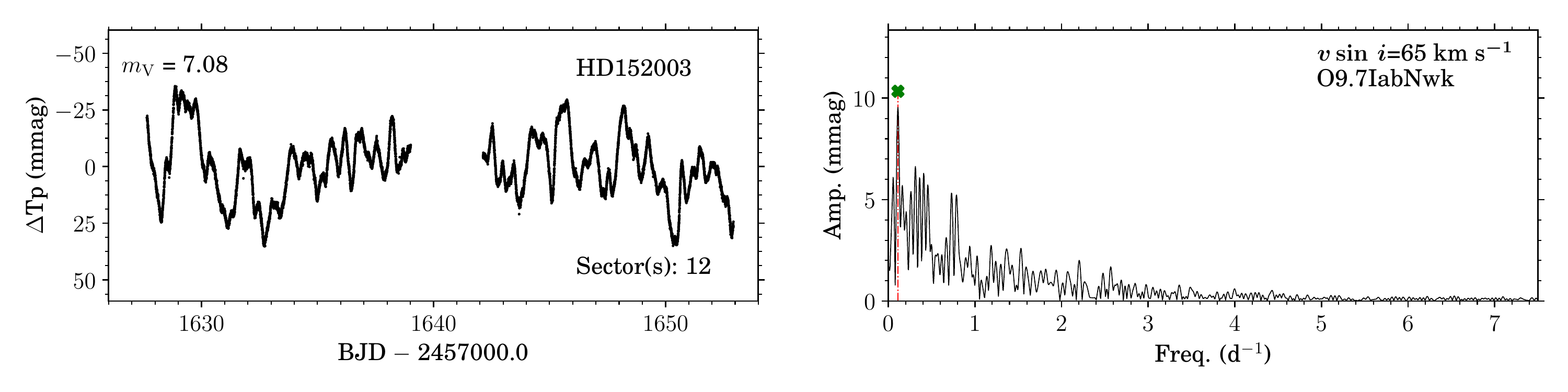}
    \caption{TESS light curve and LS-periodogram of HD\,152003. Same figure style as Fig~\ref{fig:PG_LINE_Bcep}.}
    \label{fig:LC+PG_HD152003}
\end{figure*}

HD~152003 (O9.7\,IabNwk, \citealp{Sota2014}) is a double-lined spectroscopic binary \citep{Chini2012, Sana2014}. We note variable wind emission lines in the IACOB/OWN spectroscopy. \citet{Balona1989} measured a photometric period of $1.724$~d. We note stochastic low frequency variability in the one-sector TESS light curve but do not recover the period measured by \citet{Balona1989}.

\begin{figure*}
	\includegraphics[width=2\columnwidth, scale = 1]{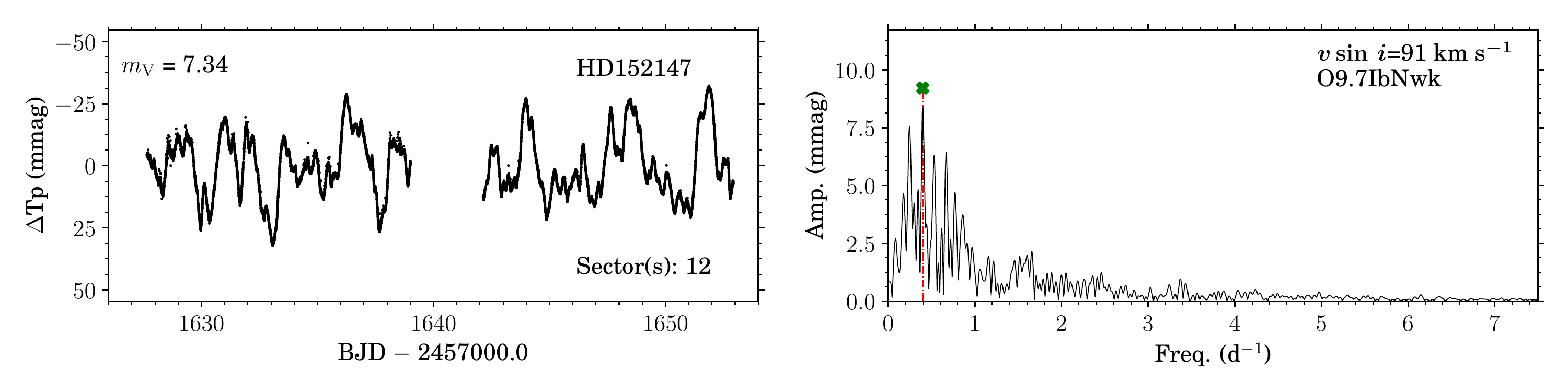}
    \caption{TESS light curve and LS-periodogram of HD\,152147. Same figure style as Fig~\ref{fig:PG_LINE_Bcep}.}
    \label{fig:LC+PG_HD152147}
\end{figure*}

HD~152147 (O9.7\,IbNwk supergiant, \citealp{Sota2014}) is a single-lined spectroscopic binary \citep{Chini2012}. We confirm this in the IACOB/OWN spectroscopy, in addition to variable wind emission. The one-sector TESS data set shows stochastic low frequency variability.

\begin{figure*}
	\includegraphics[width=2\columnwidth, scale = 1]{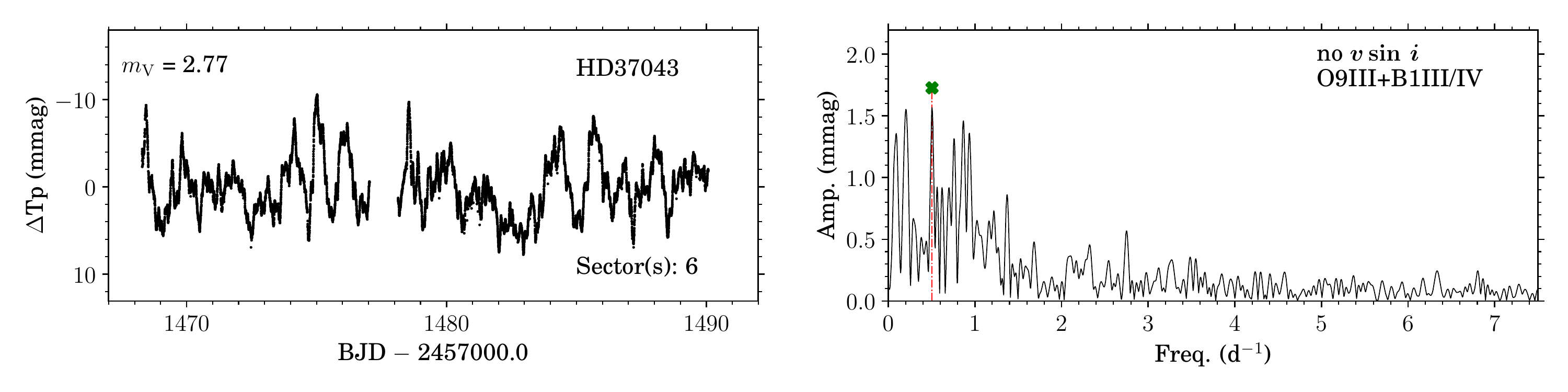}
    \caption{TESS light curve and LS-periodogram of HD\,37043. Same figure style as Fig~\ref{fig:PG_LINE_Bcep}.}
    \label{fig:LC+PG_HD37043}
\end{figure*}

HD~37043 ($\iota$~Ori) is a binary system with a primary O9\,III star and a secondary B1\,III/IV star \citep{Stickland1987}. It is currently the most massive known heartbeat star and its variability was analysed in-depth by \citet{Pablo2017} using BRITE photometry. They constrain the orbit and detect numerous tidally induced oscillations. The variability in the TESS data set occurs on similar timescales but the time-span is too short to extract individual frequencies.

\begin{figure*}
	\includegraphics[width=2\columnwidth, scale = 1]{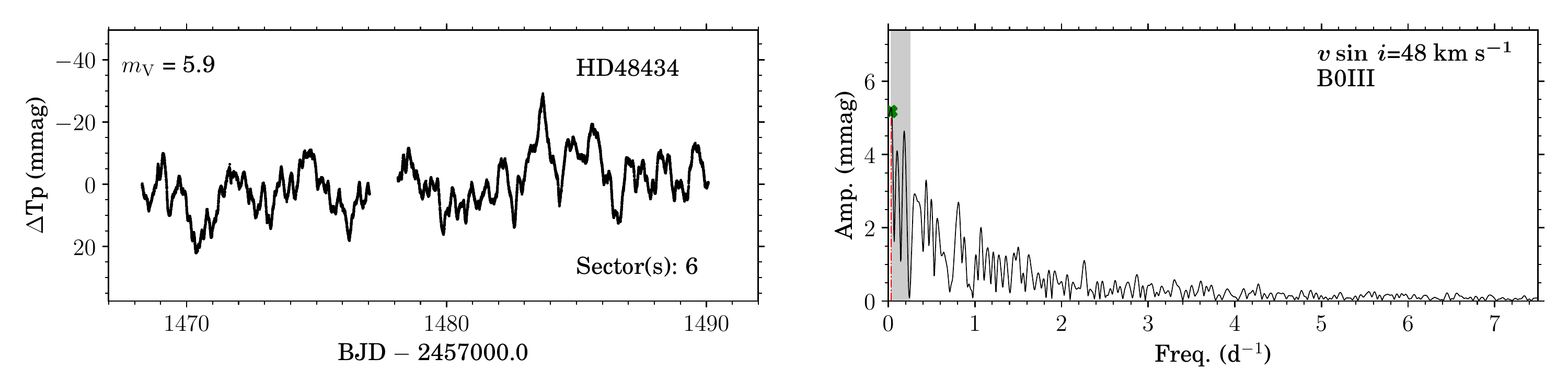}
    \caption{TESS light curve and LS-periodogram of HD\,48434. Same figure style as Fig~\ref{fig:PG_LINE_Bcep}.}
    \label{fig:LC+PG_HD48434}
\end{figure*}

HD~48434 is a B0\,III giant star \citep{Lesh1968}. The one-sector TESS data set is characterised by stochastic low frequency variability. We also note the presence of wind variability in the multi-epoch IACOB/OWN spectroscopy which may be related to the TESS photometric variability.

\begin{figure*}
	\includegraphics[width=2\columnwidth, scale = 1]{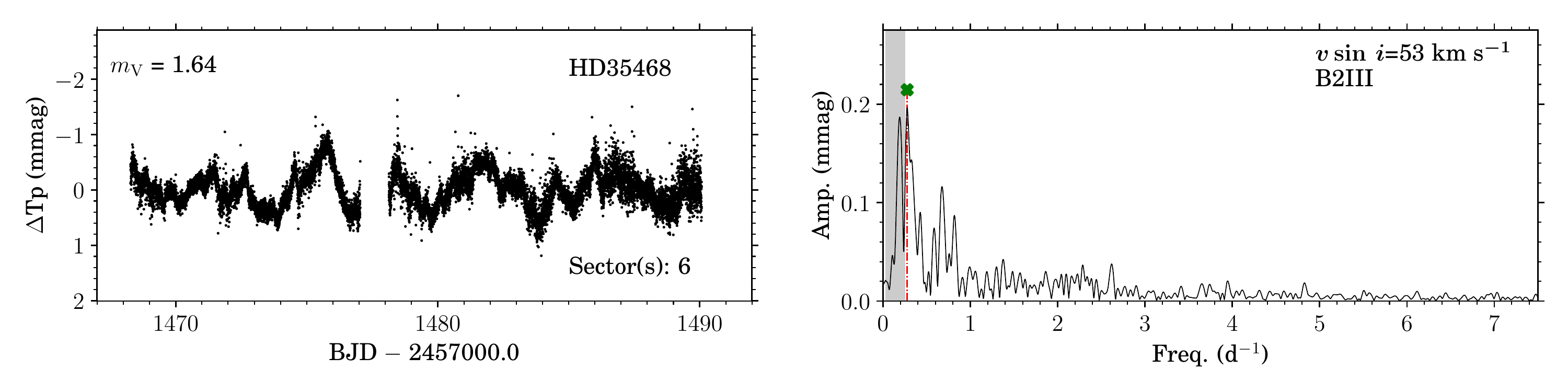}
    \caption{TESS light curve and LS-periodogram of HD\,35468. Same figure style as Fig~\ref{fig:PG_LINE_Bcep}.}
    \label{fig:LC+PG_HD35468}
\end{figure*}

HD~35468 ($\gamma$~Ori, B2\,III \citealp{Lesh1968}) known to be a low amplitude irregular variable  \citep{Krisciunas1994, Krisciunas1996}. Its BRITE light curve was used by \citet{Handler2017} as a constant reference star. We find low amplitude stochastic low frequency variability in the one-sector TESS light curve. 

\begin{figure*}
	\includegraphics[width=2\columnwidth, scale = 1]{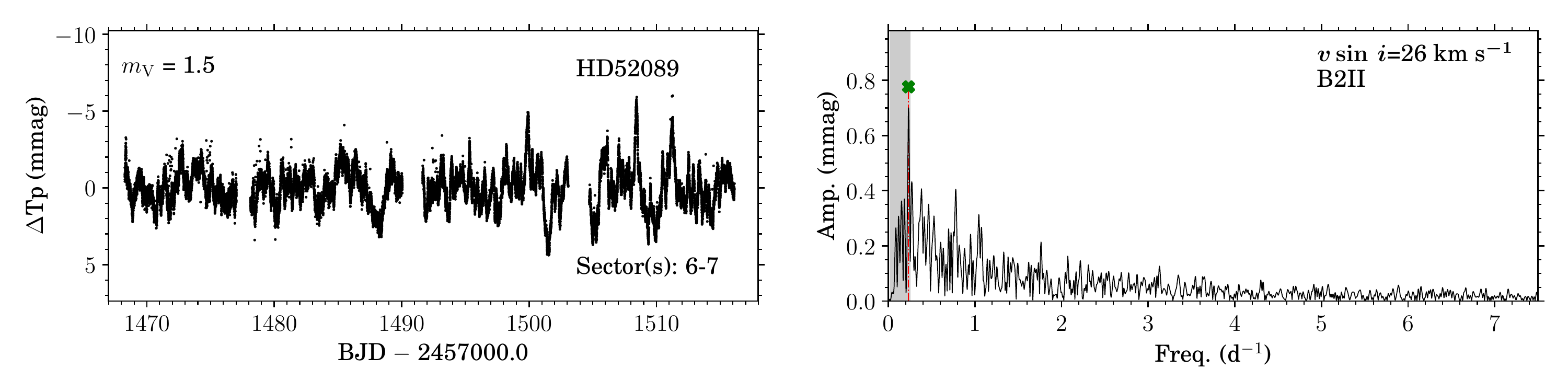}
    \caption{TESS light curve and LS-periodogram of HD\,52089. Same figure style as Fig~\ref{fig:PG_LINE_Bcep}.}
    \label{fig:LC+PG_HD52089}
\end{figure*}

HD~52089 ($\epsilon$~CMa, B2\, II,  \citealp{Morgan1955}) has a weak magnetic field \citep{Morel2008, Hubrig2014, Fossati2015, Shultz2019b}. It shows UV and X-ray emission consistent with a wind-shock model \citep{Haucke2018}. The dominant variability type in the two-sector TESS light curve is stochastic low frequency variability.

\subsection{Post-main sequence stars}\label{sec:appendix_PMS}

\begin{figure*}
	\includegraphics[width=2\columnwidth, scale = 1]{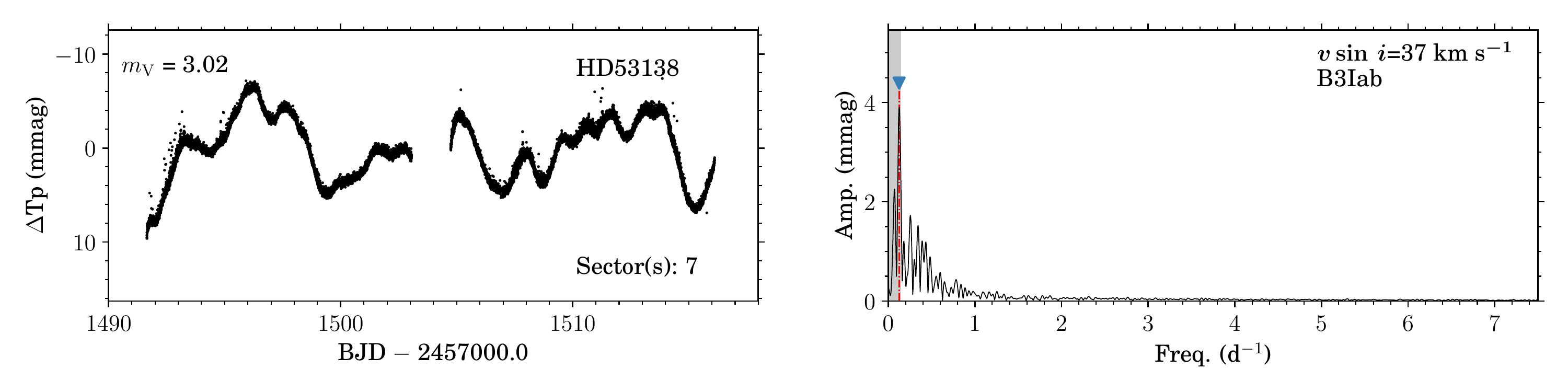}
    \caption{TESS light curve and LS-periodogram of HD\,53138. Same figure style as Fig~\ref{fig:PG_LINE_Bcep}.}
    \label{fig:LC+PG_HD53138}
\end{figure*}

HD~53138 ($o^{2}$~CMa,  B3\,Ia, \citealp{Walborn1971}) has known irregular variations in the H$_{\alpha}$ line with periods between 4 and 45~d \citep{Morel2004, Lefever2007, Haucke2018}. \citet{Lefever2007} find two periodicities in HIPPARCOS photometry ($P_{1}=24.390(1)$ and $P_{2}=3.690(1)$~d). We also note variations in the wind lines in the IACOB/OWN spectroscopy. The dominant type of variability in the one-sector TESS data set is stochastic low frequency variability. Only the dominant frequency in the periodogram is extracted as $\nu=0.1258(4)$~d$^{-1}$ (or $P=7.95(3)$~d) and it falls just on the upper boundary of the estimated rotational modulation range (v$\,\sin\,i=37$~km~s$^{-1}$).

\begin{figure*}
	\includegraphics[width=2\columnwidth, scale = 1]{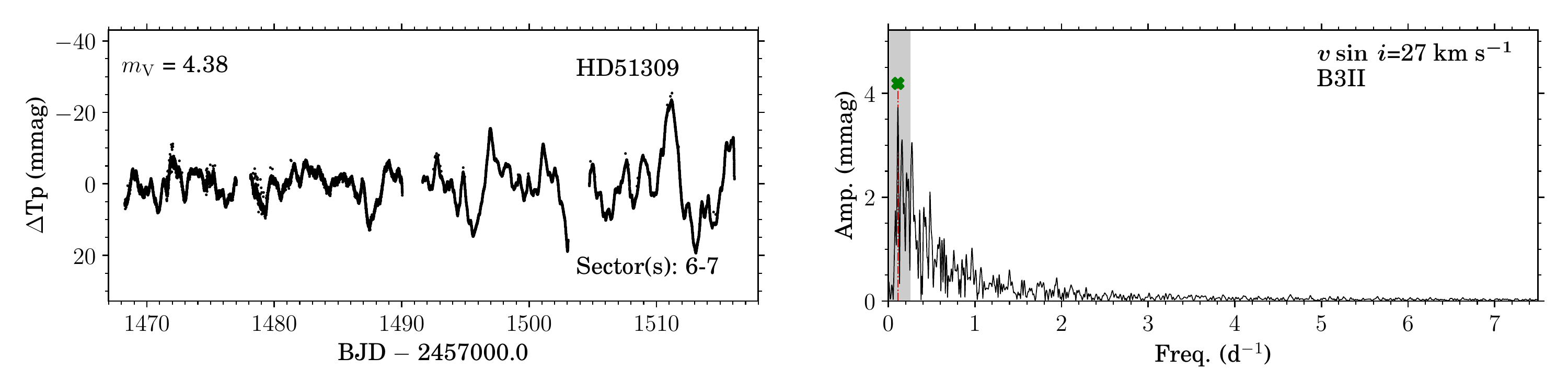}
    \caption{TESS light curve and LS-periodogram of HD\,51309. Same figure style as Fig~\ref{fig:PG_LINE_Bcep}.}
    \label{fig:LC+PG_HD51309}
\end{figure*}

HD~51309 ($\iota$~CMa, B3\,II, \citealp{Lesh1968}) is a known pulsating blue supergiant \citep{Balona1985, Stankov2005, Saio2006}. The dominant type of variability in the two-sector TESS data set is stochastic low frequency variability.

\begin{figure*}
	\includegraphics[width=2\columnwidth, scale = 1]{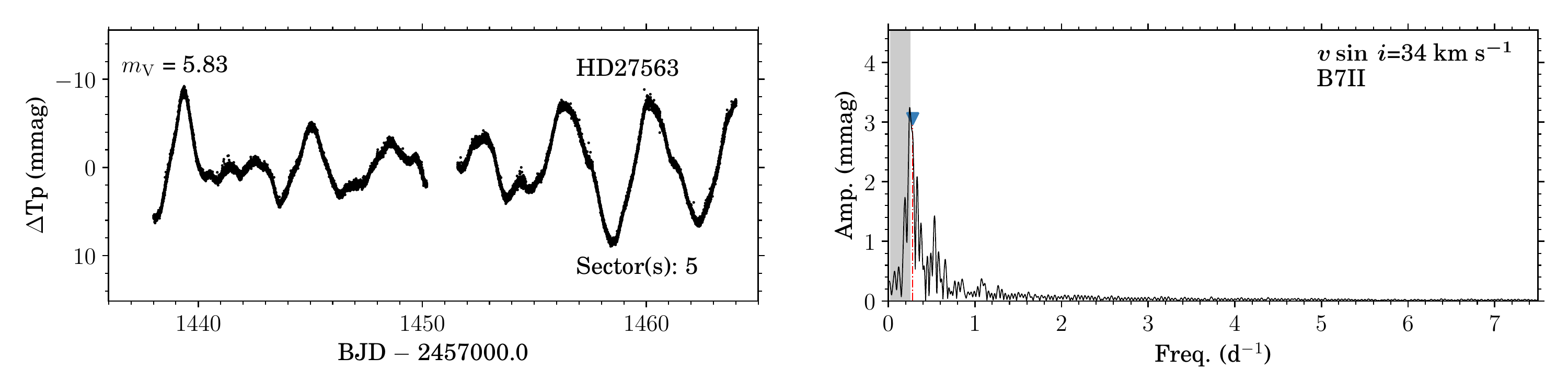}
    \caption{TESS light curve and LS-periodogram of HD\,27563. Same figure style as Fig~\ref{fig:PG_LINE_Bcep}.}
    \label{fig:LC+PG_HD27563}
\end{figure*}

HD~27563 (B7\,II, \citealp{Houk1999}) has a measured photometric period of $P\approx4$~d by \citet{Mathys1986} who classified it as multi-periodic or pseudo-periodic. Using additional photometric data \citet{Manfroid1989} suggested the presence of additional variability on multiple time scales. The one-sector TESS data set reveals stochastic low frequency variability, and one frequency $\nu=0.2820(6)$~d$^{-1}$. It is slightly higher than the estimated rotational modulation frequency range (v$\,\sin\,i=34$~km~s$^{-1}$), which could indicate either that the star is rotating fast (v$\geq450$~km~s$^{-1}$) and seen nearly pole on, or a different origin. The latter is more likely given the amplitude and shape of the light curve.

\begin{figure*}
	\includegraphics[width=2\columnwidth, scale = 1]{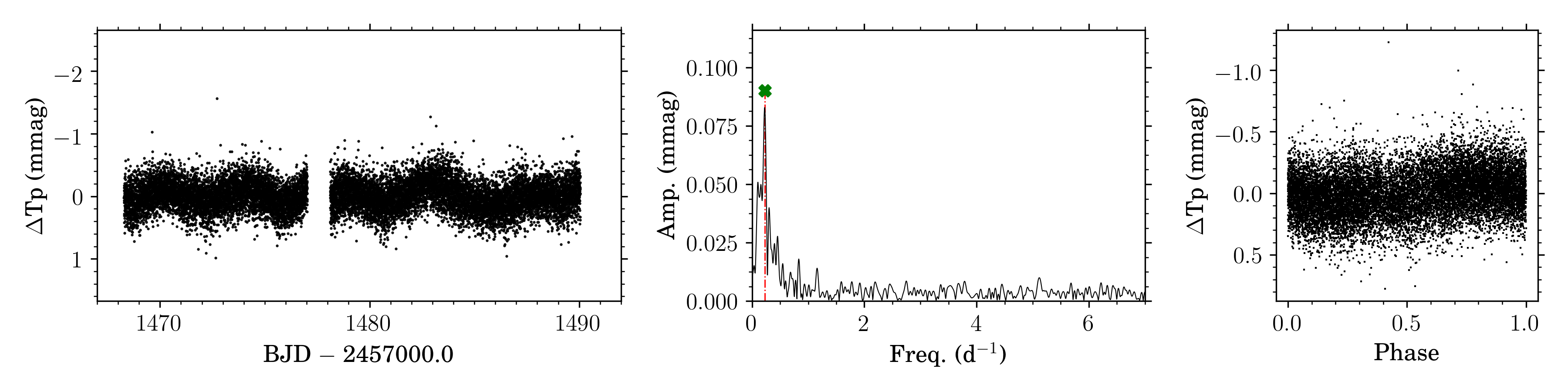}
    \caption{TESS light curve and LS-periodogram of HD\,46769. Same figure style as Fig~\ref{fig:PG_LINE_Bcep}. Here we also show a phase plot folded with the dominant frequency $\nu=0.2249(9)$~d$^{-1}$.}
    \label{fig:LC+PG_HD46769}
\end{figure*}

HD~46769 (B5\,II, \citealp{Aerts2013}) was studied in-depth by \citet{Aerts2013} using 23~d of CoRoT, detecting a dominant period of 4.84~d and several (sub-)harmonics. No significant variability was found in the residuals, and the variability was therefore explained as due to rotational modulation instead of pulsations. A low frequency power excess was measured by \citet{Bowman2019a} in the same data. We find stochastic low frequency variability in the one-sector TESS data set. To compare to the period found by \citet{Aerts2013} we provide a phase plot in Fig.~\ref{fig:LC+PG_HD46769} folded by the dominant frequency $\nu=0.2249(9)$~d$^{-1}$. 

\begin{figure*}
	\includegraphics[width=2\columnwidth, scale = 1]{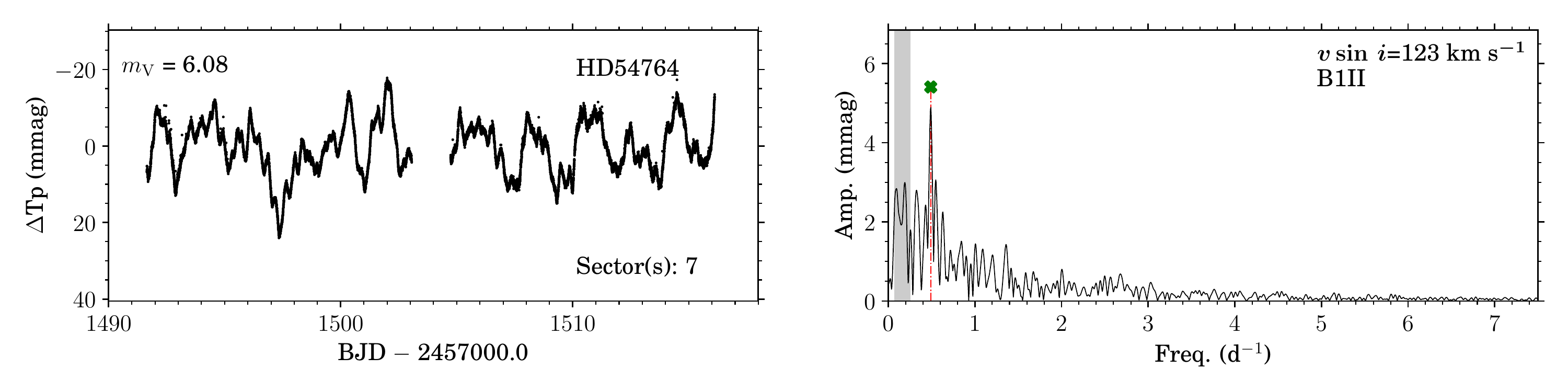}
    \caption{TESS light curve and LS-periodogram of HD\,54764. Same figure style as Fig~\ref{fig:PG_LINE_Bcep}.}
    \label{fig:LC+PG_HD54764}
\end{figure*}

HD~54764 (B1\, II, \citealp{Lesh1968}) is classified as a $\alpha$~Cygni type supergiant with a period of 2.69~d measured in HIPPARCOS data \citep{Waelkens1998}. Based on this classification \citet{Saio2006} proposed it as a potential SPBsg, a class of supergiant g mode pulsators. \citet{Lefevre2009} further notes asymmetric line profiles suggesting an undetected companion or pulsational effects. The variability in the one-sector TESS light curve is predominantly stochastic in nature.

\begin{figure*}
	\includegraphics[width=2\columnwidth, scale = 1]{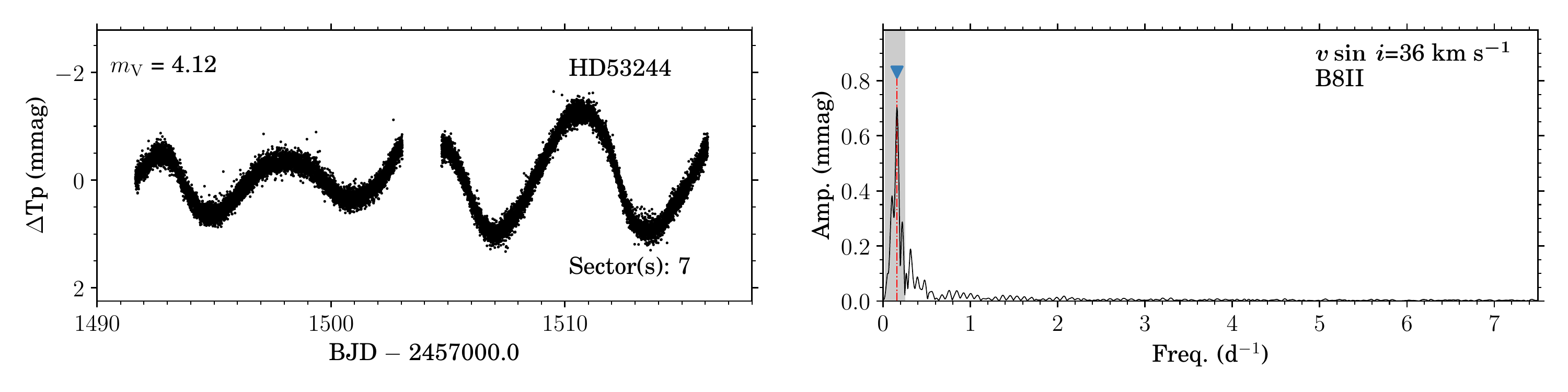}
    \caption{TESS light curve and LS-periodogram of HD\,53244. Same figure style as Fig~\ref{fig:PG_LINE_Bcep}.}
    \label{fig:LC+PG_HD53244}
\end{figure*}

HD~53244 ($\gamma$~CMa, B8\,II, \citealp{Lesh1968}) is a binary system \citep{Schneider1981, Scholler2010}. Its classified as a peculiar star because of variability in the spectroscopic Hg and Mn lines \citep{Schneider1981, Renson1991, Maza2010}. However, the presence of a magnetic field that could induce such variations was never confirmed \citep{Makaganiuk2011, Hubrig2012, Martin2018}. The one-sector TESS data set is characterised by variability due to rotational modulation with a frequency $\nu=0.1600(2)$~d$^{-1}$, or a period of $P_{\rm rot}=6.250(8)$~d. 

\begin{figure*}
	\includegraphics[width=2\columnwidth, scale = 1]{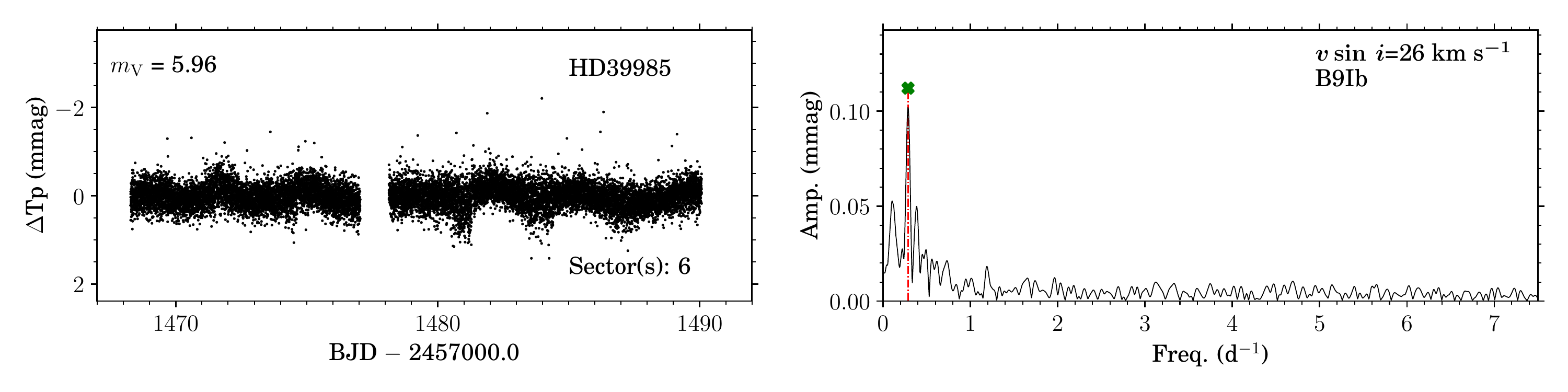}
    \caption{TESS light curve and LS-periodogram of HD\,39985. Same figure style as Fig~\ref{fig:PG_LINE_Bcep}.}
    \label{fig:LC+PG_HD39985}
\end{figure*}

HD~39985 is a B9\,Ib supergiant star \citep{Hube1970}. The one-sector TESS data set shows low amplitude stochastic variability.

\subsection{Other B main sequence stars}\label{sec:appendix_LMS}

\begin{figure*}
	\includegraphics[width=2\columnwidth, scale = 1]{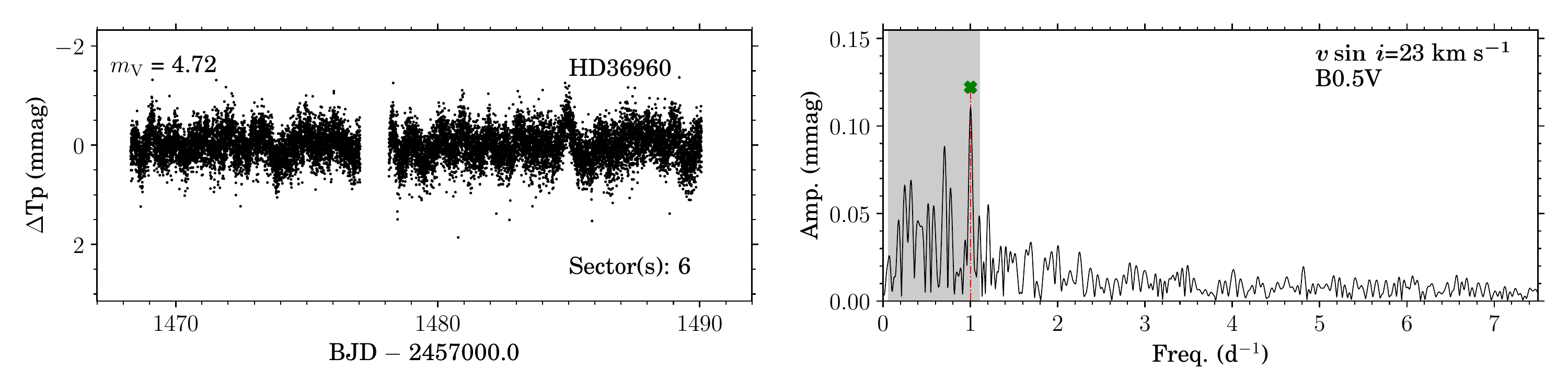}
    \caption{TESS light curve and LS-periodogram of HD\,36960. Same figure style as Fig~\ref{fig:PG_LINE_Bcep}.}
    \label{fig:LC+PG_HD36960}
\end{figure*}

HD~36960 is a B0.5\,V standard star \citep{Lesh1968}. The dominant variability in the one-sector data set is either due to stochastic low frequency variability, dense unresolved g mode pulsations, or a combination of both. We note that the TESS pixels are contaminated by HD~36959, another star in this sample. Given their similar spectral type it is difficult to disentangle their separate contributions to the light curve. As HD~36960 is the brighter star (4.72 vs. 5.53 in Johnson~V) we assign the variability to HD~36960 and mark HD~36959 as contaminated.

\begin{figure*}
	\includegraphics[width=2\columnwidth, scale = 1]{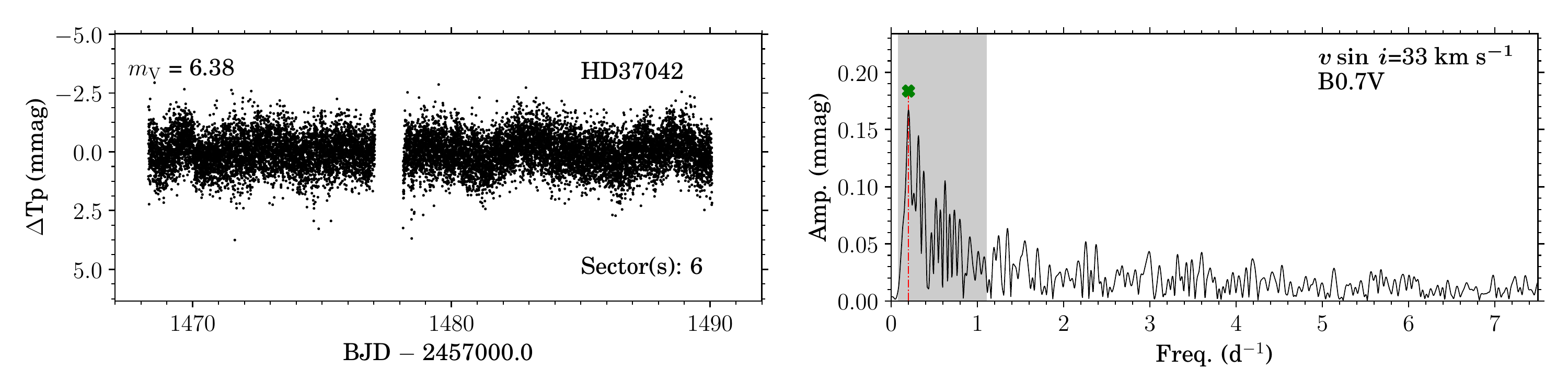}
    \caption{TESS light curve and LS-periodogram of HD\,37042. Same figure style as Fig~\ref{fig:PG_LINE_Bcep}.}
    \label{fig:LC+PG_HD37042}
\end{figure*}

HD~37042 is a B0.7\,V star \citep{SimonDiaz2010}. The one-sector TESS data set shows  stochastic low frequency variability of low amplitude.

\begin{figure*}
	\includegraphics[width=2\columnwidth, scale = 1]{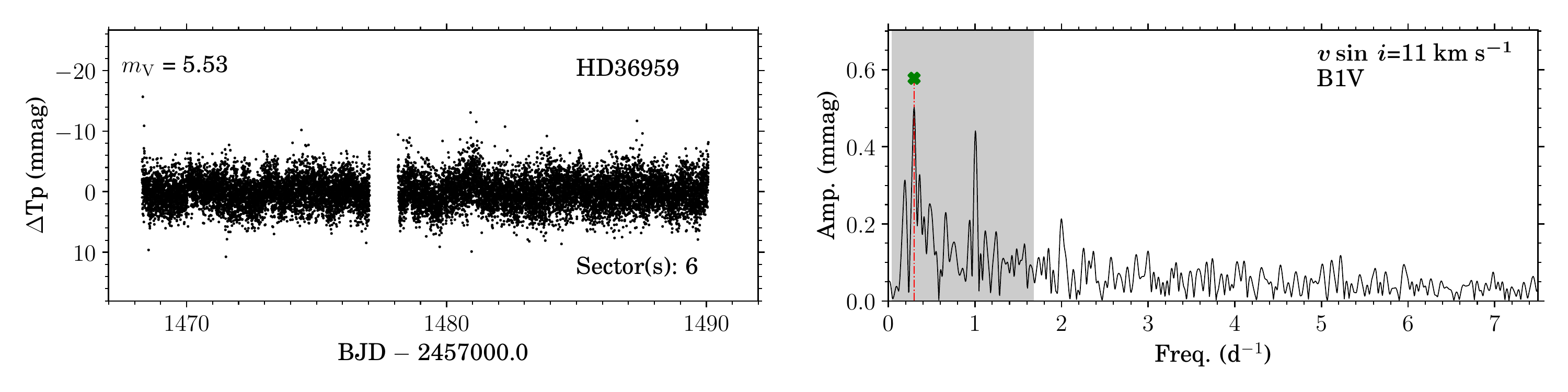}
    \caption{TESS light curve and LS-periodogram of HD\,36959. Same figure style as Fig~\ref{fig:PG_LINE_Bcep}.}
    \label{fig:LC+PG_HD36959}
\end{figure*}

HD~36959 is a B1\,V star \citep{Lesh1968}. The TESS data set is contaminated by nearby star HD~36960 discussed previously.

\begin{figure*}
	\includegraphics[width=2\columnwidth, scale = 1]{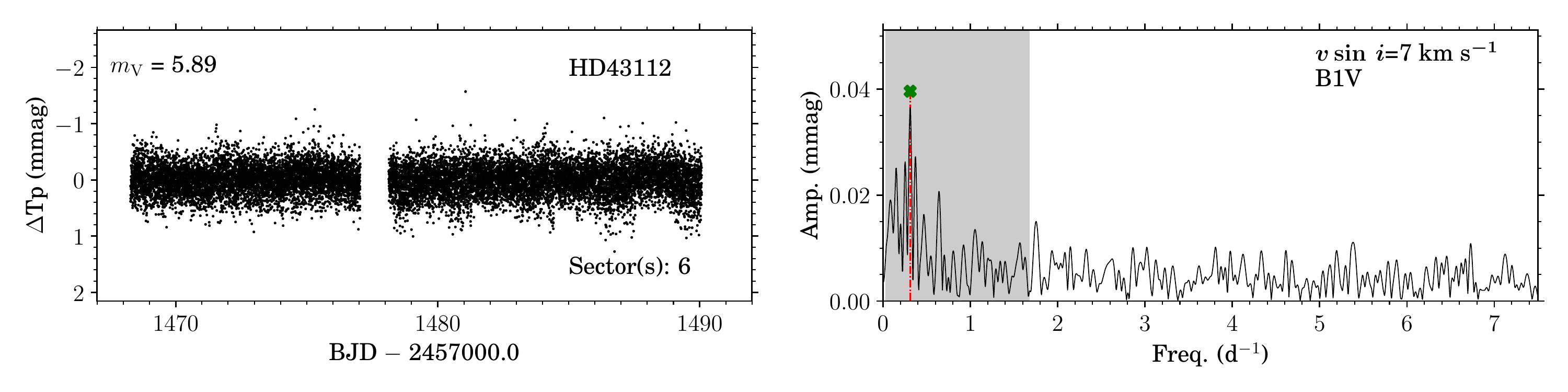}
    \caption{TESS light curve and LS-periodogram of HD\,43112. Same figure style as Fig~\ref{fig:PG_LINE_Bcep}.}
    \label{fig:LC+PG_HD43112}
\end{figure*}

HD~43112 is a B1\,V star \citep{Lesh1968}. The one-sector TESS data set is nearly constant and the stochastic low frequency variability could be instrumental in origin.

\begin{figure*}
	\includegraphics[width=2\columnwidth, scale = 1]{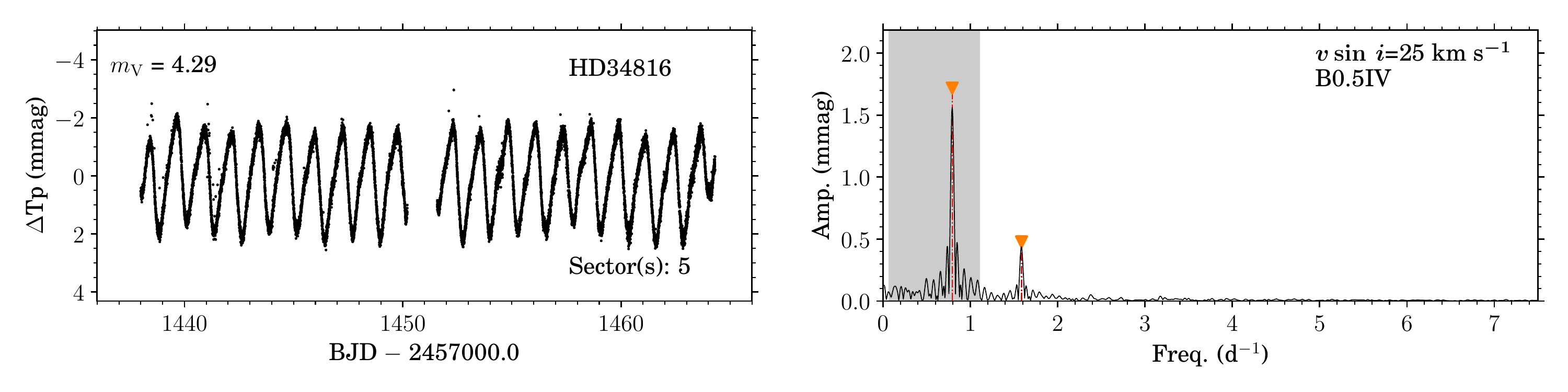}
    \caption{TESS light curve and LS-periodogram of HD\,34816. Same figure style as Fig~\ref{fig:PG_LINE_Bcep}.}
    \label{fig:LC+PG_HD34816}
\end{figure*}

HD~34816 ($\lambda$~Lep) is a B0.5\,IV star \citep{Lesh1968}. The one-sector TESS data set is characterised by rotational modulation, and two frequencies are extracted: $\nu_{1}=0.7923(2)$ and its harmonic $\nu_{2}=1.5863(3)$~d$^{-1}$ (or $P_{1}=1.2621(3)$ and $P_{2}=0.6304(1)$~d).

\begin{figure*}
	\includegraphics[width=2\columnwidth, scale = 1]{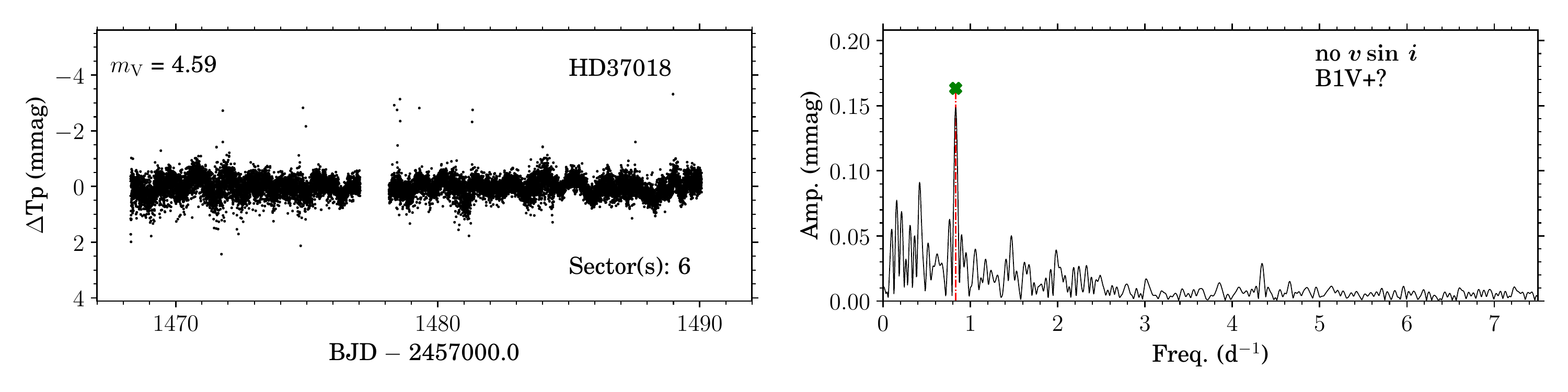}
    \caption{TESS light curve and LS-periodogram of HD\,37018. Same figure style as Fig~\ref{fig:PG_LINE_Bcep}.}
    \label{fig:LC+PG_HD37018}
\end{figure*}

HD~37018 (42~Ori, B1\,V star \citealp{Lesh1968}) is a binary star \citep{Oudmaijer2010}. We find it as SB2 in the IACOB/OWN spectroscopy. The dominant variability type in the one-sector TESS data set points to stochastic low frequency variability.

\onecolumn
\section{MESA \& GYRE inlists}\label{sec:appendix_inlists}

\subsection{MESA}

\begin{verbatim}
&star_job

  ! start a run from a saved model
    load_saved_model = .false.         

  ! save a model at the end of the run
    save_model_when_terminate = .false.
    save_photo_when_terminate = .false.

  ! display on-screen plots
    pgstar_flag = .false.

  ! history and profile columns
    history_columns_file = 'hist_columns.list'
    profile_columns_file = 'prof_columns.list'

! OPTIONS

! begin with a pre-main sequence model
    create_pre_main_sequence_model = .true.

    ! Model start up
    pre_ms_T_c = 1e5


! Opacity tables to use
    kappa_file_prefix = 'OP_a09_nans_removed_by_hand'     
    ! kappa_CO_prefix = 'a09_co'
    ! kappa_lowT_prefix = 'lowT_fa05_a09p'

! Przybilla mixture
    initial_zfracs = 8  

/ !end of star_job namelist


&controls

  !!!!!!!!!!!!!!!!!!!!!!!!!!!!!!!!!!!!!!!!!!!
  !! MIXING CONTROLS
  !!!!!!!!!!!!!!!!!!!!!!!!!!!!!!!!!!!!!!!!!!!

   use_Ledoux_criterion = .true.


 ! Do not smooth N^2
   num_cells_for_smooth_gradL_composition_term = 0

   alpha_semiconvection = 0d0
   semiconvection_option = 'Langer_85 mixing; gradT = gradr'
   thermohaline_coeff = 0d0
   alt_scale_height_flag = .true.
   MLT_option = 'Cox'
   mlt_gradT_fraction = -1
   okay_to_reduce_gradT_excess = .false.

   do_conv_premix = .true.

  !!!!!!!!!!!!!!!!!!!!!!!!!!!!!!!!!!!!!!!!!!!
  !! OVERSHOOTING CONTROLS
  !!!!!!!!!!!!!!!!!!!!!!!!!!!!!!!!!!!!!!!!!!!

  ! Minimum Diffusive mixing
    set_min_D_mix = .true.
    min_D_mix = 5d0 ! cm^2 / sec

    ! parameters specified in run_star_extras.f
    overshoot_new = .true.

    use_dedt_form_of_energy_eqn = .true.
    use_gold_tolerances = .true.

    write_profiles_flag = .false.
    photo_interval = 500000
    history_interval = 1

    pulse_data_format = 'GYRE'
    write_pulse_data_with_profile = .false.

  ! starting specifications
    !initial_mass = 15 ! specified in run_star_extras

  ! stop when the center mass fraction of he4 drops below this limit
    xa_central_lower_limit_species(1) = 'he4'
    xa_central_lower_limit(1) = 1d-3

    HB_limit = 0.95


    !!!!!!!!!!!!!!!!!!!!!!!!!!!!!!!!!!!!!!!!!!!
    !! ATMOSPHERE/WIND CONTROLS
    !!!!!!!!!!!!!!!!!!!!!!!!!!!!!!!!!!!!!!!!!!!

  ! Mass loss
    hot_wind_scheme = 'Vink'
    Vink_scaling_factor = 0.5d0 
 

    atm_option = 'table'
    atm_table = 'photosphere'



    use_eosDT2 = .false.

    !!!!!!!!!!!!!!!!!!!!!!!!!!!!!!!!!!!!!!!!!!!
    !! MESH & RESOLUTION CONTROLS
    !!!!!!!!!!!!!!!!!!!!!!!!!!!!!!!!!!!!!!!!!!!

    varcontrol_target = 1d-4
    mesh_delta_coeff = 0.4


    max_allowed_nz = 20000 ! maximum number of grid points allowed
    mesh_adjust_use_quadratic = .true.
    mesh_adjust_get_T_from_E = .true.


  ! Additional resolution based on the pressure and temperature profiles

    mesh_dlogX_dlogP_extra = 0.15             ! resol coeff for chemical gradients
    mesh_dlogX_dlogP_full_on = 1d-6           ! additional resol on for gradient larger than this
    mesh_dlogX_dlogP_full_off = 1d-12         ! additional resol off for gradient smaller than this

    mesh_logX_species(1) = 'he4'              ! taking into account abundance of He4

  ! Additional resolution near the boundaries of the convective regions
    xtra_coef_os_full_on = 1.0
    xtra_coef_os_full_off = 1.0

    xtra_coef_os_above_burn_h = 0.5d0
    xtra_dist_os_above_burn_h = 0.5d0
    xtra_coef_os_below_burn_h = 0.5d0
    xtra_dist_os_below_burn_h = 0.5d0


    xtra_coef_czb_full_on = 1.0
    xtra_coef_czb_full_off = 1.0

    xtra_coef_a_l_hb_czb = 0.5                ! resol coeff above lower hburn convective boundary
    xtra_dist_a_l_hb_czb = 0.5                ! distance above lower hburn convective boundary
    xtra_coef_b_l_hb_czb = 0.5                ! resol coeff below lower hburn convective boundary
    xtra_dist_b_l_hb_czb = 0.5                ! distance below lower hburn convective boundary

    xtra_coef_a_u_hb_czb = 0.2                ! resol coeff above upper hburn convective boundary
    xtra_dist_a_u_hb_czb = 0.5                ! distance above upper hburn convective boundary
    xtra_coef_b_u_hb_czb = 0.2                ! resol coeff below upper hburn convective boundary
    xtra_dist_b_u_hb_czb = 0.5                ! distance below upper hburn convective boundary

/ ! end of controls namelist

\end{verbatim}
\subsection{GYRE}
Example GYRE inlist.
\begin{verbatim}
&constants
/

&model
	model_type = 'EVOL'
	file = 'model.GYRE'
	file_format = 'MESA'
/


&mode
	l = 1
	m = +1
	n_pg_min = -50
	n_pg_max = -1
/

&osc
	nonadiabatic = .true.
	outer_bound = 'UNNO'
/

&num
	diff_scheme = 'MAGNUS_GL2'
	n_iter_max = 50
/

&scan
    grid_frame = 'INERTIAL'
	grid_type = 'INVERSE'
	freq_min = 0.025180
	freq_max = 3.932893
	freq_max_units = 'CYC_PER_DAY'
	freq_min_units = 'CYC_PER_DAY'
	n_freq = 600
/


&grid
    n_inner = 5
	alpha_osc = 10			! At least 5 points per oscillatory wavelength
	alpha_exp = 2			! At least 1 point per exponential 'wavelength'

/


&nad_output
	summary_file = 'output.HDF'
	freq_units = 'CYC_PER_DAY'
	summary_file_format = 'HDF'
	! Items to appear in summary file
	summary_item_list = 'M_star,R_star,L_star,l,m,n_p,n_g,n_pg,omega,freq,E_norm,W'
	
\end{verbatim}
\twocolumn
\end{appendix}

\end{document}